\documentclass[twocolumn,amsmath,amssymb,10pt,superscriptaddress,a4paper,letterpaper,fleqn,floatfix]{revtex4-1}

\usepackage{amssymb}
\usepackage{epsfig}
\usepackage{float}
\usepackage{graphicx}
\usepackage{dcolumn}
\usepackage{array}
\usepackage{bm}
\usepackage{fancyhdr}
\usepackage{longtable}
\usepackage{multirow}
\usepackage{afterpage}  
\usepackage{color}
\usepackage{rotating}
\usepackage[warnundef]{jabbrv}
\setlongtables
\usepackage{placeins}
\usepackage[colorlinks = true, linkcolor = blue, urlcolor=blue,citecolor = black, breaklinks = true]{hyperref}

\newcommand*{\bn}{$\beta$n}
\newcommand*{\bdn}{$\beta$-delayed neutron }
\newcommand*{\bne}{$\beta$n-emitter }
\newcommand*{\bdns}{$\beta$-delayed neutrons }
\newcommand*{\nbrs}{neutron-branching ratios }
\newcommand*{\nbr}{neutron-branching ratio }
\newcommand*{\Qbn}{$Q_{\beta n}$ }

\newcommand\T{\rule{0pt}{2.6ex}}       

\newcommand{\ucinq}{$^{235}$U}

\newcommand{\utrois}{$^{233}$U}
\newcommand{\puneuf}{$^{239}$Pu}
\newcommand{\npsept}{$^{237}$Np}

\newcommand{\uhuit}{$^{238}$U}
\newcommand{\puzero}{$^{240}$Pu}
\newcommand{\puhuit}{$^{238}$Pu}
\newcommand{\puun}{$^{241}$Pu}
\newcommand{\amun}{$^{241}$Am}
\newcommand{\pudeux}{$^{242}$Pu}
\newcommand{\thdeux}{$^{232}$Th}

\def\etal{\emph{et al.}\,}

\begin{document}
\setcounter{page}{1}
\title{ Development of a Reference Database for Beta-Delayed Neutron Emission}

\author{P. Dimitriou} \email[Corresponding author: ] {p.dimitriou@iaea.org}
\affiliation{NAPC-Nuclear Data Section, International Atomic Energy Agency, A-1400 Vienna, Austria }

\author{I. Dillmann} 
\affiliation{TRIUMF, Vancouver, British Columbia V6T 2A3, Canada}
\affiliation{Department of Physics and Astronomy, University of Victoria, Victoria, British Columbia V8P 5C2, Canada }

\author{B. Singh} 
\affiliation{Department of Physics and Astronomy, McMaster University, Hamilton, Ontario L8S 4M1, Canada}

\author{V. Piksaikin}
\affiliation{State Scientific Center of Russian Federation, Institute of Physics and Power Engineering, 249033 Obninsk, Russian Federation}

\author{K.P. Rykaczewski} 
\affiliation{Physics Division, Oak Ridge National Laboratory, Oak Ridge, Tennessee 37831, USA}

\author{J.L. Tain}
\affiliation{ IFIC, CSIC-Universitat de Valencia, 46071 Valencia, Spain}
\author{A. Algora}
\affiliation{ IFIC, CSIC-Universitat de Valencia, 46071 Valencia, Spain}
\author{K. Banerjee}
\affiliation{Variable Energy Cyclotron Centre, Kolkata 700064, India}
\author{I.N. Borzov}
\affiliation{National Research Centre Kurchatov Institute, Department of Nuclear Astrophysics, 1 Akademika Kurchatova pl., Moscow 123182, Russian Federation}
\affiliation{Joint Institute for Nuclear Research, Bogolubov Laboratory of Theoretical Physics, 
6 Joliot-Curie, 141980 Dubna, Moscow region, Russian Federation}
\author{D. Cano-Ott} 
\affiliation{Centro de Investigaciones Energ\'{e}ticas, Medioambientales y Tecnol\'{o}gicas (CIEMAT), Avenida Complutense 40, Madrid 28040, Spain}
\author{S. Chiba}
\affiliation{Tokyo Institute of Technology, 2-12-1-N1-9, Ookayama, Meguru-ku, Japan}
\author{M. Fallot}
\affiliation{Subatech, CNRS/in2p3, Univ. of Nantes, IMTA, 44307 Nantes, France}
\author{D. Foligno}
\affiliation{CEA, DEN, DER/SPRC/LEPh Cadarache, F-13108 Saint Paul-Lez-Durance, France}
\author{R. Grzywacz}
\affiliation{Department of Physics and Astronomy, University of Tennessee, Knoxville, Tennessee 37996, USA}
\affiliation{Physics Division, Oak Ridge National Laboratory, Oak Ridge, Tennessee 37831, USA}
\author{X. Huang}
\affiliation{China Nuclear Data Center, China Institute of Atomic Energy, Beijing 102413, China}
\author{T. Marketin}
\affiliation{Department of Physics, Faculty of Science, University of Zagreb, 10000 Zagreb, Croatia}
\author{F. Minato}
\affiliation{Nuclear Data Center, Nuclear Science and Engineering Center, Japan Atomic Energy Agency, Tokai-mura  Ibaraki-ken, Japan}
\author{G. Mukherjee}
\affiliation{Variable Energy Cyclotron Centre, Kolkata 700064, India}
\author{B.C. Rasco}
\affiliation{JINPA, Oak Ridge National Laboratory, Oak Ridge, Tennessee 37831, USA}
\affiliation{Physics Division, Oak Ridge National Laboratory, Oak Ridge, Tennessee 37831, USA}
\affiliation{Department of Physics and Astronomy, University of Tennessee, Knoxville, Tennessee 37996, USA}
\affiliation{Department of Physics and Astronomy, Louisiana State University, Baton Rouge, Louisiana 70803, USA}
\author{A. Sonzogni}
\affiliation{National Nuclear Data Center, Brookhaven National Laboratory, Brookhaven, Upton NY, USA}
\author{M. Verpelli} 
\affiliation{NAPC-Nuclear Data Section, International Atomic Energy Agency, A-1400 Vienna, Austria }
\author{A. Egorov}
\affiliation{State Scientific Center of Russian Federation, Institute of Physics and Power Engineering, 249033 Obninsk, Russian Federation}
\author{M. Estienne}
\affiliation{Subatech, CNRS/in2p3, Univ. of Nantes, IMTA, 44307 Nantes, France}
\author{L. Giot}
\affiliation{Subatech, CNRS/in2p3, Univ. of Nantes, IMTA, 44307 Nantes, France}
\author{D. Gremyachkin}
\affiliation{State Scientific Center of Russian Federation, Institute of Physics and Power Engineering, 249033 Obninsk, Russian Federation}
\author{M. Madurga}
\affiliation{Department of Physics and Astronomy, University of Tennessee, Knoxville, Tennessee 37996, USA}
\author{E.A. McCutchan}
\affiliation{National Nuclear Data Center, Brookhaven National Laboratory, Brookhaven, Upton NY, USA}
\author{E. Mendoza} 
\affiliation{Centro de Investigaciones Energ\'{e}ticas, Medioambientales y Tecnol\'{o}gicas (CIEMAT), Avenida Complutense 40, Madrid 28040, Spain}
\author{K.V. Mitrofanov}
\affiliation{State Scientific Center of Russian Federation, Institute of Physics and Power Engineering, 249033 Obninsk, Russian Federation}
\author{M. Narbonne}
\affiliation{Subatech, CNRS/in2p3, Univ. of Nantes, IMTA, 44307 Nantes, France}
\author{P. Romojaro} 
\affiliation{Centro de Investigaciones Energ\'{e}ticas, Medioambientales y Tecnol\'{o}gicas (CIEMAT), Avenida Complutense 40, Madrid 28040, Spain}
\author{A. Sanchez-Caballero} 
\affiliation{Centro de Investigaciones Energ\'{e}ticas, Medioambientales y Tecnol\'{o}gicas (CIEMAT), Avenida Complutense 40, Madrid 28040, Spain}
\author{N.D. Scielzo}
\affiliation{Physical and Life Sciences Directorate, Lawrence Livermore National Laboratory, Livermore, California 94550, USA }

\date{\today} 

\begin{abstract}
{Beta-delayed neutron emission is important for nuclear structure and astrophysics as well as for reactor applications. Significant advances in nuclear experimental techniques in the past two decades have led to a wealth of new measurements that remain to be incorporated in the databases. 

We report on a coordinated effort to compile and evaluate all the available $\beta$-delayed neutron emission data. The different measurement techniques have been assessed and the data have been compared with semi-microscopic and microscopic-macroscopic models. The new microscopic database has been tested against aggregate total delayed neutron yields, time-dependent group parameters in 6-and 8-group re-presentation, and aggregate delayed neutron spectra. New recommendations of macroscopic delayed-neutron data for fissile materials of interest to applications are also presented.

The new Reference Database for Beta-Delayed Neutron Emission Data is available online at: \url{http://www-nds.iaea.org/beta-delayed-neutron/database.html}.
}
\end{abstract}
\maketitle

\lhead{Development of a Reference Database $\dots$}          
\chead{NUCLEAR DATA SHEETS}                             
\rhead{P.~Dimitriou \textit{et al.}}                   
\lfoot{}                                                           
\rfoot{}                                                          
\renewcommand{\footrulewidth}{0.4pt}
\tableofcontents{}


\section{ INTRODUCTION}\label{sec:intro}

Neutron-rich nuclei can emit neutrons after $\beta$-decay when their decay $Q$-value is larger than the (one/two/three…) neutron separation energy: $Q_{\beta}>S_{xn}$. This decay mode is called ``$\beta$-delayed ($\beta$n) neutron emission" and was discovered in 1939 by Roberts \etal~\cite{Roberts1939a,Roberts1939b}, shortly after the discovery of fission by Meitner, Hahn, and Strassmann in 1938 \cite{Meitner1938}. Immediately after this, Bohr and Wheeler \cite{Bohr1939} improved their liquid drop model to describe these new decay mechanisms. 

The $\beta$-delayed two- and three-neutron emission ($\beta$2n and $\beta$3n) process was discovered 40 years later in $^{11}$Li \cite{Azuma1979,Azuma80}, and in 1988 the (so far) only $\beta$4n emitter, $^{17}$B, has been investigated and a tentative branching ratio of $P_{4n} =0.4(3)$ has been measured \cite{Dufour1988}.

`Delayed' in this context means, that the neutron is emitted with the $\beta$-decay half-life of the precursor $^{A}Z$, ranging from a few milliseconds for the most neutron-rich isotopes up to 80(18)~s for $^{211}$Tl, the longest-lived \bdn (\bn) precursor observed to date. These delayed neutrons have to be distinguished from the prompt neutrons evaporated immediately (in the order of 10$^{-14}$ s) after a fission event from a neutron-rich nucleus and usually form only a small fraction of the total neutrons emitted from fission ($\approx1\%$). Nevertheless, they play an important role in controlling the dynamic rate of fission in a chain-reacting assembly and hence in reactor kinetics and the safe operation of a reactor.

This important role of delayed neutrons in determining the reactor kinetics and reactivity was recognized early on and led to several measurements of delayed neutron groups \cite{Booth1939,Brostrom1939,Snell1947,DeHoffmann1948,Hughes1948}. In 1954, an extensive delayed neutron program was initiated at Los Alamos National Laboratory. One major outcome of this program was the 6-group fitting parameters by Keepin and co-workers (see Sect.~\ref{Sec:Macro-methods-total}, \cite{Keepin57}). In a least-squares fit of the delayed neutron emission curves of $^{235}$U, they found that by grouping the \bn~ emitters in six groups -- according to their half-lives -- they could reproduce the experimental data.

%

In 1990, Subgroup 6 was established by the NEA-OECD ``Working Party on International Nuclear Data Evaluation Co-operation" (WPEC-SG6) \cite{NEA-WPEC6} to assess the differences between the calculated and measured values of the reactivity scale obtained from reactor kinetics. The observed discrepancies were largely attributable to the uncertainties in the delayed neutron data used in these calculations. Their effort to address this included international benchmark measurements of the effective delayed neutron fraction, made on fast critical assemblies, which aimed to provide high-quality experimental information for $^{235}$U, $^{238}$U, and $^{239}$Pu. 

One of the outcomes of this working group was the refinement of the Keepin parameters by the laboratories in Los Alamos and Obninsk from a 6-group fit to an 8-group fit. This was achieved by separating the three longest-lived delayed neutron precursors ($^{87}$Br, $^{137}$I, and $^{88}$Br) into single groups.

Independent of the effort to determine integral delayed neutron yields for applications, there have also been several attempts to measure, compile and evaluate \bdn~data of individual precursors. Groups in Sweden~\citep{Rudstam74,Rudstam77}, USA~\cite{Greenwood1985,Greenwood1997}, and Germany~\citep{Kratz79} became active in measurements of \bdn~probabilities and spectra in the late 1970s. A large number of data was published in the following decades and subsequently compiled and evaluated by Rudstam and collaborators~\cite{Rudstam74, Rudstam77,Rudstam1980,Rudstam1993} and  Pfeiffer et al. ~\cite{Pfeiffer2002}. 

Values from these compilations and evaluations have sometimes been incorporated in ENSDF~\cite{ENSDF}, together with independent recommendations by the respective evaluator. Consequently, these values have also been transferred to evaluated data libraries such as ENDF/B~\cite{Chadwick2011,Brown2018}, JENDL~\cite{Katakura11, Katakura2015} and JEFF~\cite{jeff3.1.1, Plompen2020} decay-data libraries, as well as in nuclear astrophysics applications.
Other independent evaluations have also been performed to address specific needs in applications, such as the highly-specific evaluations of decay data for short- to very short-lived radioisotopes, including $\beta$-delayed neutron emitters, which were performed from 1996 to 1998 by Nichols~\cite{Nichols1998} for the UK-based data file. These evaluations were subsequently adopted in the JEFF decay-data libraries~\cite{Plompen2020}. 

The first comprehensive compilation and evaluation of \bdn~emission spectra was published by Brady~\cite{Brady1989a} in an effort to combine measured spectra with statistical model calculations to complete the experimental data that was sometimes scattered or limited in energy range. The resulting spectra were adopted in the ENDF/B libraries and were further supplemented with model calculations by Kawano et al.~\cite{Kawano2008}.


Until 2011 only sporadic efforts have been made 
to collect data of `microscopic' quantities related to the \bdn~emission, namely half-lives and neutron-branching ratios of very neutron-rich nuclei to reveal nuclear structure properties for nuclei far from stability. This data has also been used directly as nuclear physics input for astrophysical reaction network calculations for the "rapid neutron capture" ($r$) process in explosive astrophysical scenarios like core collapse supernovae and binary neutron star mergers, or indirectly to benchmark and improve theoretical predictions which are needed for extrapolating models towards the yet inaccessible regions of the nuclide chart (``Terra Incognita"). 

Recent experimental efforts relied largely on the existing generation of radioactive beam (RIB) facilities but with the next generation of these facilities becoming operational within this decade, a plethora of new nuclides will become accessible. The vast majority of these new nuclides on the neutron-rich side will be $\beta$n-emitters.


According to the Atomic Mass Evaluation in 2016 (AME2016) \cite{AME16}, 651 out of the 3435 known nuclei are \bn~emitters, meaning that their $Q_{\beta 1n}$ is (within mass uncertainties) positive. Theoretical models estimate that about 4000 additional nuclei should exist between the proton- and neutron dripline, most of them are on the neutron-rich side and will be \bn~emitters.

In 2011, 72 years after the discovery of \bdns, only 216 \bn~emitters had measured neutron-branching ratios: 81 nuclei in the lighter mass region ($Z<28$), 134 nuclei in the fission regions between $Z = 29-57$, and a single (unconfirmed) measurement of the $P_{1n}$ value of $^{210}$Tl \cite{Stetter1962} in the whole mass region above $Z>57$. 

Currently (as of August 2020) 653 nuclei between $^{8}$He and $^{237}$Ac have been identified as potential $\beta$-delayed one-neutron emitters ($Q_{\beta 1n}>0$ keV). However, as of August 2020 only 306 of those nuclei have measured $\beta$-delayed one-neutron branching ratios ($P_{1n}$): 114 in the light mass region $Z<28$ \cite{Birch2015}, 183 in the fission fragment region ($Z=29-57$), and 9 in the region above $Z>57$ \cite{Liang2020}. A total of 28 of these measurements provided only an upper limit for the $P_{1n}$ value, and another six provided only lower limits.

As far as delayed multi-neutron emission is concerned, 300 potential $\beta$2n-emitters have been identified, alongside 138 $\beta$3n-emitters, and 58 $\beta$4n-emitters. The number of measured multi-neutron branching ratios is however much smaller: only 31 $P_{2n}$ values have been measured (23 for $Z<28$ and eight for $Z>28$) and four $P_{3n}$ values ($^{11}$Li,$^{17}$B, $^{19}$B, and $^{23}$N). One single tentative measurement for the $P_{4n}$ value of $^{17}$B has been reported.

One of the largest and most focused recent efforts to fill this large gap between identified and measured \bdn~branching ratios is the "Beta-delayed neutrons at RIKEN" (BRIKEN) project \cite{BRIKEN2017}. Since 2016 this international collaboration has been measuring half-lives and \nbrs of almost all accessible \bn~emitters. The collaboration will conclude its experimental campaigns in 2021 and is expected to add in the near future $>$ 250 new \bdn~emitters to the existing list. The data from the experimental runs in 2016--19 is presently being analyzed, and a small fraction has already been published and included in the latest $Z>28$ evaluation \cite{Liang2020}. An overview of the BRIKEN and future campaigns can be found in Sect.~\ref{Sec:Micro-newdata}.



%

\begin{figure*}[!htb]
	\centering
\includegraphics[width=0.9\textwidth]{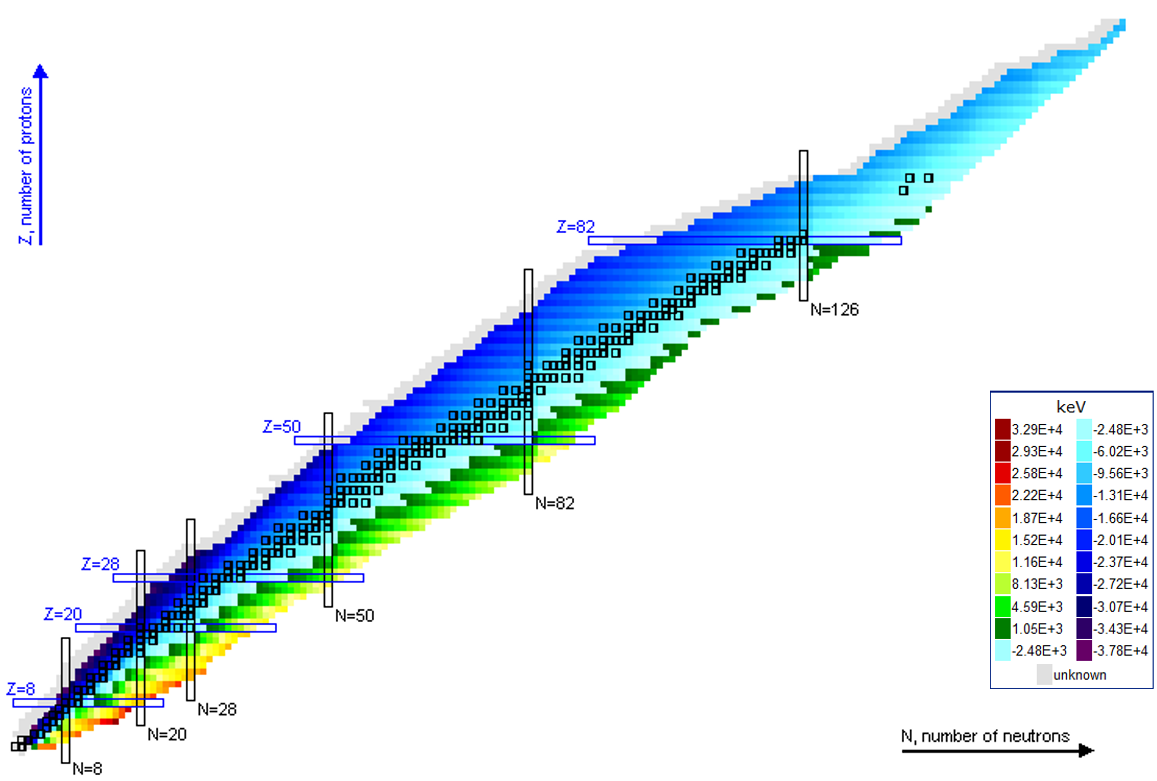}
\caption{Chart of nuclides with color-coded $Q_{\beta n}$ window (in keV, retrieved from NuDat 2.7 \cite{NuDat27}).}
\label{fig:chart-bn}
\end{figure*}
This rapidly developing landscape of measured \bn~emitters and respective application-driven needs for reliable \bn~data necessitates both, a thorough update of the existing \bn-data compilations and a review of the recommended group parameters. In response to these urgent data needs, the International Atomic Energy Agency (IAEA) coordinated an international effort to produce a `Reference Database of Beta-Delayed Neutron Emission Data'~\cite{IAEA0599, IAEA0643, IAEA0683, IAEA0735}. 

The results of this coordinated effort are reported in this paper. The methods for measurement of microscopic \bdn~data are described in Sect.~\ref{Sec:Micro-methods}, followed by the compilation and evaluation efforts (including new recommendations of standards for $P_{1n}$ measurements) and systematic descriptions in Sect.~\ref{Sec:Micro-compilation}. Comparisons of the \bdn~data (half-lives, \bdn-branching ratios and spectra) with global theoretical approaches are presented in Sect.~\ref{Sec:Micro-theory}. 

Macroscopic (integral) delayed neutron yields and spectra have been used to verify and validate the microscopic data. Measurement methods, new data, and compilation efforts are presented in Sect.~\ref{Sec:Macro-meas}. Summation and time-dependent calculations of total delayed neutron yields and spectra based on the evaluated microscopic \bdn~data are described and compared with recommended values in Sect.~\ref{Sec:Macro-Summation}. The impact of the new \bdn~data on integral reactor calculations is discussed in Sect.~\ref{Sec:Macro-Integral}. Finally, the systematics of the measured  macroscopic data are discussed in Sect.~\ref{Sec:Macro-Systematics} and new recommendations for 6- and 8-group kinetic parameters for major and minor actinides are given in Sect.~\ref{Sec:Macro-Recommended}. 

The structure of the reference database is detailed in Sect.~\ref{Sec:Database} and a summary of the main conclusions of this work is provided in Sect.~\ref{Sec:Conclusions}.

All the microscopic and macroscopic data presented herein, are available online at \url{http://www-nds.iaea.org/beta-delayed-neutron/database.html}~\cite{database}.

\section{MICROSCOPIC DATA: Methods and measurements}\label{Sec:Micro-methods}


\subsection{Delayed neutron emission probabilities}
\label{Sec:Micro-Pn}

The \nbr of a nucleus into the $\beta$1n decay channel is defined as

\begin{equation}\label{eq:P1n-def}
P_{1n}= {\frac{N_{1n}}{N_{decays}}}.
\end{equation}

where 
$N_{1n}$ stands for the 
number of decays ($N_{decays}$) that emit one neutron. 
Similar expressions hold for any other $\beta$-delayed multi-neutron decay channel. 
This quantity can be measured in various ways with different techniques, either directly by counting the number of emitted neutrons, or indirectly by counting $\gamma$-rays or other decay products. 
Special care has to be taken that the number of counted events are free of any contaminations, i.e. that the contributions from background activities are subtracted and the random noise is properly corrected for.

Eight methods for the determination of $P_{1n}$ values were extracted from the previous Rudstam evaluation \cite{Rudstam1993} and were reviewed in the IAEA CRP summary reports \cite{IAEA0599,IAEA0643,IAEA0683} for their strengths and weaknesses and, in particular, potential sources of systematic errors and uncertainties.  These methods were complemented by newer methods. Note that only a few of these methods are also suited for the measurement of multi-neutron emission probabilities. 



Throughout this article the following notation will be used: precursor ($^{A}$Z) or mother nucleus stands for the $\beta$-decaying nucleus; $\beta$-decay daughter or intermediate nucleus is the product from the $\beta$ decay of the precursor ($^{A}$Z+1), and \bdn daughter stands for the final nucleus with $^{A-1}$Z+1 for the case of one-neutron emission, $^{A-2}$Z+1 for the case of two-neutron emission, etc.

The majority of $P_{xn}$ values in the literature have been determined using methods that involve neutron counting using moderated neutron detectors. These type of counters combine a medium acting as a neutron energy moderator with a detector sensitive to thermal neutrons to maximize detection efficiency. Typical moderators are made of dense hydrogenated material, like high-density polyethylene that has replaced the paraffin wax of earlier designs. The first counters used BF$_{3}$-filled proportional tubes as a thermal neutron detector, but nowadays the more expensive $^{3}$He is preferred because of its advantages \cite{Crane1991} (see also Sect.~\ref{Sec:Micro-spectra}). This detector can be considered as an evolution of the neutron long counter \cite{Hanson1947}, where very high detection efficiencies are achieved by arranging several tubes in a 4$\pi$ geometry around the source. Apart from the large detection efficiency, another distinct advantage of this type of detector is the virtual absence of a low-energy threshold. 

Recent examples of this kind of setup applied to the measurement of $P_{xn}$ are NERO (now HABANERO) at the National Superconduction Laboratory (NSCL) at Michigan State University (USA) \cite{Pereira2010}, LOENIE at Lohengrin-Institute Laue Langevin (France) \cite{Mathieu2012}, 3Hen at HRIBF-Oak Ridge National Laboratory (USA) \cite{Grzywacz2014}, BELEN at the GSI Helmholtz Center for Heavy Ion Research in Darmstadt (Germany) and at the heavy ion accelerator laboratory of the University of Jyv{\"a}skyl{\"a} (Finland) \cite{Agramunt16}, and TETRA at ALTO - Institute de Physique Nucleaire in Orsay (IPNO, France) \cite{Testov2016}. The latest setup is BRIKEN at RIKEN Nishina Center in Wako (Japan) which is a temporary merger of the 3Hen and BELEN counters with some additional $^{3}$He tubes from RIKEN \cite{BRIKEN2017, Tolosa2018}. 

The detection efficiency of these type of detectors exhibits a dependency on neutron energy which can be a source of systematic uncertainty in the determination of $P_{xn}$ values (see  discussion in \ref{Sec:micro-bncoin}). The effect is related to the moderation process and depends on the geometry of the counter (number and spatial distribution of neutron detectors). Thus the detector design can be optimized to minimize the dependency of the efficiency on neutron energy. An example of this is the LOENIE counter \cite{Mathieu2012}: With only 18 $^{3}$He tubes a very small (17\%) but flat efficiency up to 2~MeV is achieved. 

Nowadays the general trend is to minimize the energy dependency of the efficiency curve while still maximizing the overall detection efficiency. This can be achieved by increasing the number of tubes considerably, as has been done for example in the case of the BRIKEN setup which could use up to 166 $^{3}$He tubes. The optimization of the geometry incorporating many tubes is a complicated problem that can be solved by combining Monte Carlo simulations with a parametric approach \cite{BRIKEN2017}. In this way a configuration with an efficiency in excess of 75\% below 1~MeV was obtained that drops (only) to an efficiency of 54\% at 5~MeV.

The efficiency calibration of these instruments requires the use of mono-energetic isotropic neutron sources spanning the range of neutron energies of interest to well above 5~MeV for exotic nuclei \cite{Madurga2016}. Photo-neutron ($\gamma,n$) and $\alpha$-induced reaction sources (like $^{241}$Am/$^{9}$Be or $^{24x}$Cm/$^{12}$C), as well as fission sources have been employed \cite{Reeder1977} but these are difficult to prepare or have a very broad energy spectrum. Beam-induced ($p,n$) and ($\alpha,n$) reactions have also been used \cite{Pereira2010} although it seems difficult to reach the required accuracy. 

In experiments at radioactive beam facilities \bdn emitters with well know $P_{1n}$ values and neutron spectra are used for the efficiency calibration of the setup. For sources with a broad neutron energy distribution the measured neutron detection efficiency is averaged over the respective neutron energy range. However it should be kept in mind that this is an approximation as different neutron spectra -- detected with different average efficiencies -- can have the same average energy. Thus, for these sources the measurements should be combined with detailed Monte Carlo simulations, although the accuracy of the latter in complex geometries is still under discussion.

The moderation process introduces a delay between neutron emission and detection that reaches few hundreds of $\mu$s. This poses a problem for conventional triggered event-based data acquisition systems. The use of an event gate of this length can lead to a sizable loss of data due to the acquisition dead time. This problem is overcome by modern event-less self-triggered data acquisition systems, as in the case of the most recent neutron detection setups 3Hen, BELEN, TETRA, and BRIKEN.

In the following, the eight different methods for obtaining $P_{1n}$ values as well as their advantages and disadvantages are discussed.

\subsubsection{Beta-neutron coincidences (``$\beta$--n")}\label{Sec:micro-bncoin}

In this method $\beta$'s and neutrons are counted in coincidence and compared with the number of $\beta$ counts. Thus the efficiency of the $\beta$ detector is greatly reduced, while the neutron efficiency in absolute terms is needed. This method was labelled ``n/$\beta$" in Ref.~\cite{Rudstam1993} and has been renamed here to ``$\beta$--n" to account for the proper sequence of emissions of the particles.

\begin{figure*}[!htb]
	\centering
\includegraphics[width=0.8\textwidth]{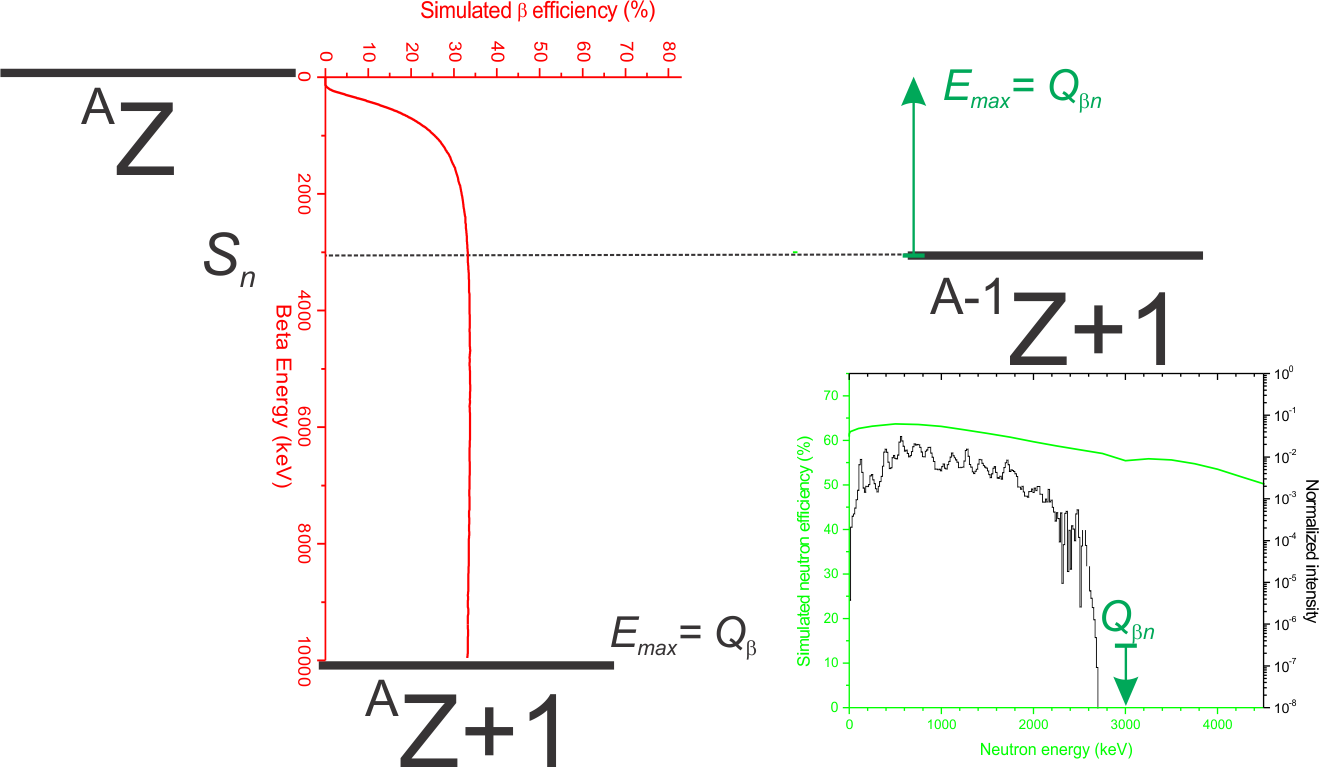}
\caption{Examples of efficiency curves for a $\beta$- and a neutron detector \cite{Caballero18} in comparison with an example neutron spectra. Note that the highest neutron energies are connected to $\beta$'s with the lowest detection efficiency due to the detector thresholds.}
\label{fig:examples}
\end{figure*}

For measuring the one-neutron emission probability ($P_{1n}$), the respective equation is

\begin{equation}\label{eq:P1n-bn}
P_{1n}= \frac{1}{\epsilon_{1n}} ~ \frac{N_{\beta 1n}}{N_\beta} ,
\end{equation}

where $\epsilon_{1n}$ is the one-neutron detection efficiency, $N_{\beta 1n}$ is the number of time-correlated $\beta$-1n events, and $N_\beta$ is the amount of detected $\beta$'s.
The main assumption in Eq. \ref{eq:P1n-bn} is that the detector efficiency for $\beta$ decays is constant for all energies and thus cancels out.
However, $\beta$ detectors have a lower threshold of typically 50--150~keV, and the respective efficiency curve shows a steep drop below $\approx$1--2 MeV (see Fig.~\ref{fig:examples}). 

In contrast, a moderated $^{3}$He-neutron detector has an (almost) constant neutron detection efficiency for energies up to 1~MeV, and then the curve drops slowly. These effects might introduce systematic errors if the calibrating isotopes have very different $\beta$- or neutron-energy spectra 
\cite{Agramunt16}. 
Examples of efficiency curves for a typical $\beta$- and moderated neutron detector used in experiments at the IGISOL facility in Jyv{\"a}skyl{\"a} are shown in Fig.~\ref{fig:examples} \cite{Caballero18}. It can be seen that the highest neutron energies are connected to low-energy $\beta$'s with the lowest detection efficiency due to the detector thresholds, and vice versa.

\begin{figure}[!htb]
	\centering
\includegraphics[width=0.45\textwidth]{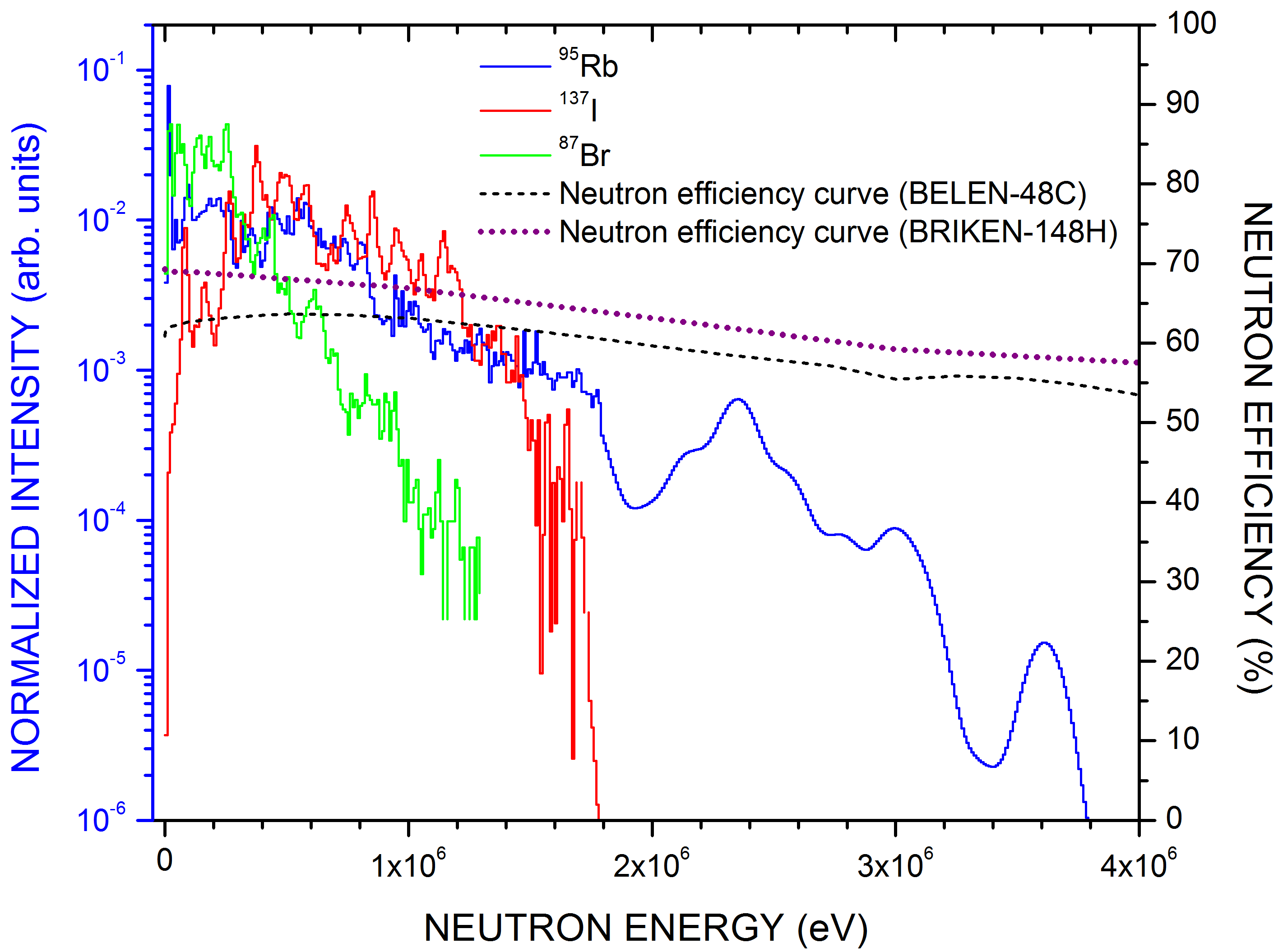}
\caption{Examples of neutron spectra for $^{87}$Br (green), $^{95}$Rb (blue), and $^{137}$I (red) in comparison with two (simulated) efficiency curves for state-of-the art neutron detectors (dashed and dotted line) used in experiments at the IGISOL facility in Jyv{\"a}skyl{\"a} \cite{Caballero18} and at RIKEN \cite{BRIKEN2017}.}
\label{fig:examples-n}
\end{figure}

An additional source of uncertainty is that the energy dependence (slope) of the neutron efficiency curve is often ignored. 
However, if the neutron energy distribution of the calibrant isotope is very different to the one of the isotope to be measured, this will induce systematic effects which need to be corrected for. For moderated neutron detectors the ``average detection efficiency" depends on the neutron energy spectra of each isotope. Fig.~\ref{fig:examples-n} shows the neutron spectra of the \bn~emitters $^{87}$Br (green line), $^{95}$Rb (blue line), and $^{137}$I (red line) as an example. For comparison, two simulated neutron detection efficiency curves for two state-of-the-art neutron detectors, BELEN \cite{Caballero18} and BRIKEN \cite{BRIKEN2017}, are plotted as dashed lines.

The correct description for the \nbr should therefore include detection efficiencies $\langle\epsilon\rangle$ which are averaged over the respective energy ranges:
\begin{equation}
P_{1n}=  \frac{\epsilon_\beta}{\langle\epsilon'_\beta~\cdot~\epsilon_n \rangle} ~ \frac{N_{\beta 1n}}{N_\beta}
\label{eq:p1ncoin}
\end{equation}

In this equation $\epsilon_\beta$ is the mean $\beta$ efficiency. $\langle\epsilon'_\beta~\cdot~\epsilon_n \rangle$ is the product of the averaged $\beta$ efficiency above the neutron separation energy weighted by the respective $\beta$ intensity spectrum and the averaged neutron efficiency weighted by the neutron spectrum \cite{Agramunt16, Caballero18}. If $\epsilon_\beta$ = $\langle\epsilon'_\beta\rangle$, then the simplified Eq.~\ref{eq:P1n-bn} is recovered. 



Even with this simplification, the method
requires the knowledge of the neutron spectra of the respective isotope to be measured (see Sect.~\ref{Sec:Micro-spectra}). Since these spectra are not available for all isotopes, 
one can use as a guidance the
spectra of calibration isotopes like the ones shown in Fig.~\ref{fig:examples-n} and listed in Sections~\ref{Sec:Eval-P1n} and \ref{Sec:Eval-Spectra}, 
selected to cover the expected neutron energy range of the isotopes of interest.

When analyzing decay curves the signals from all contributing decays need to be considered. This requires a proper knowledge of the different half-lives of all involved nuclei and possible isomeric states. 
In turn, if a half-life in a decay chain is not know, this method can also be used to extract the unknown half-lives of \bn~emitters from $\beta$- and \bn-decay curves. 
Extensions of the ``$\beta$--n" method to include multi-neutron emission increase the complexity as the neutron-emission multiplicity (x) increases. Since the neutron-detection efficiency is $<$ 100\%, a $\beta$xn decay channel will contribute to the counts $N_{\beta yn}$ with $y \leq x$ \cite{Azuma80} which must be taken into account.  It is also important to include corrections for accidental coincidences with background neutrons, in particular when the background rate is large and the $P_{xn}$ value is small. The importance of a proper background correction has been discussed recently in Refs.~\cite{Caballero18,Tolosa2018,Rasco2018}. As in the $P_{1n}$ case (Eq.~\ref{eq:p1ncoin}), the correct expressions must include variable, isotope- and decay channel-dependent $\beta$- and neutron-detection efficiencies. 

In summary, the ``$\beta$--n" method is a generic and widely employed method that is well-suited for the measurement of single- and multi-\nbrs. It is the method of choice if the rate of neutrons from the respective decay is small compared to the rate of background neutrons, since the coincidence relations will increase the sensitivity to the neutrons of interest. It is a reliable method if all pitfalls as described above are carefully considered and accounted for. The main sources of systematic uncertainty are the $\beta$ and neutron detection efficiencies that have to be simulated carefully. 






\subsubsection{Neutron and $\beta$ counting (``n,$\beta$")}
In this method the $\beta$'s and neutrons are counted simultaneously but separately (not in coincidence). This method was labelled ``n--$\beta$" in Ref.~\cite{Rudstam1993}. The $P_{1n}$ value can be extracted from the ratio of counted events corrected by the ratio of detection efficiencies:

\begin{equation}
P_{1n}= {\frac{\epsilon_\beta}{\epsilon_n}}~{\frac{N_{1n}}{N_\beta}}.
\end{equation}

The ``n,$\beta$" method requires that either both efficiencies are determined separately or that their ratio is obtained from calibration nuclei. All systematic errors on $\beta$ and neutron detection efficiencies related to the different intensity distributions and the shape of the efficiency curves discussed in the previous method apply also here and have to be included properly. The $\beta$ detection needs to be checked for potential sources of background, e.g. conversion electrons or changing noise in the detectors. This method is also quite generic and applicable to the measurement of multiple neutron emission. The main practical limitation is the neutron signal-to-background ratio that can be overcome by the use of $\beta$-n coincidences (the ``$\beta$--n" method).

\subsubsection{Relative neutron counting (``$P_{1n}$ $^{A}$Z")}
This is a neutron-only counting method and requires that in the decay chain either a descendant or the parent $^{A}$Z is a \bne with well known $P_{1n}$ value which becomes the normalization. In addition, all half-lives have to be known including that of the measured precursor which could be determined independently from $\beta$ activity curves. It is based on the analysis of the measured neutron activity curve to disentangle the different contributions. The method has been applied only in a few cases after chemical or mass separation, thus activity curves must be corrected for contaminants. In its simplest form it assumes that the neutron energy spectrum (and thus the neutron detection efficiency)  for parent and descendants are identical but eventually corrections as discussed before should be applied.

\subsubsection{Neutron counting per fission (``fiss,n")}\label{Sec:P1n-FY}

This method was labelled ``fiss" in Ref.~\cite{Rudstam1993}. Historically this is the oldest method used to measure neutron emission probabilities of individual precursors \cite{Stehney1953}. It is based on the determination of the number of \bdns per fission of a given precursor $Y^{n}_{A,Z}$ with a suitable neutron detector. This number is compared then to the fission yield $Y_{A,Z}$ of that precursor to obtain the $P_{1n}$ value:

\begin{equation}\label{eq:P1n-fiss}
P_{1n}= \frac{Y^{n}_{A,Z}}{Y_{A,Z}}
\end{equation}

Typically fission of $^{235}$U at thermal neutron energies has been used in these measurements. The main source of systematic error is the accuracy of the fission yields available at the time of the measurement. For example, in the evaluation of Rudstam \cite{Rudstam1993} the results of earlier studies were updated using the fission yields $Y_{A}$ from Wahl \cite{Wahl88} and the respective charge fraction $P_Z$. However, new mass- and charge-dependent fission yields $Y_{A,Z}$ are now available, e.g. from evaluated libraries like ENDF/B-VIII.0 and JEFF-3.1.1. Thus, a renormalization of the older data to the new fission yields standards is advisable.

This method was applied mostly to chemically separated samples \cite{Rudstam1993}, a procedure that requires a certain processing time (in the order of 1~s) and is not suited for very short-lived \bdn emitters. It is affected by systematic uncertainties of the chemical separation efficiency as well as the neutron efficiency. For older experimental data, the lack of separation of the different isotopes and the presence of contaminants could have affected the determination of the $P_{1n}$ value. Experience shows that the reliability of this method can be poor and that should be taken into account when comparing with other methods.





\subsubsection{Neutron and $\gamma$ counting (``n,$\gamma$")}

This method was called ``$\gamma$ $^{A}$Z" in \cite{Rudstam1993} and is renamed to ``n,$\gamma$" method to account for the separate counting of neutrons from the precursor  and $\gamma$'s from the daughter $^{A}$Z decay ($^{A}$Z+1 in the nomenclature in this paper). It should not be confused with the ``n--$\gamma$" method in \cite{Rudstam1993} which actually designates a ``$\gamma$,$\gamma$" method (see below). Because of the usually much longer half-live of the daughter it is convenient to measure first the neutron activity of the sample, and then the $\gamma$-activity in a separate dedicated setup \cite{Talbert1969}. 

The method requires that the absolute $\gamma$-intensity $I_{\gamma}^{abs}$ of one or more transitions in the decay of the daughter is known. Corrections for growth and decay of activities, which require a knowledge of half-lives, must be applied before calculating the $P_{1n}$:

\begin{equation}\label{eq:P1n-ng}
\frac{P_{1n}}{1-P_{1n}}= {\frac{\epsilon_\gamma \cdot I_{\gamma}^{abs}}{\epsilon_n}}~{\frac{N_{n}}{N_\gamma}}
\end{equation}

The $\gamma$-ray detection efficiency $\epsilon_{\gamma}$ is usually well known, thus the main issue is the availability and accuracy of absolute $\gamma$ intensities per decay. If the daughter decay activity is also co-produced and present in the sample, this needs to be accounted for. The method can be used also with $\gamma$-rays from subsequent descendants in the $\beta$-decay chain.

\subsubsection{Double $\gamma$ counting (``$\gamma$,$\gamma$")} \label{Sec:Micro:gammagamma}

This method involves counting of $\gamma$-rays emitted by descendants in both, the $\beta$-decay and the $\beta$xn-decay chains. The $\gamma$ activities can be labeled by their masses, which in the case of $\beta$1n emitters are A and A-1, respectively.  The method requires knowledge of the absolute $\gamma$-intensities per decay for both transitions used \cite{Okano1986}. After correcting the $\gamma$ counts by the growth and decay of activities, the possible branching-out/branching-in from/to the respective decay chains and the presence of contaminant activities in the sample, the $P_{1n}$ can be deduced from:

\begin{equation}\label{eq:P1n-gg}
\frac{P_{1n}}{1-P_{1n}}= \frac{\epsilon^{A}_{\gamma} \cdot I^{abs,A}_{\gamma}}{\epsilon^{A-1}_{\gamma} \cdot I^{abs,A-1}_{\gamma}}\frac{N^{A-1}_{\gamma}}{N^{A}_{\gamma}}
\end{equation}

The main advantage of this technique lies in the relative simplicity of detecting $\gamma$ rays as compared to detecting neutrons, but the problem is shifted to the determination or knowledge of the absolute $\gamma$ intensities per decay. The main drawback can be the low signal-to-background ratio requiring that the limited peak efficiency of $\gamma$-ray detectors is compensated by high counting statistics. Tagging with $\beta$ particles can be used to improve the sensitivity but at the cost of introducing a possible dependence on the $\beta$ detection efficiency.

If $\beta$-decaying isomers are present with half-lives similar to the ground state, $\gamma$ rays can provide a uniquely selective tool to discriminate them, and measure correctly the $P_{xn}$ values of isomeric and ground states. 

Fragmentation facilities offer new opportunities for this technique, as descendants of the nucleus of interest $^{A}$Z in the $\beta$ and $\beta$xn decay chains can be independently produced and transmitted with the beam. In this case, absolute $\gamma$ intensities for suitable transitions can be determined in the same experiment using the number of implanted ions for normalization \cite{Shearman2018}. However, the possible existence of isomers should be investigated, and the activity should be corrected according to the correlation time.
Since ions with different Z are implanted at different depths in the respective implantation detector, $\beta$ efficiency corrections might also be needed.  

The ``$\gamma$,$\gamma$" method is suitable for the measurement of multi-neutron emission branches.

\subsubsection{Ion and neutron counting (``ion,n")} \label{Sec:Micro:n-ion}

This corresponds to the ``ion" method as defined in \cite{Rudstam1993}. It is based on the direct measurement of the number of parent nuclei rather than the $\beta$'s emitted. It has been applied at mass separators by counting the number of selected ions. For example, ions can be counted by implantation on an electron multiplier that is surrounded by the neutron detector \cite{Reeder1980b}. In other arrangements two measurements are needed, one for counting the ions, where the beam is deflected to the electron multiplier, and one for counting neutrons where the beam is implanted in a collector in the middle of the neutron detector \cite{Amarel1969}.  In this case, additional systematic uncertainties must be considered. The neutron emission probability is determined as

\begin{equation}
P_{1n}= {\frac{\epsilon_{ion}}{\epsilon_n}}~{\frac{N_{1n}}{N_{ion}}}.
\end{equation}

Here $N_{ion}$ is the number of ions after correction for contaminants and background and $\epsilon_{ion}$ is the ion detection efficiency. The determination of the ion detection efficiency of the electron multiplier can be an issue as shown in \cite{Reeder1980b}.

More recently (see \cite{Winger2009} as an example), non-destructive ion counting with microchannel plates before implantation has been applied to post-accelerated mass-separated ion beams. In this example however, $\gamma$ counting of descendant decays was used instead of neutron counting to obtain the $P_{1n}$ value (``$\gamma$,ion" method).

\subsubsection{Double ion counting (``ion,ion")}\label{Sec:methods-ionion}
The ``ion,ion" method is possibly the only new method for the measurement of \bdn emission probabilities that was developed in recent years, thanks to advances in instrumentation. It is based on counting directly the number of parent nuclei and the number of daughter nuclei produced after neutron emission, circumventing the direct neutron detection. 

The method includes several different techniques like detecting recoils in a Beta-decay Paul trap (BPT) \cite{Yee13}, observing recoil tracks with an optical time-projection chamber (OTPC) \cite{Mianowski10}, or separating and counting ions with a heavy-ion storage ring \cite{Evdokimov13} or a Multi-Reflection Time-Of-Flight (MR-TOF) spectrometer \cite{Mardor2017}. With the first two methods it is also possible to extract neutron energy spectra for the one-neutron emitters via the energy of the recoil, but not in the case of multi-neutron emitters. More details can be found in Sect.~\ref{Sec:Micro-newmethods}.


\subsection{Delayed neutron spectra}\label{Sec:Micro-spectra}
Most of the available data on the spectra of $\beta$-delayed neutrons have been obtained using three different techniques: $^{3}$He spectrometers, gaseous proton recoil spectrometers, and time-of-flight spectrometers. In the following the principles, instruments, and analysis methods of the three techniques are described.

\subsubsection{$^{3}$He spectrometers}

This method is based on the detection of fast neutrons through the reaction \( \mathrm{n} + ^{3}\mathrm{He} \rightarrow \mathrm{p} + ^{3}\mathrm{H}\), which has a positive Q-value of 764~keV. In a $^{3}$He filled proportional detector the reaction leads to the appearance of a peak in the spectrum for mono-energetic neutrons. The neutron energy $E_{n}$ can be determined from the displacement of the observed peak with respect to the thermal neutron peak. For a continuum neutron energy distribution, the measured spectrum must be deconvoluted using the energy-dependent neutron detector response. 

Following the introduction of high-pressure gridded ionization chambers \cite{Shalev1973} with good energy resolution ($10-20$~keV) and efficiency, this method has been used to measure \bdn spectra of several fission systems and individual precursors \cite{Brady1989a}. One disadvantage of this type of detector is the fast drop in efficiency with neutron energy: whereas at thermal energies the reaction cross section is as large as 5316~b it drops to 0.8~b at 1~MeV. This imposes a practical limit in the maximum measurable energy of about 3~MeV. In addition, the large peak due to the thermal-epithermal neutron background limits the minimum measurable energy to about 100~keV or more.

The accurate determination of the detector response is an essential step in this method. This is typically achieved using mono-energetic neutrons from reactions like $^{7}\mathrm{Li}(\mathrm{p},\mathrm{n}) ^{7}\mathrm{Be}$. The shape of the measured response departs from that of a simple peak due to several effects. Apart from electronic and microphonic noise, the response includes the effect of $\gamma$-ray interactions (including pulse pile-up), a peak due to thermal neutron background, the effect of neutron-induced recoils in the gas, the detector-wall effect, and ballistic effects (incomplete charge integration, see \cite{Franz1977} for a detailed discussion). Information on signal risetime can be used to eliminate some of these artifacts \cite{Owen1981}, at the cost of lowering the efficiency. 

Nowadays, digital electronics allow a more detailed pulse shape analysis \cite{Chichester2012}or the use of the full response function \cite{Beimer1986}. After parametrization of the detector response as a function of $E_{n}$, several suitable algorithms can be used to unfold the response from the measured spectrum \cite{Franz1977,Rudstam1980}. The evaluation of systematic uncertainties in this method is not a simple task. Information about their magnitude can be obtained by comparison of results obtained by different groups and detector systems \cite{Reeder1980,Rudstam1980}. This issue is discussed in Sect.~\ref{Sec:Eval-Spectra}.

\subsubsection{Gaseous proton recoil spectrometers}

This method is based on the detection of protons that are elastically scattered by neutrons. The detectors used are either methane- or hydrogen-filled proportional counters at high pressure in which the gas acts at the same time as the target for the reaction and as detecting media. These detectors are insensitive to thermal neutrons and compared to $^{3}$He spectrometers have a better energy resolution below $\sim200$~keV. Because of this they are considered to be superior in this energy range \cite{Brady1989a}. They have, however, a lower efficiency which limits the practical measurable energy range up to about 1~MeV. 

At typical $\beta$n energies, struck protons are scattered isotropically and the ideal detector response is a step function. Thus the neutron energy distribution is obtained by spectrum differentiation \cite{Bennett1972} rather than by deconvolution \cite{Gold1968}. Fourier filtering can be applied to reduce the fluctuations of the extracted distribution \cite{Sloan1978}. 

However the true detector response departs from the ideal response and a number of corrections must be applied \cite{Greenwood1985}. These type of detectors are sensitive to $\gamma$-rays that contaminate the low-energy part of the spectrum ($<100$~keV), but $\gamma$ signals are  efficiently eliminated using pulse shape discrimination. Different discrimination methods have been applied \cite{Eccleston1977,Greenwood1985} and the energy-dependent impact on neutron counting must be evaluated. In addition, the response is altered by the recoils of heavier nuclei present in the gas mixture, by wall- and end effects and by position-dependent variations of the signal amplitude. The PSNS code \cite{ANL7763} allows the inclusion of all these corrections and is commonly employed to analyze the data. In general the corrections are found to be small and sometimes instead of correcting the data they are used to evaluate the size of the related systematic uncertainties \cite{Greenwood1985}. A comparison of the results obtained with this technique and with $^{3}$He spectrometers is presented in Sect.~\ref{Sec:Eval-Spectra}.

\subsubsection{Neutron energy spectroscopy with time-of-flight spectrometers}


The time-of-flight (TOF) methods involve detection of neutrons using the $\beta$-neutron delayed coincidence method. In radioactive beam experiments, the start signal for the neutron time-of-flight measurement is given by a $\beta$ detector trigger, and the stop signal is provided by a neutron detector, which is typically a fast plastic or liquid scintillator. The required time resolution for the TOF measurements is in the nanosecond range. The TOF distance and timing resolution determine the total resolution \cite{Kor09}. 

The key advantages of TOF arrays are the relatively high intrinsic efficiency for high-energy neutrons, the ability to measure neutrons without recording their full energy,  and the relatively well-understood interaction cross-sections with the scintillator which enables reliable simulations. The disadvantage of the TOF technique is the limitation of the achievable resolution for high-energy neutrons ($>$ 3~MeV) due to the $E\sim 1/TOF^2$ dependence in the typical geometry and susceptibility of the detector response to neutron scattering. Neutron-$\gamma$ discrimination can be achieved by using liquid scintillators. The high cost of the detector material prevented the use of solid organic crystals such as stilbene and para-terphenyl for making large TOF arrays. Recently developed plastic scintillators with neutron-$\gamma$ discrimination capability \cite{Zai12} are now available. 

Until recently the most successful TOF arrays used for \bdn spectroscopy were TONNERRE \cite{Buta2000}, which was used in experiments at CERN and at GANIL (France), and the TOF array developed at the NSCL (USA)~\cite{Har91}. Both were used to measure light-mass nuclei with distinct neutron transitions \cite{Morrissey1997, Sum10, Tim05}. The heaviest studied isotopes were $^{51-53}$K \cite{Per06}.  Both TOF arrays share a similar design with about 1~m flight path, and long and curved plastic scintillator bars.

The currently operational TOF arrays for decay studies are VANDLE~\cite{Mat08,Pet15} (used at various facilities in the USA, at CERN, and at RIKEN), DESCANT~\cite{Bil13,Garrett2014} at TRIUMF/ISAC in Vancouver (Canada), and MONSTER which has been built for FAIR and already used at ISOLDE and at the IGISOL facility in Jyv\"asky\"a \cite{Mar14}.  

VANDLE is a TOF array using a fully digital data acquisition system \cite{Pau14}. The focus is on a high detection efficiency with the use of straight bars made of EJ200 plastic scintillators (Eljen Technology, USA). A digital distributed trigger enables a low-energy neutron threshold down to 100~keV. 

MONSTER is a TOF array based on BC501A (St. Gobain) and EJ301 (Eljen Technology) cylindrical liquid scintillators. Both BC501 and EJ301 offer excellent neutron/$\gamma$-ray pulse shape discrimination capabilities, which allow to strongly suppress the competing $\gamma$-ray background and $\beta$-neutron detector coincidences due to cosmic rays. MONSTER is operated with a fully digital data acquisition system based on 14 bits resolution and 1 Gsample/s digitisers.


DESCANT is a 70-element array of deuterated liquid-scintillator detectors (BC-537, St. Gobain) that can be used with both the TIGRESS and GRIFFIN $\gamma$-ray spectrometers at TRIUMF-ISAC for reaction studies and the investigation of $\beta$-delayed neutron emitters \cite{Bil13,Garrett2014}.
DESCANT forms a close-packed array that replaces the forward ``lampshade" of 4 HPGe clover detectors in each of the GRIFFIN and TIGRESS $\gamma$-ray spectrometers, providing high detection efficiency for neutrons in the range of $\approx 100$~keV to 10~MeV.


The decay of $^{17}$N, a well-defined standard for neutron-branching ratio measurements (see Table~\ref{tab:standards}), can be used for energy- and detector-response calibrations in TOF experiments. The energy of the three main neutron lines are well-known from $^3$He ion chamber measurements \cite{Cut71, Mei73, Pol73, Alb76, Ohm76} and appear fully resolved in TOF measurements \cite{Buta2000, Miyatake2003}.

Alternatively, if a calibration with neutron transitions of energies higher than 2~MeV is desired, the use of $^{15}$B or $^{49}$K is possible. However, there are no high-resolution $^{3}$He-detector measurements for the decay of $^{15}$B, and only one measurement for $^{49}$K \cite{Rac84}. Several measurements of the delayed neutron emission of $^{15}$B \cite{Har91, Buta2000, Miyatake2003} and $^{49}$K \cite{Rac84, Per06} have been published using TOF arrays and show deviations up to 10\% in the reported neutron energies.

Monte Carlo simulations are a complementary tool to calibrations since they offer a large flexibility and can reach a high level of accuracy in the simulation of neutron interactions and detector properties \cite{Men14, Gar17}. They are used for extending the overall efficiency calibration of a TOF spectrometer to a broader neutron energy range than available with mono-energetic neutron sources. Moreover, Monte Carlo simulations can also be applied for determining the time-energy relation of the neutron detectors, quantifying the cross-talk and estimating the time-correlated neutron and $\gamma$-ray background due to elastic, inelastic and (n,$\gamma$) reactions. Many new measurements have been performed in the past few years and the results are presently being analyzed.

\subsection{New methods (for P$_n$ values and energy spectra)} \label{Sec:Micro-newmethods}

\subsubsection{Total absorption $\gamma$-ray spectroscopy}
\label{Sec:TAGS}

Total absorption $\gamma$-ray spectroscopy (TAGS) was first applied to $\beta$-decay studies at ISOLDE-CERN in the 1970's \cite{Duke1970} as a tool to measure accurately $\beta$-decay intensity distributions. It has been used since then to tackle a variety of problems, including the investigation of reactor decay heat evolution \cite{Algora2010}, the nuclear level densities and $\gamma$-ray strength functions far from stability \cite{Spyrou2014}, and the spectrum of reactor anti-neutrinos \cite{Rasco2016}. Recently it was proposed to use this technique to measure $P_{1n}$ values and delayed neutron spectra \cite{Rasco2017}. This is indeed an exciting possibility, the current status of the method is summarized here.

A total absorption spectrometer is a $\gamma$-ray calorimeter made with a large volume of inorganic scintillation material surrounding the source with a $\sim~4\pi$ solid-angle coverage. Such a detector is capable of absorbing the full energy of the $\gamma$-cascade following the decay, thus identifying the excitation energy of the populated state. However, due to detector inefficiency and $\beta$ penetration, this is not true for all events, and a spectrum deconvolution procedure is needed to recover the information.

Due to the large detector size, neutrons emitted after $\beta$ decay have a large probability to interact with the scintillation material through elastic collisions, inelastic scattering, and radiative capture. The latter two processes generate $\gamma$-rays that are easily detected. In principle, this is a source of contamination of the spectrum that needs to be determined and subtracted \cite{Tain2015}. 

The neutron reaction cross sections, specific to the scintillation material, determine the amount of background signals. Large differences are observed in $(n,\gamma)$ cross sections between scintillators, which is the reason for the choice of BaF$_{2}$ in the spectrometer of Ref.~\cite{Tain2015} since the neutron-induced background is minimized. 

On the other hand, Ref.~\cite{Rasco2017} proposes to turn the large capture cross section of NaI(Tl) into an advantage, allowing the measurement of neutron-emission probabilities and spectra. Beta-delayed neutrons captured in the scintillator produce a bump in the energy spectrum above $\sim 6.8$~MeV, whose shape depends on the neutron-energy distribution and the height is proportional to the $P_{1n}$ value. 

The difficulty of the method is that it relies on the Monte Carlo simulation of the transport and generation of radiation by neutrons in the detector to extract the desired information \cite{guadilla2018}. Contrary to the simulation of photon transport, the accuracy of such simulations is not guaranteed a priori and should be investigated by comparison between different codes and nuclear data libraries \cite{Men14} and, if possible, by dedicated measurements \cite{Tain2015b}. 

Alternatively, a collection of neutron emitters with well-known $P_{1n}$ values and neutron spectra can be measured with TAGS and compared with the results of the standard methods. This was done in Ref.~\cite{Rasco2017} for the case of $^{137}$I, and a good agreement was found with the $P_{1n}$ value recommended in this CRP~\cite{Liang2020} (see Table~\ref{tab:standards}). 

However, in a similar work reported in Ref.~\cite{guadilla2018}, the $P_{1n}$ value obtained for $^{137}$I was 11\% smaller, while the value for $^{95}$Rb was found to be 22\% larger than the evaluation \cite{Liang2020}. It is clear from this that further work is needed, both experimentally and from the point of view of simulations, to clarify the situation and establish the systematic uncertainties of the method. This is worth the effort -- given the sizable number of \bn~emitters that have been measured recently and will be measured in the near future with the TAGS technique.



\subsubsection{Ion-recoil methods}
\label{Sec:New-ion}

The Beta-decay Paul Trap (BPT)~\cite{Scielzo2012}, an open-geometry linear radiofrequency quadrupole (RFQ) ion trap designed for precision $\beta$-decay studies, has been used to determine \bdn (\bn) branching ratios and energy spectra from the recoil energy imparted to the daughter ions following $\beta$ decay. This novel way to perform \bn~spectroscopy circumvents the challenges associated with direct neutron detection. 

The BPT collects and suspends radioactive ions in vacuum at the center of a radiation-detector array consisting of $\Delta E$-$E$ plastic scintillator telescopes, microchannel plate (MCP) detectors, and high-purity germanium (HPGe) detectors, used to detect the $\beta$ particles, recoiling daughter ions, and $\gamma$ rays, respectively, emitted following the decay. Following $\beta$ decay, the recoil ions emerge from the trap volume essentially unperturbed by scattering and the TOF of the recoiling ion to the MCP detectors is determined. The \bn~decays can be identified because the higher-energy recoil ions characteristic of neutron emission (with kinetic energies up to tens of keV) have shorter TOFs than the other recoiling daughter ions (which have kinetic energies typically less than 500~eV). 

The \bn~branching ratios can be deduced by comparing the number of $\beta$-ion coincidences with TOFs characteristic of neutron emission to the total number of $\beta$ decays from trapped ions obtained from the number of detected: (1) $\beta$ particles emitted from the trapped precursor, (2) $\beta$-ion coincidences with longer TOFs characteristic of $\beta$-decay to bound states, and (3) $\beta$-delayed $\gamma$ rays. 

The agreement obtained for the \bn~branching ratios when using these three approaches provides additional confidence in the results. This approach was first demonstrated in Ref.~\cite{Yee13} and has subsequently been used to study mass-separated ion beams of $^{137,138,140}$I, $^{135,136}$Sb, and $^{144,145}$Cs~\cite{Czeszumska20,Wang20} delivered by the CARIBU facility at Argonne National Laboratory. 

Additional upgrades to the experimental setup can be implemented to increase the solid angle of the detector array and to improve the ion collection and cooling while minimizing the effect of the electric fields on the recoil ions. Future experiments will also benefit from increases in the intensity and purity of the low-energy beams delivered by the CARIBU facility~\cite{Hirsh2016, Savard2016}. 

The recoil-ion approach is well suited to reconstruct the kinematics of a single undetected particle through conservation of momentum and energy. The studies of $\beta$-delayed multi-neutron emission will need to incorporate auxiliary neutron detectors. 

\subsubsection{Optical Time Projection Chamber}\label{Sec:New-OTPC}

\begin{figure}[!htb]
	\centering
\includegraphics[width=0.45\textwidth]{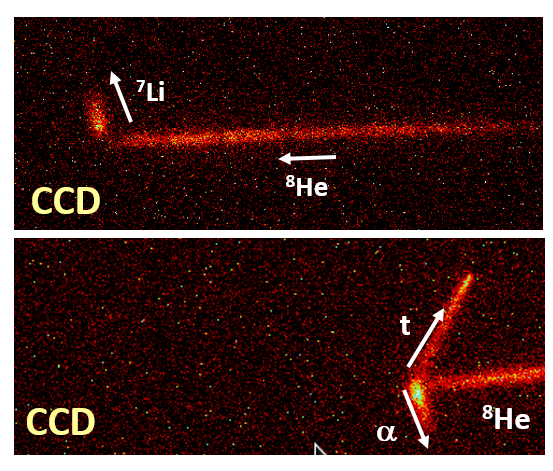}
\caption{(Top) \bdn decay of $^{8}$He into $^{7}$Li and a neutron (not visible) in the OTPC. (Bottom) $\beta$-delayed triton decay of $^{8}$He into an $\alpha$, a triton, and a neutron (not visible) in the OTPC. Pictures courtesy of Janas Zenon, University of Warzaw/ Poland.}
\label{fig:OTPC}
\end{figure}

The potential of using an optical Time Projection Chamber (OTPC)  to measure the neutron branching ratio via ion counting has been shown in \cite{Mianowski10}. The principle is the same as for the discovery of the two-proton radioactivity: a gas-filled TPC is used to produce ionization tracks of the incoming \bn-precursor and the recoiling \bn-daughter products which are then made visible via a CCD camera. Fig.~\ref{fig:OTPC} shows two different decay events of the \bn~emitter $^{8}$He. The picture shows only the incoming $^{8}$He nucleus and the recoiling $^{7}$Li, the \bdn is not visible. In the bottom of Fig.~\ref{fig:OTPC} the second decay possibility of $^{8}$He into an $\alpha$, a triton, and a neutron is shown. 

In principle one can reconstruct the energy of the emitted neutron from the length of the recoil tracks. This has been done for the $\beta$-delayed triton decay of $^{8}$He via measurement of the momenta of the recoiling charged particles ($\alpha$ and triton) \cite{Mia18}.

The main limitation comes from the generally low recoil energy and the finite resolution (diffusion effects) of the OTPC. So the technique is limited to low-mass precursors and reasonably high neutron energies. In the present setup with a gas mixture of 95\% He and 5\% N$_2$ at atmospheric pressure, a lower threshold for neutrons of $\approx 1.4$~MeV corresponding to a minimal track length of 10~mm \cite{Mia18} is required for the recoil identification. 

The OTPC method is not suitable for measuring $P_{2n}$ values (and higher) since the energy reconstruction is not possible with two outgoing (undetectable) neutrons.

\subsubsection{Measurement in a storage ring}\label{Sec:New-SR}
The $P_n$ value could also be measured via ion counting in a heavy-ion storage ring. 
This complimentary method has been proposed for radioactive beams produced at the Fragment Separator (FRS) at the GSI Helmholtz Center for Heavy Ion Research in Darmstadt/ Germany which are injected into the Experimental Storage Ring (ESR). Similarly, at the future Facility for Antiproton and Ion Research (FAIR) the \bne will be produced at the Super-FRS and injected into the Collector Ring (CR). While the \bne is circulating in the ring and decays, the measurement of the neutron-branching ratio could then be carried out via the ion counting method (see Sect.~\ref{Sec:methods-ionion}). This method has been described in Ref. \cite{Evdokimov13} but no proof-of-principle measurement has been carried out yet.

The advantage of this method is that symbiotically the mass and/or half-life of the stored ion could be measured non-destructively by time-resolved Schottky mass spectrometry \cite{Litvinov2004}. However, Schottky mass spectrometry requires ``cooling" to reduce the momentum spread of the beam. These beam cooling procedures take several seconds and thus reduce the range of accessible \bn~precursors.

Capacitive Schottky pick-up plates provide the revolution frequencies of the ions in the storage ring and allow their identification. When these ions decay, the change in mass and charge state leads to a change in the respective revolution frequencies. With Schottky spectrometry it is only possible to study the daughter nuclei that remain within the acceptance of the ring. This restricts the method in the ESR to parent-daughter pairs within an $A/q$ change of $\pm$1.5\%, but it can be complemented by particle detectors in pockets behind the dipole magnets to detect also $\beta$$^{-}$/ \bn-daughter nuclei. 

Like all described ion-counting methods this method is independent of the energy of the emitted neutron and the respective neutron detection efficiency. However, different loss mechanisms during the storage in the ring need to be carefully addressed and investigated. 

The decay of the ions can occur anywhere on the ring orbit (which is $\approx 108$~m for the ESR at GSI Darmstadt) but only those decays that occur in the long straight sections can be detected by a particle detector behind the respective dipole section. All other decays outside this section will lead to a loss of the ions. For the ESR, a multi-purpose particle detector (CsISiPHOS) has been installed and tested \cite{Najafi16}, which allows a geometrical efficiency of $\approx 20$\%. For experiments in the CR at FAIR two particle detectors will be installed and operated by the ILIMA collaboration \cite{ILIMA2017}.


\subsubsection{Ion counting using a MR-TOF}\label{Sec:New-MRTOF}
Multi-Reflection Time-Of-Flight (MR-TOF) spectrometers are becoming more and more popular as devices to measure precisely masses, determine the beam composition, or to manipulate and clean ion beams. This versatility has also led to the proposal to measure neutron-branching ratios (including multi-neutron emission branches) with a MR-TOF, parallel to the determination of masses and half-lives \cite{Mardor2017,Miyatake2018}. The setup at the Fragment Separator (FRS) at GSI Darmstadt/ Germany, the so-called FRS Ion Catcher \cite{Plass2013}, consists of a Cryogenic Stopping Cell (CSC) and a MR-TOF spectrometer.

The composition of the fast ($E > 500$ MeV/u) beam is identified with the standard set of detectors at the FRS. A beam containing only the isotopes of interest is stopped and stored in the CSC. After some decay time the ions are then extracted from the CSC into the MR-TOF for further identification and measurement. The relative intensities of \bn~precursors and respective decay daughters will then allow to extract the neutron-branching ratios.

\begin{figure}[!htb]
	\centering
\includegraphics[width=0.45\textwidth]{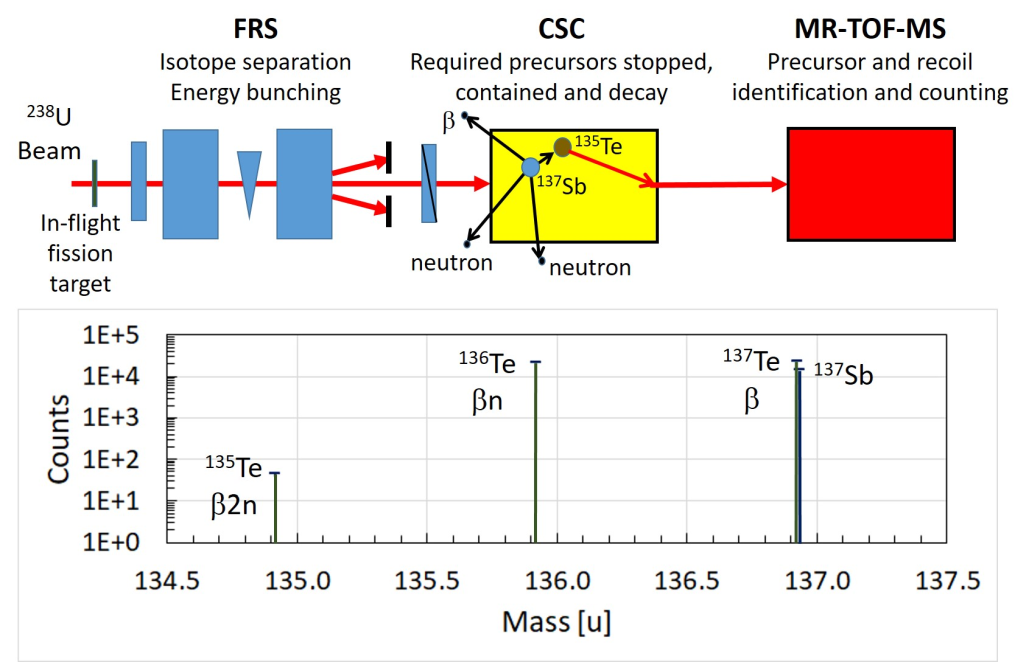}
\caption{Schematic view of the MR-TOF technique using a Fragment Separator (FRS), a Cryogenic Stopping Cell (CSC) and a MR-TOF device to count the ions. At the bottom a simulated MR-TOF mass spectrum is shown. Picture reprinted with permission from \cite{Mardor2017} (https://creativecommons.org/licenses/by/4.0/).}
\label{fig:MR-TOF}
\end{figure}

Fig.~\ref{fig:MR-TOF} gives a schematic view of the setup and a simulated MR-TOF mass spectrum for the yet unmeasured $\beta$2n decay of $^{137}$Sb. Within the next years this promising method will be tested with radioactive beams at the GSI Helmholtz Center in Darmstadt/ Germany.

\subsection{Outlook for future measurements} \label{Sec:Micro-newdata}
There has been a lot of progress in measuring \bn~emitters in the last decade, especially with precision better than 5\%. The earlier measurement campaigns were mainly driven by fission reactor studies and thus focused around the two -- low mass and high mass -- fission fragment groups. The reactor community still strives for more precise measurements of important fission fragments that are the main contributors to the delayed neutron yield $\nu_d$ (see priority list on p.~15 in Ref.~\cite{IAEA0643}) but the general scientific focus has now shifted. 

The nuclear astrophysics community is interested in measurements of the most neutron-rich, heavy ($Z >26$) nuclei that can be accessed with the new generation of radioactive beam facilities, especially those around the $N = 50$, 82, and 126 shell closures which are the progenitor nuclei that are responsible for the creation of the solar $r$-process abundance peaks at $A\approx$ 80, 130, and 195. Masses, decay half-lives, neutron-capture cross sections, and neutron-branching ratios of these neutron-rich nuclides are important nuclear physics input parameters for astrophysical network calculations, and in turn also help to improve theoretical nuclear models that are required for the prediction of even more exotic nuclei. 

\begin{figure*}[!htbp]
	\centering
\includegraphics[width=0.9\textwidth]{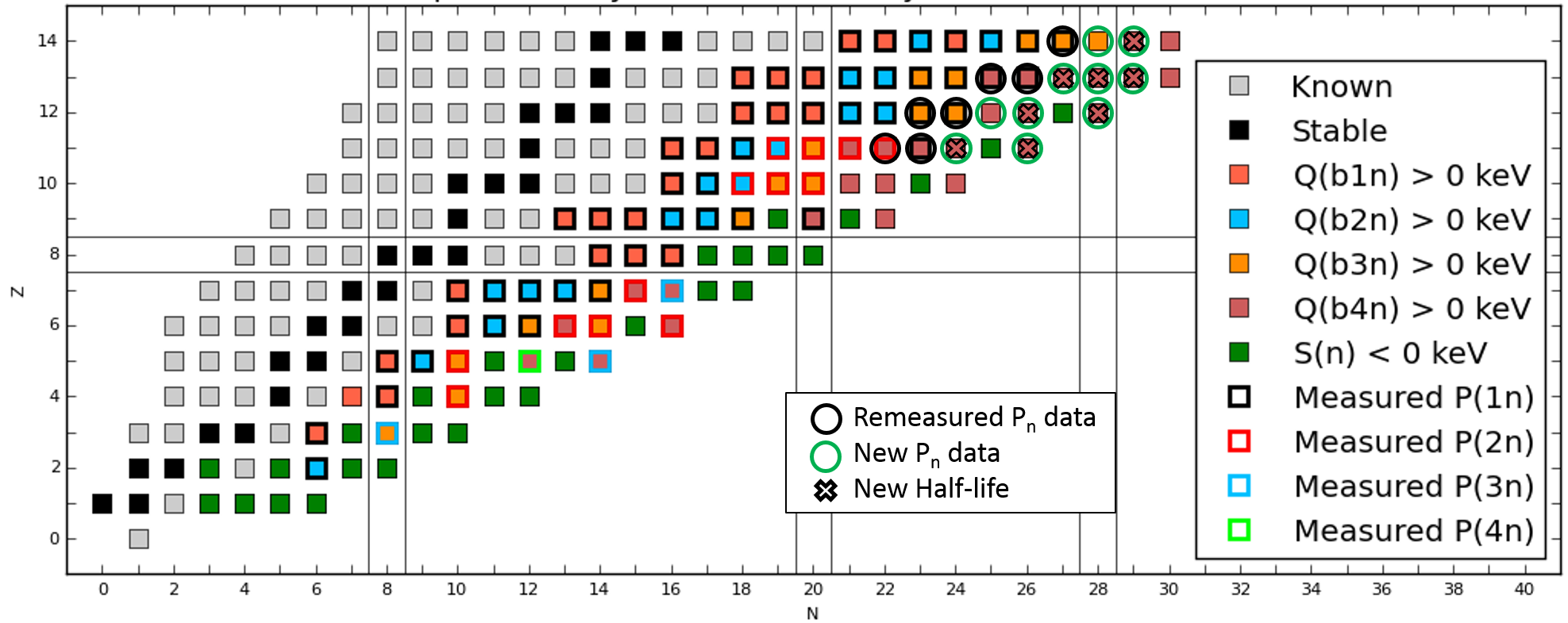}   
\includegraphics[width=0.9\textwidth]{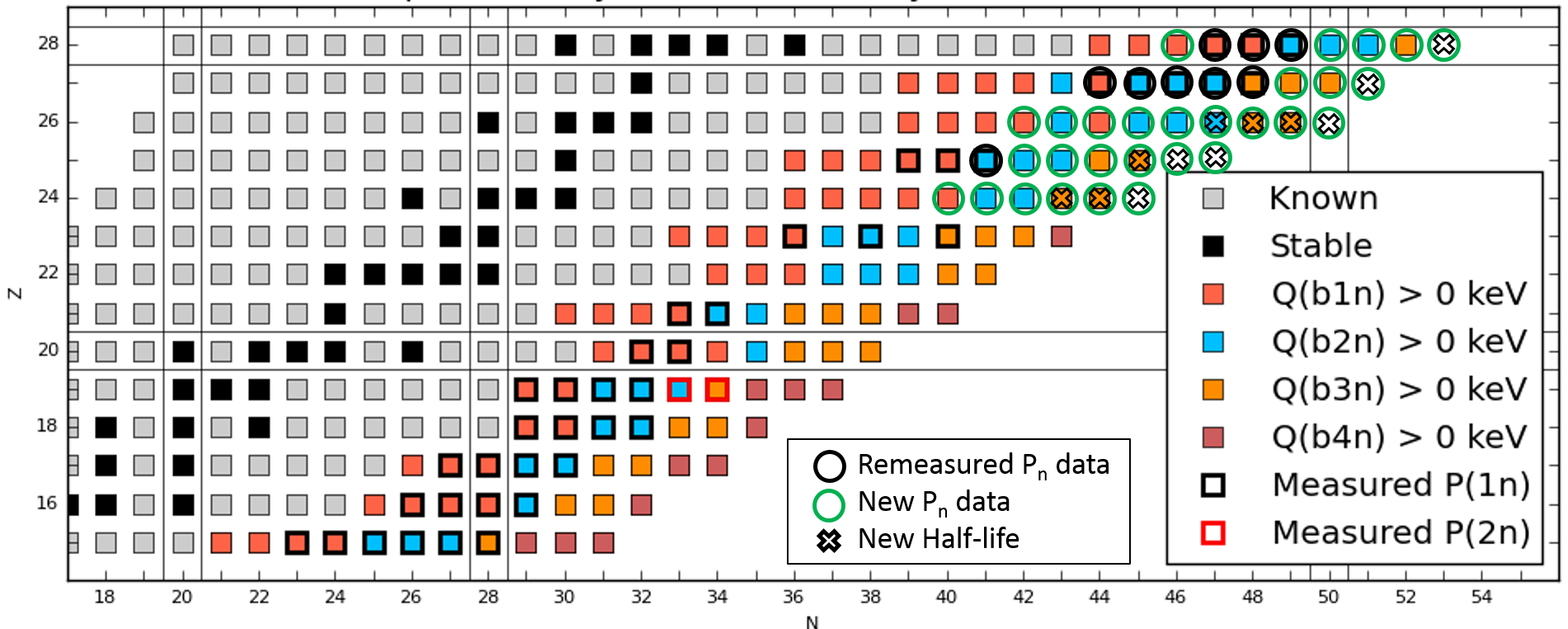}
\includegraphics[width=0.9\textwidth]{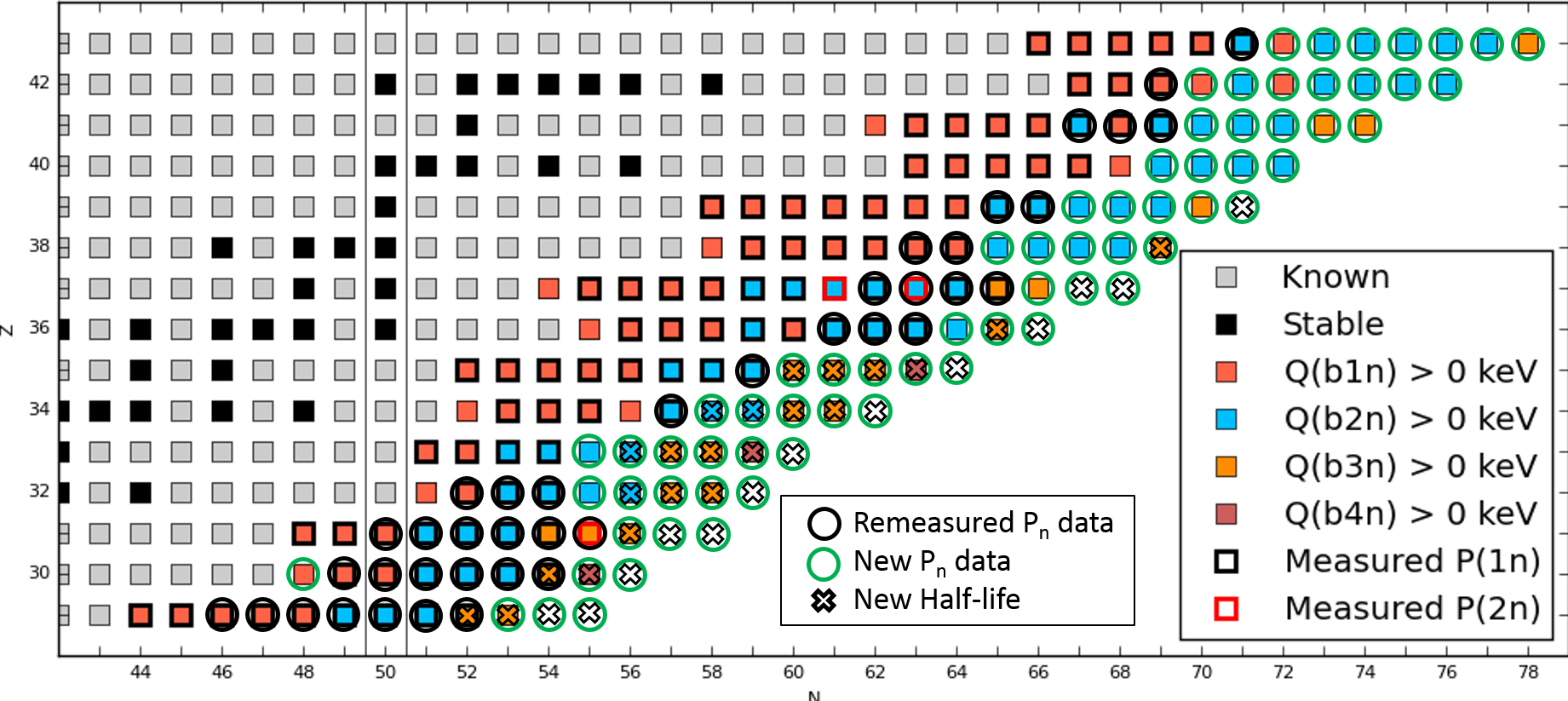}
\caption{Chart of nuclides for $Z$ = 0 -- 14 (top), $Z$ = 15 -- 28 (middle) and $Z$ = 29 -- 43 (bottom) with status for half-life and neutron-branching ratio measurements. Boxes show previously measured neutron-branching ratios, whereas circles indicate isotopes that have recently measured but have not yet been published or included in the latest evaluations \cite{Birch2015,Liang2020}. Crosses indicate isotopes for which half-lives have been measured for the first time recently.}
\label{fig:chart0-43}
\end{figure*}

\begin{figure*}[!htb]
	\centering
\includegraphics[width=0.9\textwidth]{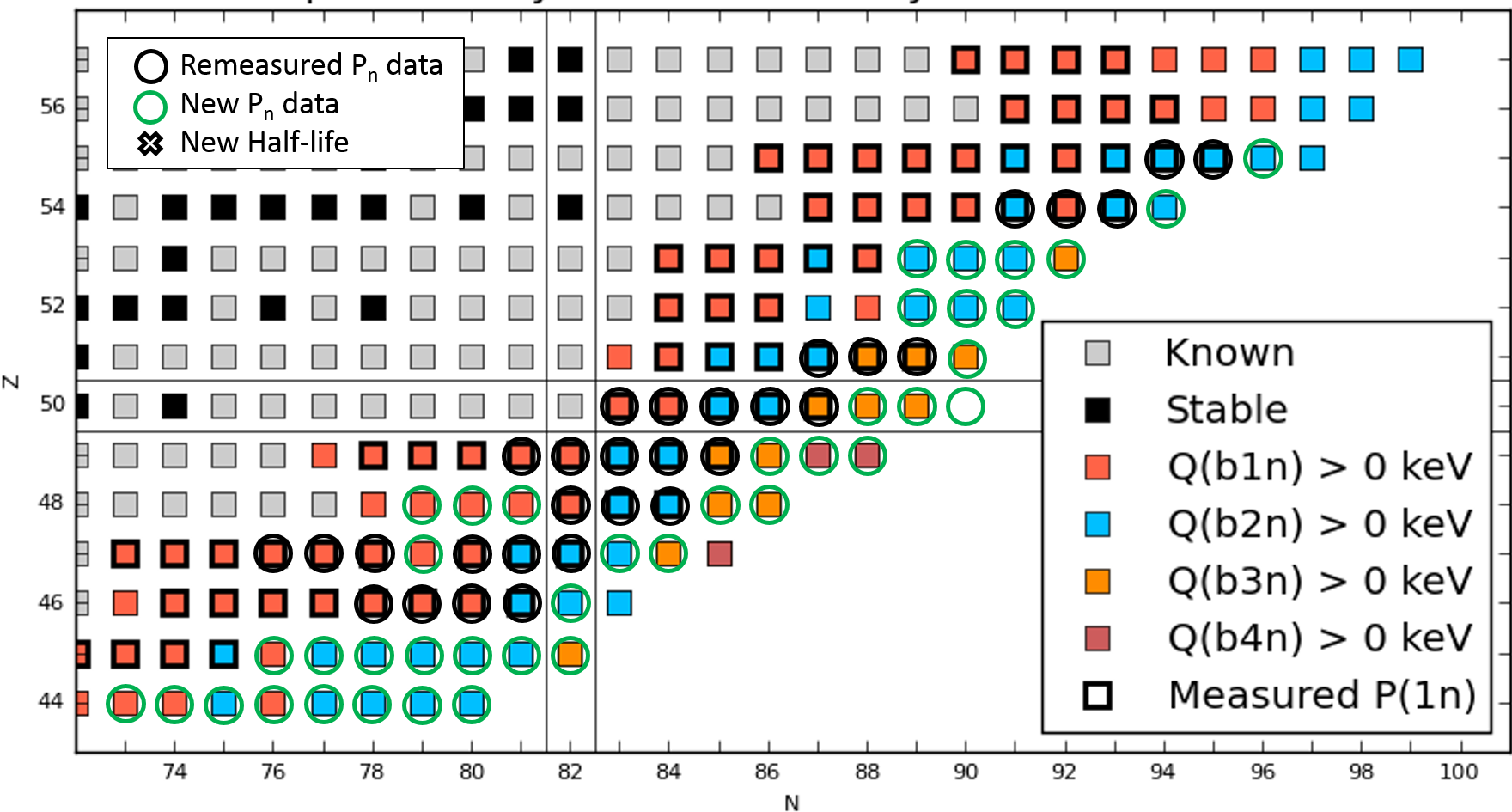}
\includegraphics[width=0.9\textwidth]{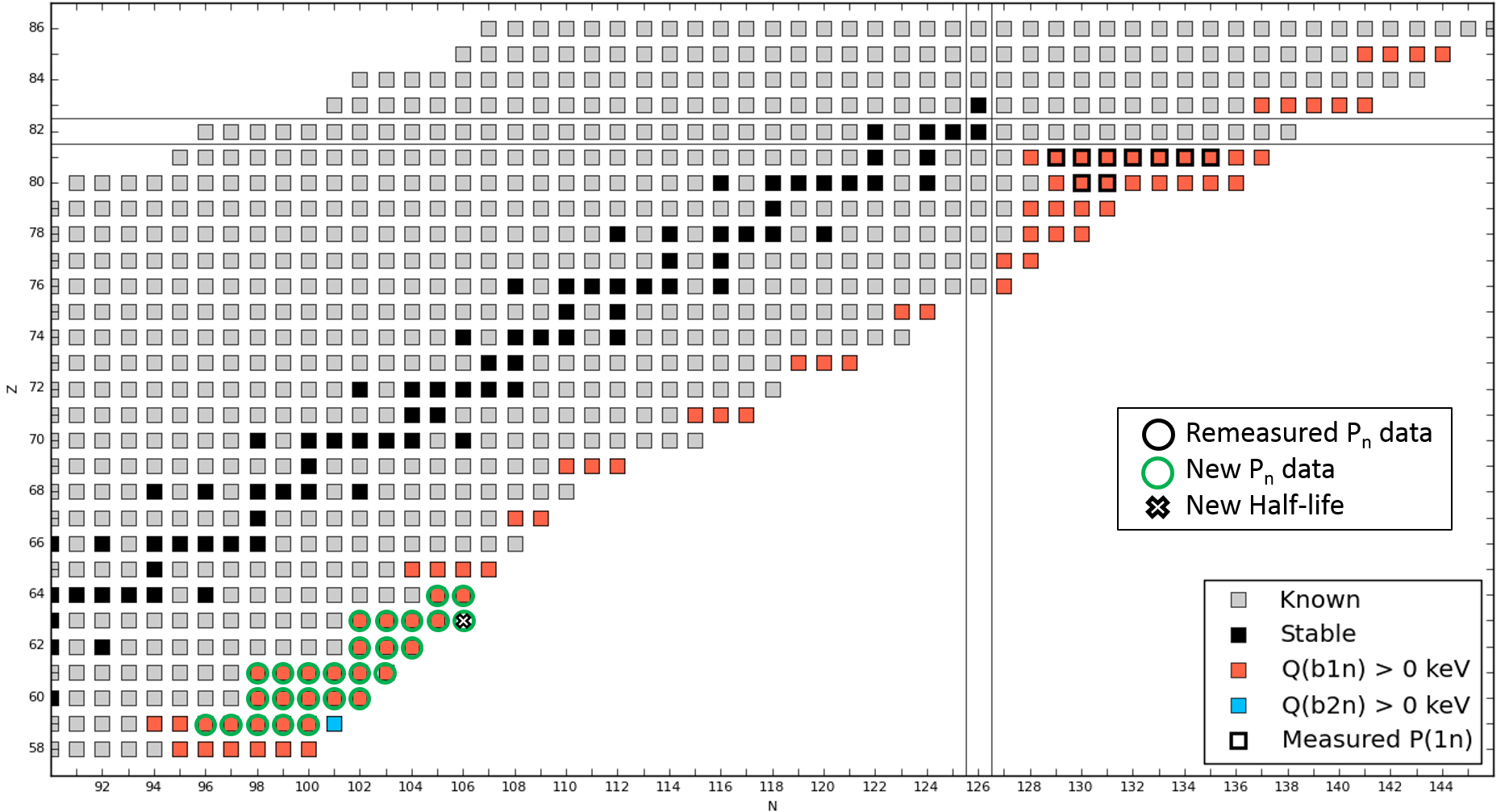}
\caption{Chart of nuclides for $Z$ = 44 -- 57 (top) and $Z$ = 58 -- 86 (bottom) with status for half-life and neutron-branching ratio measurements. Boxes show previously measured neutron-branching ratios, whereas circles indicate isotopes that have recently measured but have not yet been published or included in the latest evaluations \cite{Birch2015,Liang2020}. Crosses indicate isotopes for which half-lives have been measured for the first time recently.}
\label{fig:chart44-86}
\end{figure*}

A detailed review of the $r$-process nucleosynthesis, the astrophysical scenarios, the required nuclear data, how they can be measured at radioactive beam facilities and with what accuracy, is provided in Ref.~\cite{Horowitz2018}. Figs.~\ref{fig:chart0-43} and ~\ref{fig:chart44-86} give an overview of the status quo of \bne as of August 2020. These pictures include the vast amount of measurements that were performed during the course of the evaluation work that has been published in Refs.~\cite{Birch2015,Liang2020} (see following Sect.~\ref{Sec:Micro-compilation}), as well as the majority of the (yet unpublished) data from the BRIKEN project \cite{BRIKEN2017, Tolosa2018} which will conclude its campaign in 2021.

This decade will see the transition to the new generation of radioactive ion beam (RIB) facilities that are still under construction. Two of the main competitors for the study of the most neutron-rich \bne will be the Facility for Rare Isotope Beams (FRIB) at Michigan State University/ USA, which is planning to be operational by the end of 2021, and the Facility for Antiproton and Ion Research (FAIR) in Darmstadt/ Germany which is aiming at producing the first beams in $\approx$ 2027. Until then, the ``FAIR Phase-0" program will be operating at the existing GSI facilities, using many of the newly designed detector setups and upgrades of the accelerator facilities.

These new facilities will push the limits on what can be measured on the neutron-rich side of the chart of nuclides closer towards the driplines. Two of the key regions for future $r$-process experiments at these new facilities are the so-called ``Rare Earth Region" ($A= 160-170$) and the nuclei with $Z<75$ along the $N= 126$ shell closure. As can be seen in Fig.~\ref{fig:chart44-86} and in the recently published evaluation \cite{Liang2020}, only a small number of $\beta$-delayed neutron emitters has been measured so far for $Z>57$. Experiments in the next decade will allow  us to access these neutron-rich nuclei for the first time and achieve a better understanding of the formation of these two abundance peaks in the solar $r$-process abundance distribution.

\section{  COMPILATION AND EVALUATION}
\label{Sec:Micro-compilation}

\subsection{Compilation and evaluation methodology}
Standard procedures were followed for the evaluation of nuclear spectroscopic data as defined for the Evaluated Nuclear Structure Data File (ENSDF) and discussed at three Research Coordination meetings of the CRP \cite{IAEA0599,IAEA0643,IAEA0735}, with the addition of guidelines for half-life evaluations as formulated by A.L. Nichols and B. Singh (see p. 31--40 in Ref.~\cite{IAEA0687}). Emphasis has been placed on making sure that the $P_n$ value and half-life data have been compiled thoroughly, together with the method used for each measurement, from all the available measurements published in peer-reviewed journals as well as in certain selected secondary references. This was followed by a selection of the most reliable set of data for a nuclide which could be averaged to obtain recommended values. 

Details of the procedures and policies of evaluations for this CRP are described in two publications. A detailed paper on the compilation and evaluation of data in the $Z$ = 2 -- 28 region (cut-off date December 2014) was published by M. Birch \textit{et al.}~\cite{Birch2015} in 2015. The newly evaluated data for $Z$ = 29 -- 83 (cut-off date August 2020) can be found in J. Liang \textit{et al.}~\cite{ Liang2020}.

\subsection{New evaluated ($T_{1/2}, P_{n}$) data}
In total, as of August 2020 data for nuclide identification, half-lives and $P_{n}$ values have been extracted and evaluated for 653 neutron-rich known or potential \bdn emitters from more than 750 publications, mostly in main-stream peer-reviewed journals.

The evaluations in Refs.~\cite{Birch2015,Liang2020} included only the very first batch of data from the aforementioned BRIKEN measuring campaign at RIKEN. The upcoming publications will be included in evaluations of these mass ranges in the near future. 
The cut-off date for the data used for the calculations of macroscopic data in this CRP paper (see Secs.~\ref{Sec:Macro-Summation} and \ref{Sec:Macro-Integral}) was March 2020. The data used in these calculations can be found in the reference database~\cite{database}.

Some data that was published in the months between March 2020 and the cut-off date of the $Z>$28 evaluation was included in the published version \cite{Liang2020}. This data was not used for the calculation of macroscopic parameters in Secs.~\ref{Sec:Macro-Summation} and \ref{Sec:Macro-Integral} as already mentioned, however, the differences are minute and have a negligible impact on the calculated parameters. 

The most noteworthy addition is the inclusion of data published by the EURICA collaboration at RIKEN \cite{JWu2019}, measuring half-lives of 55 neutron-rich nuclei between $^{134}$Sn and $^{155}$La. In addition, for 13 neutron-rich nuclei the $\beta$-decay half-life was measured for the first time: $^{140-142}$Sb, $^{139-144}$Te, $^{144-146}$I, and $^{148}$Xe. This led, for example, to tiny changes in the recommended half-lives for the $P_{1n}$ standards $^{144,145,146}$Cs: 
\begin{itemize}
    \item $^{144}$Cs: 581(6)~ms $\longrightarrow$ 582(6) ms
    \item $^{145}$Cs: 321.7(11)~ms $\longrightarrow$ 321.7(10) ms
    \item $^{146}$Cs: 229.7(10)~ms $\longrightarrow$ 229.5(10) ms
\end{itemize}


\subsubsection{$Z$ = 2 -- 28 region}
Based on $Q$ values deduced from the Atomic Mass Evaluation (AME) 2016  for neutron-rich nuclides \cite{AME16,AME16a}, a total of 221 experimentally observed nuclides were identified as known or potential \bdn emitters in Ref.~\cite{Birch2015}. Of these nuclei, half-lives had been measured for 180 nuclides, $P_{1n}$ values for 114, $P_{2n}$ values for 21, and $P_{3n}$ values for only three nuclides. There is only one nucleus, $^{17}$B, where a tentative $P_{4n}$ value has been reported \cite{Dufour1988}. In addition there were 9 isomeric values for $P_{1n}$ values reported. For about 50\% of the cases, $P_{xn}$ values are available from only one measurement. 

Since the cut-off date for this part of the evaluation was in 2015, work on an update to include recent publications has started by the Canadian community. So far, four new nuclides have been identified as \bne, and two decay half-lives and three new $P_{2n}$ values were measured.


\subsubsection{$Z$ = 29 -- 57 region}
The latest evaluation comprised the heavier mass region and was recently published \cite{Liang2020} with a cut-off date of August 2020. A total of 318 nuclei
were identified as \bdn emitters, based on $Q$ values deduced from the AME2016 \cite{AME16,AME16a}.  This region is of utmost importance due to the production of most of these nuclides in the fission of uranium and other actinides, and therefore of relevance in nuclear power reactors.  Out of these 318 nuclei, half-lives have been measured for 284 nuclides, $P_{1n}$ values for 183 nuclides, and $P_{2n}$ values for only eight nuclides. So far no $P_{3n}$ or $P_{4n}$ values have been determined in this mass region.  

The data for isomers are generally poorly determined and often the half-lives and $P_{n}$ values are mixed with the data for ground states. For a large number of cases $P_{1n}$ values are available from only one measurement, and for several others only upper limits are reported. However the data for the main isotopes contributing to fission yields seem to be in good order.  

\subsubsection{$Z$ = 58 -- 87 region}
In this mass region 110 experimentally known nuclides were identified as \bdn emitters based on $Q$ values from the AME2016 \cite{AME16,AME16a}.  Half-lives are known for only 54 of these, and $P_{1n}$ values for only 9 nuclides. Measurements for six Tl isotopes ($^{211-216}$Tl) and two Hg isotopes ($^{210,211}$Hg) were reported only recently  \cite{Caballero2016,Caballero2017a}, while a tentative value of P$_{1n}$ for $^{210}$Tl was reported in a secondary publication in 1961 \cite{Stetter1962}.  


\subsection{P$_{1n}$ Standards}\label{Sec:Eval-P1n}
The selection of certain radionuclides as standards for P$_{1n}$ follows criteria which are given in Refs. \cite{Birch2015, Liang2020}. The main criteria for ``good" standards are: 
\begin{itemize}
\item Easy production at various facilities (large quantities, clean beams, no isomers).
\item Measured using reliable methods outlined in Sect.~\ref{Sec:Micro-Pn} in four or more independent experiments.
\item Consistent results with overall uncertainty of less than 5\%. 
\end{itemize}
Exceptions are $^{49}$K, $^{82}$Ga, and $^{146,147}$Cs, where the overall uncertainty is 5--10\%. $^{147}$Cs has only three independent measurements, but these measurements are very robust and consistent. 

In summary, the following nuclides are recommended as P$_{1n}$ standards, covering different $Z$ regions up to $Z$=55: $^{9}$Li, $^{16}$C, $^{17}$N, $^{49}$K, $^{82}$Ga, $^{87}$Br, $^{88}$Br, $^{94}$Rb, $^{95}$Rb, $^{137}$I, $^{145}$Cs, $^{146}$Cs, and $^{147}$Cs. The recommended half-lives and $P_{1n}$ values for these standards can be found in Table~\ref{tab:standards}.


\begin{table}[!htb]
\caption{Table with $P_{1n}$ standards and their corresponding half-lives as recommended from this CRP \cite{Birch2015, Liang2020}.}
\label{tab:standards}
\begin{tabular}{c|c|c|c} \hline\hline
Nuclide & Half-life (s) & $P_{1n}$ (\%) & Ref.\\ 
\hline
$^{9}$Li  & 0.1782(4)  &  50.5(10) & \cite{Birch2015}\\ 
$^{16}$C  & 0.7546(80)  &  99.28(12) & \cite{Birch2015}\\
$^{17}$N  &  4.171(4) &  95.1(7) & \cite{Birch2015}\\
$^{49}$K  & 1.263(50)  &  86(9) & \cite{Birch2015}\\
$^{82}$Ga  &  0.601(2) & 22.7(20)  & \cite{ Liang2020}\\ 
$^{87}$Br  & 55.64(15)  &  2.53(10) & \cite{Liang2020}\\
$^{88}$Br  & 16.29(8)  & 6.72(27)  & \cite{Liang2020}\\
$^{94}$Rb  &  2.704(15) & 10.39(22)  & \cite{Liang2020}\\
$^{95}$Rb  &  0.378(2) & 8.8(4)  & \cite{Liang2020}\\ 
$^{137}$I  &  24.59(10) & 7.63(14)  & \cite{Liang2020}\\
$^{145}$Cs  & 0.582(6)  &  13.5(6) & \cite{Liang2020}\\
$^{146}$Cs  &  0.3217(10) &  14.3(8) & \cite{Liang2020}\\
$^{147}$Cs  & 0.2295(10)  & 28.5(20)  & \cite{Liang2020}\\
\hline\hline
\end{tabular}
\end{table}

\subsection{Systematics for P$_{1n}$ values}\label{Sec:Eval-Syst}
In several studies \cite{Amiel1970,KHF1973,McCutchan2012,Miernik2013} correlations between known values of $P_{1n}$ and other $\beta$ decay gross properties ($T_{1/2}$, $Q_\beta$, $S_n$) have been described to estimate unknown $P_{1n}$ values. The basic correlation between $P_{1n}$ and $\beta$-decay properties is:

\begin{equation}
 P_n = \frac{\int_{S_n}^{Q_\beta}S_\beta(E)
    f(Z, Q_\beta - E)~dE}
    {\int_{0}^{Q_\beta}S_\beta(E)f(Z, Q_\beta - E)~dE},
    \label{eq:Pn}
\end{equation}

where $S_\beta(E)$ is the $\beta$-decay strength function and $f(Z, Q_\beta - E)$ is the Fermi integral. The integrations is performed over the available $\beta$ energy window, $Q_\beta - E$.

One of the earliest semi-empirical expressions originates from Amiel \textit{et al.} in 1970 \cite{Amiel1970}. Already then it was known that the neutron emission probability depends on the available $Q_{\beta n}$ energy window, the level density, and the competition of neutron and $\gamma$ emission. A plot of the log $P_{1n}$ vs. log $Q_{\beta n}$ resulted in a linear dependence with slope $m$. This led to the conclusion that the dependency of the P$_{1n}$ value on the $Q_{\beta n}$ window is dominant over any influence from the level density.

The evaluated half-lives and $P_{1n}$ values in this CRP were analyzed by the three empirical approaches proposed by K.L. Kratz and G. Herrmann \cite{KHF1973} (Kratz-Herrmann formula, KHF), an improved KHF systematics developed by E.A. McCutchan \textit{et al.}~\cite{McCutchan2012}, and a level-density dependent systematics by K. Miernik \cite{Miernik2013}. These estimates can be used as guidance for future experiments close to the known region, as well as to cross-check if outliers in existing data appear. They are however of limited use for $r$-process abundance calculations since they do not reach out far enough in $A$.

For $Z$ = 2 -- 28 region, global fits with the Kratz-Herrmann and McCutchan \textit{et al.} systematics were presented in Ref. ~\cite{Birch2015}. More detailed systematic plots using all the three approaches are presented for the $Z$ = 28 -- 57 region in Ref.~\cite{Liang2020}, where a new approach of fitting the McCutchan \textit{et al.} systematics by individual $Z$ has also been presented for most of the elements in this region.  It appears that this approach has better potential for predicting unknown $P_{1n}$ values. Unfortunately, there is not enough data in the $Z$ = 58 -- 87 region to attempt fits with the systematics.

\subsubsection{Kratz-Herrmann formula}
In the Kratz-Herrmann formula (KHF) \cite{KHF1973} the $\beta$-strength function is assumed to be constant above a certain cut-off energy $C$, and to be zero below that energy. It also assumes that the Fermi integral might evolve as a function of the $Q_{\beta n}$ value. This leads to the following description:

\begin{equation}
{P_{1n}} \sim a~ {\left(\frac{Q_{\beta n}}{Q_{\beta} - C}\right)^b}, \label{eq:KHF}
\end{equation}

where $a$ and $b$ are free parameters (fitted to the respective data range), and $C$ is the cut-off parameter. The values for the cut-off parameter depend on the proton/neutron number: 
\begin{itemize}
\item $C = 0$ for even-even nuclei. 
\item $C = 13/\sqrt{A}$ for even-odd/odd-even nuclei. 
\item $C = 26/\sqrt{A}$ for odd-odd nuclei.
\end{itemize}

For the latest fit parameters and a log-log plot of $P_n$ as a function of $\left(\frac{Q_{\beta n}}{Q_{\beta} - C}\right)^b$, the reader is referred to Ref.~\cite{Liang2020}.

\subsubsection{McCutchan systematics}
The considerable scatter and large reduced $\chi^2$ in the plots of the KHF systematics questions the reliability of predictions made employing this systematics. Thus, an improved systematics was developed by McCutchan \textit{et. al.} \cite{McCutchan2012} which considers both, the \bdn emission probability as well as the half-life. As the half-life is inversely proportional to the decay $Q$ value, the ratio $P_{1n}$/$T_{1/2}$ is considered as a function of the $Q_{\beta n}$ value:

\begin{equation}
\frac{P_{1n}}{T_{1/2}} \sim c~ Q_{\beta n}^d, 
\label{eq:McC}
\end{equation}

where $c$ and $d$ are fitted parameters.

\subsubsection{Improved systematics along $Z$ chains}
While the $P_{1n}/T_{1/2}$ systematics from McCutchan show an improvement over the KHF, there is still considerable scatter. "Isotopic" systematics along a given $Z$ can provide a much better picture for unknown $P_{1n}$ values. In Ref.~\cite{Liang2020} this method was used for nuclides in the region $Z=29-57$ which had three or more isotopes with a measured value for their $P_{1n}$ and $T_{1/2}$, respectively.

The improved systematics along a given $Z$ shows linear behavior when plotted in a log-log scale. Despite some scatter in this presentation, extreme outliers can indicate isotopes that require further investigation, e.g. the isotopic chains of As, Pd, and Ba \cite{Liang2020}.

\subsubsection{Systematics based on the Effective Density Model}
The phenomenological effective level density function model by Miernik \cite{Miernik2013} assumes that the $\beta$ = strength function is proportional to the level density $\rho(E)$ fed by $\beta$-decay:

\begin{equation}
S_\beta (E) ~\propto~ \rho(E) = \frac{\textrm{exp}(a_d\sqrt{E})}{E^{2/3}}
\end{equation}

The parameter $a_d$ is fitted to existing data, so that the derived formula can be used across an entire mass surface. 

In order to calculate $P_n$ using this systematics, one needs to first find the $a_d$ for the nuclides of interest. Then the $\beta$-decay energy Q$_\beta$ of the parent and the neutron separation energy $S_n$ can be used within the integration of Eq.~\ref{eq:Pn}. The Fermi integral can be calculated precisely with available tables, and the integration of Eq.~\ref{eq:Pn} can be carried out numerically. 

An attempt to extend this model to multi-neutron emission (up to P$_{3n}$) can be found in \cite{Miernik2014}. The assumption is that the neutrons are emitted sequentially, and that the many-body channels are negligible. The limitation of this description is also that only Gamow-Teller transitions are considered and that the neutron-$\gamma$ competition is approximated with a "threshold method". 

Numerical values for the fit parameters and more details are given in Refs.~\cite{Miernik2013,Liang2020}.

\subsection{Delayed neutron spectra for individual precursors}\label{Sec:Eval-Spectra}

Delayed neutron spectra have been measured for less than 50 \bn~emitters. These include 34 fission products for which good quality data exists which was obtained using $^{3}$He and gaseous proton recoil spectrometers and measured by several groups at different installations. As it was pointed out in the late 1970's \cite{Kratz79} discrepancies were observed between the different sets of data in particular at low neutron energies. This lead to an attempt to benchmark the different methods at different laboratories \cite{Owen1981} and triggered new measurements at other facilities \cite{Reeder1980, Greenwood1985}.

An evaluation of the available neutron spectra for the 34 individual fission products was carried out during the thorough and comprehensive work presented in the PhD Thesis of M. C. Brady in 1989 \cite{Brady1989a}. The list of isotopes included in the evaluation is given in Appendix D of that work. Most of the experimental data was coming from $^{3}$He spectrometer measurements carried out by the Studsvik group (G. Rudstam \etal), the Mainz group (K.-L. Kratz \etal) and the Pacific Northwest Laboratory (PNL) group (P. Reeder \etal). Proton recoil data from the Idaho National Engineering Laboratory (INEL) group (R. C. Greenwood \etal) was also included. Additional data was published by this group in 1997 \cite{Greenwood1997} and thus was not included in the Brady evaluation. 

The data from the Studsvik group was provided with uncertainties. The uncertainty  for the other data sets was inferred from generic assessments by the respective authors. Brady \cite{Brady1989a} stated that the Mainz and Studsvik data are in "fair agreement" above few hundred keV, and are considered to be of better quality than PNL data. Probably because of this she gave preference to a single data set instead of averaging different data sets, when available. 

In general, due to the smaller uncertainties and broader energy range, preference was given to $^{3}$He data from the Mainz group. If available, the proton recoil data from INEL replaced the $^{3}$He data below 200~keV. The selected data were extended by theoretical model calculations to cover the full \Qbn window. This procedure lead to a renormalization of the experimental data. The recommended data is presented in graphical form (figures 12 to 45 in Ref.~\cite{Brady1989a}). 

The originally evaluated data from Brady was incorporated into the ENDF/B-VI decay data sub-library \cite{Brady89}. Uncertainties are not included explicitly in the recommended data of \cite{Brady1989a} neither can they be retrieved from ENDF/B.  In addition an evaluation of possible systematic errors from the comparison of different data sets was not carried out in Brady's work. The lack of realistic uncertainties for individual precursor spectra diminishes its helpfulness. For example, it does not allow for an evaluation of uncertainties on aggregate spectra calculated by the summation method. In addition it is difficult to define reference \bdn spectra for calibration or inter-comparison purposes. This issue is further discussed below.

\subsubsection{Digitization of delayed neutron (DN) energy spectra} \label{Sec:Eval-digitize}
Digitization of several \bdn spectra has been done using the GSYS2.4.3 code and its later version 2.4.7~\cite{GSYS2.4}. This code is routinely and extensively used for the digitization of cross section plots for the EXFOR database~\cite{Otuka2014}.  The quality of the digitization was checked by plotting the extracted data and comparing the plot with the corresponding figure in the paper. 

The extracted datasets of all the figures were sent to the IAEA Nuclear Data Section for further checking by an expert staff scientist. All the data files were finally prepared for inclusion into the Reference Database for delayed neutrons~\cite{database}.

\paragraph{$Z$ = 2 -- 28 region:} 
Beta-delayed neutron spectra from $^{8}$He \cite{Bjornstad1981}, $^{9}$Li \cite{Nyman1990,Hirayama2015}, $^{11}$Li \cite{Azuma1979,Aoi1997,Morrissey1997,Hirayama2004}, $^{14}$Be \cite{Aoi1997}, $^{15}$B \cite{Buta2000,Miyatake2003}, $^{16}$C \cite{Grevy2001}, $^{17}$B \cite{Yamamoto1997}, $^{17}$N \cite{Buta2000,Miyatake2003,Grevy2001}, $^{18}$N \cite{Lou2007}, $^{21}$N \cite{Lou2008,Li2009}, and $^{27,28}$Na \cite{Ziegert1981} were digitized. Note that some of these are time-of-flight spectra, e.g. for $^{9}$Li from Ref. \cite{Nyman1990}.

\paragraph{$Z$ = 29 -- 57 region:} 
Spectra are available from different groups:

\begin{itemize}
\item R.C. Greenwood and A.J. Caffrey \cite{Greenwood1985}: Spectra of delayed neutrons for the isotope-separated, fission product precursors $^{93-97}$Rb and $^{143-145}$Cs were digitized independently by two groups over an energy region of $\approx$ 10 -- 1300~keV. The data was transformed to an equidistant scale with an energy bin of 1~keV for use in the summation method.

\item P.L. Reeder \etal \cite{Reeder1980}: Measured energy spectra of delayed neutrons from $^{93-95}$Rb and $^{143}$Cs were digitized.

\item M.C. Brady \cite{Brady1989a}: Several measured delayed neutron spectra obtained from different experimental groups (mainly Mainz, Studsvik, and groups at the TRISTAN online separator) were analyzed in this work as raw and adjusted spectra. This work includes $^{79-81}$Ga, $^{85}$As, $^{87-98}$Rb, $^{129,130}$In, $^{134}$Sn, $^{135}$Sb, $^{136}$Te, $^{137-141}$I, and $^{141-147}$Cs. Both the raw as well as the adjusted spectra from this work were digitized. 

\item R.C. Greenwood and K.D. Watts \cite{Greenwood1997}: Delayed neutron spectra of $^{87,88,90}$Br, $^{137-139}$I, and $^{136}$Te were presented in this paper. All were digitized, and were subsequently transformed to equidistant scale with an energy bin of 1~ keV for use in the summation method.
\end{itemize}


\subsubsection{Assessment of spectra as a reference for calibration purposes}\label{Sec:Ref-Spectra}
Defining \bdn reference spectra would be an important asset in practical work. These could serve to benchmark new setups and methods for the measurement of \bdn spectra. It will also serve to study the neutron energy dependency of the efficiency in devices aiming to determine neutron emission probabilities. The lack of an evaluation of uncertainties, in particular systematic uncertainties, introduces ambiguities in the results.

\begin{figure}[!htb]
 \begin{center}
 \includegraphics[width=0.45\textwidth]{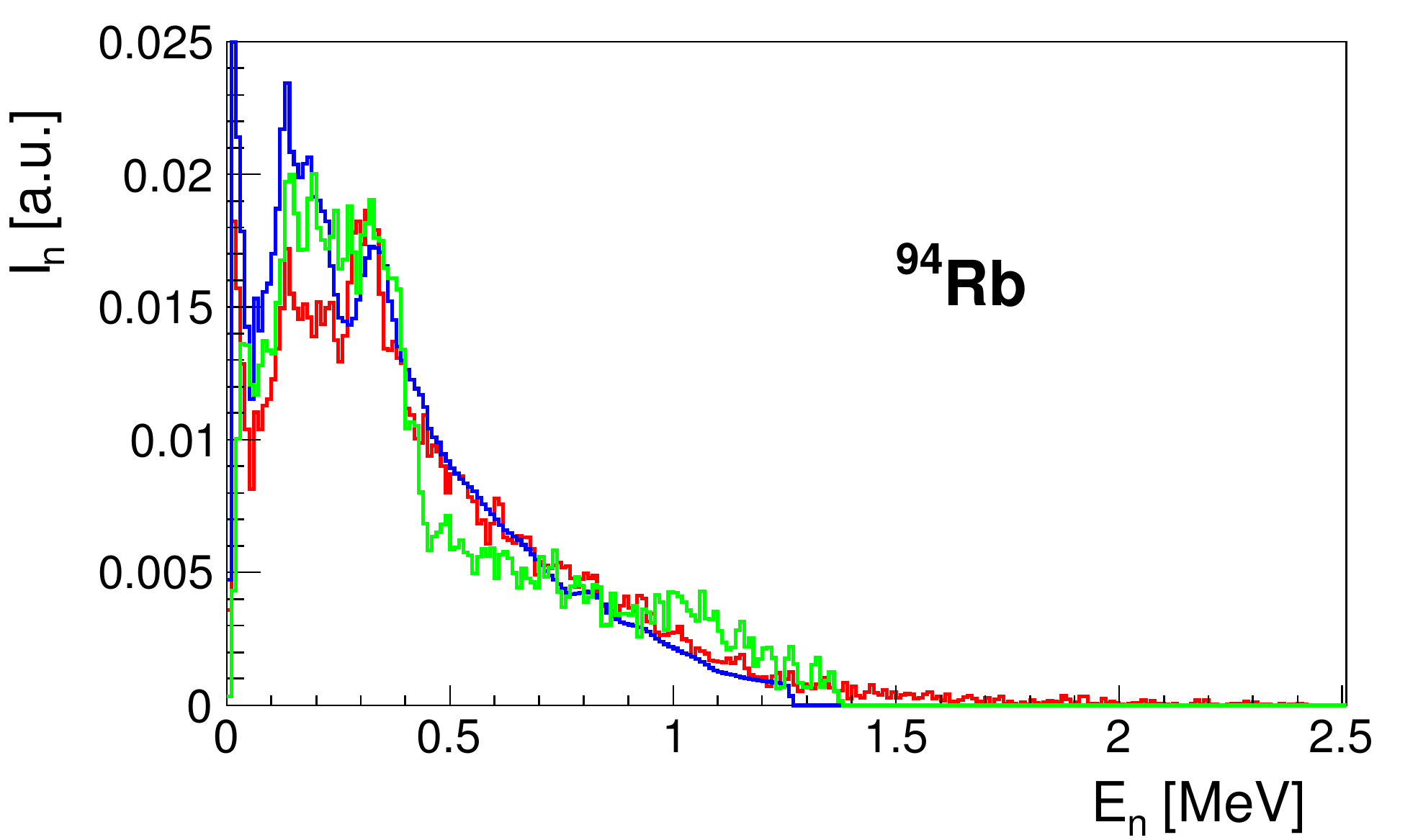}
 \includegraphics[width=0.45\textwidth]{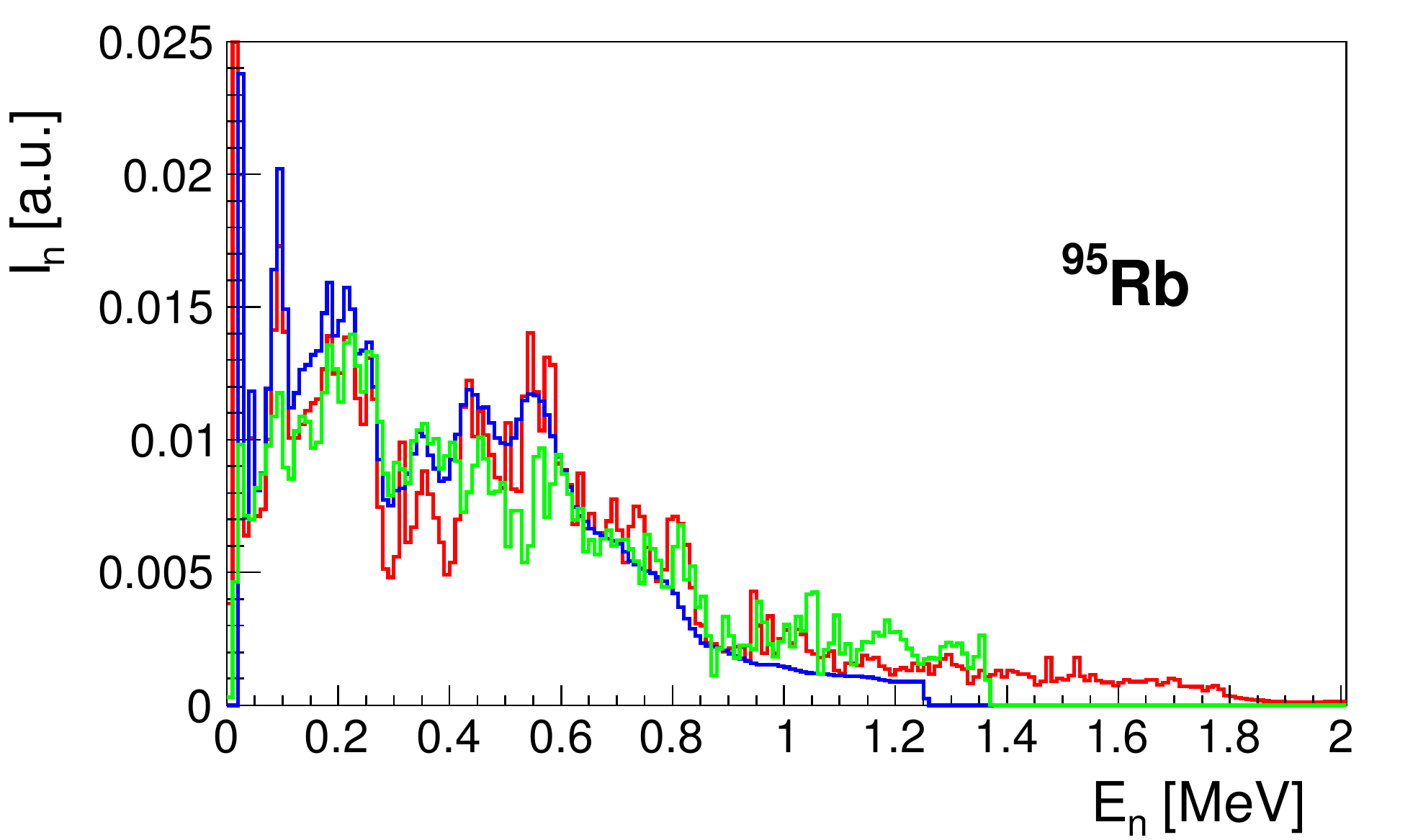}
 \end{center}
 \caption{Comparison of measured \bdn spectra for $^{94}$Rb (top panel) and $^{95}$Rb (bottom panel). Red line: Data retrieved from ENDF/B-VII.1 database. Blue line: Data from Greenwood \etal (INEL) \cite{Greenwood1985}. Green line: Data from Reeder \etal (PNL) \cite{Reeder1980}}
 \label{Fig:94Rb95RbSpec}
\end{figure}

An attempt was made within the CRP to asses the size of systematic uncertainties by comparison of spectral data obtained by different groups and different methods. However, it was realized that with the available information it is not possible to make a quantitative assessment. To illustrate the difficulty of the task, the cases of $^{94}$Rb and $^{95}$Rb are shown in Fig.~\ref{Fig:94Rb95RbSpec}, which have reasonably large $P_{1n}$ values and are easily produced in fission reactions. 

In the evaluation work of Brady \cite{Brady1989a}, $^{3}$He spectrometer data was available for both nuclei, provided directly by the Mainz group and the Studsvik group. Data was available also from the PNL group \cite{Reeder1980}, but was disregarded because it had lower statistics and a smaller energy range. Preference was given to Mainz data over Studsvik data for both isotopes, but no information is given about the differences between both data sets. Below 200~keV the Mainz data was replaced by proton-recoil data from INEL \cite{Greenwood1985}. 

Figure \ref{Fig:94Rb95RbSpec} displays the resulting evaluated spectra as retrieved from the ENDF/B-VII.1 library \cite{Chadwick2011} (same spectra are available in ENDF/B-VIII.0~\cite{Brown2018}). It is compared with the full INEL spectrum obtained by digitization of the graphical representation in \cite{Greenwood1985}. The data from PNL is also included in the figure, and was obtained by digitization of figures in \cite{Reeder1980}. The three data sets are histogramed with a 10~keV bin width and are normalized to the counts in the region from 0.2 to 1.2~MeV. As can be seen there are notable differences. When comparing values integrated over a range of 100~keV, to minimize the effect of different energy resolution and statistics, the discrepancies reach up to 30--50\%. The drop observed in the PNL data at around 0.5~MeV is likely due to a sudden rise of their efficiency that the authors quote as of uncertain origin but was also observed by other groups. Later this rise of efficiency has been identified as a spurious effect of neutron resonances excited in iron which is generally present in the experimental setup \cite{Tengblad1987}. 

It is not easy to arrive to a conclusion regarding the size of systematic errors from this information alone. In particular the loss of the primary information handled by Brady is unfortunate. In addition, it seems that the systematic errors of the different techniques are not fully understood. A proper evaluation work will require new well-controlled measurements that should profit from the advances in radioisotope production methods and instrumentation made in recent years.

For practical applications the interested user should be warned about possible large uncertainties in the shape of \bdn spectra retrieved from the ENDF/B-VIII.0~\cite{Brown2018} and ENDF/B-VII.1~\cite{Chadwick2011} libraries. Whenever possible the effect of spectra obtained from different measurements should be studied. It should be noted also that the experimental spectra in the ENDF/B data base are augmented with model calculations to cover the full \Qbn energy window. These calculations use FRDM+QRPA $\beta$-strength distributions \cite{Moller2003} and Hauser-Feshbach neutron branching ratios \cite{Kawano2008} to obtain the neutron spectra as described in the following sections. The model spectra additions are easily identified due to the smooth character of their distribution.

\section{  THEORY OF MICROSCOPIC DATA}
\label{Sec:Micro-theory}

In this section we compare the new evaluated CRP ($T_{1/2}, P_n$) data described in Sect.~\ref{Sec:Micro-compilation} with ``state-of-the-art" microscopic models. The comparisons are limited to \textit{global self-consistent $\beta$-decay models} that take into account both the allowed Gamow-Teller (GT) and first forbidden (FF) decays on equal footing and can be used to calculate both half-lives and \bdn~emission probabilities. Self-consistent models have the advantage of being founded on first principles. Their parameters are determined from a limited set of sample nuclei and are then kept unchanged for the whole nuclear chart. Thus, they are considered more reliable for extrapolations to extreme $N/Z$ ratios. However, before they are used for predicting nuclear properties in unexplored mass regions, their accuracy has to be tested against existing experimental data.
The aim of this section is to ascertain whether these global models are robust in the description of the evaluated CRP data which include new results from ongoing large-scale experiments at RIB facilities worldwide. This will then mean they can be used in extrapolations to mass regions where experimental data are either unavailable or inaccessible. 

Beta-decay properties are influenced by various different factors. First of all, the competition between the GT and high-energy FF transitions accelerates the $\beta$-decay, while in the region of the ground-state spin inversion such concurrence may produce isotopic irregularities in the half-lives and $P_n$ values. Second, the effects of many particle-many hole (np-nh) configurations substantially enriches and softens the $\beta$-decay strength distribution. Third, a local fragmentation of the $\beta$-strength due to deformation may change nuclear $\beta$-decay characteristics. In transitional nuclei a co-existence of spherical and deformed shapes may also be of importance. 

In the comparisons, we use two self-consistent global models which treat GT and FF transitions consistently, but do not take into account complex np-nh configurations or deformation effects. For completeness we also include the results of the recently updated microscopic-macroscopic model of Ref.~\cite{Moller2019} which are taken from the publicly available supplementary tables (\url{https://t2.lanl.gov/nis/molleretal/publications/ADNDT-BETA-2018.html}). 

The three models are listed below:
\begin{itemize}
\item Fayans Energy Density Functional (EDF)~\cite{Fayans2000} combined with continuum quasiparticle random-phase approximation (CQRPA). A common abbreviation DF+CQRPA is used below for different versions of the density functional, namely DF3 from ~\cite{Borzov1996,Borzov2003,Borzov2005} as well as 
the extended 
DF3a which was developed within the CRP to compare with the new evaluated data.
\item Relativistic Hartree--Bogoliubov+QRPA~\cite{Marketin2016} based on the D3C$^{*}$ functional (RHB+RQRPA)~\cite{Marketin2007}. This model was also extended within the CRP to perform global calculations.

\item Microscopic-macroscopic model of Ref.~\cite{Moller2019} based on the latest version of the finite-range droplet model FRDM12 published in Ref.~\cite{Moeller2016} combined with the BCS+RPA `(Q)RPA' description for the GT decays and a phenomenological approach based on the `Gross theory' for first forbidden (FF) decays. The \bdn~emission properties are estimated by considering the competition between neutron and $\gamma$ emission in the daughter nucleus using the Hauser-Feshbach statistical model (HF). This approach `FRDM12+(Q)RPA+HF' is different from the sharp cut-off model used in the previous calculations~\cite{Moller2003} whereby multi-\bdn~emission~emission was assumed to occur at energies above the corresponding neutron separation. 
\end{itemize}

In addition to the above models, which are used in isotopic chains and global comparisons of both  $T_{1/2}$ and $P_{xn}$ values, we include the results from the finite amplitude method (FAM)~\cite{Nakatsukasa2007} in selected comparisons of half-lives. The FAM is a mean-field model used 
up to now solely for calculations of half-lives, however,  it provides a  breakthrough approach for treating deformed nuclei~\cite{Mustonen2014,Mustonen2016,Shafer2016} with the Skyrme functional SkO'~\cite{Bender2001}, therefore is worth comparing with the CRP data for deformed nuclei.
    
A few attempts have been made to include complex np-nh configurations within the self-consistent approach in the study of the fine structure of the $\beta$-decay strength functions and integral $\beta$-decay properties. The preferred models are the latest versions of the interacting shell-model (SM)~\cite{Zhi2013,Yoshida2018} which include the allowed GT and FF decays. Also ``beyond the QRPA approximation" models based on the EDF approach have been developed and applied to selected nuclei near the closed shells. Practical schemes of this kind of models are the phonon-phonon coupling model (PPC) within the FRSA approximation ~\cite{Severyukhin2014,Severyukhin2017,Sushenok2018} and particle-vibration coupling models (PVC)~\cite{Niu2015,Niu2018,Robin2016} of the $\beta$-decay strength function. However, these models are still limited to the allowed GT approximation. Among the models with complex configurations, \bdn~emission branching ratios have been tackled so far only in the SM frameworks ~\cite{Zhi2013,Yoshida2018} and FRSA ~\cite{Severyukhin2017,Sushenok2018}. We have included published results from the SM ~\cite{Zhi2013, Yoshida2018}, FRSA ~\cite{Severyukhin2014,Severyukhin2017,Sushenok2018}, and PVC models ~\cite{Niu2015,Niu2018,Robin2016} in the comparison of half-lives of $^{78}$Ni in this report.

Finally, we mention the self-consistent models for spherical and deformed nuclei in the GT approximation ~\cite{Borzov2000,Minato2016,Yoshida2013,Yoshida2019,Sarriguren2014,Martini2014} which have been used in selected isotopic regions. Ref.~\cite{Minato2016} in particular is dedicated to the next generation of JENDL evaluated decay data, including $\beta$-decay half-lives, \bdn~branching ratios, as well as neutron energy spectra. A detailed review of the self-consistent models of $\beta$ decay can be found in Refs.~\cite{Borzov2020,Borzov2006,Borzov2017}.

Sect.~\ref{Sec:Micro-theory-A} discusses the extensions to the two self-consistent models that were performed within the CRP. Comparisons of the models with the evaluated CRP ($T_{1/2}, P_n$) data are shown in Sect.~\ref{III} for:
(i) isotopic chains of nuclides in the fission mass region, (ii) isotopic chains of nuclides relevant to nuclear astrophysics and (iii) global comparisons with respect to the ($T_{1/2}, P_n$) data and mass number $A$. Finally, theoretical DN spectra are compared with experimental spectra from Brady's thesis~\cite{Brady1989a} in Sect.~\ref{Sec:Micro-theory-spectra}. 

\subsection{Self-consistent QRPA-based models}\label{Sec:Micro-theory-A}

The DF+CQRPA and RHB+QRPA models have been extended to perform large-scale calculations of $\beta$-decay half-lives and delayed multi-neutron emission probabilities within the CRP. These extensions are briefly described in the following.

\subsubsection{DF+CQRPA}
The $\beta$-decay strength function is calculated in the extended finite Fermi systems theory (FFST) \cite{Migdal1967} based on the energy density functional (EDF) theory which allows for fully self-consistent description of the ground state properties of nuclei with pairing. Treating the effects of momentum and energy dependence of the functional and effective forces on an equal footing leads one to the Fayans EDF \cite{Fayans2000} with the fractional-linear dependence on normal density. The surface and pairing components of the EDF contain the gradient of the density. Such an ansatz is more general than the standard density dependence of the Skyrme functional that is a result of effective 3N correlations.
Within the present CRP the ground-state and $\beta$-decay characteristics for more than 200 (quasi-) spherical nuclei were re-calculated with the DF3a functional \cite{Tolokonnikov2010} which differs from the previous version DF3 by stronger effective tensor-like components. It gives the same quality of description of the ground-state properties as the DF3 functional and a better description of $\beta$-decay of heavy nuclei \cite{Madurga2012}.

Based on such an approach a framework that allows for large-scale continuum QRPA calculations of the GT and FF $\beta$-decay properties (DF3a+CQRPA) was developed in Refs.~\cite{Borzov2003,Borzov2005}. For large-scale $\beta$-decay calculations the effective approximation is used for the spin-isospin effective NN interaction. 
In the one-particle-one-hole (ph) channel it consists of the Landau- Migdal interaction and one-$\pi$ and $\rho$-meson exchange terms (modified for nuclear medium effects). 
For nuclei with pairing correlations, the density-dependent, $A$-dependent, zero-range $T=1$ ground-state pairing is used. For the spin-isospin effective NN interaction in the particle-particle (pp) channel (isoscalar $T=0$ effective interaction or dynamic pairing)  a mass-independent zero-range interaction with constant strength is assumed.

Allowed and first-forbidden transitions are treated on equal footing in terms of the full set of multipole operators that depend on the space and spin variables.  
The relativistic vector operator $\alpha$ and axial charge operators $\gamma_5$ are reduced to their non-relativistic 
space-dependent counter parts via the relations for conservation of vector current (CVC) and partial conservation of axial current (PCAC).  An advantage of such a scheme is that it is convenient to use in the full ph-basis continuum pnQRPA framework ~\cite{Borzov2003, Borzov2005}. 

The correlations beyond the QRPA are included by re-scaling the spin-dependent multipole responses by the same energy-independent quenching factor $Q=(g_{A}/G_{A})^2$.
The one-pion component of the residual interaction is quenched by the same factor $Q$.
The axial-vector coupling constant $|g_A/G_A|= 1.0$ is used both for GT and FF decays, and the amplification of the $\gamma_5$ component of the spin-dipole operator is accounted for as in Ref.~\cite{Warburton1991}. 
The energy-dependent effect of np-nh configurations can be included via the spreading width $\Gamma_{\downarrow}$ 
\cite{BertschBroglia1994}.

In the case of odd-$A$ and odd-odd nuclei (hereafter called ``odd" nuclei), their degenerate ground states are treated through averaging the corresponding multiplet components via equal filling approximation. The interaction of the odd nucleon with a ``core" and rearrangement effects are properly taken into account ~\cite{Borzov1996}. 

In 2016, the DF+CQRPA framework was extended by fixing the odd nucleon in a given state. In this way, the DF3a+CQRPA model can account for possible ground-state spin inversion effects which have been shown to impact the $\beta$-decay observables \cite{Borzov2017,Borzov2018}.
Though the present version of the model is basically limited by the spherical approximation, it should be noted that the deformation caused by the unpaired nucleon is properly taken into account in the FFST equations for the odd-odd and odd-$A$ nuclei.


\subsubsection{Relativistic Density Functional+QRPA}
Large-scale calculations of $\beta$-decay half-lives and $\beta$-delayed neutron-emission probabilities were performed in Ref.~\cite{Marketin2016} with the main goal of applying the results in heavy-element nucleosynthesis and comparing them with the new CRP ($T_{1/2}, P_n$) database. The model was based on the relativistic nuclear energy density functional (RNEDF), which constructs the nuclear ground state from the nucleon mean-field and a minimal set of exchange mesons together with the electromagnetic field. The ground-state properties were obtained from the relativistic Hartree-Bogoliubov model which correctly describes the pairing in open-shell nuclei. In the calculation, the D3C$^{*}$ density functional, which includes momentum-dependent terms~\cite{Marketin2007}, was used together with the pairing part of the Gogny D1S interaction for the description of the pairing correlations. 
For the isoscalar proton-neutron (pairing-like) NN-interaction, a two-Gaussian form was used following its introduction in Ref.~\cite{Engel1999}. Unlike the DF+CQRPA approach, which is free from fitting the parameters to the `output' half-lives, the strength of the interaction was adjusted to reproduce the experimental half-lives available at the time.

The FF transitions were  treated at the same level as the allowed GT transitions. 
The same value of the axial-vector coupling constant $|g_{A}/G_{A}|=1.0$ was used for all transitions, both allowed and first-forbidden, while the amplification of the $\gamma_5$ operator was not included.
For odd-$A$ and odd-odd nuclei, the ground state energy was computed by constraining the average number of particles to be odd which resulted in an \textit{even} RHB state. Its energy differs from the \textit{true} RHB state by the energy of the odd quasi-particle (or both in the case of odd-odd nuclei). This approach was tested on $Q_{\beta}$-values for even-$Z$ and odd-$Z$ isotopic chains and was found to be in satisfactory agreement with the data.

The limitations of the model are: (i) the model considers nuclei to be spherical, disregarding the possible effects of deformation on the transition spectra, and (ii) the model determines the ground state and excitations at zero temperature which has implications for applications to the $r$-process nucleosynthesis since the latter takes place at finite temperatures. Both approximations are necessary to reduce the total computational cost of the large-scale calculations.


\subsection{Comparison with new evaluated ($T_{1/2}, P_n$) data for important $\beta$-delayed neutron emitters} \label{III}

The evaluated CRP ($T_{1/2}, P_n$) data described in Sect.~\ref{Sec:Micro-compilation} are compared with the previously listed self-consistent global models \cite{Borzov2003,Marketin2016} and microscopic-macroscopic results \cite{Moller2019} for light and heavy fission products. The CRP data can be downloaded from the IAEA online database~\cite{database}. The selected isotopic chains contain the most important \bne associated with the highest fission yields, as recommended by the evaluated libraries ENDF/B-VIII.0 ~\cite{Brown2018}, JEFF-3.1.1 ~\cite{jeff3.1.1} and JENDL/DDF-2015 \cite{Katakura2015}. In order to assess the quality of the description of the $\beta$-decay characteristics, we analyze both the T$_{1/2}$ and P$_{xn}$ values.

 \subsubsection{Light fission products}\label{V}
\paragraph{Arsenic isotopes}
The $\beta$-decay characteristics of the As isotopic chain ($Z$ = 33) are compared with the calculations from the new DF3a functional within the DF3a+CQRPA model, as well as with the RHB+QRPA \cite{Marketin2016} and FRDM12+(Q)RPA+HF \cite{Moller2019} models in Fig.~\ref{Fig:HL-P1n}. As can be seen in the figure, the DF3a+CQRPA calculation in the GT approximation overestimates the evaluated half-lives by up to a factor of 10. Inclusion of the FF decays and the experimental ground-state spin (J$^\pi = 5/2^-$) leads to a slight underestimation (by up to 19\%) of the evaluated half-lives, while the RHB+QRPA and FRDM12+(Q)RPA result in an underestimation by up to a factor of 8. The deformation of As isotopes is rather moderate and the quadrupole moments are estimated to be $\beta_2\approx 0.1-0.15$ \cite{Stone2014}. The difference in the results of spherical and deformed calculations of the half-lives is not significant.

A sudden decrease of the $P_{1n}$ value when crossing $N$ = 50 is correlated with the onset of the FF transitions. At $N$ = 52, 53 a qualitative agreement is observed between DF3a+CQRPA and FRDM12+(Q)RPA+HF, though the experimental $P_{1n}$ values  
are up to a factor of 3 higher. However, the experimental data do not support a sharp increase of the $P_{1n}$ value at $N$ = 54 as seen in the FRDM12+(Q)RPA+HF and RHB+RQRPA models.

\paragraph{Bromine isotopes}
The CRP half-lives and $P_{1n}$ values for Br isotopes ($Z$ = 35, Fig.~\ref{Fig:HL-P1n}) are compared with results obtained from the DF3a+CQRPA and RHB+RQRPA models. Both frameworks show a behavior that is typical of the dominance of the GT decays.
The deformation in the Br isotopes is estimated to be higher than in As isotopes  ($|\beta_2|\approx$ 0.15 -- 0.2) \cite{Stone2014}. However it is observed that the T$_{1/2}$ given by the spherical DF3a+CQRPA and RHB+RQRPA are closer to the experimental data than those from the deformed FRDM12+(Q)RPA model. The latter underestimates the T$_{1/2}$ by up to a factor of 25 at $A < 93$ and shows a sudden increase of T$_{1/2}$ at $A = 94$. At the same time, the $N$ dependence of $P_{1n}$ values in FRDM12+(Q)RPA+HF is in better agreement with the evaluated data, while the DF3a+CQRPA calculation describes the general trend of experimental half-lives and $P_{1n}$ values.


In Table~\ref{table:T_Br-I}, the evaluated half-lives and $P_{1n}$ values for Br isotopes are compared with those obtained from the DF3a+CQRPA (spherical option) and RHB+RQRPA models, as well as with the results from the FRDM+RPA model of Ref.~\cite{Pfeiffer2002} that uses the spherical GT approximation (``QRPA-2"). The latter model (based on FRDM of Ref.~\cite{Moeller1995}) was used to estimate the total (integral) DN yields in the evaluation of Wilson and England~\cite{Wilson02}. Finally, the deformed FRDM12+(Q)RPA calculations  that include the FF decays within the Gross theory \cite{Moller2019} are also displayed. 

As can be seen in the table, DF3a+CQRPA describes the general trend of the isotopic distribution  of the half-lives and  $P_{1n}$ values for $N<58$. It can be used for estimating the total (integral) DN yield, 
as the cumulative fission yields of the DN emitters amplify the contribution of this region (see Sect.~\ref{Sec:Macro-summation-yields-basic}). 

\paragraph{Molybdenum isotopes}
The influence of deformation on the $\beta$-decay properties is illustrated for the Mo isotopes ($Z$ = 42) in Fig.~\ref{Fig:HL-P1n}. Here, $\beta_2$ is $\approx -0.3$ for $110 < A < 118$ while for $A \approx 126$ it is $\beta_2 \approx 0$, as estimated from the quadrupole moments ~\cite{Stone2014}.

As can be seen from Fig.~\ref{Fig:HL-P1n}, the $T_{1/2}$ in Mo isotopes obtained from the spherical DF3a+CQRPA and RHB+RQRPA model are in agreement with the CRP data. They are also close to the results from the deformed FRDM12+(Q)RPA and FAM models in the $N = 72 - 82$ region.  However, the FRDM12+(Q)RPA shows some irregularities in the left wing of the isotopic chain corresponding to stronger deformation ($\beta_2 \approx -0.35$) for $N < 72$, as well as in the (quasi-)spherical region $N > 82$. At the same time, the $A$-dependence of the half-lives in the spherical models is quite smooth at $N > 82$.  It should be noted that a detailed study in Ref.~\cite{Sarriguren2010} has shown some sensitivity of the Mo half-lives due to the oblate, prolate or spherical shape.

\paragraph{Rubidium isotopes}
For the strongly deformed Rb isotopes ($Z$ = 37) with $\beta_2 \approx 0.35$ \cite{Stone2014}, the spherical RHB+RQRPA describes the experimental half-lives except for $^{92-94}$Rb but does not explain the $P_{1n}$ values for the important \bdn~emitters $^{95,96}$Rb (Fig.~\ref{Fig:HL-P1n}). The deformed FRDM12+(Q)RPA model tends to underestimate the half-lives in the region of strong deformation for $^{92-98}$Rb, nevertheless it provides a good description of the $P_{1n}$ data.



\subsubsection{Heavy fission products}\label{VI}
\begin{figure*}[!hp]
\begin{centering}
  \includegraphics [width=1.0\linewidth] {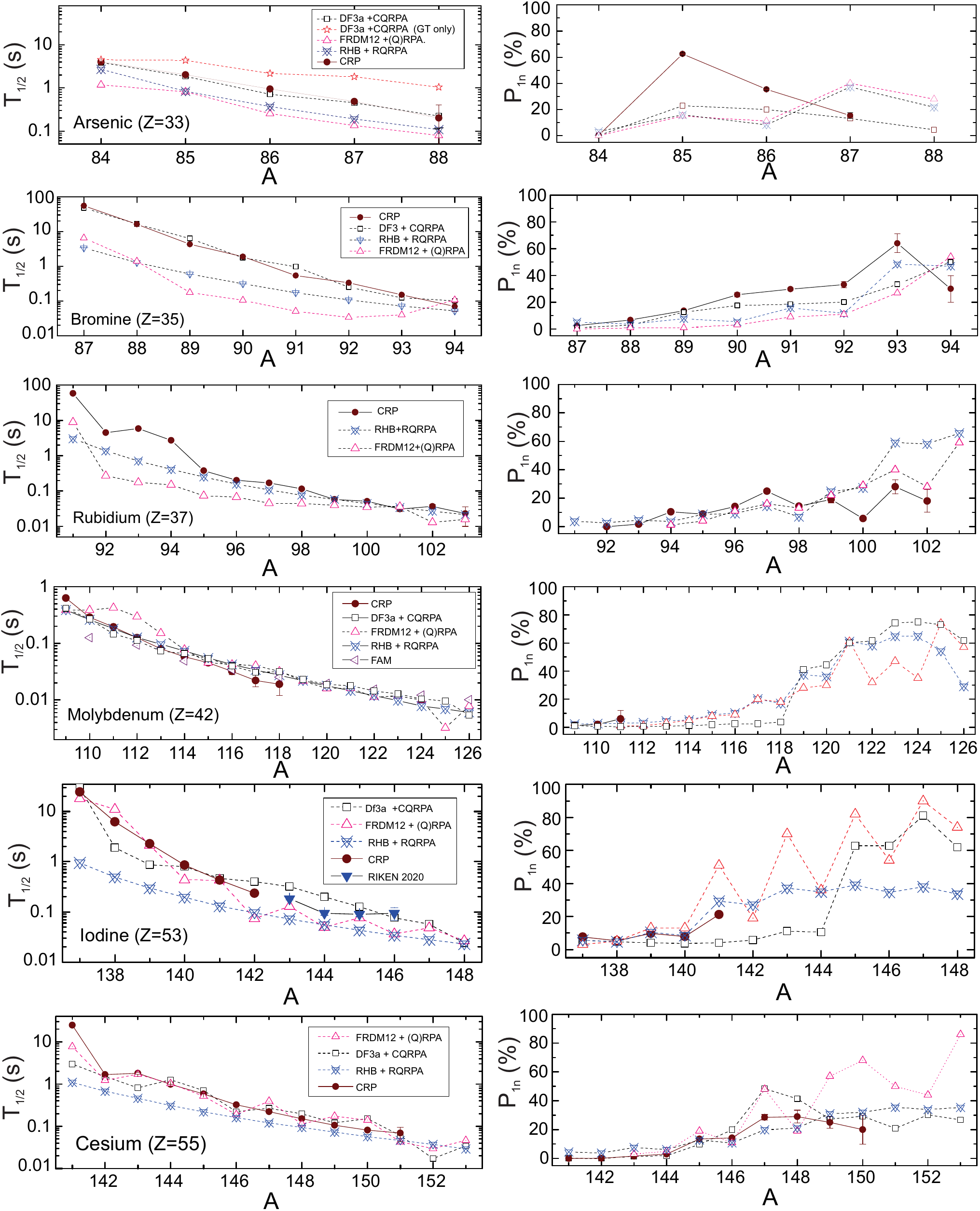}
  \caption{(Left panels) Predictions for half-lives from various theoretical models for As, Br, Rb, Mo, I, and Cs isotopes compared with the evaluated CRP data. For $^{143-146}$I also the latest RIKEN data from Ref.~\cite{Wu2020} is shown. (Right panels) Predictions for $P_{1n}$ values for the same isotopes from various theoretical models compared with the evaluated CRP data. For more information, see text.} 
  \label{Fig:HL-P1n}
\end{centering}
\end{figure*}

\paragraph{Iodine isotopes}
In Fig.~\ref{Fig:HL-P1n}, the half-lives and $P_{1n}$ values for the iodine chain ($Z$ = 53) are shown, including the important \bdn~emitters $^{137-139}$I which are the main contributors to the total (integral) DN yields for thermal neutron-induced fission of $^{235}$U. The spherical DF3a+CQRPA and the deformed FRDM12+(Q)RPA describe the half-lives reasonably for these moderately deformed nuclei with $\beta_2 \approx 0.10-0.15$ \cite{Stone2014}. 

As the FRDM+BCS+RPA framework adopted in Ref.~\cite{Moller2019} does not retain the SO(8)-symmetry of the QRPA, it infers a notable spurious odd-even staggering in half-lives and $P_n$ values. 
The RHB+RQRPA produces a very smooth $N$ dependence and systematically underestimates the half-lives but interestingly agrees best with the experimental $P_{1n}$ values.

In the DF3a+CQRPA calculation two effects are important: a ground-state spin inversion and contribution of the FF decays. The first one leads to a stabilization of the half-lives at $N$ = 88 -- 90 ($^{141-143}$I) and the second induces a reduction in the $P_{1n}$ values. Notice that a stabilization of the half-lives of $^{144-146}$I was observed in the recently published RIKEN data \cite{Wu2020} (shown in Fig.~\ref{Fig:HL-P1n} and in Table~\ref{table:T_Br-I}).

\paragraph{Cesium isotopes}
For the isotopic chain of the Cs \bdn~emitters ($Z$ = 55, Fig.~\ref{Fig:HL-P1n}), the deformation is $\beta_2 \approx 0.15-0.2$ 
~\cite{Stone2014}. Both, deformed FRDM12+(Q)RPA and spherical self-consistent models, underestimate the $T_{1/2}$ at the beginning of the chain ($A$ = 141) where nuclei have a lower deformation. 

The evaluated half-lives at $A \approx 150$ indicate a kind of plateau like the one observed in the iodine isotopes and in self-consistent calculations. The evaluated $P_{1n}$ values at $A>146$ also show a stabilization effect.
This could be related to the ground-state spin inversion at $N>90$ \cite{Moon2017}.  

As in the iodine case discussed before, the RHB+RQRPA seems to reproduce the experimental $P_{1n}$ values the best yet underestimates the half-lives.
On the other hand, the DF3a+CQRPA and FRDM12+(Q)RPA show a behavior that reflects the competition of GT and FF decays in the region up to $N=92$.
For $N>92$ ($A>147$), the DF3a+CQRPA and RHB+RQRPA describe the experimental $P_{1n}$ values due to increasing contribution from the FF decays. Thus, for the I and Cs isotopic chains, deformation is still not 
a decisive factor for the mass dependence of the $P_{1n}$ values. It can be seen that for these isotopes the spherical approaches, namely DF3a+CQRPA and RHB+RQRPA, provide a level of accuracy comparable to the highly parametrized deformed FRDM12+(Q)RPA framework.

In Table~\ref{table:T_Br-I}, the DF3a+CQRPA half-lives and $P_{1n}$ values for iodine isotopes are compared with the spherical option of the FRDM+RPA model (``QRPA-2") ~\cite{Pfeiffer2002} and the recent deformed option of the FRDM12+(Q)RPA ~\cite{Moller2019}.  As can be seen, the FRDM+RPA, which was used in the evaluation of Wilson and England~\cite{Wilson02} for estimating the total (integral) DN yields, often overestimates the experimental $P_{1n}$ values.  As it does not reproduce the experimental half-lives, the corresponding isotopes are not correctly assigned to the I-VI group of the 6-group model that is used to describe the total (integral) DN activity (see Sect.~\ref{Sec:Macro-Recommended} for more on group constants). 

\begin{table*}[!htb]
\caption{Evaluated and theoretical $T_{1/2}$ and $P_{1n}$ values for bromine ($Z$= 35) and iodine ($Z$=55) isotopes. ``FRDM+RPA" is the model labelled ``QRPA-2" in Ref.~\cite{Pfeiffer2002}. $^*$Latest data from Ref.~\cite{Wu2020}. }\label{table:T_Br-I}
\begin{tabular}{c|c|c|c|c|c|c|c|c|c|c}
\hline\hline
    & \multicolumn{2}{|c}{CRP evaluation } &\multicolumn{2}{|c}{DF3a (this work)} & \multicolumn{2}{|c}{FRDM + RPA \cite{Pfeiffer2002} } & \multicolumn{2}{|c}{FRDM12 + (Q)RPA + HF \cite{Moller2019} } & \multicolumn{2}{|c}{RHB + RQRPA  \cite{Marketin2016} } \\ 
    & $T_{1/2}$~(s) & $P_{1n}~(\%)$ & $T_{1/2}$~(s) & $P_{1n}~(\%)$ & $T_{1/2}$~(s) & $P_{1n}~(\%)$ & $T_{1/2}$~(s) & $P_{1n}~(\%)$ & $T_{1/2}$~(s) & $P_{1n}~(\%)$ \\
\hline
$^{87}$Br	& 55.64(15) & 2.53(10) & 47.730 & 0.99 & 37.25&	1.129 &	6.543 &	0.0 & 3.407 & 5.3 \\
$^{88}$Br	& 16.29(8) &	6.72(27) &	16.550 &	2.9 &	104.99 &	28.599 &	1.400 &	1.0 & 1.292 & 3.7 \\
$^{89}$Br	& 4.338(22) &	13.7(6)&	6.290 &	12.6&	10.76&	41.746 &	0.177 &  1.0 & 0.597 & 7.8 \\
$^{90}$Br   &	1.911(10) &	25.6(15)&	1.770 &	17.5&	17.33&	99.800 &	0.106 &	3.0 & 0.313 & 5.4 \\
$^{91}$Br   &	0.544(10)	& 29.8(8)&	0.980 &	18.5&	0.762&	73.365 &	0.051 &	9.0 & 0.175 & 15.8 \\
$^{92}$Br   &	0.334(14) &	33.1(25)&	0.250 &	20.1&	0.054&	8.457 &	0.034 &	11.0 & 0.108 & 11.7 \\
$^{93}$Br   &	0.152(8) &	64(7)&	0.120 &	33.4 &	0.221&	100 &	0.040 &	27.0 & 0.072 & 48.5 \\
$^{94}$Br   &	0.070(20) &	30(10)&	0.100 &	50.1 & 	0.034&	14.221 &	0.108 & 53.0 & 0.052 & 46.9 \\
\hline
$^{137}$I   & 24.59(10) & 7.63(14)  & 29.28 & 11.8 & 3365.424 &  98.980 & 18.099  &  3.0 & 0.948 & 5.5 \\       
$^{138}$I    & 6.251(31) & 5.30(21)  & 5.1 & 1.5 & 9020.949 & 69.824 & 11.038 &  5.0 & 0.496 & 4.6 \\
$^{139}$I    & 2.280(11) & 9.74(33)  & 0.87 &  3.95 & 58.145 & 99.988 & 2.133 &  13.0 & 0.299 & 10.3 \\
$^{140}$I    &  0.86(4) &  7.88(43)  & 0.79 & 3.69  & 	17.216	 & 99.728 & 0.433 &  13.0 & 0.194 & 8.5 \\
$^{141}$I    &  0.43(2) & 	21.2(30)   & 0.46 & 19.5 & 2.347 & 100 & 0.428 & 51.0 & 0.131 & 29.3  \\
$^{142}$I    &  0.235(11)$^*$ & 	-    & 0.401 & 5.61 & 1.400 & 99.952 & 0.074 & 19.0 & 0.096 & 27.0 \\
$^{143}$I    &  0.182(8)$^*$ & 	-	 & 0.319	 & 11.1	 & 0.150	 & 77.120	 & 0.128	 & 70.0 & 0.072 & 37.2 \\
$^{144}$I    & 0.094(8)$^*$ & 	-	 & 0.198	 & 10.4	 & 0.058	 & 29.379	 & 0.049	 & 36.0 & 0.055 & 34.9 \\
$^{145}$I    & 0.0897(93)$^*$ & 	-	 & 0.126	 & 61.9	 &  -	 &  -	 & 0.076	 & 82.0 & 0.043 & 39.1 \\
$^{146}$I    & 0.094(26) $^*$ & 	-	     & 0.0762	 & 62.7	 &  -	 &  -	 & 0.037	 & 54.0 & 0.034 & 34.6 \\
\hline \hline
\end{tabular}
\end{table*}
It can be concluded that for fission products a complex interplay of many factors influences the $\beta$-decay properties. The local fragmentation of the $\beta$ strengths due to deformation is not necessarily the main driving force.
What is important is an appropriate description of the ground-state properties: total energy release and single-particle energies. The competition of the GT and high-energy FF transitions, effects of complex np-nh configurations and possible ground-state spin inversion also have a strong impact on the results. In transitional nuclei, a co-existence of spherical and deformed shapes may also play an important role. 

Summarizing,  the self-consistent models give a robust description of the $\beta$-decay properties of fission products. The quality of description is comparable to that of the highly-parametrized microscopic-macroscopic FRDM12+(Q)RPA+HF model. As it will be shown in the next section, the self-consistent models are preferable due to their predictive power.

\subsection{Comparison for isotopes relevant to nuclear astrophysics}

In this section, the global models are compared with the CRP ($T_{1/2}, P_n$) data
for very neutron-rich isotopes which are 
at the edge of the current possibilities of ongoing RIB experiments and yet are indispensable for the modeling of the $r$-process nucleosynthesis in nuclear astrophysics. 

We focus on two reference semi-magic isotopic chains: Ni ($Z=28$,  $\pi$1$f_{7/2}$ orbital is filled) near the crossing of the neutron major shell $N=50$ and  Sn ($Z=50$, $\pi$1$g_{9/2}$ orbital is filled) around $N=82$. The  reason for choosing these isotopic chains is that $\beta$-decay properties of isotopes before and after the major neutron shell crossings offer a sensitive test for nuclear structure models. 
The theoretical results are compared with the available CRP data and the half-lives from the latest RIKEN experiment \cite{Wu2020}. 


Throughout the comparisons we show that the mass dependence of total half-lives and $P_{xn}$ is sensitive to a structural indicator $\%FF$ that represents the ratio of the FF (or GT) decays over the total rate:
\begin{equation}
\%FF={\lambda}_{\rm FF}/{\lambda}=(T_{\rm GT}-T)/T_{\rm GT}.    
\end{equation}
 
Here, $\lambda$ and ${\lambda}_{{\rm FF}}$ are the total rate and rate of the FF transitions only, respectively, $T$ and $T_{{\rm GT}}$ are the total half-life and the half-life of the GT transitions only, respectively. Naturally, $\%FF$ reflects selection rules and, in particular, a reduction of the unique FF compared to the non-unique FF. 

\subsubsection{$\beta$-decay properties in the $^{78}$Ni region}

For the reference doubly-magic nucleus $^{78}$Ni ($N$= 50, $Z$= 28)
and the isotopes in the region (both with $Z<28$ and $Z \ge 28$), the high-energy (low $E_{x}$) GT transitions  mostly define the total half-lives at $N<50$, while the FF decays give less-important contribution to the total $\beta$-decay rates. 
Eventually, for nuclei with occupied proton 1$f_{7/2}$ orbital beyond the $N=50$ neutron shell, the high transition-energy FF decays start to compete with the GT decays, although the former continue to make a substantial contribution to the total half-life as their transition energies increase with $N-Z$.

\begin{figure}[htb]
\begin{centering}
\includegraphics[width=\linewidth]{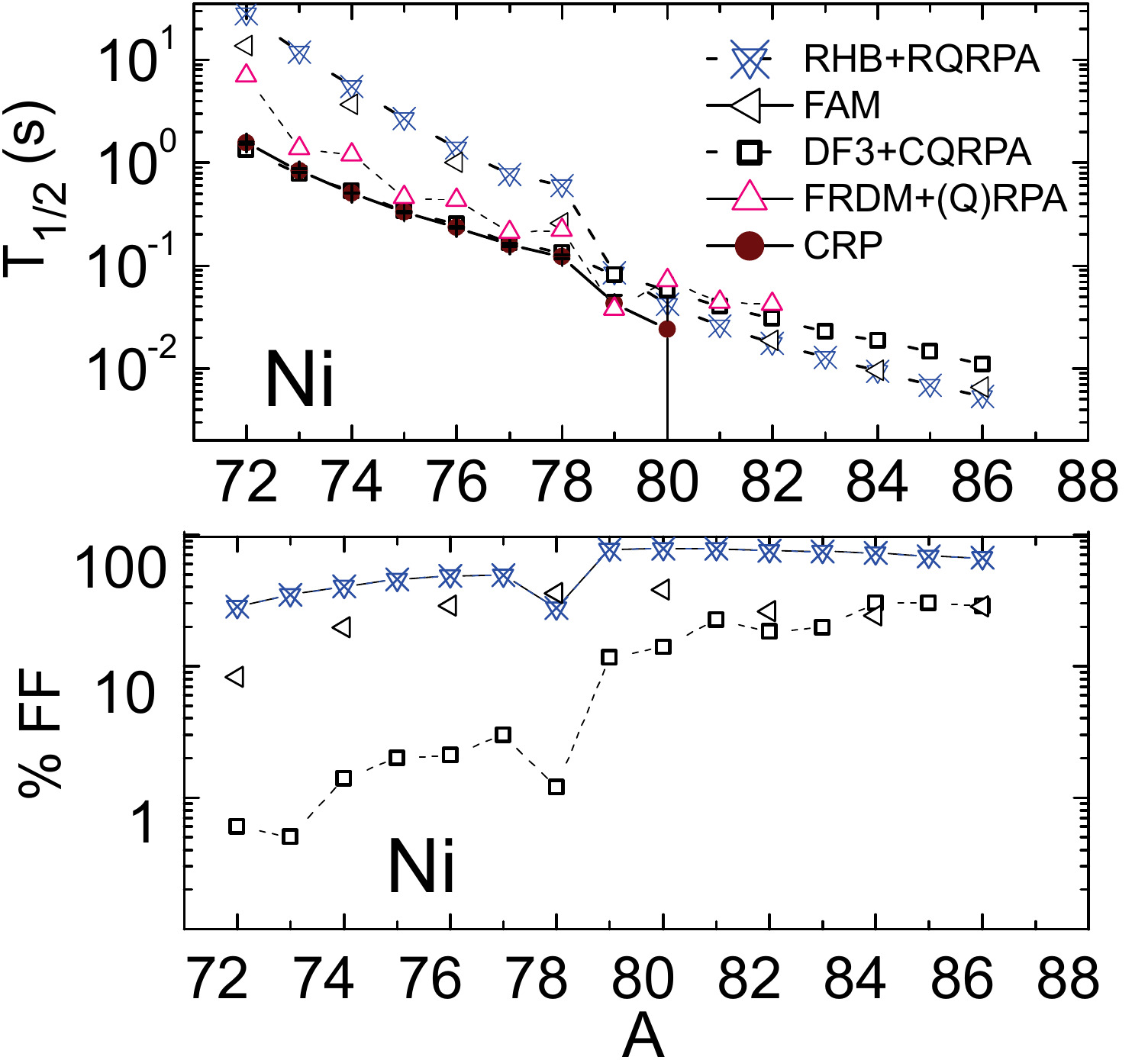}
\caption{a) (Upper panel) Half-lives for the Ni isotopes ($Z$= 28) calculated from DF3a+CQRPA (present work), RHB+RQRPA \cite{Marketin2016}, FAM (even-even only) \cite{Mustonen2014}, and FRDM12+(Q)RPA \cite{Moller2019} compared with the evaluated half-life data \cite{Birch2015}; 
b) (Lower panel)  Contributions of the first-forbidden transitions to the total decay rate ($\%FF$ value).} \label{Fig_Ni}
\end{centering}
\end{figure}

The total half-lives for the Ni chain obtained from DF3a+CQRPA~\cite{Borzov2003,Borzov2017}, RHB+RQRPA~\cite{Marketin2016}, FRDM12+(Q)RPA~\cite{Moller2019}, and
FAM~\cite{Mustonen2014} are compared with evaluated data in Fig.~\ref{Fig_Ni}. The FRDM12+(Q)RPA and DF3a+CQRPA results are close to the experimental data, though an 
odd-even staggering is observed in the FRDM12+(Q)RPA.

For the spherical RHB+RQRPA as well as for the deformed FAM, the deviation from the evaluated half-lives at $N<50$ is quite pronounced. For the doubly-magic nucleus $^{78}$Ni, they predict 258~ms and 605~ms, respectively, compared to the experimental value of $T_{1/2}$=122.2(51)~ms \cite{Xu2014,Birch2015}.

In order to understand the origin of these deviations, it is necessary to analyze the contribution of the FF (or GT) decays to the total rate ($\%FF$ value, Fig.~\ref{Fig_Ni}).
For Ni isotopes  with neutron numbers $N \le 50$, as the $\pi$1$f_{7/2}$ orbital is filled, the GT decays ($\nu$1$f_{5/2}$, $\pi$1$f_{5/2}$),
($\nu$1$p_{1/2,3/2}$, $\pi$1$p_{1/2,3/2}$) overwhelm the hindered FF-unique decay ($\nu$1$g_{9/2}$, $\pi$1$f_{5/2}$). Naturally, the $\%FF$ values obtained from the
DF3a+CQRPA for $A<78$ are low and the drop of the $\%FF$ values at $A=78$ is mostly related to the competition with the GT transitions. For the FAM and RHB+QRPA models, the calculated $\%FF$ values for $A \le 78$ are higher by 30 and 50\%, respectively. However, high $\%FF$ values for $A<78$ are not supported by the selection rules and the available decay schemes.

Thus, an overestimation of the half-lives for $A<78$ by RHB+RQRPA and FAM may stem from the predicted balance between the GT and FF strengths.
For $N>50$, the non-unique decays (2$d_{5/2}$, 2$s_{1/2}{\rightarrow}2p_{1/2,3/2}, 1f_{5/2}$) contribute to the decay process, yet the $\%FF$ values calculated from DF3a+CQRPA reach
only 25\% for $A=86$, which is quite close to the FAM prediction. For $A>78$, all calculations give similar half-lives although the RHB+QRPA gives different $\%FF$ values.

It turns out that the differences in the total $Q_{\beta}$ values, multi-neutron emission thresholds, and quasi-particle levels predicted by the different models are very important. The different phase-(sub)-spaces defined by these quantities are reflected in the $\beta$-strength distributions and are translated into the half-lives and delayed multi-neutron emission probabilities. 

\begin{figure}[htb]
\begin{centering}
\includegraphics [width=\linewidth]{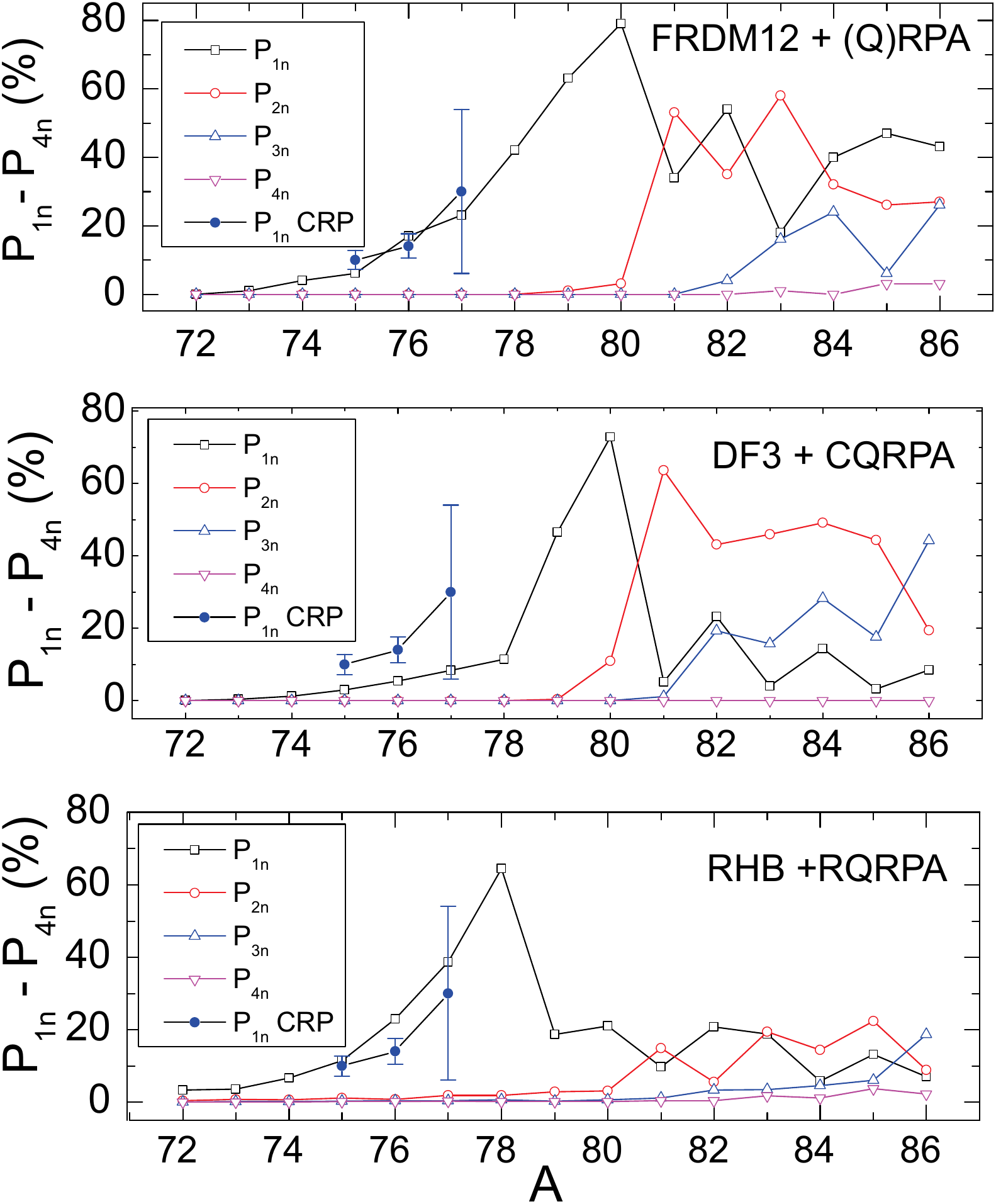} 
\caption{Delayed multi-neutron emission probabilities ($P_{1n}$--$P_{4n}$) for Ni ($Z=28$) isotopes calculated with FRDM12+(Q)RPA+HF \cite{Moller2019}, DF3a+CQRPA~(present work), and RHB+RQRPA~\cite{Marketin2016}. These models are compared to the evaluated $P_{1n}$ data for $^{75-77}$Ni as published in Ref.~\cite{Birch2015}.
 }
\label{Fig_Ni2}
\end{centering}
\end{figure}

The $P_{xn}$ for the Ni chain obtained from the DF3a+CQRPA, RHB+RQRPA, and FRDM12+(Q)RPA+HF models are compared in Fig. \ref{Fig_Ni2}. It can be seen that in the RHB+RQRPA calculations, the peak of the one-neutron emission probability occurs at $^{78}$Ni, two mass units earlier compared to the other two model calculations, and that the two-neutron emission probability mass behavior also differs. 
However, the three experimental $P_{1n}$ data points for $^{75-77}$Ni are very nicely reproduced by the FRDM12+(Q)RPA+HF and the RHB+RQRPA model, whereas the DF3a+CQRPA model seems to underpredict these isotopes.

\begin{figure}[!htb]
\begin{centering}
\includegraphics [width=0.9\linewidth]{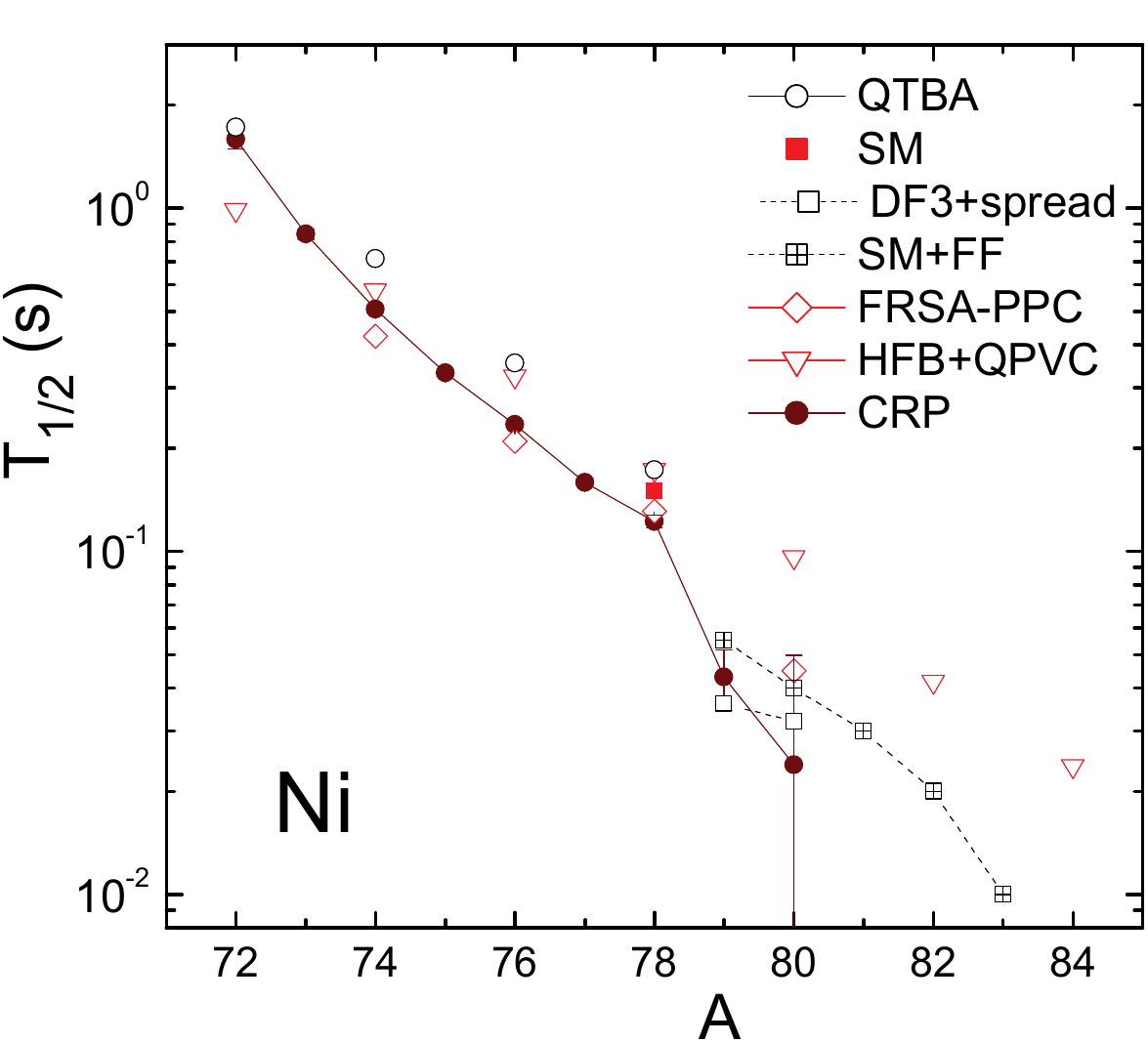}
\caption{Half-lives for Ni isotopes calculated from QTBA \cite{Robin2016}, FRSA+QRPA+PPC \cite{Sushenok2018} and HFB+QRPA+PVC \cite{Niu2018} in comparison with with the evaluated data from Ref.~\cite{Birch2015}.
Also shown are calculations from the SM \cite{Zhi2013}, NuShellX \cite{Alshudifat2016} and DF3a+CQRPA with spreading \cite{Borzov2017}.}
  \label{Fig_th.11}
\end{centering}
\end{figure}

Finally, in Fig.~\ref{Fig_th.11} we compare the half-lives for the Ni chain with the recently developed ``beyond the pnQRPA" models, such as the FRSA+PPC \cite{Severyukhin2014} which takes into account the phonon-phonon coupling, the PVA models with the particle-vibration coupling (PVC) ~\cite{Colo1998, Niu2015, Niu2018}, and the ``quasi-particle time blocking approximation" (QTBA) \cite{Robin2016}. Hybrid shell-model results from ~\cite{Alshudifat2016} are also included in the figure. 

The calculations in Fig.~\ref{Fig_th.11} have been performed within the GT approximation which is more justified for $A<78$. 
In the pnQRPA+PVC developed in the recent paper~\cite{Niu2018}, the experimental half-lives for the Ni isotopes are reasonably well described assuming a rather strong $T=0$ pairing (except for $A=80$).  
The corresponding 1p-1h QRPA calculation in the same framework has given a factor of about 10-15 longer half-lives ~\cite{Niu2018}. Thus the renormalization due to the PVC turns out to be very strong leading to QRPA+PVC half-lives that are only by a factor 1.5 shorter for $^{72}$Ni and by a factor 1.5 longer for $^{78}$Ni.

A similar strong renormalization due to the PVC effect is obtained in the QTBA approach~\cite{Robin2016}. Again, the pure pnQRPA half-lives obtained without assuming phonon-phonon coupling in the QTBA are much higher than the corresponding quantities obtained from the FRSA~\cite{Sushenok2018} which includes phonon-phonon coupling. Thus, the renormalization factor due to the effective interactions in these approaches differs significantly (by about 1--2 orders of magnitude). 
Most probably this difference is due to the fact that the QTBA model uses experimental $Q_{\beta}$ values instead of self-consistently derived ones.

The above discussion is important if one considers the potential of these models: once the contribution of FF decays is included, they could be implemented in global calculations of $\beta$-decay half-lives and $\beta$-delayed neutron branching ratios. It has already been shown that ``hybrid" models, i.e.  DF3a+CQRPA including the complex configurations through the spreading width \cite{Borzov2017} and the shell-model NuShellX with the FF decays added from the DF3a+CQRPA \cite{Alshudifat2016}, can give an overall reasonable description of the experimental half-lives in Ni for $A>78$. 

Only a simultaneous account of the FF decays and np-nh configurations in these models allows one to describe the ``sudden shortening" of the half-lives after crossing the $N=50$ shell that was observed in the RIKEN data \cite{Xu2014}. It is therefore reasonable to expect that the self-consistent `beyond pnQRPA' models have the potential to describe all these effects. In this respect, it is important to include the contribution of the first-forbidden decays in the present (spherical)  `beyond pnQRPA' schemes. 

On the other hand, an effort to include the contribution of the complex configurations into the deformed Finite Amplitude Method (pnFAM) \cite{Mustonen2014} would result in a universal approach that could be applied to the whole nuclear chart.

\subsubsection{$\beta$-decay properties in the $^{132}$Sn region}

For the isotopes beyond the major neutron shell $N=82$ in the $^{132}$Sn region, the concurrence of GT and FF decays has much in common with the $^{78}$Ni region. We show that the contribution of GT and FF transitions to the $\beta$-decay rates differs for the $Z<50$ and $Z \ge 50$ isotopes. The intensive GT decays in the $Z<50$ nuclei mostly correspond to the ($\nu$1$g_{7/2}$, $\pi$1$g_{9/2}$) configuration. The high-energy GT decays contribute strongly to the total half-lives of the nuclei with $Z<50$, $N<82$.  After crossing the $N=82$ neutron shell, the high-energy FF decays which are mainly related to the ($\nu$1$h_{11/2}$, $\pi$1$g_{9/2}$) configuration are active. 
In contrast to what is observed in nuclei in the $Z \ge 28$ region, the high-energy GT transitions for $Z>50$ nuclei are hindered, as the p1$g_{9/2}$ orbital is blocked. At the same time, the high energy FF transitions ($\nu$2$f_{7/2}$, $\pi$1$g_{7/2}$), ($\nu$1$f_{7/2}$, $\pi$2$d_{5/2}$) come into effect and mostly define the total half-life.

\begin{figure}[htb]
\includegraphics [width=\linewidth]{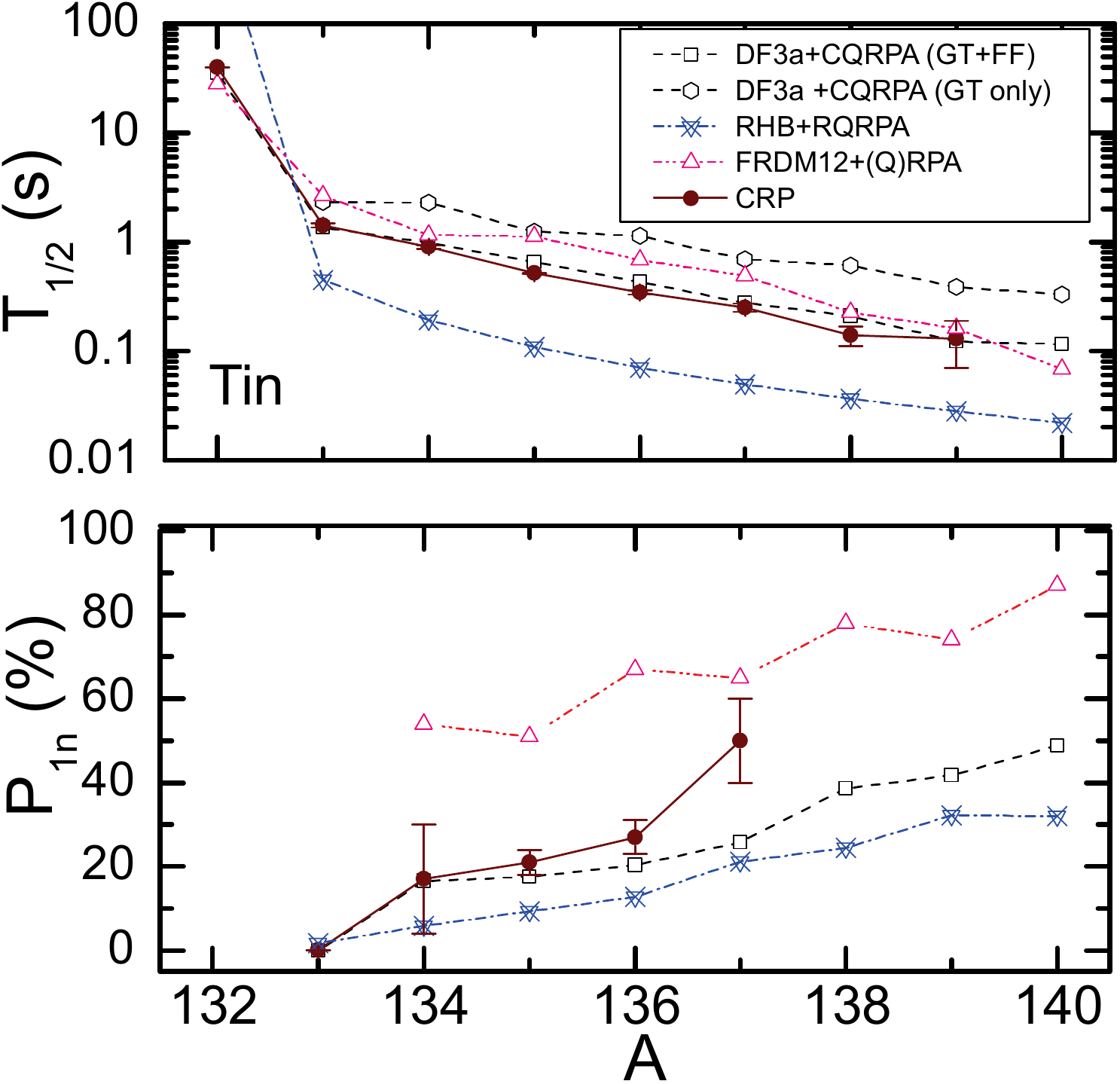}
\caption{Half-lives (upper panel) and $P_{1n}$ values (lower panel) for Sn isotopes ($Z=50$). Calculated values from DF3a+CQRPA \cite{Borzov2017}, RHB+RQRPA \cite{Marketin2016}, FAM \cite{Mustonen2014}, and FRDM12+(Q)RPA \cite{Moller2019} are compared with the evaluated ($T_{1/2}, P_{1n}$) data. The half-life for $^{132}$Sn (no $\beta$n emitter, 39.7(8)~s) was taken from ENSDF.}
\label{Fig_th.12}
\end{figure}

\begin{figure}[htb]
\includegraphics [width=0.8\linewidth]{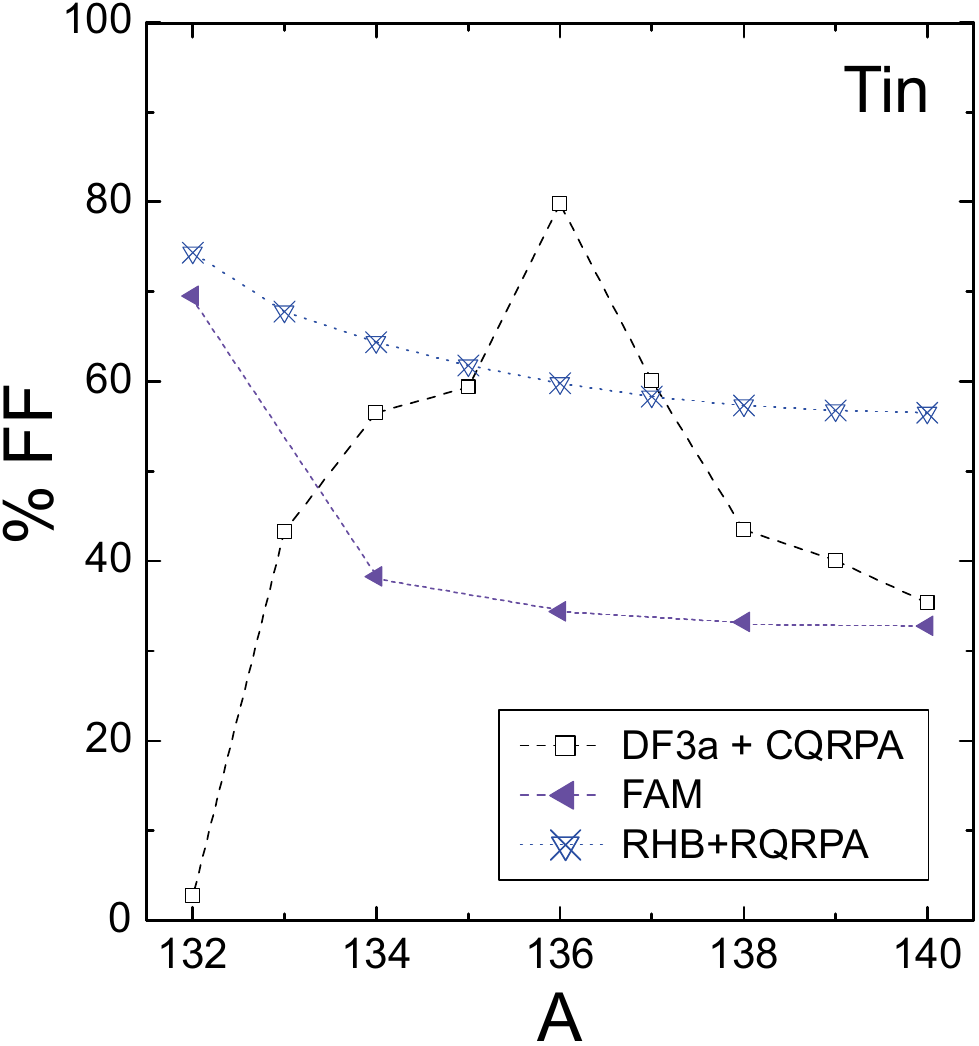}
\caption{Contributions of the FF transitions to the total decay rate ($\%FF$ values) for Sn isotopes obtained from DF3a+CQRPA (present work), RHB+RQRPA \cite{Marketin2016}, and FAM \cite{Mustonen2014}.}
 \label{Fig_th.13}
\end{figure}

Fig.~\ref{Fig_th.12} shows the comparison of the half-lives obtained from DF3a+CQRPA, RHB+RQRPA, FRDM12+(Q)RPA, and FAM with the evaluated CRP data. For completeness, we include $^{132}$Sn in the plot even though it is not a \bne. Within the GT-only approximation, the DF3a half-lives are up to a factor 5 longer than the data. Taking into account the FF decays sets the balance between the GT and FF strengths as shown in Fig.~\ref{Fig_th.13} and allows for an 
excellent description of the experimental half-lives for the Sn isotopic chain.

A rather high percentage of FF decays of 60--75\% calculated from the RHB+RQRPA may be the reason for the strong underestimation of the half-lives of $^{133-140}$Sn by a factor of 4--5. 
On the other hand the FRDM12+(Q)RPA half-lives overestimate the experimental data by a factor of $\approx 2$ while the DF3a+CQRPA (GT+FF) calculations
are quite close to the experimental values.



Note that for the reference doubly-magic nucleus $^{132}$Sn the very low $\%FF$ value predicted by DF3a+CQRPA calculations (Fig.~\ref{Fig_th.13}) is consistent with the decay scheme observing 99\% intensity for the decay to the GT state at 1.325 MeV with a $\log(ft)=4.2$. 
Accordingly, the DF3a calculates a half-life of 35.7~s, compared to
the evaluated
value of 39.7(5)~s from ENSDF. In contrast, the RHB+ RQRPA predicts 610~s and the FAM a value of 820~s.

The corresponding $P_{1n}$ values  (Fig.~\ref{Fig_th.12}) obtained with RHB+RQRPA and DF3a+CQRPA agree in general with the mass dependence of the $\%FF$ values predicted by these models, however they underestimate the CRP data for $A>133$. 
The FRDM12+(Q)RPA+HF \cite{Moller2019} overestimates the $P_{1n}$ values considerably which is consistent with
a gradually increasing share of the FF decays at $N>82$ due to the increasing contribution of the $\Delta{J}=0$ g.s.-to-g.s. transitions at $N=83-86$.
The stabilization of the $P_{1n}$ values at $N>86$ predicted by RHB+RQRPA and DF3a+CQRPA reflects a decrease of the $\%FF$ values for $N=87-90$ ($^{137-140}$Sn) due to the competition from the open high-transition energy GT decays.

As was already mentioned,  the $\beta$-decay characteristics of the semi-magic Ni and Sn isotopic chains are highly sensitive to the details of the nuclear structure aspects of the models, in particular the energy-density functional used. As we move further away from the closed shells, these characteristics become less sensitive to the details of the models. For example in the Pd and Cd isotopes that have pairing in both neutron and proton sectors there is good agreement between the self-consistent DF3a+CQRPA and RHB+RQRPA calculations \cite{Borzov2017, Borzov2006, Marketin2016}.

To summarize, from the comparison of theory and evaluated data for a fairly broad range of nuclei
shown in this section, the recommended theoretical approach is the quasiparticle random phase approximation (pnQRPA) based on the energy density functional (EDF) approach. Recently developed self-consistent models have been successfully used in this CRP to calculate the half-lives and \bdn~rates of fission products and nuclei near the major closed shells that are important for $r$-process modeling.  
Compared to the old semi-empirical models (``Gross theory" versions) and the semi-microscopic FRDM-based (Q)RPA framework for the GT decays that is augmented by a statistical model for the FF decays, the self-consistent $\beta$-decay models have the obvious advantage of being well-founded on first principles. By definition, the parameters of the self-consistent global models should be kept the same when extrapolating to different mass regions across the nuclear chart. 

Global models based on the energy density functional approach are not only used in $\beta$-decay studies but also for a consistent description of a range of nuclear properties in a broad mass region, such as binding energies, radii, magnetic moments, fission barriers as well as nuclear responses to different probes. Consequently, there is a wide variety of data available to constrain the EDF parameters and one can avoid using the ``output" half-lives and $P_n$ values for such a purpose. Self-consistent global models are thus a reliable and universal instrument for the description of the middle-heavy, heavy and superheavy nuclei which are in the scope of the present paper.

Finally, it is worth mentioning the rapidly developing ``beyond the pnQRPA" models such as FRSA+phonon-phonon coupling~\cite{Severyukhin2014,Severyukhin2017},
the PVA models with particle-vibration coupling \cite{Colo1998} and the ``quasi-particle time blocking approximation" (QTBA) \cite{Robin2016}.
These approaches can potentially describe the $\beta$-decay strength function over a wide range of masses across the nuclear chart with a quality comparable to the multi-configurational shell-model.  
For example, the ``sudden acceleration" of the $\beta$-decay after crossing the $N=50$ shell, that was recently discovered in the Ni isotopic chain at RIKEN \cite{Xu2014}, has been described by assuming np-nh configurations within the DF3a+CQRPA approach using quasi-particle spreading. This confirms that such a phenomenon can only be explained by considering FF decays and complex configurations simultaneously.

\subsection{Global comparisons of theoretical results}

Global comparisons between the CRP evaluated data and the self-consistent models DF3a+CQRPA, RHB+QRPA, as well as the microscopic-macroscopic model FRDM12+(Q)RPA(+HF) are presented in this section.  Large-scale calculations of the $\beta$-decay energy releases ($Q_{\beta}$, $Q_{\beta n}$), half-lives ($T_{1/2}$) and delayed neutron (DN) emission probabilities ($P_{n}$) for hundreds or even thousands of spherical, near-spherical and  deformed  nuclei have been performed so far within several models. Nowadays these models provide a reliable data input for the modeling of abundances in the astrophysical $r$ process. They have also been used as predictions for the ongoing large-scale experimental campaigns at RIB facilities worldwide (see, e.g. Refs.~ \cite{Xu2014,Caballero2016,Lorusso2015,Wu2020}).

\begin{figure*}[!htb]
\centering
\includegraphics[width=0.8\linewidth]{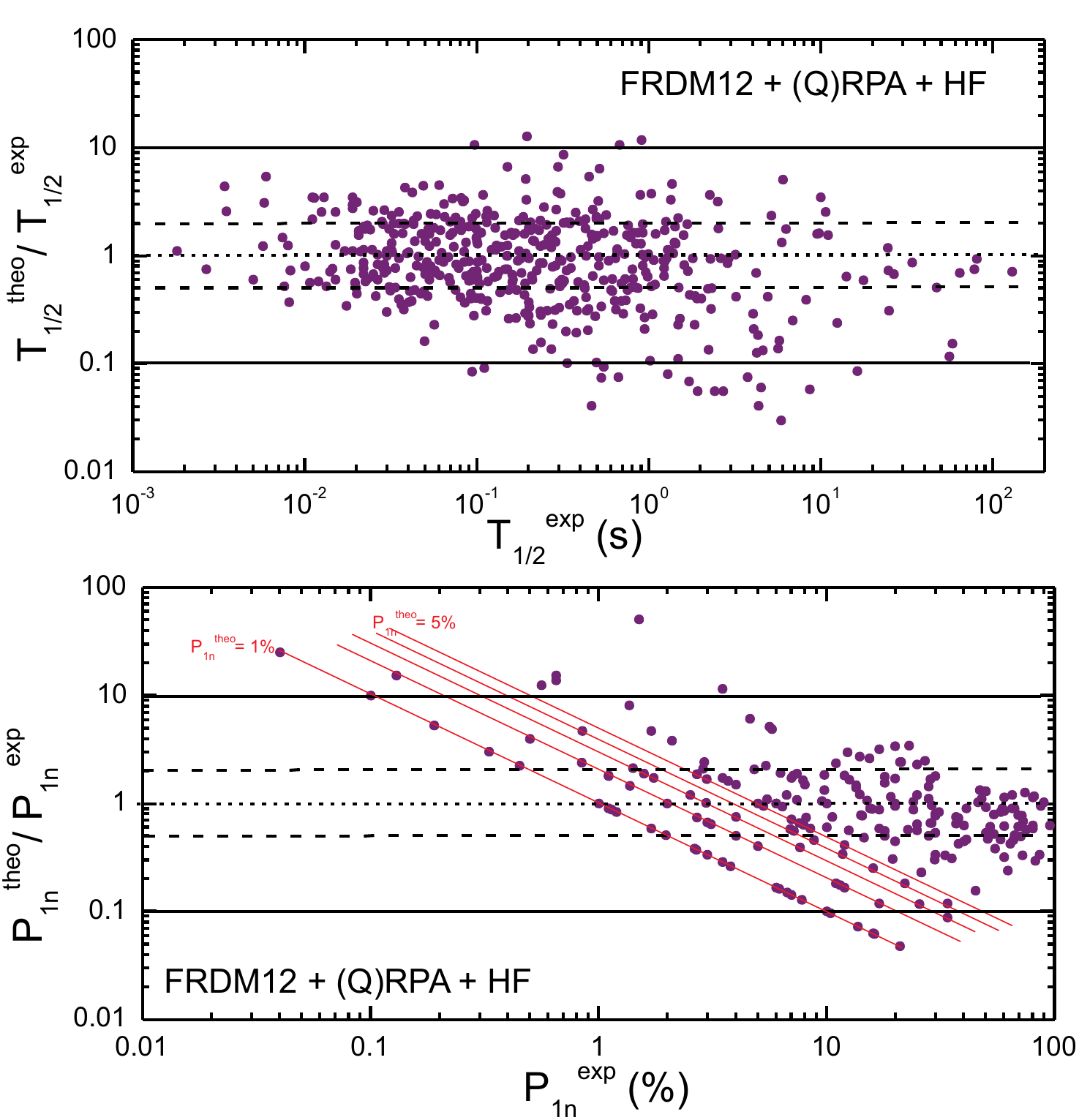}
\caption{Ratios of theoretical and experimental results for the half-lives (upper panel) and  $P_{1n}$ values (lower panel) calculated using the FRDM12+(Q)RPA+HF formalism~\cite{Moller2019}, plotted versus the corresponding experimental values. Diagonally aligned nuclei in the lower plot are nuclei that have the same theoretical $P_{1n}$ value. Lines for $P_{1n}$= 1, 2, 3, 4, and 5\% are shown to guide the eye.}
\label{fig:th_global_moeller}
\end{figure*}

In Fig.~\ref{fig:th_global_moeller}, the ratio of half-lives and $P_{1n}$ values from FRDM12+(Q)RPA \cite{Moller2019} and the evaluated data are plotted with respect to the evaluated values.
It is obvious that the half-lives and neutron-emission probabilities of almost all $\beta$n-emitter considered here are reproduced within one order of magnitude. 

In the case of half-lives, the majority is reproduced within a factor of two or better, denoted by the black dashed lines. It is also noticeable that the scatter around the ratio of 1 decreases with decreasing experimental half-life, which is a common pattern. Shorter half-lives are characterized by larger $Q$-values, and a larger portion of the decay strength is included within the $Q_{\beta}$ window. As a result the half-lives of shorter-lived nuclei are less sensitive to the deviations of the model from the true strength function. 

In the case of the comparison with experimental $P_{1n}$ values in the bottom part of Fig.~\ref{fig:th_global_moeller} several nuclei can be found aligned on diagonal lines. 
This surprising feature is due to theoretical $P_{1n}$ values that appear only as integer values in the FRDM12+(Q)RPA+HF model. Lines for $P_{1n}$= 1,2,3,4, and 5\% are shown in red to guide the eye. For cases where the experimental values are very small, e.g. lower than 1\%, any deviation of the model predictions from the true energies and strengths of the transitions will, as a consequence, result in relatively large deviations of the total emission probabilities from the observed values. 

\begin{figure*}[!htb]
\centering
\includegraphics[width=0.8\linewidth]{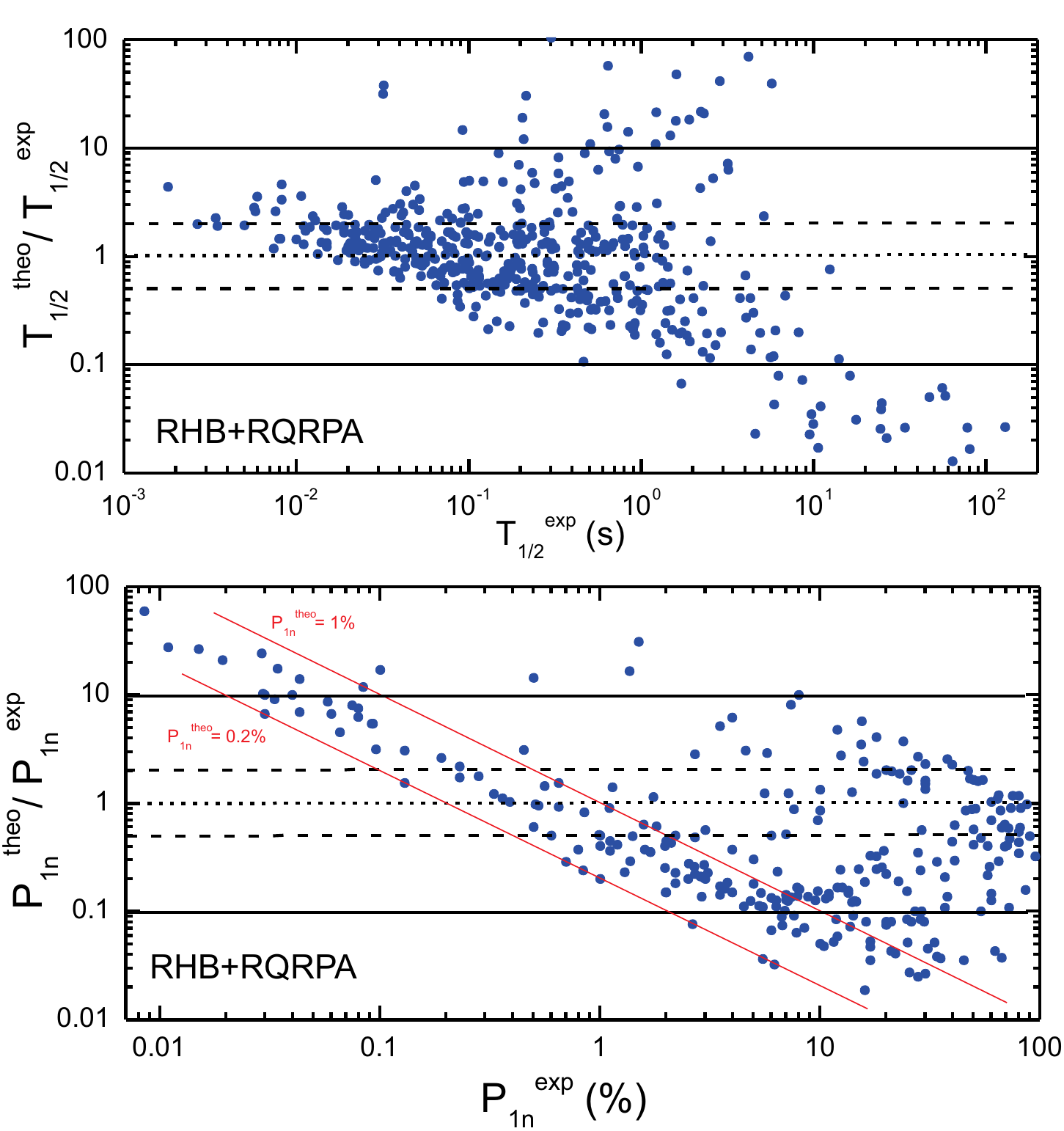}
\caption{Ratios of theoretical and experimental results for the half-lives (upper panel) and  $P_{1n}$ values (lower panel) calculated using the relativistic RHB+RQRPA formalism~\cite{Marketin2016}, plotted versus the corresponding experimental values. Diagonally aligned nuclei in the lower plot are nuclei that have the same theoretical $P_{1n}$ value. Lines for $P_{1n}$= 0.2 and 1\% are shown to guide the eye.}
\label{fig:th_global_rqrpa}
\end{figure*}

Figure~\ref{fig:th_global_rqrpa} shows the same as Fig.~\ref{fig:th_global_moeller}, but for theoretical results obtained using the RHB+RQRPA model \cite{Marketin2016} which was extended for nucleosynthesis calculations and for use in comparisons in this CRP. The general behaviour of the model is quite similar to the FRDM12+(Q)RPA+HF shown above, but with some notable differences. 
In the case of half-lives, the relativistic approach manages to provide, on average, a somewhat better description for half-lives $<$200~ms as indicated by the smaller scatter of ratios around the value of 1. However, for longer half-lives $>$10~s the model seems to systematically underestimate the experimental half-lives. In the case of neutron emission probabilities, the same diagonal trend as for the FRDM12+(Q)RPA+HF model in Fig.~\ref{fig:th_global_moeller} can be seen, due to the fact that many nuclei have similar theoretical $P_{1n}$ values around 1--2\%.





\begin{figure*}[!htb]
\centering
\includegraphics[width=0.8\linewidth]{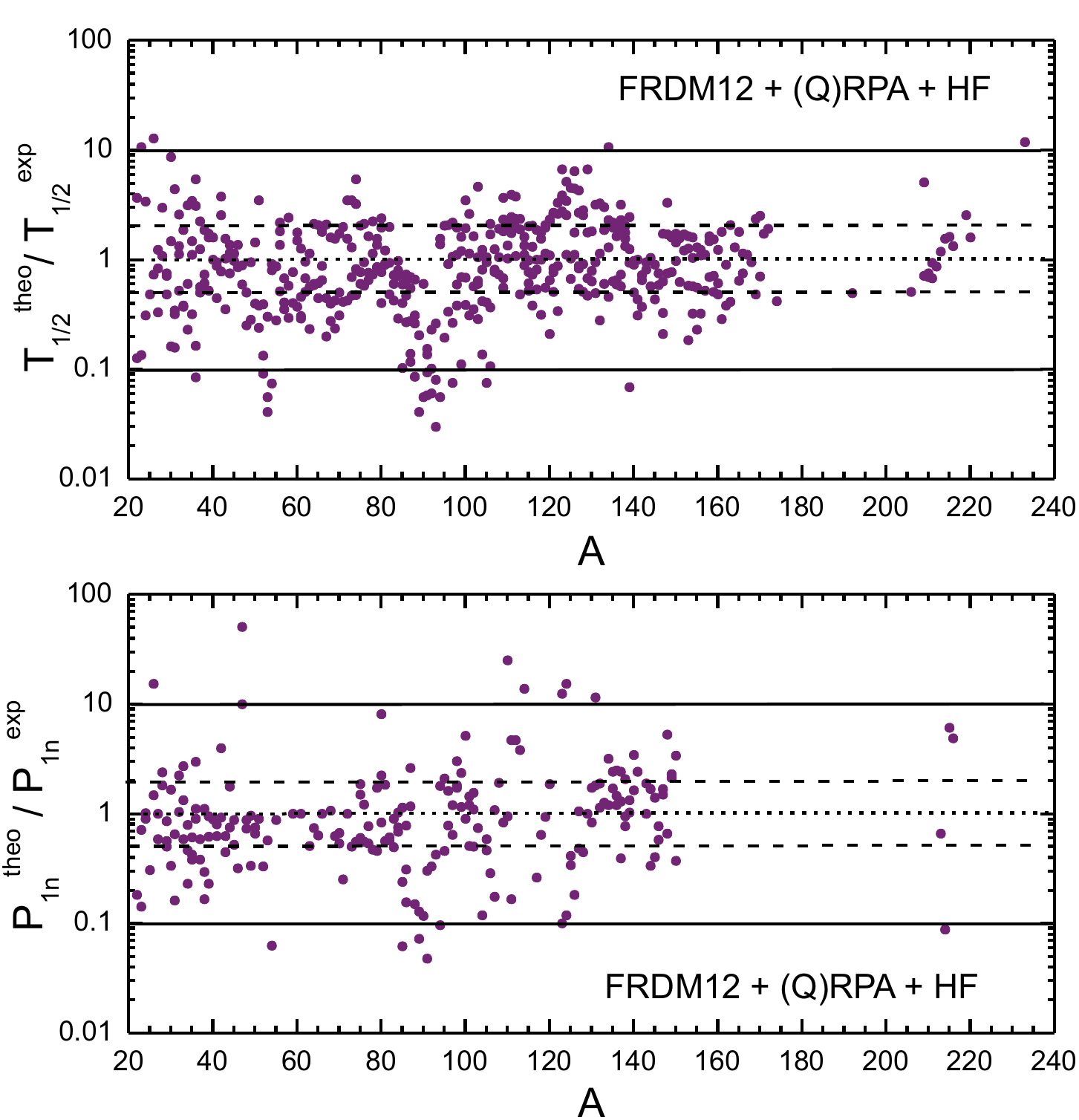}
\caption{Ratios of theoretical and experimental results for the half-lives (upper panel) and total $P_{1n}$ values (lower panel) calculated using the FRDM12 +(Q)RPA +HF formalism~\cite{Moller2019}, plotted versus the atomic mass number.}
\label{fig:th_global_moeller_mass}
\end{figure*}

\begin{figure*}[!htb]
\centering
\includegraphics[width=0.8\linewidth]{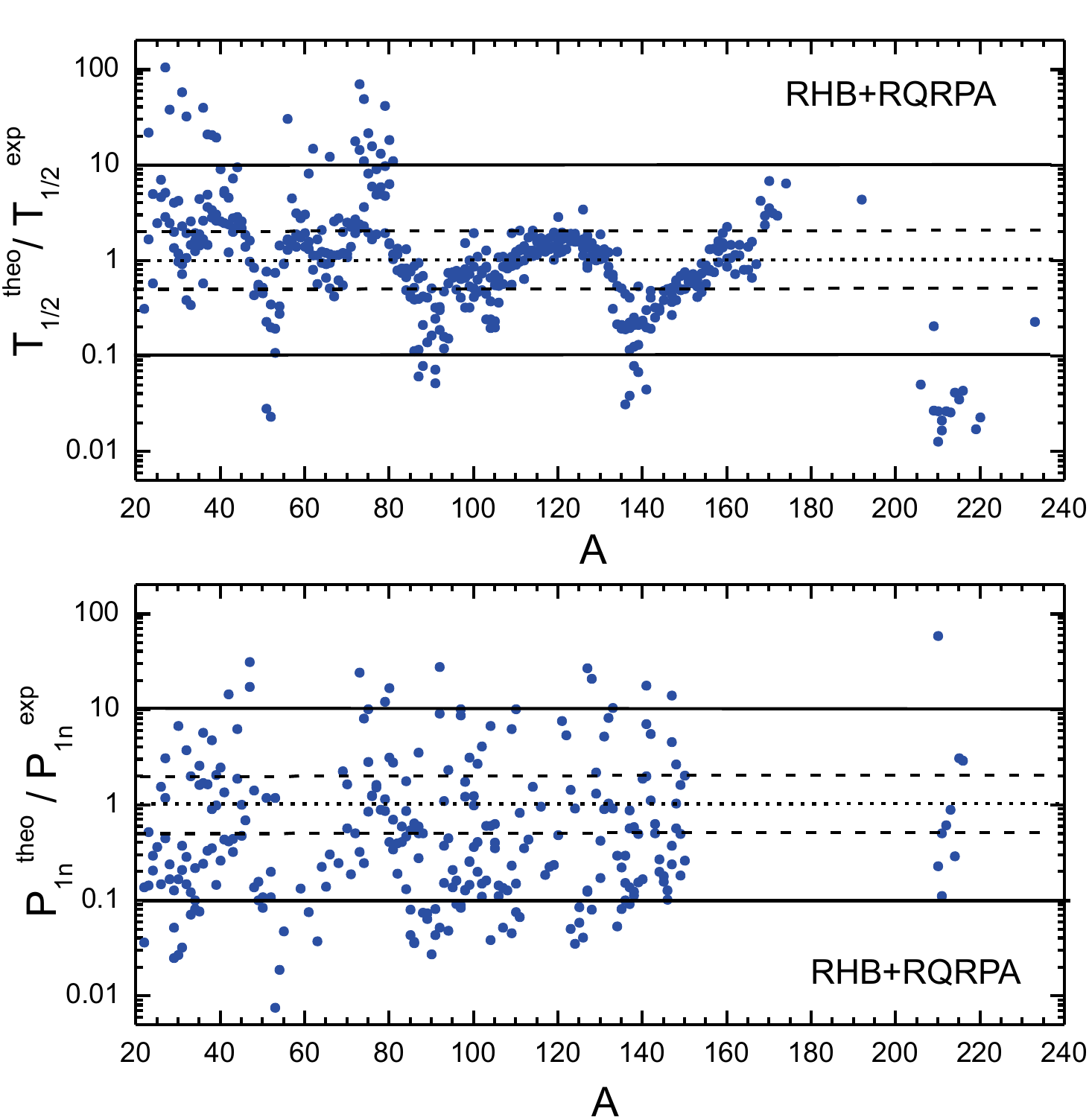}
\caption{Ratios of theoretical and experimental results for the half-lives (upper panel) and  $P_{1n}$ values (lower panel) calculated using the relativistic RHB+RQRPA formalism~\cite{Marketin2016}, plotted versus the atomic mass number.}
\label{fig:th_global_rqrpa_mass}
\end{figure*}

Figures~\ref{fig:th_global_moeller_mass} and \ref{fig:th_global_rqrpa_mass} present the same results as in the previous two figures, but now the ratios are plotted versus the atomic mass number $A$, in order to explore the impact of shells on the results. The FRDM12+(Q)RPA+HF formalism provides a very robust description of the data, without showing any visible strong dependence of the model predictions on the atomic mass number (except for a notable dip in the deformed $A\approx 90-95$ region). This is true both for the half-lives and for the one-neutron emission probabilities. 

However, the results obtained with the relativistic RHB+RQRPA model (Fig.~\ref{fig:th_global_rqrpa_mass}) show significant effects of the nuclear shell structure on the quality of the results in the description of the $\beta$-decay half-lives. These results display a clear arch between two mass regions, from $A \approx 90$ to $A \approx 140$, which is consistent with the fact that the model does not take nuclear deformations into account. Therefore, the model produces deviations in the regions of the nuclear chart between closed shells where nuclei have non-negligible ground-state deformations. 
However, the $\beta$-decay half-lives are well reproduced within a factor of 2 (although longer than measured) for the mass region $A=100-130$.

\begin{figure*}[!htb]
\centering
\includegraphics[width=0.9\linewidth]{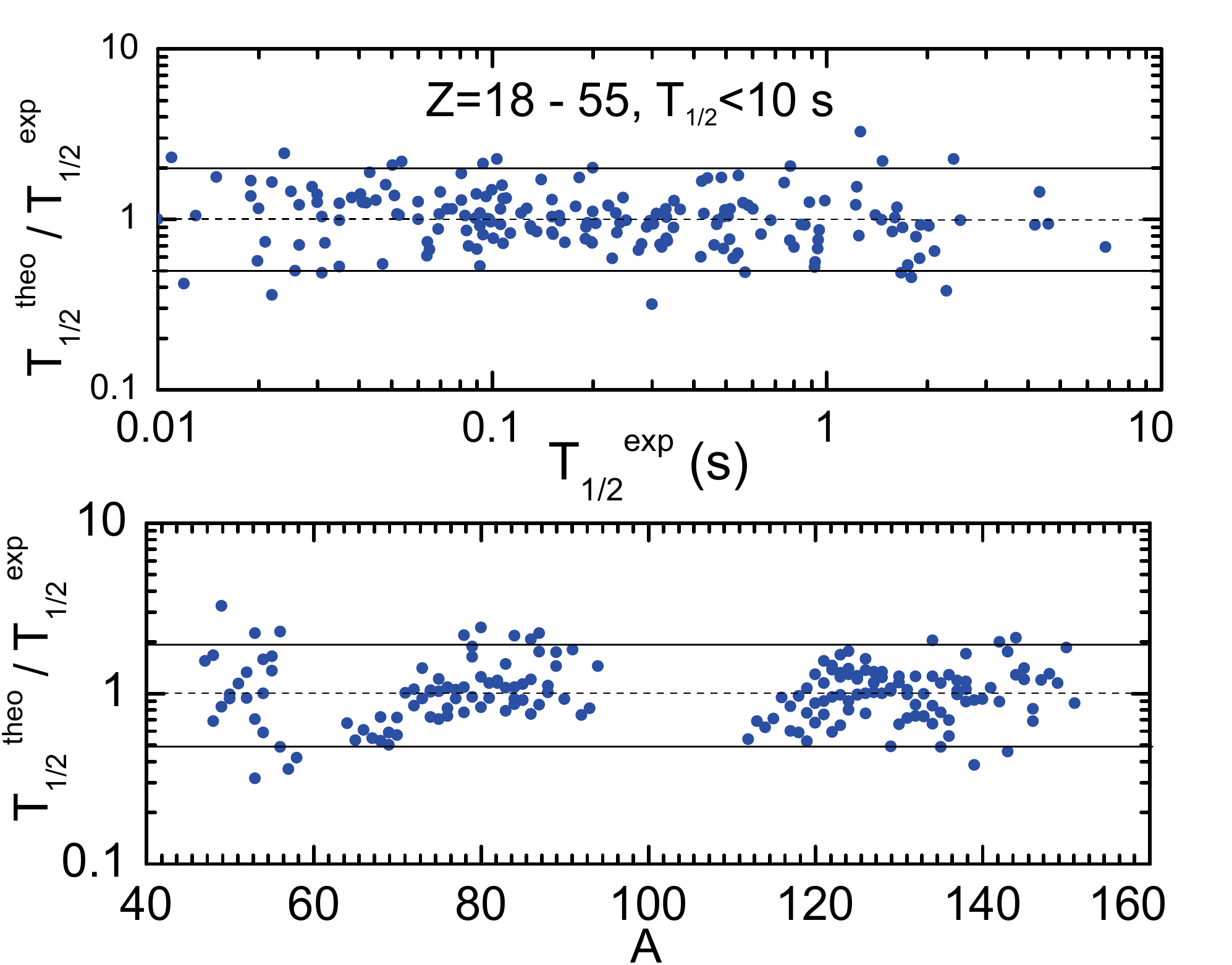}
\caption{Ratio of theoretical and experimental values of half-lives of (quasi-) spherical nuclei with $Z = 18 - 21,\, 25 - 35$, and $44 - 55$,  calculated using the DF3a+CQRPA formalism and plotted versus the evaluated experimental half-lives in the top panel and versus mass $A$ in the bottom panel.}
\label{fig:th_global_df3_thalf}
\end{figure*}

In the upper panel of  Fig.~\ref{fig:th_global_df3_thalf}, the half-lives calculated within
the spherical DF3a+CQRPA approach are plotted with respect to the evaluated values. The figure includes (near-) spherical nuclei with $Z=25-35$ and $Z=44-55$
regions with $T_{1/2}$
ranging from 10 ms to 10 s. This constraint is a consequence of the strength function formalism which assumes that
QRPA results are better for short-lived nuclei with relatively large
Q$_{\beta}$ energy window. Also, the limitation of the comparison to spherical and near-spherical nuclei makes it easier to understand the origin of the deviations from the experimental half-lives. Fig.~\ref{fig:th_global_df3_thalf} shows that most of the half-lives are in agreement within a factor of two with the evaluated data. 

In the lower panel of Fig.~\ref{fig:th_global_df3_thalf}, 
the ratio of the theoretical and experimental half-lives are plotted as a function of mass number $A$. 
The results indicate that the data is in general reproduced within a factor of 2 over the whole mass region, with no obvious discrepancies at shell closures or for deformed nuclei around $A\approx$85 and 120.


In a number of  cases, however, such as in K and Sb isotopes, the deviation has been shown to be related to ground-state spin inversion rather than to deformation \cite{Borzov2017, Borzov2018}.

\begin{figure*}[!htb]
\centering
\includegraphics[width=0.9\linewidth]{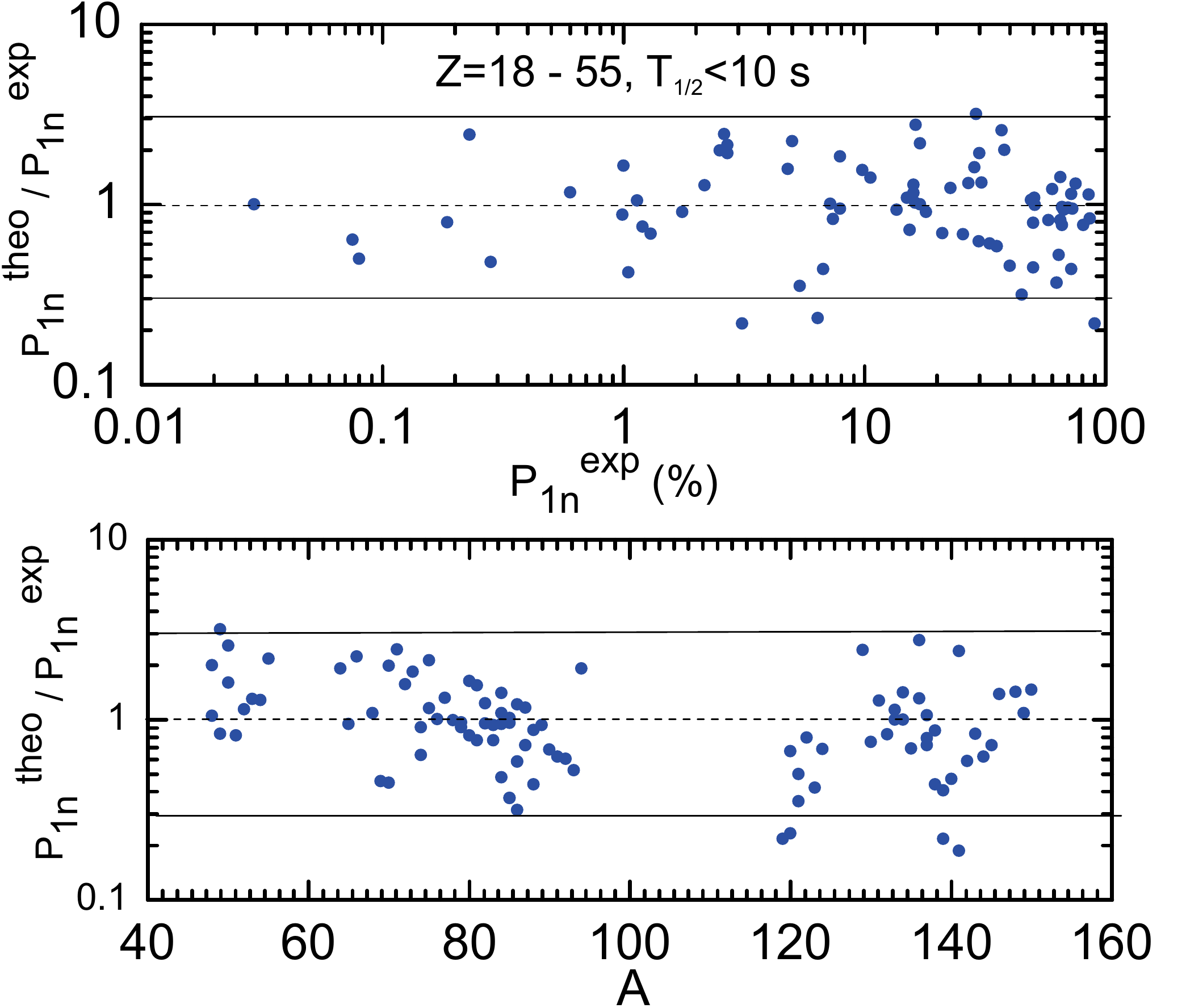}
\caption{Ratio of theoretical and experimental values of $\beta$-delayed one-neutron emission probabilities ($P_{1n}$) of (quasi-)spherical nuclei with $Z = 18 - 21,\, 25 - 35$ and $44 - 55$, calculated using the DF3a+CQRPA formalism and plotted versus the evaluated experimental $P_{1n}$ values in the top panel and the mass $A$ in the bottom panel.
}
\label{fig:th_global_df3_p1n}
\end{figure*}

Similarly, the ratio of neutron emission probabilities of (near-)spherical nuclei is shown
in Fig.~\ref{fig:th_global_df3_p1n} as a function of $A$. Overall, the experimental data are reproduced within the same factor of two. Again the exceptions to this trend are weakly deformed nuclei in the $Z = 31-35$, $A \approx 85$ and $Z = 44-45$, $A \approx 120$ regions which have rather small
$P_{1n}$ values. In these cases, neglecting the spreading of the $\beta$ strength due
to deformation leads to a systematic underestimation of the $P_{1n}$ values. 

It is worth stressing that the $\beta$-decay strength function can be modified by
including the quasiparticle-phonon coupling. The sensitivity of the neutron emission
probabilities to the strength distribution in the vicinity of the neutron emission thresholds
$S_{xn}$ is extremely high. Consequently, in the one-particle-one-hole pnQRPA approach, the accuracy of global predictions for the $P_{1n}$ values is lower than for half-lives.

\begin{figure*}[!htb]
\includegraphics[width=0.9\textwidth]{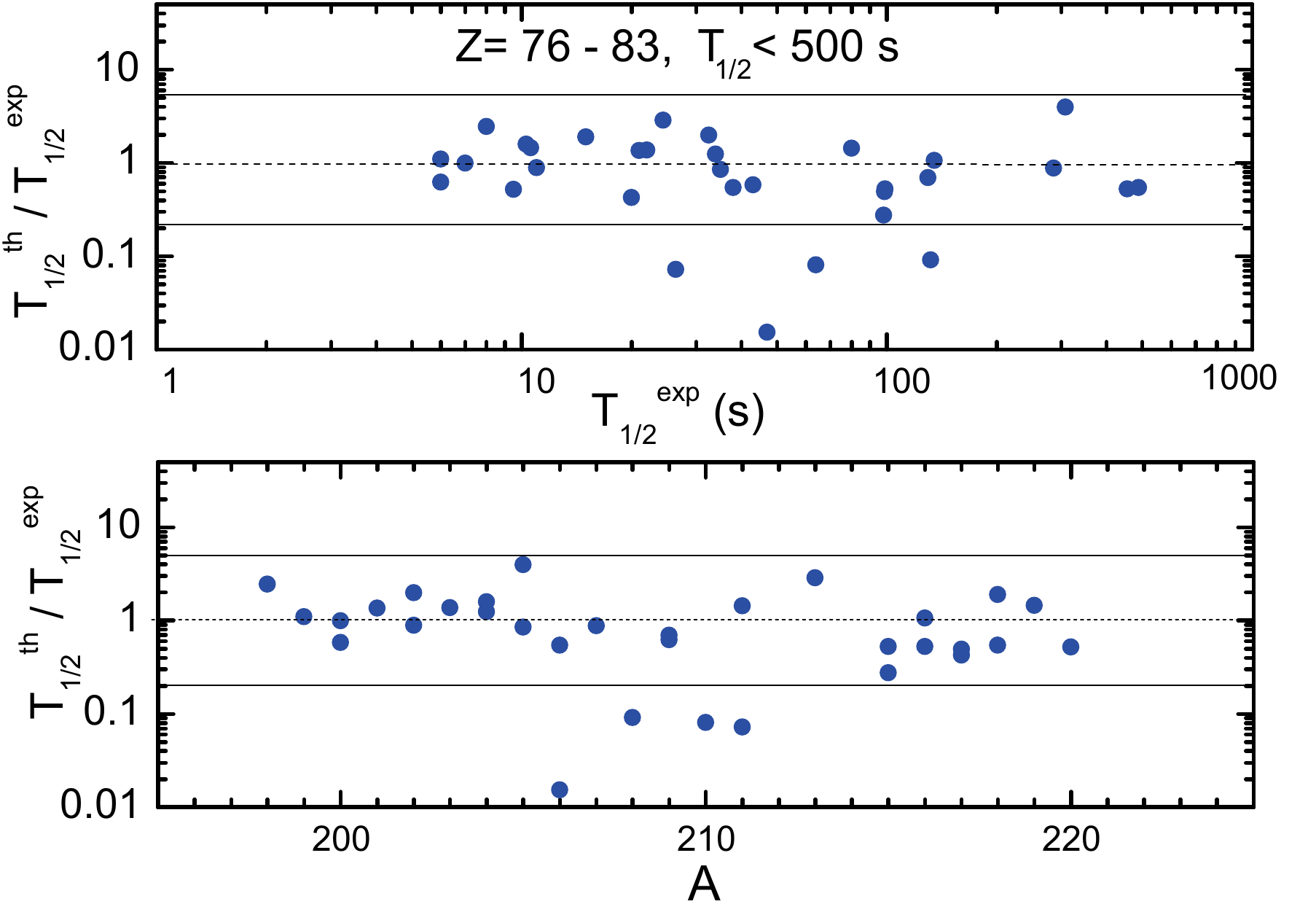}
\caption{Same as in Fig.~\ref{fig:th_global_df3_thalf} but for heavy ($Z = 76 - 83$, Os to Bi) spherical isotopes calculated using the DF3a+CQRPA formalism \cite{Borzov2011,Caballero2016,Caballero2017a}.
The experimental data is taken from the recent evaluation \cite{Liang2020}.}
\label{fig:th_global_df3_he1}
\end{figure*}

In Fig.~\ref{fig:th_global_df3_he1}, the ratio of DF3a+CQRPA predictions to experimental half-lives \cite{Caballero2016,Caballero2017a} is shown for the heavier Os to Bi nuclei ($Z$ = 76--83). As one approaches the $N = 126$ shell closure, nuclei are either spherical or have a small ground-state deformation. For
the bulk of the nuclei included in the DF3 calculations, we observe a larger spread of the half-lives of up to a factor of 5.  For Hg isotopes around $A = 210$ with relatively low Q$_{\beta}$ values, the DF3a+CQRPA underestimates the half-lives by one and RHB+RQRPA by two
orders of magnitude, respectively (see Ref.~ \cite{Caballero2016}). Both DF3a+CQRPA and RHB+RQRPA underestimate the experimental half-life of $^{206}$Au
\cite{Morales2015,CaballeroFolch2017} by a factor of about 30 -- 60. 

\begin{figure}[!htb]
\includegraphics[width=0.45\textwidth]{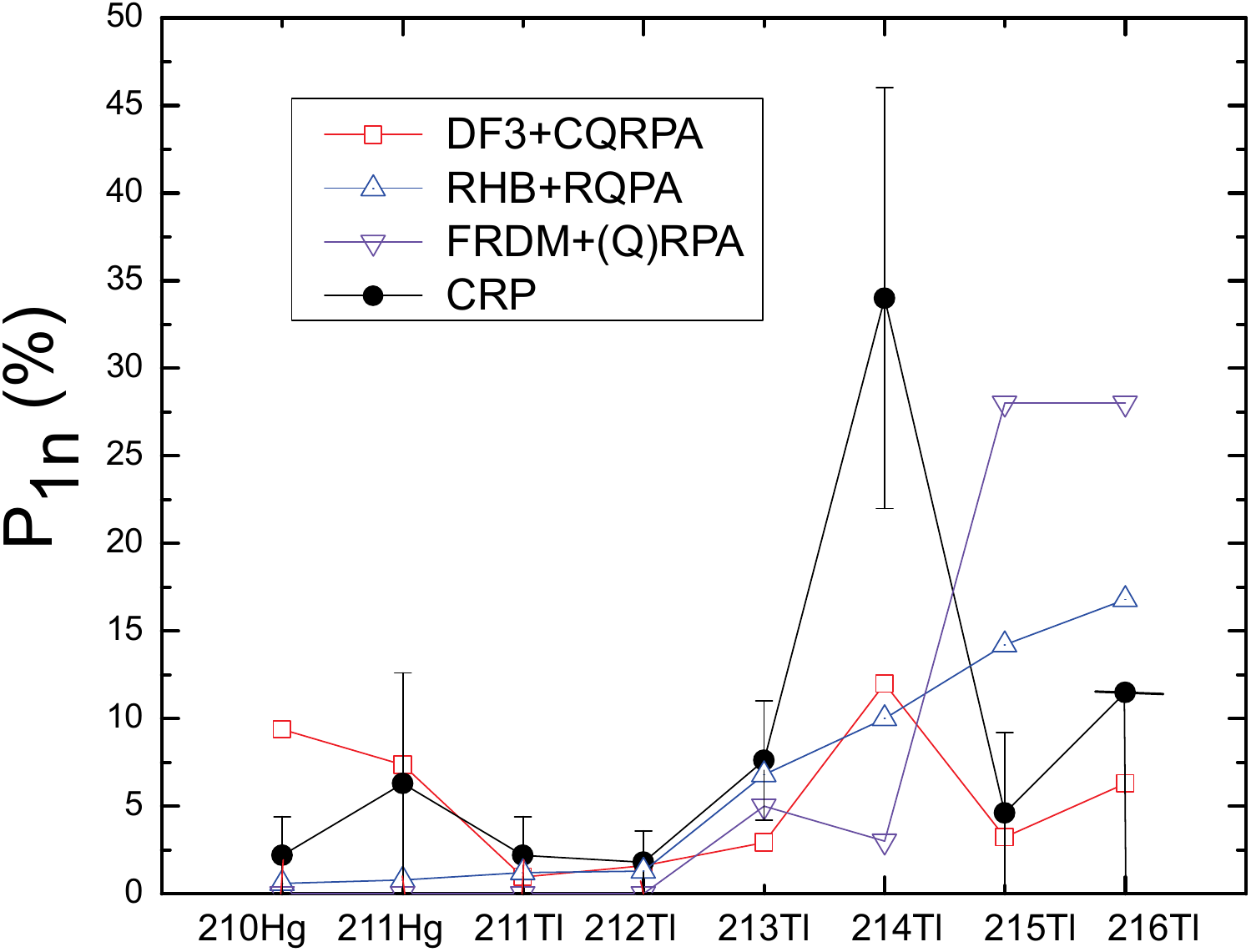}
\caption{Theoretical and experimental values for the $\beta$-delayed one-neutron emission probabilities of heavy (quasi-) spherical Hg and Tl isotopes ($Z = 80, 81$), calculated using the DF3a+CQRPA \cite{Borzov2011,Caballero2016,Caballero2017a}, RHB+RQRPA \cite{Marketin2016}, and FRDM12+(Q)RPA+HF models \cite{Moller2019}.
The experimental data is taken from the evaluation in Ref.~\cite{Liang2020}.}
\label{Fig_th_he_Pn}
\end{figure}

A comparison of the existing eight experimental $P_{1n}$ values for nuclei beyond the $N=126$ shell closure \cite{Caballero2016,CaballeroFolch2017} is shown in Fig.~\ref{Fig_th_he_Pn}. Note, however, that these are the heaviest nuclei for which $P_{1n}$ values have been measured so far. As can be seen, the experimental uncertainties are still too large for any detailed comparison of the different model performances.



It is worth mentioning that global comparisons of $\beta$-decay half-lives have also been performed within the deformed self-consistent QRPA model of Ref.~\cite{Martini2014}. The model uses the GT approximation and a global D1M Gogny interaction. Only the half-lives of 145 spherical and deformed even-even nuclei for which the experimental half-lives are known were included in the comparison. The calculated ratio $T_{1/2}^{theo}/T_{1/2}^{exp}$ is quite homogeneous with respect to the mass number $A$ and, more particularly, the quadrupole deformation parameter $\beta_2$. The treatment of deformation reduces the overall deviation
from the data to basically the same factor of 10 which was obtained for spherical nuclei in
Refs.~\cite{Borzov2003, Marketin2016}. Including deformation in the model is therefore expected to lead to a marked improvement in the RHB+RQRPA half-lives for deformed nuclei and similar levels of accuracy that were obtained for spherical nuclei in Refs.~\cite{Borzov2003, Marketin2016}. 

As far as 
the comparison of different microscopic models with the CRP data is concerned, we observe that regardless of the treatment of deformation all models using a concept of the $\beta$-decay strength function have difficulty describing nuclei close to the valley of $\beta$ stability with low Q{$_\beta$} values. The RHB+RQRPA calculations, for instance, tend to underestimate the half-lives for very heavy nuclei, around $A \approx 210$ by more than an order of magnitude \cite{Caballero2016, Caballero2017a}. This generic feature is also true for the microscopic-macroscopic model and ``Gross theory". 

Neutron-rich nuclei near the closed shells are the main playground for developments in the microscopic theories. The theoretical predictions of their $\beta$-decay properties -- while being a crucial test for any nuclear structure theory -- are also critical for planning experiments at RIB facilities. 
Additionally, these nuclei are also key for a better understanding of the astrophysical $r$ process. 



From the global comparisons we deduce that DF3a+CQRPA \cite{Borzov2003}, RHB+RQRPA \cite{Marketin2016}, and FRDM+(Q)RPA(+HF) \cite{Moller2019} models give overall a comparable description of the $\beta$-decay properties. Note that a somewhat smaller spread in the DF3a+CQRPA half-life calculations (Fig.~\ref{fig:th_global_df3_thalf}) with respect to the RHB+RQRPA (Fig.~\ref{fig:th_global_rqrpa_mass}) is partially a result of the limited comparison of (quasi-)spherical-only nuclei used in the DF3a calculations. An important source of deviations from the experimental data may come from the description of the ground-state properties. 

In particular, the underestimation of the total decay phase-space $Q_{\beta}$, as well as $Q_{\beta{xn}}$ sub-spaces may have a destructive impact on the predicted $\beta$-decay observables. This is especially true for the $P_{xn}$-values,
as these observables are more sensitive to the different "systematic" uncertainties of the theoretical strength functions.  

Finally, with regards to the predictive power of the models, if we consider the smaller number of parameters involved in the self-consistent models, as well as their universality, we can conclude that the predictions of these models are more reliable than those of the ``Gross theory" and FRDM-based models whose parameters are determined near the stability line. 




\subsection{Delayed neutron spectra calculations}\label{Sec:Micro-theory-spectra}

The microscopic-macroscopic approach described in the previous sections was also used in the calculations of delayed neutron (DN) spectra of fission products in Ref.~\cite{Kawano2008}. The FRDM95~\cite{Moeller1995} was combined with the (Q)RPA model~\cite{Moller2003,Moeller1990} to estimate $\beta$-strength functions which were then used as input in the Hauser-Feshbach statistical model (HF) to obtain DN branching ratios and spectra. The HF calculations require several nuclear ingredients. For neutron and $\gamma$-ray transmission coefficients, the Koning-Delaroche global optical potential~\cite{Koning2003} and the generalized Lorentzian E1 strength function~\cite{Kopecky1993} were adopted, respectively,
with a parameter set obtained from the Reference Input Parameter Library (RIPL-2)~\cite{Belgya2006}. For the nuclear level densities, they used RIPL-2 for discrete levels and the Gilbert-Cameron level density formula~\cite{Gilbert1965} at high energies. Spin and parity selection rules for the decay of the precursor nuclei to daughter nuclei were observed. 

The microscopic-macroscopic approach (FRDM+(Q)RPA+HF) to calculating \bdn~spectra is important for various applications as it can predict the spectra for nuclei located on the neutron-rich side of the nuclear chart for which experimental data are not available. Furthermore, as has been discussed in Sect.~\ref{Sec:Micro-theory}, it provides an improved description of the decay process leading to DN emission as it considers nuclear structure effects such as nuclear shell effects, level structures, and the spin-parity selection rules. The \bdn spectra calculated by~\cite{Kawano2008} are now adopted in the ENDF/B-VIII.-0 decay-data library~\cite{Brown2018}.

Within the CRP, we performed similar calculations of DN spectra by using a Hartree-Fock-BCS model (SHFBCS) -- instead of the FRDM -- combined with QRPA to obtain the $\beta$-decay properties. 
The DN emission part was calculated using the HF model~\cite{Minato2016}. 
In this SHFBCS+QRPA+HF model, the pairing correlation is treated by the BCS approximation with a zero-range volume-type force. The ground and excited states were self-consistently calculated with the SkO' Skyrme force~\cite{SkO'} in the cylindrical coordinate, 
where the axial symmetry was assumed. A zero-range volume-type pairing was used in the pp-channel ($T = 0$ pairing), and the strength parameters were adjusted to obtain the best possible description of available half-life data for each isotopic chain. Only allowed GT transitions are considered in the $\beta$-decay channel in Ref.~\cite{Minato2016}.
Odd-mass nuclei were treated in the same way as the even-mass nuclei, and the blocking and polarization effects caused by coupling between valence particle(s) and core nuclei were not considered. The HF statistical model implemented in the code CCONE~\cite{Iwamoto2013} was used to calculate the competition between DN and $\gamma$ decays. The global neutron optical potential of Ref.~\cite{Koning2003} and enhanced generalized Lorentzian (EGLO) model~\cite{Kopecky1993,Capote2009}
were used to calculate the neutron and $\gamma$ transmission coefficients, respectively. For level densities, we used RIPL-3~\cite{Capote2009}
for discrete levels and the formula of Gilbert and Cameron with Mengoni--Nakajima parameter set~\cite{Mengoni1994} for higher energies.

\begin{figure}[!htb]
\centering
\includegraphics[width=\linewidth]{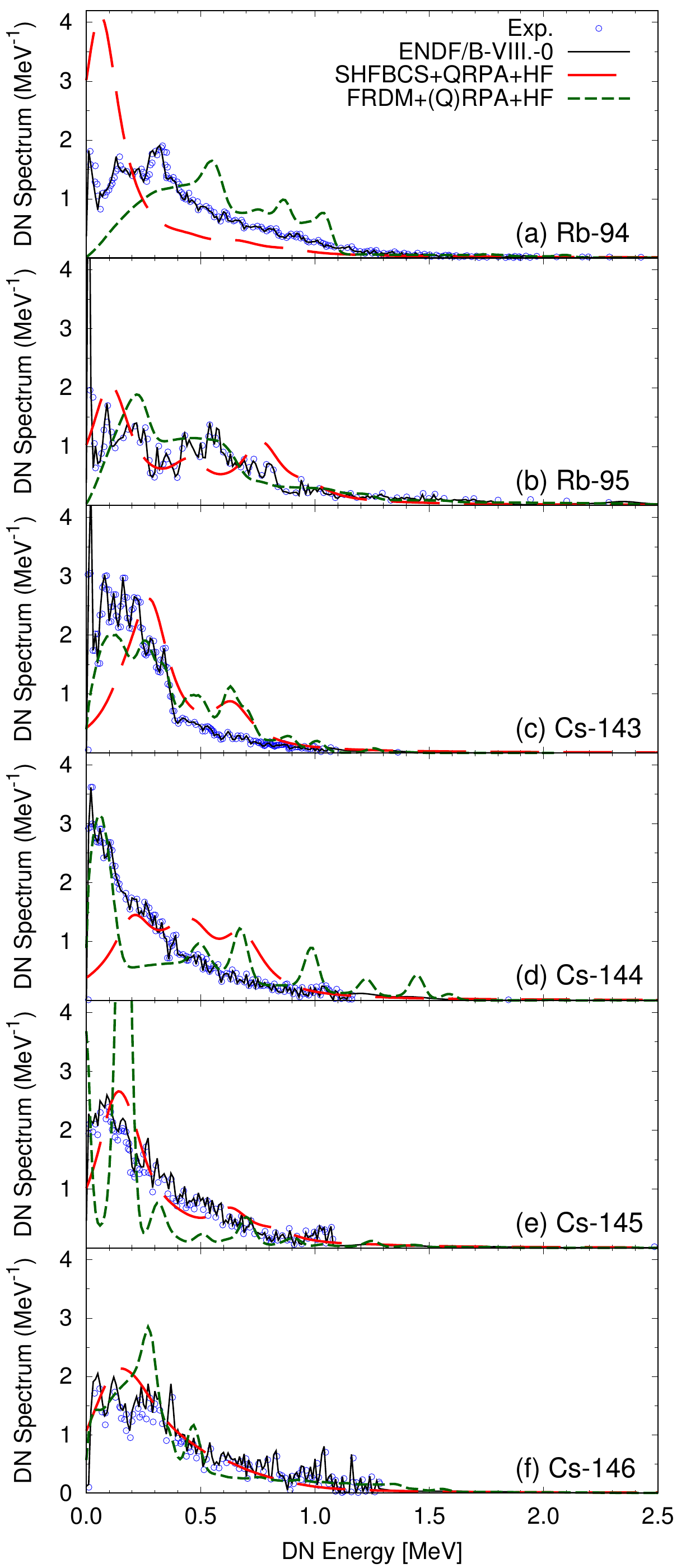}\\
\caption{\bdn spectra for $^{94,95}$Rb and $^{143-146}$Cs calculated with SHFBCS+QRPA+HF ~\cite{Minato2016} and FRDM+(Q)RPA+HF~\cite{Kawano2008}. The results are compared with experimental data available in~\cite{Brady1989a} and the evaluated spectra in ENDF/B-VIII.0~\cite{Chadwick2011}.}
\label{fig:th_spectra1}
\end{figure}

Figure~\ref{fig:th_spectra1} shows the \bdn spectra obtained for Rb and Cs isotopes from SHFBCS+QRPA+HF and FRDM+(Q)RPA+HF.
The results are compared with the evaluated data of ENDF/B-VIII.0~\cite{Brown2018} and the experimental data obtained from Ref.~\cite{Brady1989a} that were described in Sect.~\ref{Sec:Eval-digitize}.

For FRDM+(Q)RPA+HF, the \bdn spectra of $^{94}$Rb and $^{145}$Cs deviate from experimental data, while those of $^{95}$Rb and $^{143,144,146}$Cs are reproduced approximately. The SHFBCS+QRPA+HF model cannot describe the \bdn spectra of $^{94}$Rb and $^{144}$Cs, while those of the other nuclei are approximately reproduced.
It is worth noting that some of the \bdn spectra are reproduced reasonably well by these theoretical models, even though QRPA approaches are unable to estimate the nuclear excited states within an uncertainty of a few hundreds keV.
The deviations from experimental data are mainly attributed to the calculated $\beta$-strength function since both approaches use similar HF implementations.
For example, the \bdn spectrum of $^{145}$Cs is reasonably reproduced with SHFBCS+QRPA+HF but not with FRDM+(Q)RPA+HF, which implies that the $\beta$-strength function of SHFBCS+QRPA+HF is more suitable for this nucleus than the FRDM+(Q)RPA+HF strength function. This conclusion is based on the assumption that the particle evaporation process is correctly described. 

Similar systematic detailed comparisons of experimental and calculated DN spectra that duly consider the uncertainties in the measured spectra and the ingredients of the models are expected to lead to further improvements of the theoretical $\beta$-decay models in the future. However, as already mentioned in Sect.~\ref{Sec:Eval-digitize}, it is still not clear how to quantify the experimental uncertainties in the measured spectra of Ref.~\cite{Brady1989a}, therefore, until that is possible, caution should be exercised when comparing with these experimental DN spectra.

The published and digitized experimental DN spectra are available on the IAEA online database~\cite{database}.

\section{ MACROSCOPIC DATA: Methods and Measurements}
\label{Sec:Macro-meas}
In the following sections we deal with macroscopic delayed neutron data produced as a result of neutron-induced fission of major and minor actinides. These data are of utmost importance for studies on kinetic response and safe operation of fission reactors. 

\subsection{Methods }
\label{Sec:Macro-methods}
\subsubsection{Total delayed neutron yields and time-dependent parameters}
\label{Sec:Macro-methods-total}

The direct methods used for measuring the macroscopic delayed-neutron characteristics from neutron-induced fission of heavy elements are based on irradiating the investigated sample by neutrons and subsequently measuring the appropriate delayed-neutron observables: the time dependence of delayed-neutron activities or the delayed-neutron energy spectrum. Neutron sources used for this purpose are mainly the neutron flux in nuclear reactors \cite{Waldo81, Saleh97, Charlton97, Besant77, Benedetti82, Synetos79, Rose57, Gudkov89}, critical assemblies \cite{Keepin57, Loaiza97,Loaiza98}, or electrostatic accelerator-based nuclear reactions: Li(p,n), T(p,n), T(d,n), D(d,n) \cite{Masters69, Krick72, Cox68, Cox70, Cox74, Piksaikin06, Fieg72}. All experimental methods are based on the assumption that the delayed-neutron activity arising from the irradiation of the fissionable sample can be represented by the linear superposition of the exponential decay modes, each with its own decay constant 
\begin{align}\label{v.1}
\frac{dN(t)}{dt} & =A\sum_{i=1}^{m}(1-e^{-\lambda_{i}\cdot t_{irr}})\cdot a_i \cdot e^{-\lambda_{i}\cdot t}\, ,\\
A &=\epsilon_n\cdot \sigma_f\cdot \phi\cdot N_f\cdot \nu_d \, \nonumber,
\end{align}
where A is the saturation delayed-neutron activity, $a_i$ and $\lambda_i$ the relative abundance and decay constant of the i-th delayed neutron group, respectively, $t_{irr}$ the irradiation time, m the number of delayed neutron groups each of which is associated with a group of delayed-neutron precursors with similar decay constants, $\epsilon_n$ the efficiency of the neutron detector, $\nu_d$  the total delayed-neutron yield per one fission, $\phi$ the neutron flux (1/cm$^{2}$s), $\sigma_f$ the fission cross section (barn), $N_f$ the number of atoms in a fissionable sample. 

A comprehensive study of the delayed-neutron emission properties conducted by Keepin \textit{et al.} \cite{Keepin57} on $^{232}$Th, $^{233}$U, $^{235}$U, $^{238}$U, $^{239}$Pu, and $^{240}$Pu lead to the development of a 6-group model of the temporary delayed-neutron characteristics ($a_i,T_i$), where $T_i$ = ln2/$\lambda_i$. This model has been successfully used by the reactor community for many years. Only recently, an alternative 8-group model based on a set of decay constants that are universal for all fissioning systems was proposed and justified by Spriggs \textit{et al.}~\cite{Spriggs99,Spriggs02}.

In the case of instantaneous irradiation (prompt burst irradiation, $\lambda_{i}t_{irr}<<1$), Eq.~\eqref{v.1} becomes
\begin{equation}\label{v.2}
\frac{dN(t)}{dt}=A\cdot t_{irr}\sum_{i=1}^{m}\lambda_i\cdot a_i\cdot e^{-\lambda_{i}\cdot t}\, .
\end{equation}
From Eqs.~\eqref{v.1} and \eqref{v.2} it becomes clear that the accuracy of the delayed-neutron parameters $\nu_d$ and ($a_i,T_i$) extracted from measured data depends to a large extent on parameters such as the accumulated statistics, the fission rate in the sample $R_s=\sigma_{f}\cdot\phi\cdot{N}_f$, the transportation time of the sample from the irradiation position to a neutron detector and the neutron background. The quality of the obtained delayed-neutron data is also related to the incident neutron energy because the temporary delayed-neutron parameters ($a_i,T_i$) are known to have a noticeable energy dependence when expressed in terms of the average half-life of delayed neutron precursors $\langle T\rangle$ (\cite{Piksaikin02a}, \cite{Piksaikin02b}). 

\paragraph{Nuclear reactor-based methods.} 

The most important advantage of the nuclear reactor-based method is the high intensity of the neutron flux. With such a high-intensity flux it is possible to use very small samples of fissionable nuclides (up to micrograms) as was demonstrated by Waldo \textit{et al.}  \cite{Waldo81}. However, the long time it takes to transport the sample in the reactor measurements does not allow one to resolve all 6 groups of delayed neutrons. Furthermore, given that the energy of the primary neutrons can take a wide range of values or maybe is unspecified in many of the reactor experiments, it is not straightforward to use these data in the evaluation of delayed-neutron parameters. 

As a rule, a single irradiation procedure is used in the nuclear reactor-based experiments. The time dependence of the delayed neutron activity registered by a multichannel analyzer can be obtained by integration of Eq.~\eqref{v.1} (\cite{Piksaikin06,Piksaikin02a}. 
\begin{align}\label{v.3}
& \int_{t_k}^{t_{k+1}} \frac{dN(t)}{dt}dt =\frac{dN(t_k)}{dt}\cdot \Delta t_k =  \nonumber \\ 
 & A\cdot\sum_{i=1}^{m}(1-e^{-\lambda_i \cdot t_{irr}})\cdot \frac{a_i}{\lambda_i}\cdot (1-e^{-\lambda_i\cdot \Delta t_k})\cdot e^{-\lambda_i\cdot t_k} \nonumber \\ & + B\cdot \Delta t_k\, ,
\end{align}

where ($dN(t_{k})/dt)\cdot\Delta t_k$ is the number of counts registered in the time channel $t_k$ with width $\Delta t_k$, $B$ the intensity of neutron background. This decay curve is used to estimate the temporal delayed-neutron parameters ($a_i,T_i$), the saturation activity $A$, the intensity of the background B, and the covariance matrix of the estimated parameters with the help of the least squares method (LSM). The obtained delayed-neutron decay curve, the values of the estimated parameters ($a_i,T_i$) and the background $B$ are used to calculate the \textit{total delayed neutron yield}  \cite{Synetos79,Piksaikin02a,Piksaikin02c,Roshchenko06,Piksaikin13}.

One possible type of irradiation is the instantaneous (prompt burst) irradiation (see Eq.~\eqref{v.2}). In this case, a measured delayed-neutron decay curve can be obtained by the integration of Eq.~\eqref{v.2}
\begin{align}\label{v.4}
& \int_{t_k}^{t_{k+1}} \frac{dN(t)}{dt}dt =\frac{dN(t_k)}{dt}\cdot \Delta t_k =  \nonumber \\ 
 & A\cdot t_{irr}\sum_{i=1}^{m} a_i\cdot (1-e^{-\lambda_i\cdot \Delta t_k})\cdot e^{-\lambda_i\cdot t_k} + B\cdot \Delta t_k\, .
\end{align}

It is this type of irradiation that was used by Keepin \textit{et al.} \cite{Keepin57} for the measurement of the decay curves which were used for the estimation of short-lived group parameters (from 3 through 6) and the determination of the total delayed neutron yield. The Godiva reactor (a bare spherical $^{235}$U assembly) was used as neutron source. In addition, the total delayed-neutron yield was determined with the help of the following approximation \cite{Keepin57}.
\begin{align}\label{v.5}
& \sum_{t_1}^{t_2} \frac{dN(t_k)}{dt}\cdot \Delta t_k - B\cdot (t_2 - t_1) = \nonumber \\
 & A\cdot t_{irr} \sum_{i=1}^{m} a_i\cdot \left(e^{-\lambda_i\cdot t_1} - e^{-\lambda_i\cdot t_2}\right) = \nonumber \\
& A\cdot t_{irr} \sum_{i=1}^{m} a_i = \nu_d\cdot\epsilon\cdot R_s\cdot t_{irr}\, ,
\end{align}

where $t_{irr}$ is the time of the neutron pulse, $R_s\cdot t_{irr}$ the total amount of fissions in a sample, and the sum of the relative abundances is equal to 1. 

\paragraph{Accelerator-based delayed neutron experiments.} 

The neutron sources used in this method are the monoenergetic neutron fluxes from the accelerator-based nuclear reactions Li(p,n), T(p,n), T(d,n)$^4$He, D(d,n)$^3$He. The main advantage of the accelerator-based method is the possibility to investigate the energy dependence of delayed neutron properties in a wide range of incident neutron energies from thermal to 20 MeV \cite{Maksyutenko58, Cox68, Cox70, Cox74, Masters69, Krick72, Fieg72,Piksaikin06}. Eqs.~\eqref{v.1} and \eqref{v.2} reflect one single cycle of measurement which includes the irradiation of the sample, the decay, and the counting of delayed-neutron activity. However, in accelerator-based experiments a sample undergoes a number of cycles with the purpose of increasing the statistics. The general equation for the determination of the total delayed-neutron yields and temporal delayed-neutron characteristics ($a_i,T_i$) on the basis of the delayed-neutron decay curve accumulated during n cycles is given by~(\cite{Piksaikin06}, \cite{Piksaikin02d}, \cite{Piksaikin98}, \cite{Piksaikin02a}, \cite{Roshchenko06a}, \cite{Piksaikin13})
\begin{align}\label{v.6}
& \frac{dN(t_k)}{dt}\cdot \Delta t_k  =  \notag\\ & A\cdot \sum_{i=1}^{m} T_{hi}\cdot \frac{a_i}{\lambda_i}\cdot (1-e^{-\lambda_i\cdot \Delta t_k})\cdot e^{-\lambda_i\cdot t_k} + B\cdot \Delta t_k\,,
\end{align}
where $T_{hi}$ determines the history of irradiation
\begin{align}
& T_{hi}  = (1-e^{-\lambda_i\cdot t_{irr}}) \times \notag \\
& \Biggl( \frac{n}{1-e^{-\lambda_i\cdot T}}-e^{-\lambda_i\cdot T}\cdot\left(\frac{1-e^{-n\cdot\lambda_i\cdot T}}{(1-e^{-\lambda_i\cdot T})^2}\right) \Biggr)\,, \notag \\ 
& A=\epsilon\cdot\sigma_f\cdot\phi\cdot N_f\cdot\nu_d\,, \notag
\end{align}
with $n$ the number of cycles, $T$ the period of one cycle (irradiation, decay, counting, delay). 

This equation is used to estimate the values of the saturation activity $A$, the temporal delayed-neutron parameters ($a_i,\,T_i (i = 1, ..., m)$) and the background $B$ on the basis of the decay curves measured with different irradiation times $t_{irr}$. The $A$ and $B$ values and group parameters of delayed neutrons are estimated using the iterative least-squares method \cite{Piksaikin02a}. The decay curves obtained with short irradiation times were used to estimate the parameters of the short-lived groups of delayed neutrons, and the decay curves corresponding to long exposure times were used to estimate the parameters of long-lived groups. The result of the analyses of data from one run of measurements comprising $n$-cycles of irradiation and counting is the set of the relative abundances of delayed neutrons $a_i$, the half-lives of their precursor $T_i$, and the covariance matrix of the group parameters \cite{Piksaikin02e}. 

\paragraph{Modulated accelerator beam technique.} 

\begin{figure}[!htb]
	\centering
\includegraphics[width=\linewidth]{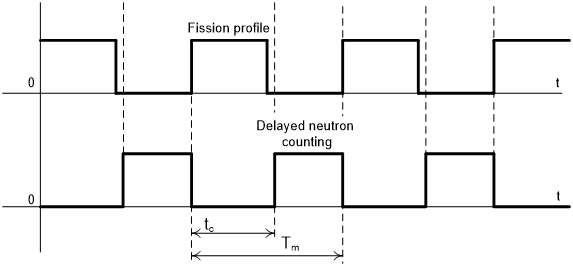}
\caption{Time sequences in the modulated neutron-source techniques \cite{Masters69}. Delayed neutron counting is made in time intervals $(T_m-t_c)$.}
\label{fig:5_1}
\end{figure}

In delayed neutron measurements, a short irradiation time (see Eq.~\eqref{v.2}) simplifies the interpretation of the experimental data obtained in the total delayed yield measurements. This idea helped to develop an efficient method for the determination of the energy dependence of the total delayed-neutron yield from fission of several heavy nuclei \cite{Masters69}. This technique is based on a modulated neutron-source generated by a Cockcroft-Walton accelerator which utilizes the D(d,n) reaction for producing 3.1 MeV and the T(d,n) reaction for 14.9 MeV neutrons. Figure~\ref{fig:5_1} shows the time sequences used for the modulated neutron-source technique. 

A long-counter neutron detector operating in anti-synchronism with the accelerator beam registers delayed neutrons in the time interval ($T_m-t_c$). Interpretation of the experimental data for this method is based on the equation for the \textit{non-cumulative} DN decay curve measured after the $n$-th cycle of irradiation-counting sequences
\begin{align}\label{v.7}
& \frac{dN(t_k)}{dt}\cdot \Delta t_k = \notag\\
& A\cdot\sum_{i=1}^{m} T_{hi}\cdot \frac{a_i}{\lambda_i}\cdot (1-e^{-\lambda_i\cdot \Delta t_k})\cdot e^{-\lambda_i\cdot t_k} + B\cdot \Delta t_k \, ,
\end{align}
\begin{align} 
T_{hi}  = (1-e^{-\lambda_i\cdot t_{irr}}) \cdot\left(\frac{1-e^{-n\cdot\lambda_i\cdot T_m}}{1-e^{-\lambda_i\cdot T_m}}\right) \notag
\end{align}

where the saturation activity is
$A = \epsilon_n\cdot\sigma_f\cdot\phi\cdot{N}_f\cdot\nu_d$
, and $T_m$ the modulation period.
After reaching the equilibrium ($n \rightarrow\infty$) in the short time periodic irradiation ($\lambda_i T_m << 1$) with the period of modulation $T_m$, the summation of delayed-neutron counts in the counting time ($T_m-t_c$) can be expressed as follows:
\begin{align}\label{v.8}
\sum_{t_1}^{t_2}\frac{dN(t_k)}{dt}\cdot \Delta t_k - B\cdot(t_2-t_1) & = \notag\\
A\cdot\sum_{i=1}^{m} (1-e^{-\lambda_i\cdot t_{irr}})\cdot\frac{a_i}{\lambda_i}\cdot \frac{(e^{-\lambda_i\cdot t_1}-e^{-\lambda_i\cdot t_2})}{(1-e^{-\lambda_i\cdot T_m})} &= \notag\\
A\cdot t_{irr}\frac{T_m-t_c}{T_m}\,
\end{align}

where $t_1$ is the time when the delayed-neutron counting $t_c$ starts, and $t_2$ is the period of modulation $T_m$ (see Fig.~\ref{fig:5_1}). Another advantage of this method as compared to the nuclear reactor neutron-source techniques is a minimum of scattering and thermalizing material present in the vicinity of the sample. The total delayed-neutron yield was measured from the fission of $^{232}$Th, $^{233}$U, $^{235}$U, $^{238}$U, and $^{239}$Pu \cite{Masters69} nuclides. Later, Krick and Evans \cite{Krick72}, using the same methodology, measured the total delayed-neutron yield as a function of the neutron energy from the fission of $^{233}$U, $^{235}$U, $^{238}$U, $^{239}$Pu, and $^{242}$Pu in the energy range 0.1 - 6.5 MeV. It should be noted that the on-beam arrangement of a neutron detector can lead to a degradation of its counting characteristics. Special measures should be taken to find and eliminate this effect (see paragraph devoted to measurements with the T(d,n)$^{4}$He neutron source). 

\paragraph{The neutron detector in delayed-neutron measurements.} 

The neutron detector most commonly used in delayed-neutron measurements is an assembly of $^{3}$He or boron counters distributed in a polyethylene or paraffin moderator. A moderator allows one to shift the neutron energy to the region of high registration efficiency. The efficiency of the neutron detector is determined using calibrated neutron sources (Pu-Be, Am-Li, $^{252}$Cf) or the monoenergetic neutrons produced from $^{7}$Li(p,n), D(d,n), $^{51}$V(p,n) nuclear reactions measured at a charged-particle accelerator \cite{Piksaikin97}.

\paragraph{Fission rate determination in accelerator and reactor-based methods.} The main method used for this purpose in nuclear reactor experiments is spectroscopy of gamma rays emitted from fission products. Keepin \textit{et al.} \cite{Keepin57} determined the total number of fission in the samples by standard counting of the 67 h $\beta$-activity from $^{99}$Mo. Benedetti \textit{et al.} \cite{Benedetti82} determined the fission rate in the samples by measuring the gamma activity of fission products ($^{103}$Ru, $^{131}$I, $^{140}$Ba/$^{140}$La) that were induced in the samples after irradiation.
In the accelerator-based technique, the fission rate in the sample is determined by measuring the neutron flux with the help of a fission chamber that is placed in the immediate vicinity of the sample. In the modulated neutron-source method, the fission rate in the fissionable sample is determined by measuring the count rates in two fission chambers sandwiching the sample (\cite{Masters69}, \cite{Krick72}). In the case of accelerator-based methods with \textit{off-beam arrangement of the neutron detector}, the fission chambers are placed in front of and behind the sample along the beam line of the accelerator (\cite{Piksaikin06},\cite{Piksaikin02c}, \cite{Piksaikin99}, \cite{Roshchenko06}).

\paragraph{The effect of the concomitant neutron source D(d,n)$^3$He in delayed- neutron experiments with an accelerator neutron source T(d,n)$^4$He.} 

In experiments using the T(d,n)$^4$He reaction as the neutron source, one has to take into account the effect of a concomitant neutron source which originates from the implantation of deuteron ions in the backing of the tritium target. As the number of implanted ions grows, the intensity of the concomitant neutron source D(d,n)$^{3}$He increases. This feature of the T(d,n)$^{4}$He reaction makes it difficult to interpret the experimental data obtained in the respective experiments. A special method has been developed to measure the intensity of the D(d,n)$^{3}$He neutron source in correlation with the ion charge accumulated on a tritium target \cite{Piksaikin06a}. This method allows one to account for the effect of the concomitant neutron source and has been used in measurements of the relative abundances and periods of delayed neutrons for neutron-induced fission of $^{232}$Th, $^{233}$U, $^{235}$U, $^{236}$U, $^{238}$U, $^{239}$Pu, $^{237}$Np, and $^{241}$Am in the energy range from 14 to 18 MeV. Details of the procedure are presented in (\cite{Piksaikin06a}, \cite{Piksaikin07}).

\paragraph{The effect of degradation of the neutron detector counting rate characteristic in an intense field of high energy neutrons.} 

The delayed-neutron decay curve recorded in measurements of relative abundances and periods of delayed neutrons from the fission of heavy nuclides induced by neutrons from the T(d,n)$^{4}$He reaction was found to undergo noticeable degradation in the first several seconds after switching off the ion beam. A special experiment has been conducted to measure the counting property of the neutron detector after its irradiation by neutrons from the T(d,n)$^{4}$He reaction. In this experiment, the time dependence of the count rates from the Am-Li neutron source was measured immediately after the irradiation of the neutron detector by a T(d,n)$^{4}$He neutron flux of different intensities in the energy range 14-18 MeV. The obtained data were used to correct the measured decay curves of the delayed neutron activity in the measurements of the relative abundances and periods from fission of $^{232}$Th, $^{233}$U, $^{235}$U, $^{236}$U, $^{238}$U, $^{239}$Pu, $^{237}$Np, and $^{241}$Am by neutrons in the energy range from 14 to 18 MeV. Details of the procedure are presented in \cite{Piksaikin07}.

\subsubsection{Delayed neutron integral spectra}\label{Sec:Macro-method-spectra}

There are two main methods used for the investigation of integral delayed-neutron spectra. These are neutron spectrometry using $^{3}$He \cite{Evans79} and proton-recoil detectors \cite{Greenwood1985,Greenwood1997,Fieg72} and the time-of-flight method \cite{Tanczyn88}. The time-of-flight method gives an excellent resolution at low energy, however this resolution worsens rapidly at energies above a few hundred keV. As an example, the full width at half maximum (FWHM) of a TOF system with flight path 50 cm and time resolution 3 ns is 1.6, 17.3, and 82.6 keV at energies 50, 300 and 900 keV, respectively. The proton-recoil spectrometer resembles the time-of-flight method: the resolution is good at low energies, while deteriorating at high energies. The full width at half maximum of a proton-recoil spectrometer with 2.5 atm of H$_2$ for the same energies mentioned above, is 4.2, 12.6 and 25.6, respectively. The $^{3}$He-spectrometer seems to be the optimal tool over the whole energy range. The resolution of this type of spectrometer is about 12-30 keV for energies ranging from 10 keV to 3 MeV. A detailed description of the techniques used for measurements of delayed-neutron spectra including unfolding procedures are given in a comprehensive review by Das \cite{Das94}. Since the publication of this review, experimental work on composite delayed-neutron spectra has been carried out only at IPPE using an accelerator-based neutron beam \cite{Piksaikin17}. By applying the least-squares method to the composite spectra measured in several delayed time intervals following fission~\cite{Vilani92}, one can unfold the group energy spectra $\chi_i(E_n$). The main purpose of obtaining the experimental group spectra is to use them to calculate the effective fraction of delayed neutrons $\beta_{eff}$ \cite{Santamarina12}.

\subsection{New measurements and compilation }\label{Sec:Macro-meas-new}

Since the last evaluation of macroscopic delayed-neutron parameters was performed in the framework of WPEC-SG6 \cite{NEA-WPEC6,DAngelo02}, experimental activities related to macroscopic delayed neutron data have mainly been carried out at IPPE (Obninsk). These activities have been devoted to the measurement of the energy dependence of the total delayed neutron yield and the relative abundances of delayed neutrons and half-lives of their precursors for neutron-induced fission of $^{232}${Th}, $^{233}$U, $^{235}$U, $^{236}$U, $^{238}$U, $^{239}$Pu, $^{237}$Np, and $^{241}$Am in the energy range from thermal to 18 MeV~\cite{Piksaikin02a,Piksaikin02c,Roshchenko06,Piksaikin13}. In addition, the integral energy spectra of delayed neutrons from thermal neutron-induced fission of $^{235}$U have been measured \cite{Piksaikin17}. The relative abundances and half-lives of delayed neutrons from the relativistic proton-induced fission of $^{238}$U have also been measured \cite{Egorov17}. All the data related to neutron-induced fission have been compiled in the macroscopic section of the database developed by this CRP~\cite{IAEA0643} under the guidance of the Nuclear Data Section. The database is an extension of the summary of delayed neutron parameters by Spriggs and Campbell \cite{Spriggs02a} and the compilation by Tuttle \cite{Tuttle75} and comprises the total delayed neutron yields, and the relative abundances and half-lives of delayed neutrons in the 6- and 8-group models with covariance and correlation data. The main features of the new IPPE macroscopic delayed-neutron data and the earlier measured data are presented in the following sections. 

\subsubsection{Energy dependence of the relative abundances and half-lives of delayed neutrons}\label{Sec:Macro-meas-energy}
The energy dependence of the relative abundances $a_i$ and half-lives $T_i$ of delayed neutrons have been measured at the IPPE electrostatic accelerators for neutron-induced fission of $^{232}$Th, $^{233}$U, $^{235}$U,$^{236}$U, $^{238}$U, $^{239}$Pu, $^{237}$Np, and $^{241}$Am in the energy range from thermal to 18 MeV. Monoenergetic neutron beams were generated by means of the nuclear reactions T(p,n), D(d,n), T(d,n). The method used in these experiments is described in Section~\ref{Sec:Macro-methods-total}. Efforts were made to improve the experimental and data processing procedures by a) shortening the transportation time, b) extending the delayed-neutron counting time, and c) averaging the temporal delayed-neutron parameters obtained in different experimental runs to improve the accuracy of the delayed neutron temporal parameters ($a_i,T_i$). The IPPE measurements in the high energy range (14 -- 18 MeV) revealed two plausible reasons for the large discrepancies observed in the delayed-neutron temporal parameters: (i) the deterioration of the counting efficiency of the neutron detector in a high-intensity neutron flux and ii) the contamination of the accelerator target by neutrons produced by the concomitant D(d,n)$^{3}$He neutron source in measurements using the T(d,n)$^{4}$He reaction. 

\begin{figure*} 
\centering
\includegraphics[width=\textwidth]{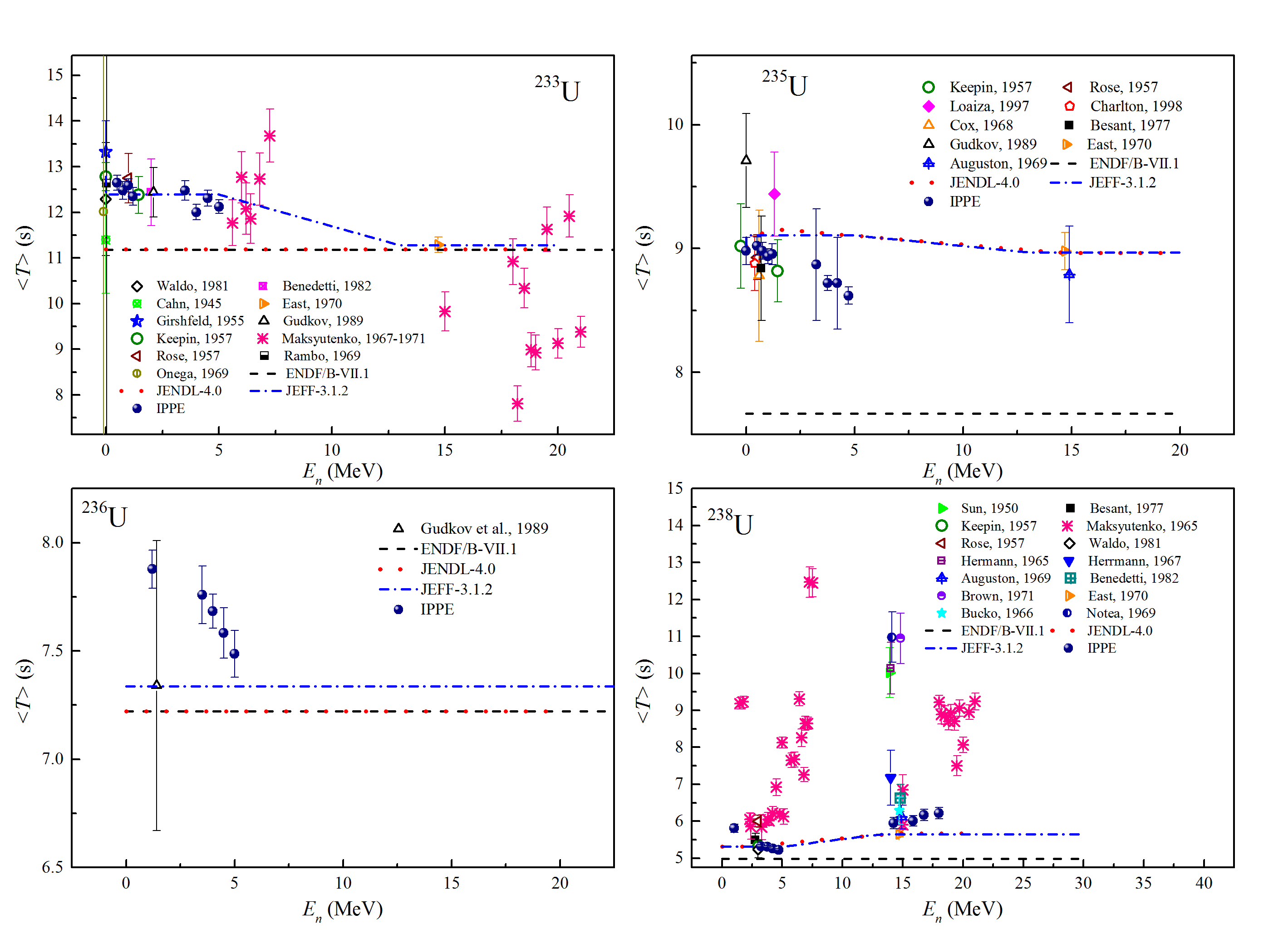}
\caption{The energy dependence of the average half-life of delayed-neutron precursors from neutron-induced fission of $^{233}$U, $^{235}$U, $^{236}$U, $^{238}$U. The references of the data can be taken from the compilation by Spriggs and Campbell \cite{Spriggs02a} except for the IPPE data~\cite{Piksaikin02a,Piksaikin02e,Piksaikin04,Piksaikin07,Piksaikin13,Isaev98}.}
\label{fig:5_2}
\end{figure*}

\begin{figure*} 
	\centering
\includegraphics[width=\textwidth]{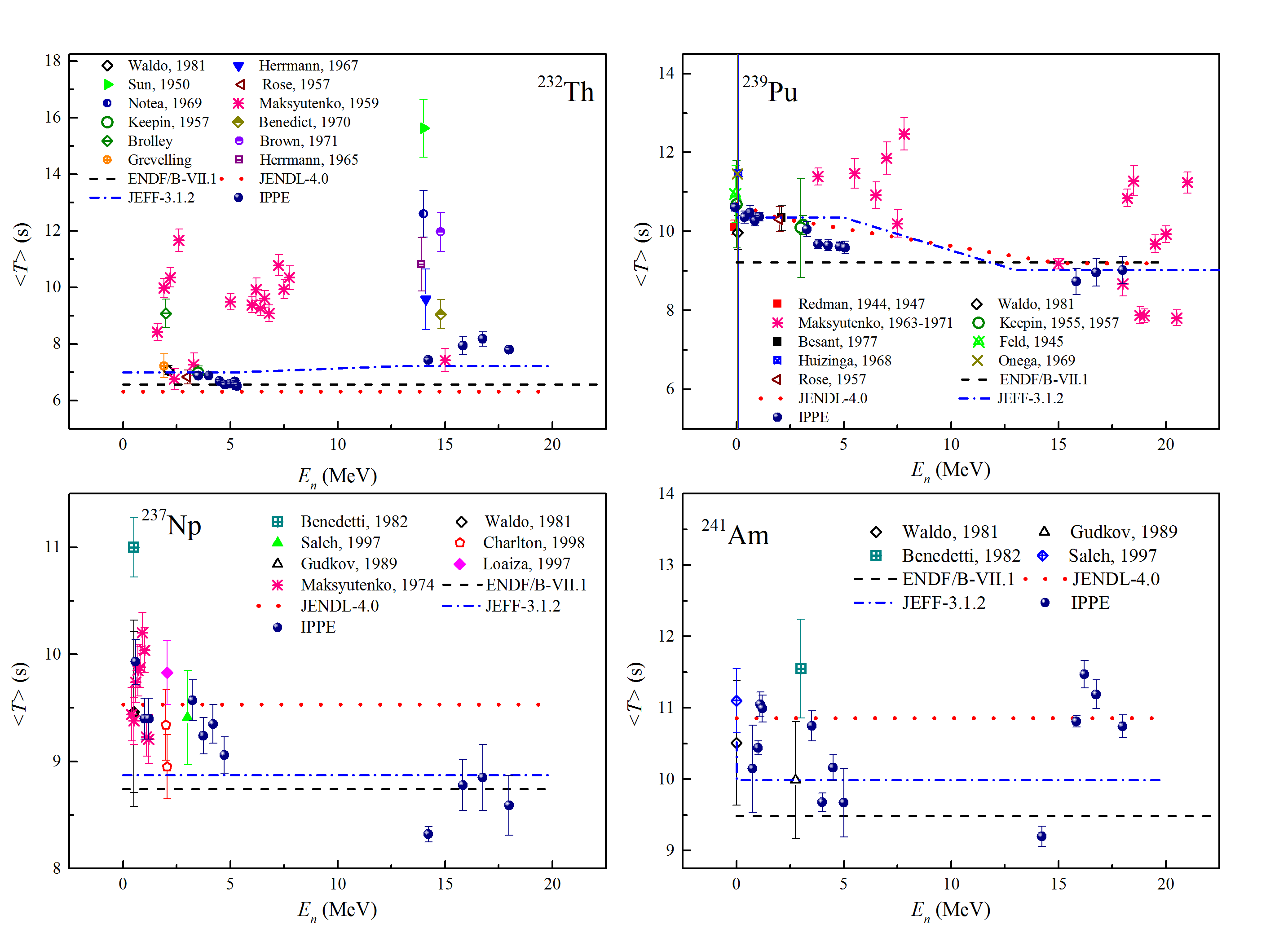}
\caption{The energy dependence of the average half-life of delayed-neutron precursors from neutron-induced fission of $^{232}$Th, $^{237}$Np, $^{239}$Pu, $^{241}$Am. The references of the data can be taken from the compilation by Spriggs and Campbell \cite{Spriggs02a} except for the IPPE data \cite{Piksaikin98,Piksaikin02a,Piksaikin02d,Piksaikin11,Piksaikin13,Gremyachkin17,Gremyachkin17a,Roshchenko06a,Roshchenko10}.}
\label{fig:5_3}
\end{figure*}

The comparison of the delayed-neutron parameters ($a_i,T_i$) obtained in the IPPE measurements with other published data is shown in Figs.~\ref{fig:5_2} and \ref{fig:5_3}. The quantity that is compared is the average half-life $\langle T  \rangle$~of the delayed-neutron precursors introduced by Piksaikin \textit{et al.}~\cite{Piksaikin02b}. Since $\langle T  \rangle$~is the combination of the delayed neutron parameters ($a_i,T_i$) obtained in the least-squares fitting procedure and represents an average property of the system, it is not affected by the correlations which inevitably exist between the delayed-neutron parameters. It should be noted that the $\langle T  \rangle$~value depends neither on whether one considers N as individual precursors or delayed-neutron groups of precursors, nor on the time boundaries of the delayed neutron groups. 

\begin{figure} 
	\centering
\includegraphics[width=\linewidth]{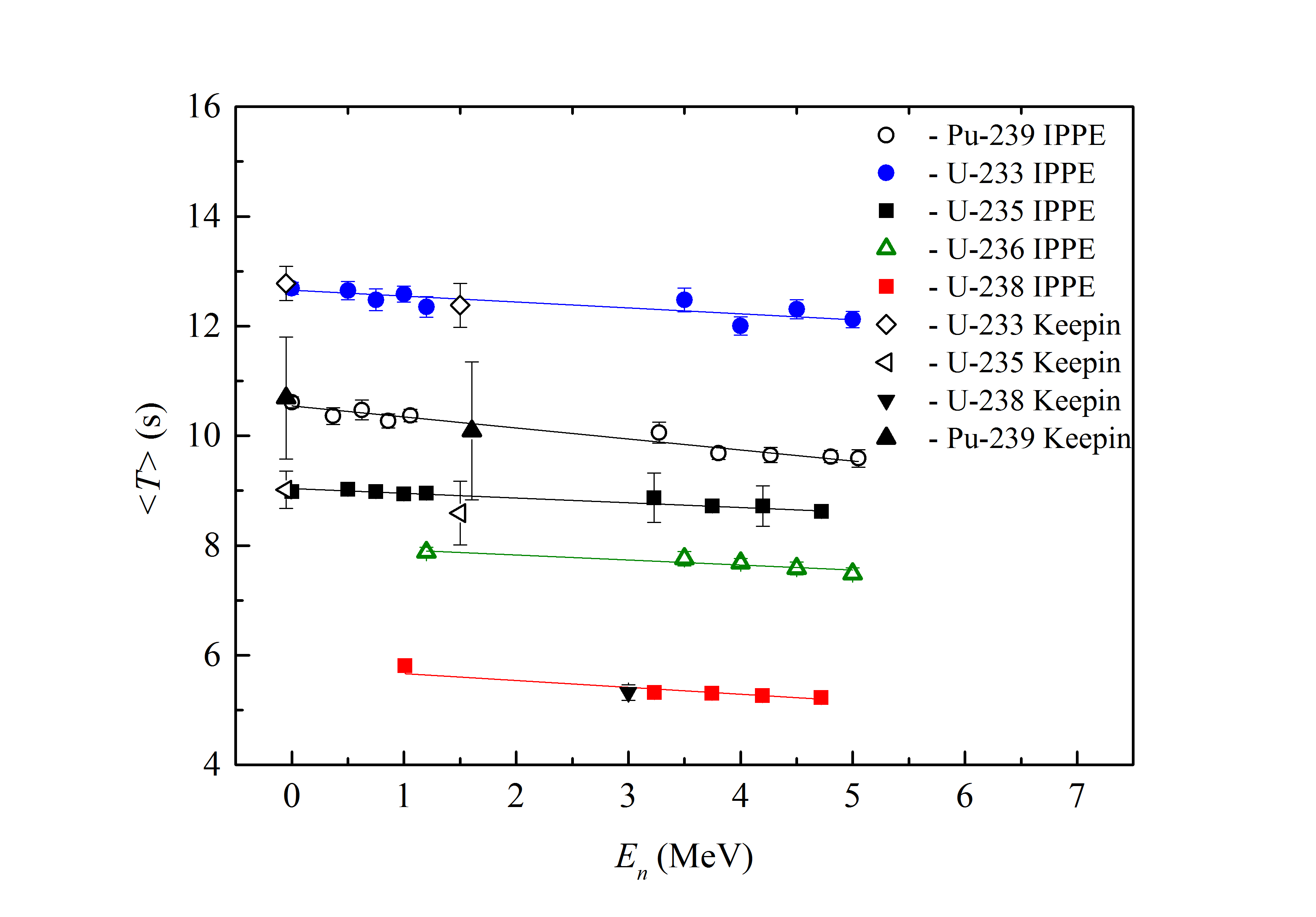}
\caption{The energy dependence of the average half-life of delayed-neutron precursors from neutron-induced fission of $^{233}$U, $^{235}$U, $^{236}$U, $^{238}$U and $^{239}$Pu based on the data from IPPE~\cite{Piksaikin02a,Piksaikin13} and Keepin \textit{et al.}~\cite{Keepin57}. The data of Maksyutenko \textit{et al.}~\cite{Spriggs02a} presented in Figs.~\ref{fig:5_2}-\ref{fig:5_3} were not considered.}
\label{fig:5_4}
\end{figure}

It can be seen from Figs.~\ref{fig:5_2} and~\ref{fig:5_3} that the average half-life of delayed-neutron precursors for all fissioning systems under consideration decreases with increasing energy of primary neutrons in the energy range from thermal to 5 MeV. The real scale of the variation in the relative abundances and half-lives of delayed-neutrons can be estimated in terms of the average half-life of delayed-neutron precursors with the help of the regression analysis of the IPPE data shown in Fig.~\ref{fig:5_4} for the uranium isotopes $^{233}$U, $^{235}$U, $^{236}$U, $^{238}$U and $^{239}$Pu. Results of the regression analysis of the energy dependence $\langle T(E_n)\rangle = A+B\cdot{E}_n$ are presented in Table~\ref{tab:5.1}.

\begin{table} 
\caption{Results of the regression analysis of the energy dependence $\langle T(E_n)\rangle = A+B\cdot{E}_n$.}
\label{tab:5.1}
\begin{tabular}{c|c|c} \hline \hline
Isotope&	Intercept value A &	Slope value B \\
       &       (s)            & (s/MeV) \\ \hline
\T $^{233}$U &	12.63$\pm $0.06	&-0.107$\pm $0.028	\\
$^{235}$U &	9.04$\pm $0.01	&-0.086$\pm $0.005	\\
$^{236}$U & 8.08$\pm $0.08	&-0.114$\pm $0.020	\\
$^{238}$U \footnote{The energy range for analyses of $^{238}$U data was 3.2-5 MeV.}&	5.56$\pm $0.05&	-0.070$\pm $0.012\\
$^{239}$Pu &	10.54$\pm $0.04&	-0.200$\pm $0.014\\ \hline \hline
\end{tabular}
\end{table}

The temporal parameters of delayed neutrons presented in this paragraph have been used in the evaluation of the relative abundances of delayed neutrons and half-lives of their precursors (see Section~\ref{Sec:Macro-Recommended}). 

\subsubsection{Energy dependence of total delayed neutron yield}\label{Sec:Macro-meas-new.2}
\begin{figure*} 
	\centering
\includegraphics[width=\linewidth]{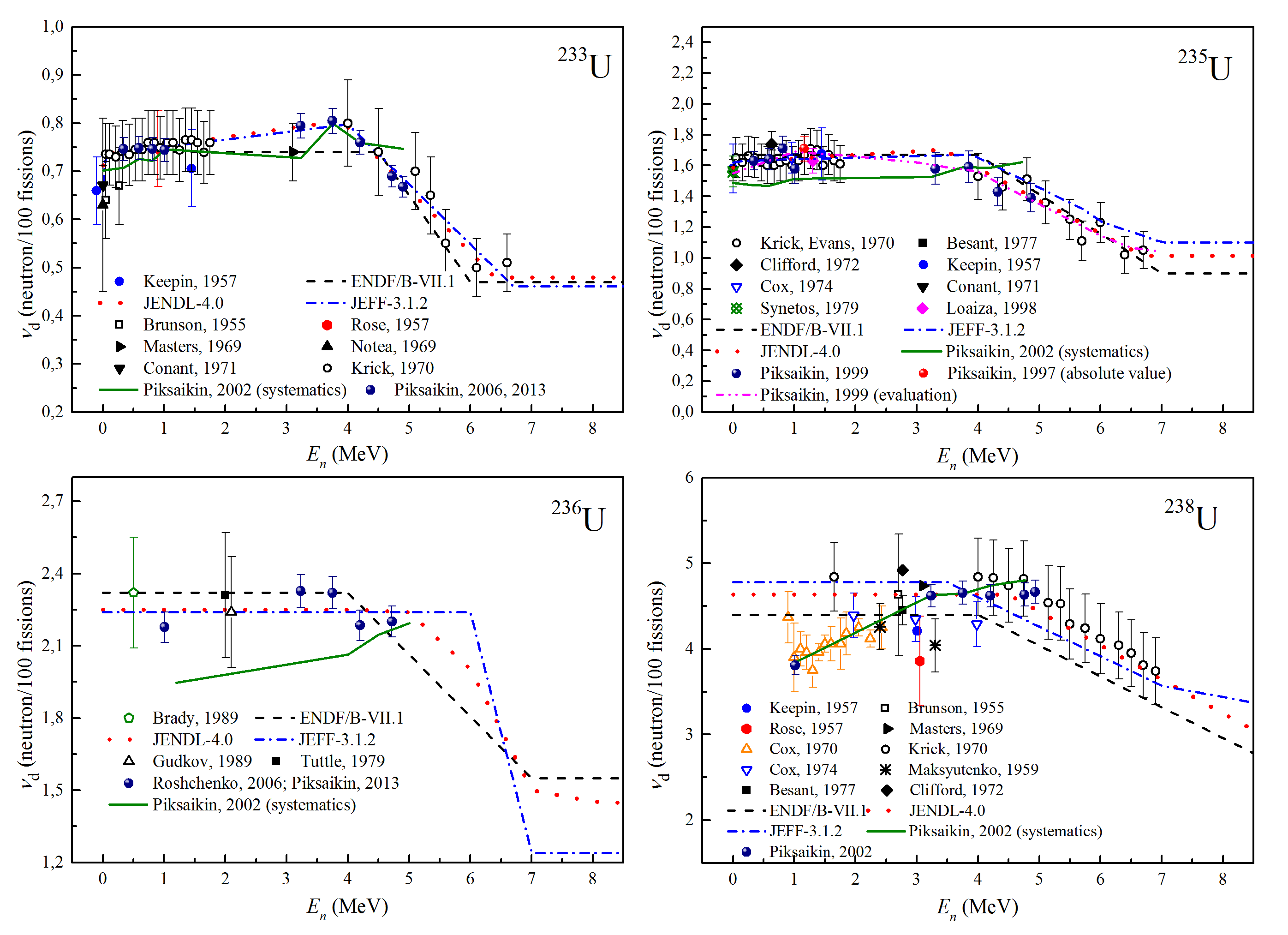}
\caption{The energy dependence of the total delayed-neutron yield from neutron induced fission of $^{233}$U, $^{235}$U, $^{236}$U, $^{238}$U. The references can be taken from the evaluation by Tuttle \cite{Tuttle75} except for the IPPE data by Piksaikin \textit{et al.} \cite{Piksaikin02b,Piksaikin02c,Piksaikin13,Piksaikin97a,Piksaikin99a} and Roshchenko \textit{et al.} \cite{Roshchenko06}.}
\label{fig:5_5}
\end{figure*}

\begin{figure*} 
	\centering
\includegraphics[width=\linewidth]{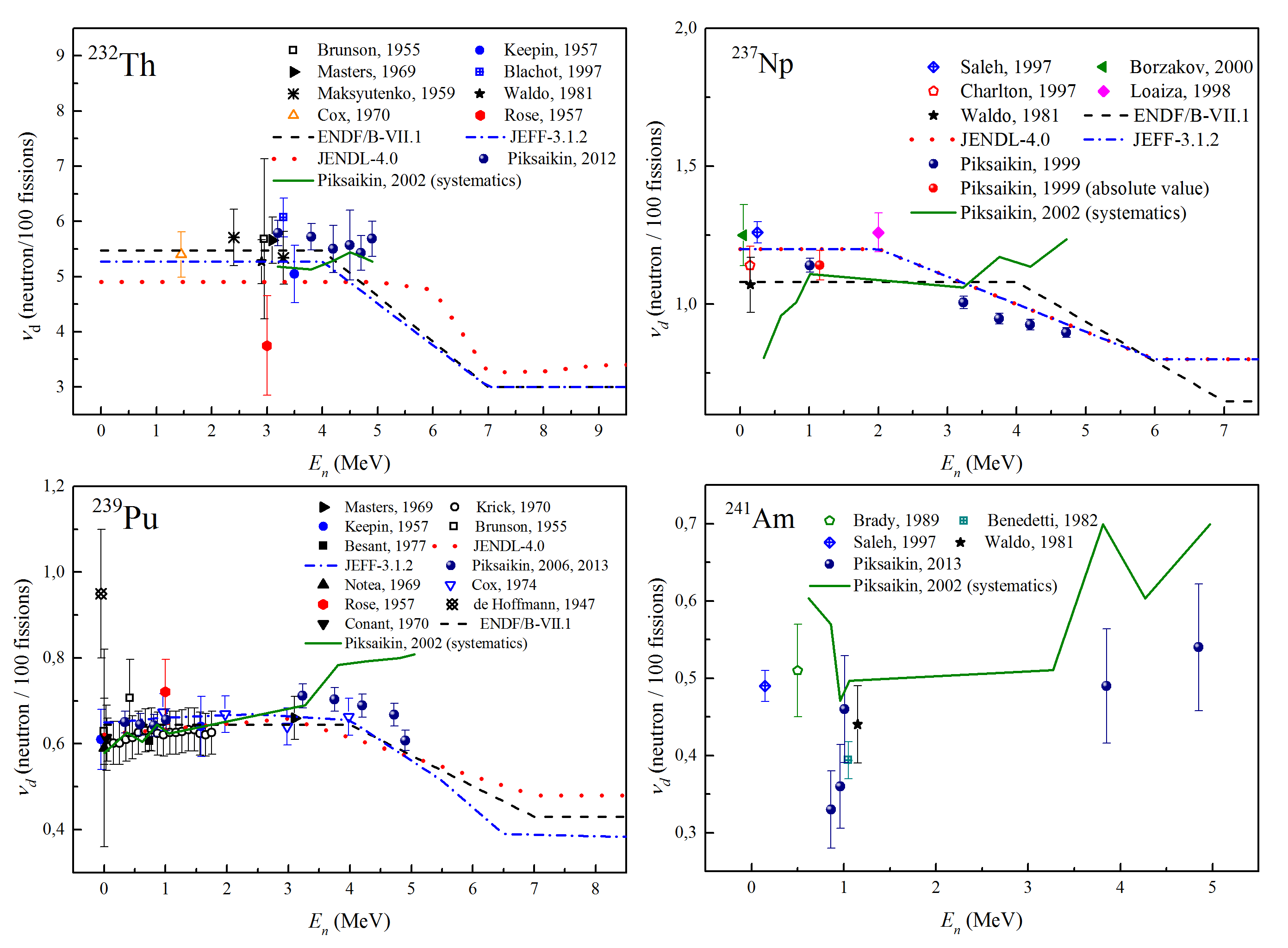}
\caption{The energy dependence of the total delayed-neutron yield from neutron-induced fission of $^{232}$Th, $^{237}$Np, $^{239}$Pu, $^{241}$Am. The references can be taken from the evaluation by Tuttle \cite{Tuttle75} except for the IPPE data of Piksaikin \textit{et al.} \cite{Piksaikin02b,Piksaikin13,Piksaikin12,Piksaikin97a,Piksaikin99a}.}
\label{fig:5_6}
\end{figure*}

The energy dependence of the total delayed-neutron yield from neutron-induced fission of $^{232}$Th, $^{233}$U, $^{235}$U, $^{236}$U, $^{238}$U, $^{239}$Pu, $^{237}$Np and $^{241}$Am in the energy range 0.3 -- 5 MeV measured at the IPPE accelerator facility is presented in Figs.~\ref{fig:5_5} and ~\ref{fig:5_6}. 

Most of the data have been obtained after the last recommendations were made for the major nuclides $^{235}$U, $^{238}$U and $^{239}$Pu \cite{DAngelo02}. The IPPE data are compared with both the earlier measured data and the evaluated data from the ENDF/B-VII.1~\cite{Chadwick2011}, JEFF-3.1.1~\cite{jeff3.1.1}, and JENDL-4.0~\cite{Shibata2011} data libraries. The total delayed-neutron yields obtained on the basis of the systematics and correlation properties of delayed neutrons $\nu_d(E_n)=a\cdot\langle T(E_n)\rangle^b$ \cite{Piksaikin02b} are also presented in Figs.~\ref{fig:5_5} and ~\ref{fig:5_6}.

The IPPE data \cite{Piksaikin02b,Piksaikin02c,Roshchenko06,Piksaikin13,Piksaikin97a,Piksaikin99a} are in agreement with the corresponding data measured by Krick \textit{et al.} \cite{Krick72} for target nuclides $^{233}$U, $^{235}$U, $^{238}$U and $^{239}$Pu. The energy dependence of both sets of data for the nuclides under consideration is well reproduced by the data analysis on the basis of the correlation properties of delayed neutrons $\nu_d(E_n)=a\cdot\langle T(E_n)\rangle^b$ at least in the energy range up to 3-4 MeV \cite{Piksaikin02b}. According to the WPEC-SG6 recommendations \cite{DAngelo02} based mainly on the measurements of the effective delayed neutron fraction $\beta_{eff}$ at thermal and fast assemblies, the energy dependence of the total delayed-neutron yield for $^{235}$U and $^{239}$Pu does not exceed 1$\%$. The energy dependence of $\nu_d$ in a 1 MeV interval calculated on the basis of Krick's data \cite{Krick72} shows the following relative increase: $^{233}$U - 1.98$\%$ ($\pm$0.6$\%$), $^{235}$U - 0.52 $\%$ ($\pm$1$\%$), $^{239}$Pu - 2.06 $\%$ ($\pm$0.5$\%$). In Table~\ref{tab:5.2} these data are compared with the corresponding data by Keepin \cite{Keepin57} and values calculated on the basis of the systematics $\nu_d(E_n)=a\cdot\langle T(E_n)\rangle^b$ \cite{Piksaikin02b}. As can be seen in the table both the experimental data and the data obtained on the basis of the systematics show noteworthy energy dependence of the total delayed neutron yield. The difference between the thermal and fast energy $\nu_d$ values obtained for $^{235}$U and $^{239}$Pu by the summation calculation on the basis of the JEFF-3.1.1 fission yield data~\cite{jeff3.1.1} is 24.5$\%$ and 21.2$\%$, respectively \cite{Gremyachkin15}.

\begin{table} 
\caption{Comparison of the energy dependence of the total delayed neutron yield from fission of $^{233}$U, $^{235}$U, $^{239}$Pu.}
\label{tab:5.2}
\begin{tabular}{c|c|c|c} \hline \hline
	& \multicolumn{3}{c}{$d\nu_d / \nu_d /1$ MeV $(\%)$} \\ \hline
Isotope	&Krick \textit{et al.} &	Keepin & IPPE~\cite{Piksaikin02b} \\
        &~\cite{Krick72} & ~\cite{Keepin57} & from $\nu_d(En)=a\cdot\langle T(En)\rangle^b$ \\ \hline
\T $^{233}$U &	1.98$\pm $0.6 &	4.80 &	1.84$\pm $0.62  \\
$^{235}$U &	0.52$\pm $1.0 &	2.47 &	2.04$\pm $1.31  \\
$^{239}$Pu&	2.06$\pm $0.5 &	3.13 &	6.26$\pm $1.69  \\ \hline \hline
\end{tabular}
\end{table}

Thus, this strong indication of an energy dependence of $\nu_d$ for $^{233}$U, $^{235}$U, and $^{239}$Pu observed in direct measurements is in contrast to the results obtained with help of the $\beta_{eff}$ methods \cite{DAngelo02}. More work is needed to determine the energy dependence of the total delayed neutron yields reliably. The most efficient and straightforward way to do this is to conduct relative measurements with accuracy $\pm1\%$ \cite{dAngelo2002a}. Apart from this, the reason why the summation method gives such a large difference between thermal and fast $\nu_d$ values also needs to be explored \cite{Gremyachkin15}. 

\subsubsection{Integral energy spectra of delayed neutrons from fission of $^{235}$U by thermal neutrons}\label{Sec:Macro-meas-spectra}

\begin{figure} 
	\centering
\includegraphics[width=\linewidth]{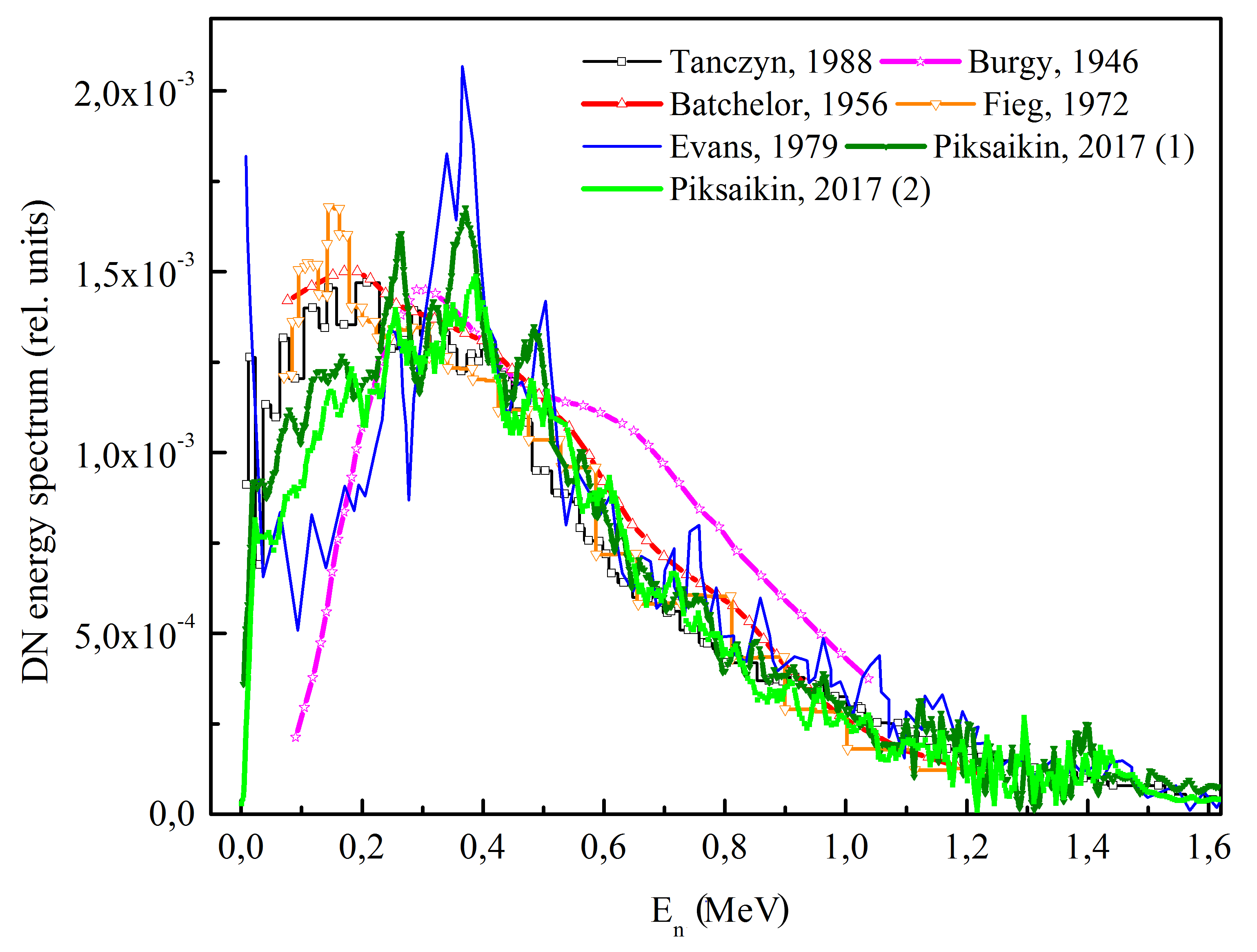}
\caption{Quasi-equilibrium spectrum of delayed neutrons from the fission of $^{235}$U by epi-thermal neutrons~\cite{Piksaikin17}. IN IPPE spectrum (1): counting time is 1.88 s after irradiation of 120 s. In IPPE spectrum (2): counting time is 0.88 s after irradiation of 20 s. Quasi-equilibrium spectra by other authors have been calculated with the help of the group spectra obtained by decomposition of appropriate aggregate spectra in different time windows. The data references can be found in the review of Das~\cite{Das94}.}
\label{fig:5_7}
\end{figure}

A comprehensive analysis of the composite delayed neutron (DN) spectra measurements can be found in the review by Das \cite{Das94}. This review includes also the energy spectra from individual precursors measured by \cite{Kratz79}, \cite{Rudstam74,Rudstam77} and \cite{Greenwood1985,Greenwood1997}, and the spectra of 235 nuclides calculated with the help of the evaporation model. Recently the integral energy spectra were measured from the epithermal neutron induced fission of $^{235}$U at the IPPE accelerator based neutron beams. The measurements were made to emphasize particular groups of delayed neutron precursors using the high-resolution $^3$He-spectrometer of FNS-1 type and applying different irradiation and counting time intervals. Two sets of measurements were performed. The first one was made with an irradiation time interval of 120 s, which was followed by measurements of delayed neutron spectra in a sequence of time intervals as follows: 0.12-2 s, 2-12 s, 12-22 s, 22-32 s, and 32-152 s after the end of irradiation. The second set of measurements was made with an irradiation time interval of 20 s and measurements of delayed neutron spectra in the time intervals: 0.12-1 s, 1-2 s, 2-3 s, 3-4 s, and 4-44 s after the end of irradiation. About 3000 irradiation-counting cycles have been made within these measurements. The time sequence of irradiation and counting intervals used made it possible to consider some of the measured spectra as quasi-equilibrium. The comparison of these quasi-equilibrium spectra with the corresponding spectra from other experiments are presented in Fig.~\ref{fig:5_7}.

It can be seen from Fig.~\ref{fig:5_7} that the IPPE \cite{Piksaikin17} near-equilibrium spectrum has more pronounced peak structure compared to the corresponding experimental data obtained by other authors. The peak structure observed in the data of Evans \textit{et al.} \cite{Evans79} is similar to the peak structure of the IPPE data but the low-energy part of the spectra (below 300 keV) is about 30\% lower than both the IPPE data and the data obtained by the summation method using the microscopic spectra from individual precursors. This could be explained by possible errors in the determination of the efficiency of the $^3$He-spectrometer or the correction that accounts for the neutron flux attenuation in the shielding used for rejection of gamma rays \cite{Piksaikin17}. Another important feature of the present near-equilibrium spectrum is the lower intensity of DN below 200 keV as compared to the earlier experiments with the exception of the data of Evans \textit{et al.} \cite{Evans79}. Comparison of the IPPE integral energy spectra measured at different time intervals after the end of the irradiation with the spectra calculated on the basis of microscopic DN data is presented in Section~\ref{Sec:Macro-Integralspec}. More details of the experimental procedure employed in the IPPE experiments can be found in \cite{Piksaikin17}.

\subsubsection{ALDEN: new measurements of delayed neutron data}\label{Sec:Macro-meas-future}
Apart from the experimental activities at IPPE (Obninsk), recently new experiments have been conducted at the Institut Laue-Langevin (ILL) to measure aggregate delayed-neutron data. The ALDEN (Average Lifetime of DElayed Neutrons) project consists of a series of experiments designed to measure delayed neutron group abundances with the aim of improving the data and providing realistic uncertainties and correlations for the most important fissioning systems involved in reactor applications ($^{235}$U, $^{238}$U, $^{239}$Pu, $^{241}$Pu) \cite{Foligno19}. 

The first experimental campaign of ALDEN
was held at the end of 2018 at ILL (Institut Laue-Langevin)~\cite{Foligno19}. The ILL reactor produces a very intense and well-characterized thermal neutron flux. The experiment consisted of irradiating the fissile target and letting it decay naturally. Delayed neutrons were measured by LOENIE, a $^{3}$He detector, optimally designed to have an efficiency that is independent of the incident neutron energy~\cite{Foligno19}. LOENIE is composed of a cylindrical polyethylene block with a central hole for the fissile material and 16 smaller holes for the $^{3}$He counters. An airtight fission chamber containing a deposit of fissile material was used as target in the experiment. With the fission chamber one estimates the fission rate during the irradiation phase, which is then used to normalize the delayed-neutron activity curve.  

During the decay phase, the $^{3}$He counters register delayed neutrons as a function of time. This time-dependent activity curve was used to derive the groups abundances in the 8-group model. Several irradiation cycles were performed, each with a different irradiation duration in order to saturate a specific group of precursors. 

\begin{figure} 
	\centering
    \includegraphics[width=\linewidth]{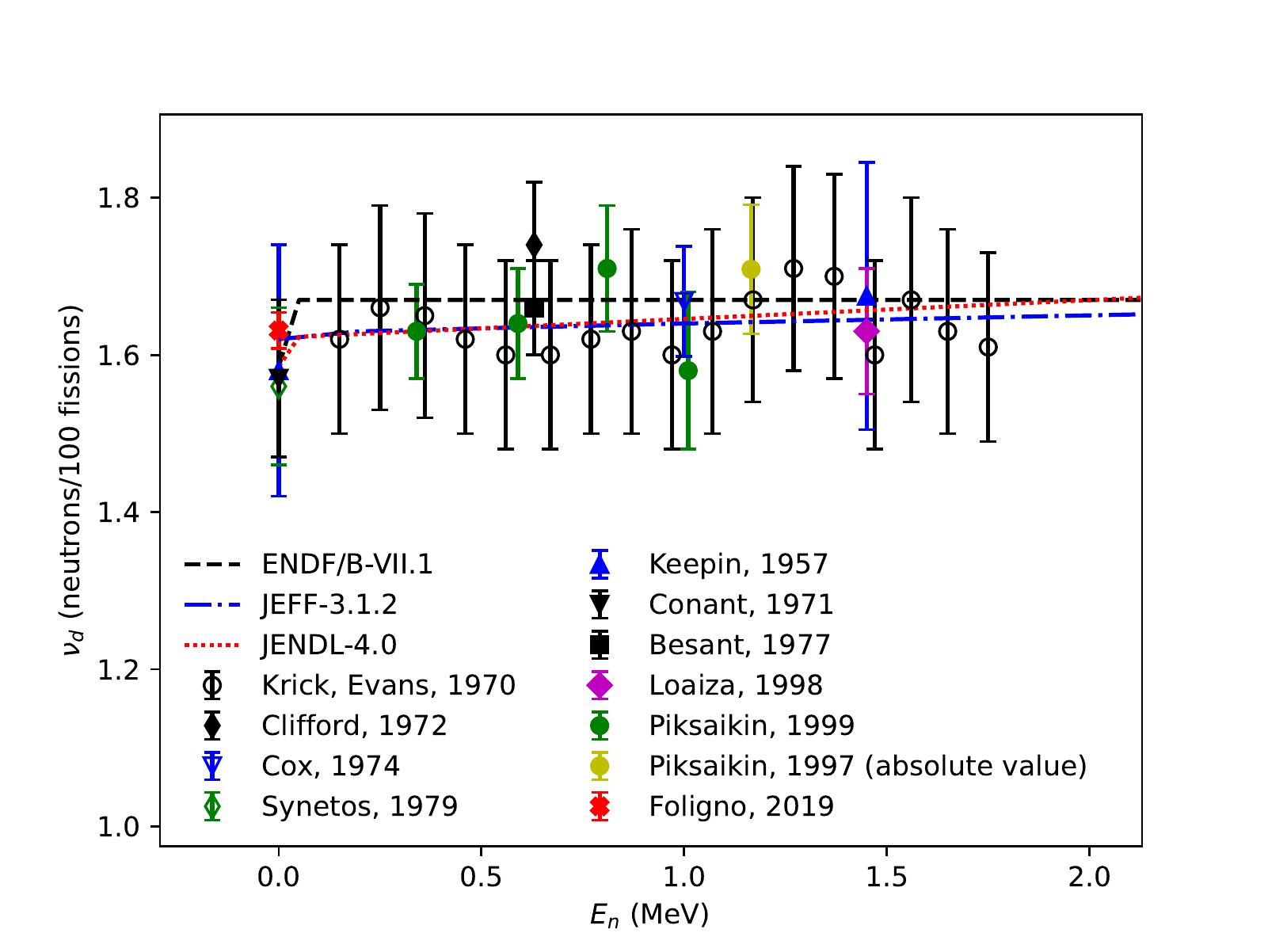}
\caption{The energy dependence of the total delayed-neutron yield from neutron induced fission of $^{235}$U. The red cross comes from the 2018 ALDEN campaign~\cite{Foligno19}. The other experimental data are taken from the evaluation of Tuttle \cite{Tuttle75} except for the IPPE data which are from by Piksaikin \textit{et al.} \cite{Piksaikin97a,Piksaikin99a}.}
\label{fig:nud_Foligno}
\end{figure}

The result obtained for thermal neutron-induced fission of $^{235}$U from this first experimental campaign in 2018 is compared with other experimental data and evaluated libraries in Fig.~\ref{fig:nud_Foligno}. More ALDEN measurements for the other major actinides are planned to take place in the near future.

\subsection{Estimation of energy dependence of total delayed neutron yields}\label{Sec:Macro-incident-energy}
As discussed in Section~\ref{Sec:Macro-meas-new.2}, 
the incident-energy dependence of the total delayed neutron yield has been measured extensively by the IPPE group in the past twenty years \cite{Piksaikin97, Roshchenko06, Piksaikin02c, Piksaikin06, Piksaikin13}. On the other hand, the theoretical study of the energy-dependence of the total delayed neutron yield has not been pursued that actively. 
The evaluated libraries, such as ENDF/B-VIII~\cite{Brown2018}, JEFF-3.1.1~\cite{jeff3.1.1} and JENDL-4.0~\cite{Shibata2011}, either perform an interpolation of the existing data or use the systematics proposed by Tuttle~\cite{Tuttle79}, Benedetti~\cite{Benedetti82}, Waldo~\cite{Waldo81}, and Manero \cite{Manero72}. As a result, the evaluated values lie on a structureless curve. 

The dependence of total delayed neutron yields on the incident neutron energy results from two factors. 
One is the dependence of fission product yields on the incident neutron energy. 
Fission yield distributions are usually determined by the potential energy surface of fissioning nuclei, which varies with incident neutron energy and excitation energy of the compound nuclei. 
The other factor
is the competition with open channels other than ($n,f$). In particular, ($n,n'f$) and ($n,2nf$) channels have a large impact on the energy dependence of the total delayed neutron yields. Similarly, other channels accompanying multiple particle emissions like ($n,3nf$) also affect the energy dependence of delayed neutron emission. 

In Ref.~\cite{Minato18}, a new method for evaluating the incident neutron energy dependence of total delayed neutron yields using least square fitting was presented. 
The model adopted the incident neutron energy dependence of fission product yields, the multi-chance fission, and the evaluated decay data of Ref.~\cite{Birch2015} ($Z$ = 2 -- 28). 
As mentioned in Section~\ref{Sec:Micro-compilation}, new evaluated ($T_{1/2},P_{n}$ data for $Z > 28$ nuclei were produced within the CRP \cite{Liang2020} in addition to Ref.~\cite{Birch2015}. New calculations of the incident neutron energy dependence have thus been performed using the new CRP data of section~\ref{Sec:Micro-compilation}, and the results are presented in the following.

%
In the model of ~\cite{Minato18}, the most probable charge $Z_p$ is expressed by the following equations;
\begin{equation}
Z_p(A,E_n)=\alpha Z_p^{0}(A)+\beta c(A)E_n+\gamma E_n^2.
\label{energdep}
\end{equation}
The function $c(A)$ accounts for the fact that the energy dependence of $Z_p$ changes smoothly for light fragments and relatively rapidly for heavy fragments~\cite{Nethaway74,Roshchenko06b}. 
The parameter $Z_p^0(A)$ is taken from the evaluation of England and Rider~\cite{England94}. 

To take into account the odd-even effect observed in fission yield distributions, the method proposed by Madland and England~\cite{Madland76} was adopted, where $(1+C_1)$, $(1-C_1)$, $(1-C_2)$, and $(1+C_2)$ are multiplied by the fission yields of even-even, odd-odd, odd-even, and even-odd nuclei, respectively. The renormalization is carried out after applying the multiplication factors. The energy dependence of the odd-even effect is approximated as
\begin{equation}
C_k(E_n)=C_k^0\frac{2}{1+\exp\left(\xi E_n\right)}
\label{pairing}
\end{equation}  
where $C_k^0$ ($k=1,2$) are taken from Madland and England's evaluation~\cite{Madland76}. The values of $\alpha, \beta$, $\gamma$ in Eq.~\eqref{energdep} and $\xi$ in Eq.~\eqref{pairing} are determined by a least squares fitting to experimental data on total delayed neutron yields for $^{233,235,236,238}$U and $^{239-242}$Pu.  More details about the method can be found in Ref.~\cite{Minato18}.

For $^{240,241,242}$Pu, the parameter $\gamma$ was set equal to zero because there are 
not enough available experimental data points for all the parameters to be determined uniquely.
The least square fitting was thus performed only for $\alpha, \beta$, and $\xi$ for the above-mentioned actinides. The parameters obtained from fitting the new evaluated CRP data are listed in Table~\ref{parameters}, The differences observed between the values in  this table and those in Table~3 of \cite{Minato18} reflect the impact of the new evaluated CRP data ($T_{1/2},P_{n}$) from Section~\ref{Sec:Micro-compilation}.
It should be noted that in this work, the total delayed neutron yield of $^{234}$U (i.e. from $^{234}$U($n,f$)) taken from JENDL-4.0 is multiplied by a factor of 1.2. This increase improves the results obtained for the delayed neutron yields of $^{235}$U and $^{236}$U at energies where the second-chance fission $^{235}$U($n,n'f$) or the third-chance fission $^{236}$U($n,2nf$) occurs.
This prescription is validated by the large ambiguity affecting the original total delayed neutron yield value of $^{234}$U of JENDL-4.0 as it is obtained from Tuttle's systematics~\cite{Tuttle79}.

\begin{table}
\centering
\caption{Parameters determined by the least square fitting to the experimental data.}
\begin{tabular}{c|crcc}\hline \hline
Nuclide  & $\alpha$ & \multicolumn{1}{c}{$\beta$} & \multicolumn{1}{c}{$\gamma$} & $\xi$ \\
\hline
U-$233$& $1.00380$ & $ 0.855$ & $-5.083\times 10^{-4}$ & $1.272$ \\
U-$235$& $0.99994$ & $ 1.520$ & $-8.631\times 10^{-4}$ & $1.124$ \\
U-$236$& $1.00028$ & $-0.195$ & $\,1.596\times 10^{-3}$ & $1.924$ \\
U-$238$& $1.00411$ & $-1.922$ & $\,4.101\times 10^{-3}$ & $0.948$ \\
\hline
Pu-$239$& $1.00013$ & $-0.010$ & $6.816\times 10^{-4}$ & $0.530$ \\
Pu-$240$& $1.00166$ & $0.326$ & $0$ & $0.379$ \\
Pu-$241$& $0.99951$ & $0.794$ & $0$ & $0.741$ \\
Pu-$242$& $1.00373$ & $0.414$ & $0$ & $1.554$ \\ \hline \hline
\end{tabular}
\label{parameters}
\end{table}

\begin{figure}[!htb]
\centering
\includegraphics[width=0.90\linewidth]{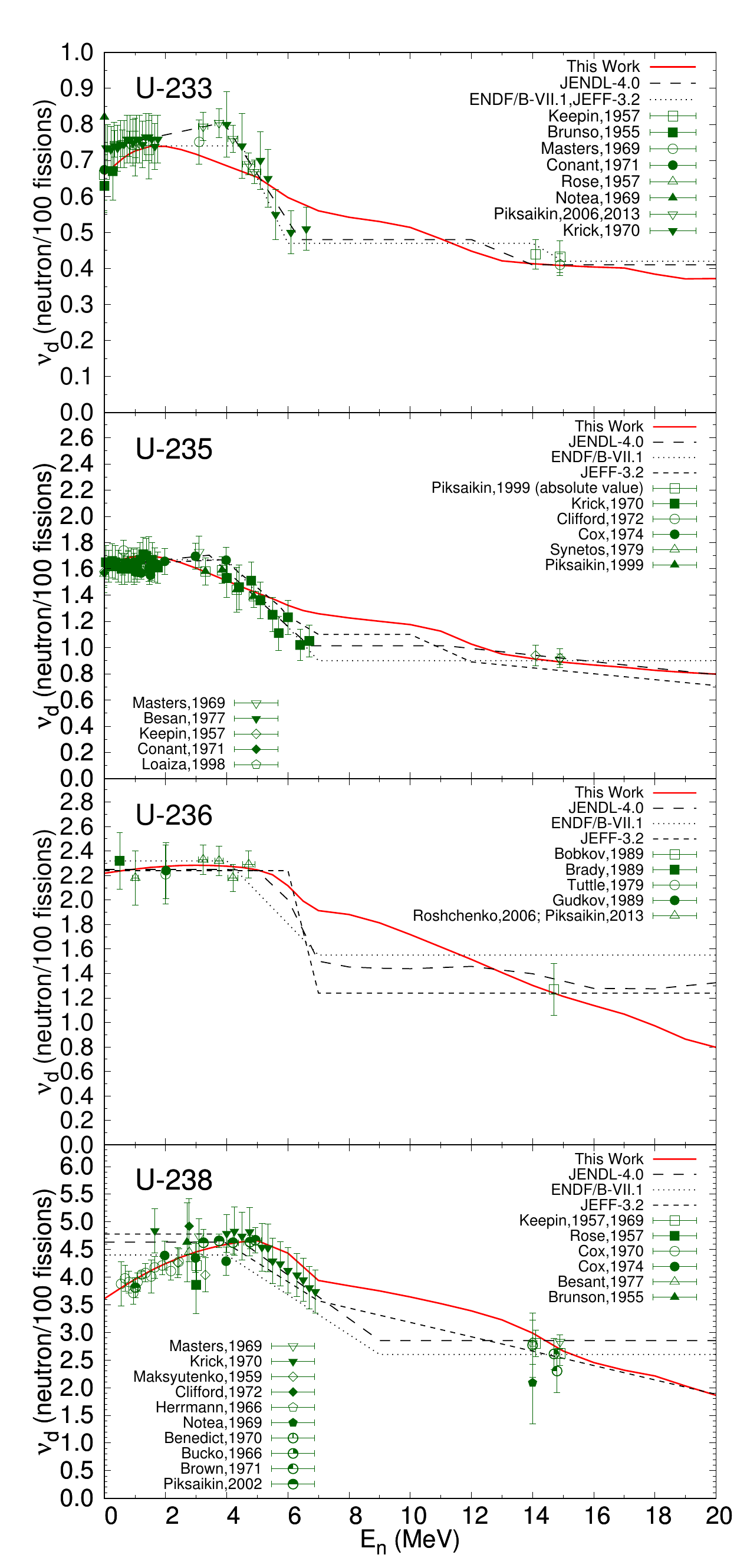}
\caption{Delayed neutron yield of uranium isotopes with $A=233$ to $238$. The data are taken from Bobkov \textit{et al.}~\cite{Bobkov89}, Loaiza \textit{et al.}~\cite{Loaiza98}, Brady \textit{et al.}~\cite{Brady89}, Gudkov \textit{et al.}~\cite{Gudkov89}, and the evaluation by Tuttle~\cite{Tuttle75,Tuttle79}, while the IPPE data are taken from Piksaikin \textit{et al.} \cite{Piksaikin02b,Piksaikin02c,Piksaikin13,Piksaikin97a,Piksaikin99a} and Roshchenko \textit{et al.} \cite{Roshchenko06}.}
\label{Fig.5.2.1}
\end{figure}

The total delayed neutron yields for uranium isotopes of $A=233,\,235,\,236,\,238$ are shown in Fig.~\ref{Fig.5.2.1} together with the experimental data and the evaluated values from JENDL-4.0~\cite{Shibata2011}, ENDF/B-VII.1~\cite{Chadwick2011} and JEFF-3.1.1~ \cite{jeff3.1.1} libraries. The energy dependence of the experimental data for $^{233,235,236,238}$U is reproduced reasonably well with the new parameterization. 


The present results agree roughly with the evaluated data, however, several differences are noted: First, the new results give a smooth curve up to about 4 MeV, while the evaluated data give a straight line. Second, they are systematically higher than the evaluated data in the energy region from 8 to 10 MeV. At energies above 15 MeV, the new results are lower than the evaluated data for $^{236,238}$U. At these high energies, measurements of total delayed neutron yields are required to improve the evaluated data.

\begin{figure}[!htb]
\centering
\includegraphics[width=0.90\linewidth]{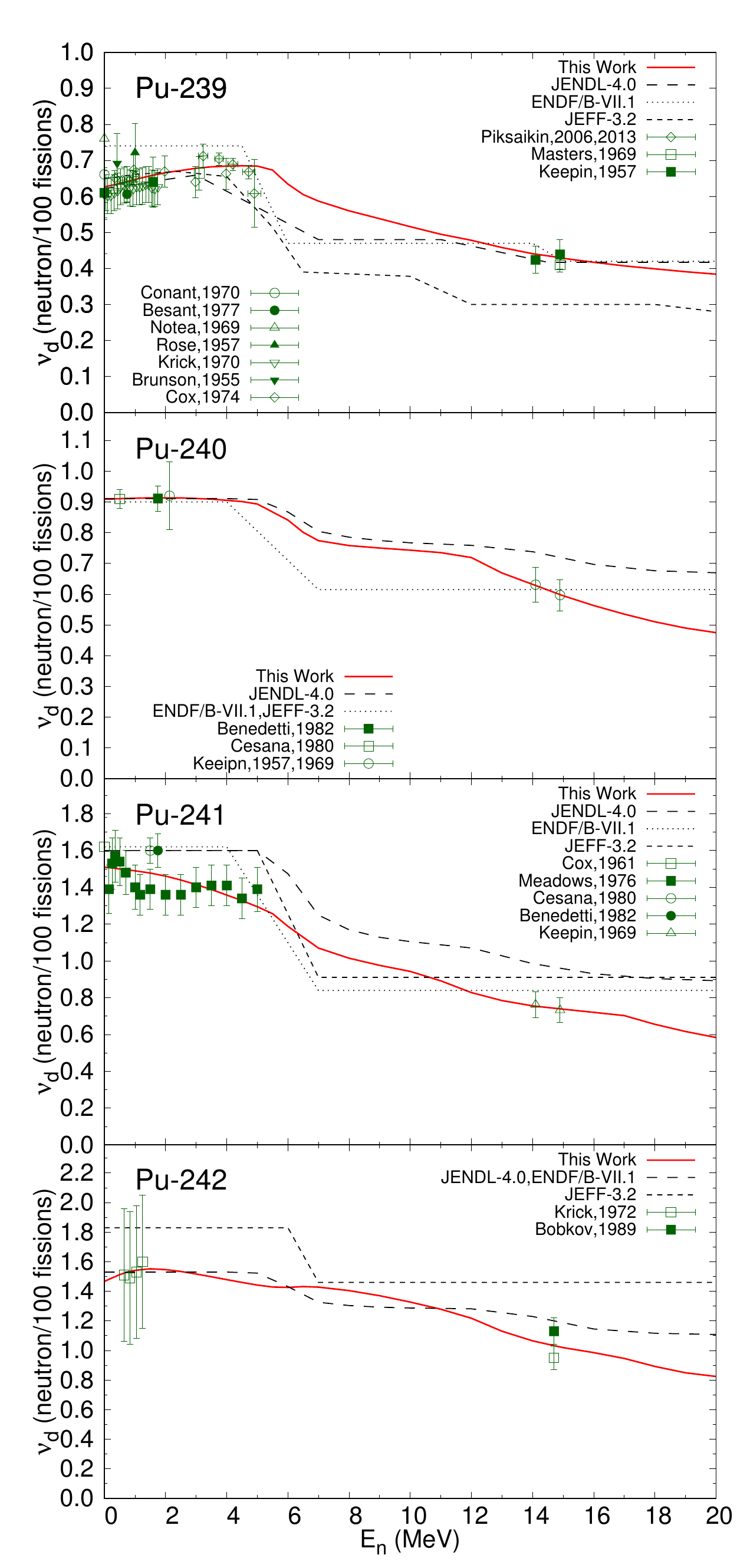}
\caption{Delayed neutron yield of plutonium isotopes with $A=239$ to $242$. The data are taken from Bobkov \textit{et al.}~\cite{Bobkov89}, Benedetti \textit{et al.}~\cite{Benedetti82}, Cesana \textit{et al.}~\cite{Cesana80,Cesana80b}, and the evaluation by Tuttle~\cite{Tuttle75}, while the IPPE data are taken from Piksaikin \textit{et al.} \cite{Piksaikin13}.}
\label{Fig.5.2.2}
\end{figure}

Figure~\ref{Fig.5.2.2} shows the total delayed neutron yields for plutonium isotopes of $A=239,\,240,\,241,\,242$. As seen in the uranium isotopes, the energy dependence of the total delayed neutron yields is well reproduced. 
When compared with the evaluated data, the present results are closest to JENDL-4.0 up to around 12 MeV, but tend to be lower above 14 MeV except for $^{239}$Pu. 
However, there are fewer available experimental data for the plutonium isotopes $^{240,241,242}$Pu compared to the uranium isotopes. 
As in the case of the uranium isotopes, measurements of total delayed neutron yields at energies close to the second-chance fission threshold are required, in order to make a reliable extrapolation up to 20 MeV.

\begin{figure}[!htb]
\centering
\includegraphics[width=0.99\linewidth]{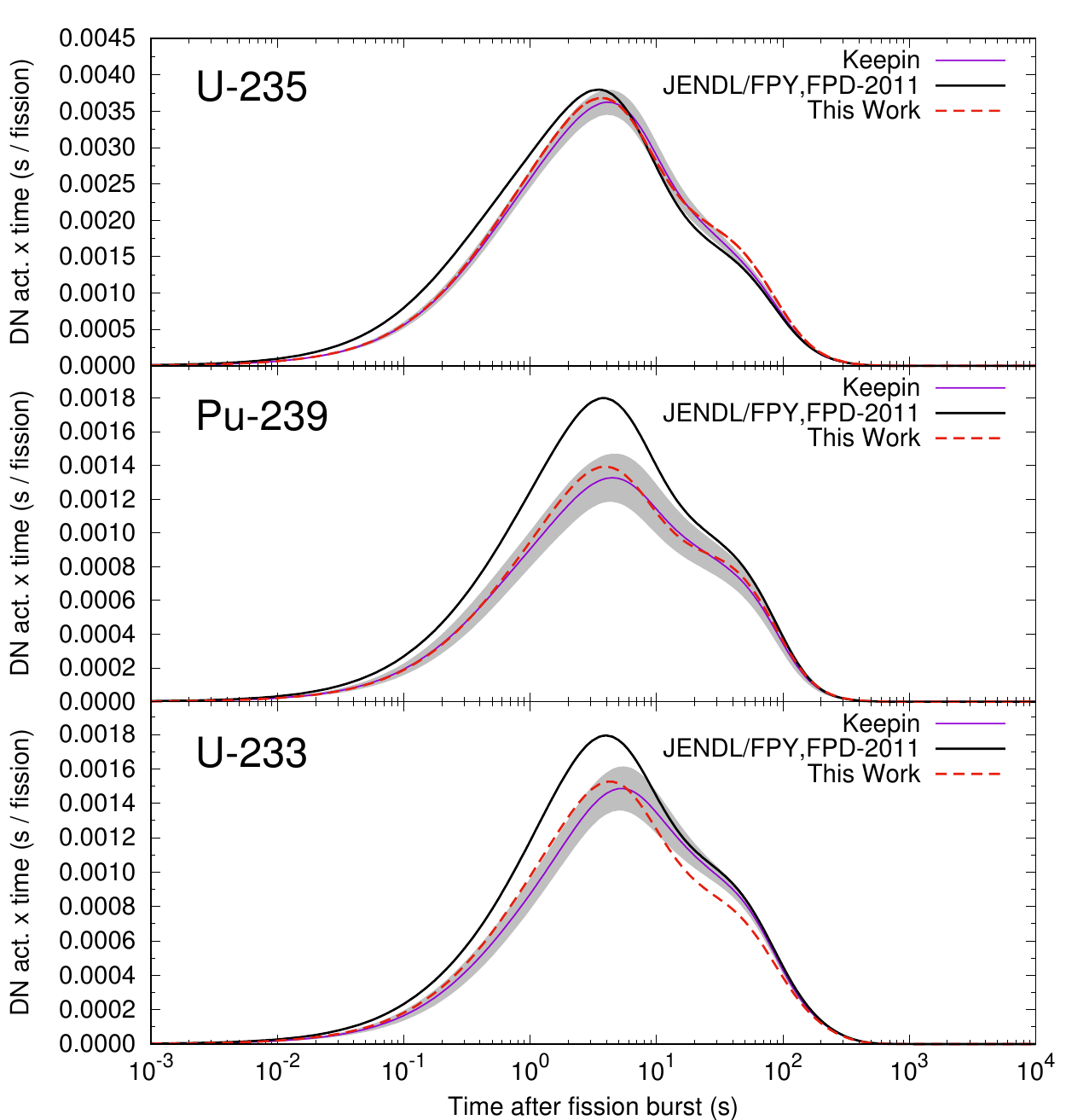}
\caption{Delayed neutron activities multiplied by time $t$ after neutron radiation for thermal neutron fission of $^{235}$U (top), $^{239}$Pu (middle), and $^{233}$U (bottom). Instant neutron radiation is assumed. The present calculation indicated by the dashed lines are compared with the summation calculation of JENDL/FPY-2011 and FPD-2011 (JENDL/FPY,FPD-2011) \cite{Katakura11} (solid line) and the results of Keepin's six group \cite{Keepin57} (errors are shown by the shade).}
\label{Fig.5.2.3}
\end{figure}

Figure~\ref{Fig.5.2.3} shows the delayed neutron activities for thermal neutron fission of $^{235}$U, $^{239}$Pu and $^{233}$U. The activities are plotted as a function of time after fission burst, $t$, and the figures compare the results obtained with the current model with Keepin's six group constants~\cite{Keepin57} and the summation calculations using
the fission product yields from the JENDL/FPY-2011 library and the radioactive decay data from the JENDL/FPD-2011 library (JENDL/FPY,FPD-2011)~\cite{Katakura11}. For $^{235}$U, the present calculations and JENDL/FPY,FPD-2011 reproduce Keepin's data reasonably. However, JENDL/FPY,FPD-2011 overestimates the delayed neutron activity at times $t<5$ sec. For $^{239}$Pu and $^{233}$U, the delayed neutron activities of JENDL/FPY,FPD-2011 overestimate those of Keepin's six group evaluation~\cite{Keepin57} significantly, while the present calculations are rather close to Keepin's results within errors. 

\begin{figure}[!htb]
\includegraphics[width=0.99\linewidth]{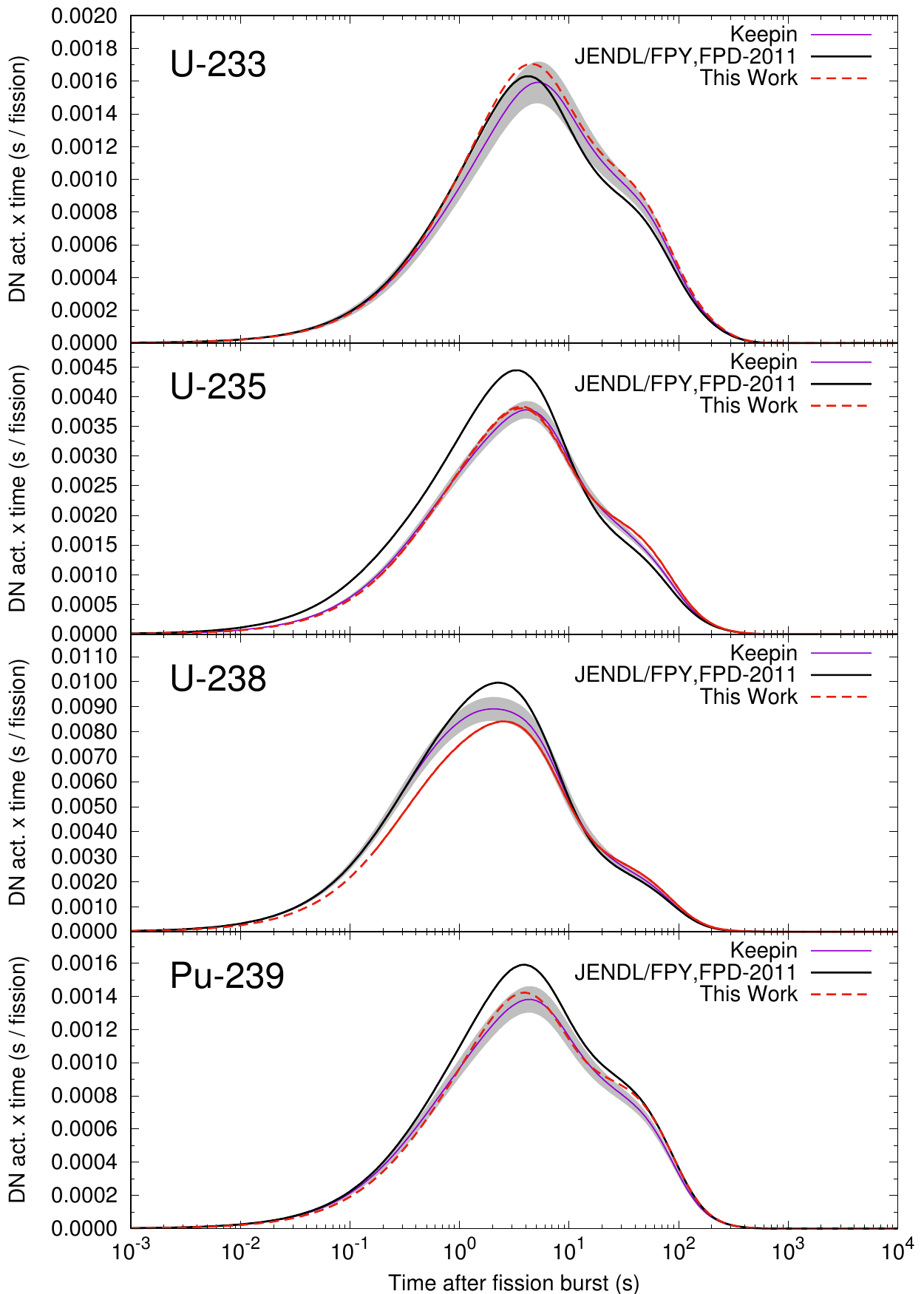}
\caption{Same as Fig. \ref{Fig.5.2.3}, but for fast neutron fission of $^{235}$U, $^{238}$U, $^{233}$U, and $^{239}$Pu.}
\label{Fig.5.2.4}
\end{figure}

Figure~\ref{Fig.5.2.4} shows the delayed neutron activities for the fast neutron fission of $^{235}$U, $^{238}$U, $^{233}$U, $^{239}$Pu and $^{240}$Pu. Similar to the results observed for thermal neutron fission, the delayed neutron activity obtained with the new model calculations is close to Keepin's six group evaluations within errors. JENDL/FPY,FPD-2011 overestimates Keepin's data significantly except for $^{233}$U.

To conclude, the incident neutron energy dependence of the delayed neutron yields of uranium and plutonium isotopes was re-estimated with the model of Ref.~\cite{Minato18} using the new evaluated CRP data introduced in Sect.~\ref{Sec:Micro-compilation}. The present model reproduces the experimental total delayed neutron yields reasonably well. Moreover, the delayed neutron activities at thermal and fast neutron fissions evaluated by Keepin's six group model are also well reproduced. 

However, new evaluations of fission yield data which are able to reproduce both decay heat and delayed-neutron emission data simultaneously, using the summation method, are urgently needed.  New measurements of total delayed neutron yields at different incident neutron energies are also important for the evaluation of both fission yields as well as delayed neutron yields.

\section{ SUMMATION CALCULATIONS}
\label{Sec:Macro-Summation}

Summation and time-dependent calculations of total delayed neutron yields, delayed neutron spectra and delayed neutron activities using the CRP evaluated microscopic ($T_{1/2}, P_n$) data are compared with recommended data and evaluated libraries in this section. The aim is to assess the new CRP data against a broad range of available macroscopic data.

\subsection{A numerical ($T_{1/2},P_{n}$) data file for practical applications} \label{Sec:Macro-summation-file}

A numerical file of definitive $T_{1/2}$ and $P_n$ values has been created for use in summation and other calculations for applications spanning fission reactor technologies, anti-neutrino spectra and nuclear astrophysics. The file is based exclusively on the evaluated data produced by the CRP as described in Sect.~\ref{Sec:Micro-compilation}~\cite{database}.
These data are purely experimental without any input from systematics or theory. It is left to the user to decide how to treat discrepant values (when compared with theory or systematics) or missing values due to lack of experimental data.

Furthermore, the ``non-numerical" features of the ($T_{1/2}, P_{n}$) tables, such as limits or approximate values, have been replaced with purely numerical values, and asymmetric limits have been symmetrized. A list of all the modifications implemented in the numerical file with respect to the tables available online is given below:

\begin{itemize}
\item upper limits were symmetrized, i.e. if $P_{1n} < 0.2$ then the adopted value in the numerical file would be $P_{n} = 0.1 \pm 0.1$. 

\item lower limits are 'symmetrized' as follows:
$P_{n} > x$ becomes $P_{n}=y \pm z$ so that $x=y-z$ and $100=y+z$. When the latter cannot be enforced, then the lower limit is doubled with an uncertainty of 50\%, i.e. $y=2x$ and $z=x$, with $y+z=3x \le 100$.

\item asymmetric uncertainties were symmetrized following the prescription: 

\begin{minipage}{0.3\textwidth}
\begin{equation*}
x(^a_b) \rightarrow y\pm z;\, y=x+\frac{a-b}{2},\, z = \frac{a+b}{2}
\end{equation*}
\end{minipage}

\item approximate values were assigned 50 $\%$ uncertainty, i.e. $P_{1n} \approx 0.2$ was replaced by $P_{1n} = 0.2 \pm 0.1$

\item when multiple activities are measured but cannot be distinguished, i.e. when a $P_{n}$ is given for both the ground state and isomeric states, then the latter is split equally among the states while the uncertainty for each state remains the same as the original recommended one. As an example, one takes for $^{80}$Ga a recommended value for both, the ground state and the isomeric state , of $P_{1n}$ = 0.90(7)\% . In the numerical file, the value $P_{1n}$ = 0.45(7)\% was adopted for the ground state and isomeric state, respectively. If the original uncertainty is larger than the split $P_{1n}$ value, one can adopt a 100$\%$ uncertainty for the split value. For example, in the case of $^{116}$Rh, where $P_{1n}<2.1$\% for both ground state and isomeric state, the limit is symmetrized to give $P_{1n}= 1.05 \pm 1.05$ \% and the resulting $P_n$ is split among the two states with 100$\%$ uncertainty, i.e. as $P_{1n}= 0.525\pm 0.525$ \% for the ground state and isomeric state, respectively.

\item there are a few cases ($^{38,39}$Al, $^{43}$P) where the reported measured $P_{n}$ value can take unphysical values within the reported uncertainty range. In such cases the uncertainty was adjusted so that $P_{n}$ lies within the physical limits ($0\le P_n \le 100$ \%). 

\item zero values for Q$_{\beta xn}$ and $P_{xn}$ have the following meaning: (i) there are no measurements available, (ii) the decay mode is forbidden. 

\item nuclides that have only been identified in measurements and have no measured $T_{1/2}$ or $P_n$ value are not included in the numerical file.

\end{itemize}

The format of the file is as follows:

\noindent Nuclide name, Z, A, liso (index indicating order of states, i.e. for ground state liso = 0, for 1st isomeric state liso = 1, etc.), Level Energy, $\Delta$ Level Energy, $\% \beta-$, $\Delta \% \beta-$, $Q_{\beta n1}$, $\Delta Q_{\beta 1n}$, $Q_{\beta n1}$, $\Delta Q_{\beta 2n}$, $Q_{\beta n3}$, $\Delta Q_{\beta 3n}$, $T_{1/2}$, $\Delta T_{1/2}$, $P_{1n}$, $\Delta P_{1n}$, $P_{2n}$, $\Delta P_{2n}$, $P_{3n}$, $\Delta P_{3n}$.

The numerical file has been used in the summation calculations described in the following sections with the aim of verifying and validating the new evaluated $P_n$ tables produced by the CRP. It is available for downloading at the IAEA Beta-delayed neutron database~\cite{database}. 
A more detailed description of the contents of the IAEA Beta-delayed neutron database is given in Section~\ref{Sec:Database}.

\subsection{Total delayed neutron yields  }\label{Sec:Macro-summation-yields}
%
%

\subsubsection{Basic summation calculations}\label{Sec:Macro-summation-yields-basic}

In this section, summation calculations of the total delayed neutron yield $\nu_d$ performed with the CRP data ($T_{1/2}, P_n$) are compared with other decay data libraries and with recommended values.

\paragraph{Comparison with evaluated libraries.}

The first step in verifying and testing the new evaluated CRP ($T_{1/2},P_n$) values using the numerical file was to perform a basic inter-comparison of summation calculations of the  delayed neutron yields among the CRP participants. Total delayed neutron yields (DN) were calculated for thermal and fast neutron-induced fission of major and minor actinides by four independent groups, namely CEA-Cadarache (Foligno), CIEMAT (Cano), JAEA (Minato) and SUBATECH-Nantes (Fallot).

The calculations are based on the summation method whereby the total delayed neutron yield $\nu_d$ is defined as

\begin{equation}
\nu_d=\sum_i N_{ni}\cdot CFY_i
\label{eq:summation}
\end{equation}

where $N_{ni}$ is the average delayed neutron multiplicity emitted by the precursor $i$ and is given by $N_n=1\cdot P_{1n}+2\cdot P_{2n}+3\cdot P_{3n}...$, $P_{xn}$ are the delayed neutron emission probabilities $P_{xn}$ and $CFY_i$ are the cumulative fission yields for precursor i. The above equation holds for equilibrium irradiation where the production and decay rates of the precursor nuclides are proportional to their cumulative fission yields.  

Assuming the $N_n$ and $CFY$ quantities in the above expression are independent, the uncertainty of the total DN yield is given by

\begin{equation}
\sigma^2_{\nu_d} = \sum_i N^2_{ni}\cdot \sigma^2_{CFY_i} + \sum_i \sigma^2_{N_ni} \cdot CFY^2_i 
\end{equation}

where $\sigma_{N_ni}$ and $\sigma_{CFY_i}$ are the uncertainties of the delayed neutron multiplicities $N_{ni}$ and cumulative fission yields $CFY_i$ of the individual precursors.

In the following calculations, the cumulative fission yields $CFY_i$ are taken from the JEFF-3.1.1 fission yield library~\cite{jeff3.1.1} which has been widely tested in a broad range of applications, including in summation calculations of DN yields, and has proved to perform well. The $\beta-decay$ and \bdn data, ($T_{1/2}, P_n$), are taken from the new CRP tables. Nuclides for which there are no $P_n$ values in the new CRP tables are not considered as \bdn emitters, i.e. $P_n=0$. The remaining decay data ($\gamma$ decays-IT), when used in the basic summation calculations or time-dependent calculations, are taken from ENDF/B-VIII.0~\cite{Brown2018}. This combination of input data is hereafter referred to as CRP+ENDF/B-VIII.0. The calculations are compared with total DN yields obtained by combining $CFY_i$ from JEFF-3.1.1 and decay data from the ENDF/B-VIII.0 decay data sub-library. This comparison enables us to observe the direct impact of the new CRP data ($T_{1/2}, P_n$) on the total DN yields as all the other data inputs are the same. It is also interesting to explore the differences between the new improved CRP+ENDF/B-VIII.0 decay data sub-library and the other available evaluated decay-data sub-libraries. Therefore, additional calculations were performed by combining the $CFY_i$ from JEFF-3.1.1~\cite{jeff3.1.1} with the decay data from JEFF-3.1.1~\cite{jeff3.1.1} and JENDL/DDF-2015~\cite{Katakura2015} sub-libraries. Note that the release of JEFF-3.3 library occurred after most of the work carried out by the CRP was completed. Since JEFF-3.3 and JEFF-3.1.1 do not differ with respect to the DN data, it was decided that for the purposes of the CRP work it was sufficient to present the results that had already been obtained with JEFF-3.1.1 as they would in practice be identical with those obtained with JEFF-3.3.

The results of the summation calculations for the major actinides $^{235}$U, $^{238}$U, $^{239,241}$Pu are given in Table~\ref{tab:TotalYields}. Perfect agreement is observed among the four different summation calculations which confirms that the calculations performed by the different groups within this project are equivalent and hence reliable. We do not need to report the results of all four groups hereafter, but just the results of the CRP. The values in Table~\ref{tab:TotalYields} are considered as reference values for the given combination of decay data and fission product yields library.

\begin{sidewaystable*} 
\caption{Total delayed neutron yields ($\nu_d$) for thermal and fast neutron-induced fission of major actinides obtained from the summation method using three different decay data libraries (CRP+ENDF/B-VIII.0, ENDF/B-VIII.0, JEFF-3.1.1, JENDL/DDF-2015) in combination with the cumulative fission yields CFY from JEFF-3.1.1. Four groups participated in the inter-comparison, namely CEA (France), CIEMAT (Spain), JAEA (Japan), and NANTES (France). The numbers in brackets are the relative uncertainties neglecting correlations.}\label{tab:TotalYields}
\label{tab:nubars}
\begin{minipage}{\textwidth}
\begin{tabular}{lllll||llll}
\multicolumn{9}{c}{CRP+ENDF/B-VIII.0 decay data}\\ \hline \hline
 & \multicolumn{4}{c}{thermal} & \multicolumn{4}{c}{fast} \\ \hline
\T $\nu_d$ & CEA & CIEMAT & JAEA & NANTES & CEA &CIEMAT & JAEA& NANTES  \\ \hline
\T $^{235}$U &  0.01673 (5$\%$) & 0.01673 (5$\%$) &0.01673 (5$\%$) & 0.01673 (5$\%$)&   0.01883 (5$\%$)   & 0.01883 (5$\%$) & 0.01883 (5$\%$)  & 0.01883 (5$\%$)  	\\
$^{238}$U &       &        &      & 		&   0.04604 (3$\%$)     & 0.04604 (3$\%$)  & 0.04604 (3$\%$) & 0.04604 (3$\%$) \\
$^{239}$Pu & 0.00680 (7$\%$) & 0.00680 (7$\%$) & 0.00680 (7$\%$)  &0.00680 (7$\%$) & 0.00756 (7$\%$)    & 0.00756 (7$\%$) & 0.00756 (7$\%$)  & 0.00756 (7$\%$) \\
$^{241}$Pu &0.01397 (4$\%$)& 0.01397 (4$\%$) & 0.01397 (4$\%$)   &0.01397 (4$\%$) 	& 0.01462(5$\%$)    & 0.01462(5$\%$)  & 0.01462(5$\%$) & 0.01462(5$\%$)	\\
\hline\hline
\end{tabular}
\end{minipage}

\begin{minipage}{\textwidth} 
\begin{tabular}{lllll||llll}
 \multicolumn{9}{c}{ENDF/B-VIII.0 decay data}\\ \hline \hline
 & \multicolumn{4}{c}{thermal} & \multicolumn{4}{c}{fast} \\ \hline
\T $\nu_d$ & CEA & CIEMAT & JAEA & NANTES & CEA &CIEMAT & JAEA& NANTES  \\ \hline
\T $^{235}$U & 0.01609 (5$\%$) & 0.01609 (5.$\%$) &  0.01609 (5$\%$)  & 0.01609 (5$\%$) & 0.01799 (5$\%$)  &      0.01799 (5$\%$) & 0.01799 (5$\%$) &  0.01799 (5$\%$)	\\
$^{238}$U &     &        &      & 		& 0.04492 (3$\%$)  & 0.04492 (3$\%$) & 0.04492 (3$\%$)   &  0.04492 (3$\%$)	\\
$^{239}$Pu &  0.0066 (7$\%$)    &  0.0066 (6$\%$)& 0.0066(7$\%$) &0.0066 (7$\%$)	&0.00722 (7$\%$)   & 0.00722 (7$\%$)&0.00722 (7$\%$)    &0.00722 (7$\%$)  	\\
$^{241}$Pu & 0.01374 (5$\%$) & 0.01374 (5$\%$)   &0.01374 (5$\%$)      & 0.01374 (5$\%$) &  0.01407 (5$\%$)   &  0.01407 (5$\%$)  & 0.01407 (5$\%$)   &0.01407 (5$\%$) 	\\
\hline\hline
\end{tabular}
\end{minipage}

\begin{minipage}{\textwidth}
\begin{tabular}{lllll||llll}
 \multicolumn{9}{c}{JEFF-3.1.1 decay data} \\ \hline \hline
 & \multicolumn{4}{c}{thermal} & \multicolumn{4}{c}{fast} \\ \hline 
\T $\nu_d$ & CEA & CIEMAT & JAEA & NANTES & CEA &CIEMAT & JAEA& NANTES  \\ \hline
\T $^{235}$U & 0.01477 (5$\%$)      &   0.01477 (5$\%$)   & 0.01477 (5$\%$)    & 0.01477 (5$\%$) & 0.01698 (5$\%$)   &   0.01698 (5$\%$) & 0.01698 (5$\%$)   &  0.01698 (5$\%$)	\\
$^{238}$U &       &        &      & 		& 0.04037 (3$\%$) & 0.04037 (3$\%$)  & 0.04037 (3$\%$)   & 0.04037 (3$\%$) \\
$^{239}$Pu & 0.00605 (7$\%$)     & 0.00605 (7$\%$)  & 0.00605 (7$\%$)    & 0.00605 (7$\%$) &0.00675 (7$\%$) &   0.00675 (7$\%$)& 0.00675 (7$\%$)   & 0.00675 (7$\%$) 	\\
$^{241}$Pu & 0.01232 (5$\%$) &  0.01232 (5$\%$) & 0.01232 (5$\%$) & 0.01232 (5$\%$) & 0.0129 (5$\%$)  & 0.0129 (5$\%$)  & 0.0129 (5$\%$) &0.0129(5$\%$) 	\\
\hline\hline
\end{tabular}
\end{minipage}

\begin{minipage}{\textwidth} 
\begin{tabular}{lllll||llll}
 \multicolumn{9}{c}{JENDL/DDF-2015 decay data} \\ \hline \hline
 & \multicolumn{4}{c}{thermal} & \multicolumn{4}{c}{fast} \\ \hline
\T $\nu_d$ & CEA & CIEMAT & JAEA & NANTES & CEA &CIEMAT & JAEA& NANTES  \\ \hline
\T $^{235}$U & 0.01637 (6$\%$) & 0.01637 (6$\%$) & 0.01637 (6$\%$) & 0.01637 (6$\%$) & 0.01849 (6$\%$)  &      0.01849 (6$\%$) & 0.01849 (6$\%$)&  0.01849 (6$\%$)	\\
$^{238}$U &     &        &      & 		& 0.04511 (4$\%$)  & 0.04511 (4$\%$) & 0.04511 (4$\%$) &  0.04511 (4$\%$)	\\
$^{239}$Pu &  0.00666 (7$\%$)    &  0.00666 (7$\%$)& 0.00666 (7$\%$)& 0.00666 (7$\%$)  &0.00739 (7$\%$)   & 0.00739 (7$\%$)&0.00739 (7$\%$)3&0.00739 (7$\%$)  	\\
$^{241}$Pu & 0.01378 (5$\%$)    & 0.01378 (5$\%$)   &0.01378 (5$\%$) & 0.01378 (5$\%$) &  0.01436 (5$\%$)   &  0.01436 (5$\%$)  & 0.01436 (5$\%$) &0.01436 (5$\%$) 	\\
\hline\hline
\end{tabular}
\end{minipage}
\end{sidewaystable*}

The results displayed in Table~\ref{tab:TotalYields} show that overall the CRP $P_n$ data lead to larger values of the total DN yields ($\nu_d$) compared to the other evaluated decay data libraries ENDF/B-VIII.0, JEFF-3.1.1 and JENDL/DDF-2015. The observed differences are larger with respect to JEFF-3.1.1, which generally gives the lowest values of total DN yields among all three evaluated libraries. In the case of fast fission of $^{238}$U and thermal fission of $^{241}$Pu, the CRP+ENDF/B-VIII.0 total DN yields are in agreement with the ENDF/B-VIII.0 and JENDL/DDF-2015 results. 

The 20 most important contributors to the total DN yields ($\nu_d$) displayed in Table~\ref{tab:TotalYields} are listed in Table~\ref{tab:nubar-contribution}. The first column includes the percentage contribution of the precursors resulting from the CRP ($T_{1/2}, P_n$) data, while the neighboring columns include the contributions of the same precursors obtained when using the ENDF/B-VIII.0, JEFF-3.1.1 and JENDL/DDF-2015 decay data, respectively. Both ENDF/B-VIII.0 and JENDL/DDF-2015 have adopted $P_n$ values from ENSDF~\cite{ENSDF}, however, depending on the date they were retrieved from the ENSDF database, the values may differ among each other and with respect to the current ENSDF data value. JEFF-3.1.1 uses $P_n$ values from an evaluation by Nichols~\cite{Nichols1998} as well as from ENSDF.

From Table~\ref{tab:nubar-contribution} we see that overall, the differences between the contributions of the top 20 precursors among the various libraries are less than $10\%$. Some notable cases are discussed below: 

$^{137}$I is the main contributor to $\nu_d$ for nearly all the major actinides and incident neutron energies. The $^{137}$I contribution is enhanced (on average by a few $\%$) when using the CRP data and is mainly responsible for the small increase in the total DN yields observed in Table~\ref{tab:TotalYields}. The increased relative contribution of $^{137}$I is related to a small increase in the CRP $P_n$ value for this precursor, which is now $P_{n} = 7.63(14) \%$, compared to the value of $7.14(23) \%$ used in ENDF/B-VIII.0 and JENDL/DDF-2015. The increase is more pronounced with respect to JEFF-3.1.1 which has an adopted value of $P_n=6.5(4) \%$ from \cite{Nichols1998}.

The precursor $^{85}$As is among the top twelve contributors in the CRP list with $P_n = 62.5(10)\%$. It also figures high in the ENDF/B-VIII.0 and JENDL/DDF-2015 contributors lists due to the large value $P_n = 59.4(24) \%$ adopted in both files, from ENSDF. The difference between the CRP contribution and JEFF-3.1.1 is considerably larger (over $100\%$) due to the relatively lower value of $P_n = 22(3) \%$ adopted in JEFF-3.1.1~\cite{Nichols1998}. 

$^{91}$Br is among the top 20 contributors in the CRP list with $P_n = 29.8(8) \%$, which is similar to the JEFF-3.1.1 value $P_n=20(2) \%$ taken from~\cite{Nichols1998} and JENDL/DDF-2015 $P_n=20(3)$ taken from ENSDF. In ENDF/B-VIII.0, however,  this precursor does not have a delayed-neutron decay branch, therefore it does not contribute to the delayed neutron activity.

$^{136}$Te appears in the top 20 list of the CRP calculations although it has a relatively small $\beta$-delayed neutron fraction of $P_n = 1.37(5) \%$. Similar results are obtained with ENDF/B-VIII.0 and JENDL/DDF-2015 which have both adopted the value of $P_n = 1.311(5)$ from ENSDF. On the other hand, this precursor does not have a delayed-neutron branch in the JEFF-3.1.1 decay data library. 

The contribution of $^{89}$Br to the CRP list is lower by more than $13\%$ in comparison to JEFF-3.1.1. This difference is observed across several of the major actinides, and is related to the small difference in $P_n$ values, with $P_n=13.7(6)$ being the CRP value and $P_n=14.1(4)$~\cite{Nichols1998} the value in JEFF-3.1.1. ENDF/B-VIII.0 and JENDL/DDF-2015 contributions are identical which is due to the fact that they have adopted the same value $P_n=13.8(4)$ from ENSDF.

The above-mentioned differences in the individual precursor $P_n$ values may have a non-negligible impact on the DN group spectra which are estimated from summations over the energy spectra of individual precursors weighted by their relative contributions to the total DN yield, in addition to the differences observed in the total delayed neutron yields. 

The uncertainties associated with the new total DN yields in Table~\ref{tab:TotalYields} are fairly small compared to the large uncertainties affecting the cumulative fission yields CFYs. However, this is expected considering that the estimation of the uncertainties ignores the effect of correlations which are particularly strong between the CFY uncertainties. A more rigorous treatment of the propagation of uncertainties including correlations would yield different results.

\afterpage{
\begin{sidewaystable*}
\caption{The twenty most important contributors to $\nu_d$ for thermal and fast neutron-induced fission of the major actinides $^{235}$U, $^{238}$U, $^{239}$Pu and $^{241}$Pu using the CRP+ENDF/B-VIII.0 decay data (CRP). In columns 'ENDF/B', 'JEFF' and 'JENDL', we give the contributions of these precursors obtained when using ENDF/B-VIII.0, JEFF-3.1.1 and JENDL/DDF-2015 decay-data libraries, respectively. All the contributions are given in $\%$.}\label{tab:nubar-contribution} 
\begin{minipage}{\textwidth}
\begin{tabular}{lcccc|lcccc|lcccc}\hline  \hline
  \multicolumn{5}{c}{$^{235}$U thermal} & \multicolumn{5}{c}{$^{235}$U fast} &\multicolumn{5}{c}{$^{238}$U fast} \\ 
\T Nuclide & CRP & ENDF/B & JEFF & JENDL & Nuclide&CRP & ENDF/B &JEFF&  JENDL&Nuclide&CRP & ENDF/B& JEFF & JENDL \\  \hline
$^{137}$I	&	16.3	&	15.8&	15.7	&	15.6	&	$^{94}$Rb	&	13.5	&	14.2&	14.5&	13.9&	$^{137}$I	&	9.3	&	8.9	&	9.0	&	8.9	\\
$^{89}$Br	&	11.1	&	11.6&	13.0	&	11.4	&	$^{137}$I	&	12.9	&	12.7&	12.2&	12.3&	$^{94}$Rb	&	7.6	&	7.9	&	8.4	&	7.8	\\
$^{94}$Rb	&	9.3	&	9.8		&	10.2	&	9.6	&	$^{89}$Br	&	11.5	&	12.1&	13.1	&	11.8	&	$^{90}$Br	&	6.7	&	6.7	&	7.3	&	6.7	\\
$^{90}$Br	&	7.4	&	7.6		&	8.1	&	7.5	&	$^{90}$Br	&	8.3	&	8.5		&	8.8	&	8.3	&	$^{89}$Br	&	6.3	&	6.5	&	7.4	&	6.4	\\
$^{88}$Br	&	7.3	&	7.4		&	8.2	&	7.3	&	$^{88}$Br	&	8.2	&	8.4		&	9.0	&	8.1	&	$^{85}$As	&	5.5	&	5.4	&	2.2	&	5.3	\\
$^{85}$As	&	5.3	&	5.3		&	2.1	&	5.2	&	$^{98m}$Y	&	5.9	&	6.1		&	6.5	&	5.9	&	$^{139}$I	&	5.4	&	5.7	&	6.2	&	5.8	\\
$^{138}$I	&	4.7	&	5.1		&	5.3	&	5.0	&	$^{85}$As	&	5.0	&	5.0		&	2.0	&	4.8	&	$^{135}$Sb	&	5.0	&	5.7	&	4.5	&	5.6	\\
$^{98m}$Y	&	4.1	&	4.2		&	4.6	&	4.1	&	$^{95}$Rb	&	5.0	&	5.1		&	5.4	&	5.0	&	$^{91}$Br	&	4.8	&	0.0	&	3.7	&	3.3	\\
$^{139}$I	&	3.5	&	3.7		&	4.0	&	3.8	&	$^{138}$I	&	3.9	&	4.3		&	4.3	&	4.2	&	$^{138}$I	&	4.6	&	5.0	&	5.3	&	4.9	\\
$^{95}$Rb	&	3.5	&	3.6		&	3.8	&	3.5	&	$^{93}$Rb	&	3.6	&	3.2		&	3.5	&	3.1	&	$^{95}$Rb	&	4.2	&	4.3	&	4.7	&	4.2	\\
$^{93}$Rb	&	3.4	&	3.1		&	3.4	&	3.0	&	$^{87}$Br	&	3.1	&	3.3		&	3.4	&	3.2	&	$^{88}$Br	&	2.8	&	2.8	&	3.2	&	2.8	\\
$^{87}$Br	&	3.2	&	3.5		&	3.6	&	3.4	&	$^{99}$Y	&	3.0	&	2.7		&	2.9	&	3.0	&	$^{145}$Cs	&	2.4	&	2.7	&	2.9	&	2.7	\\
$^{91}$Br	&	2.7	&	0.0		&	2.1	&	1.9	&	$^{91}$Br	&	2.5	&	0.0		&	1.8	&	1.7	&	$^{96}$Rb	&	2.2	&	2.1	&	2.4	&	2.1	\\
$^{99}$Y	&	2.2	&	2.0		&	2.2	&	2.2	&	$^{139}$I	&	2.2	&	2.4		&	2.5	&	2.4	&	$^{99}$Y	&	2.0	&	1.8	&	2.0	&	2.0	\\
$^{135}$Sb	&	2.1	&	2.4		&	1.9	&	2.4	&	$^{143}$Cs	&	1.4	&	1.5		&	1.6	&	1.5	&	$^{98m}$Y	&	2.0	&	2.0	&	2.2	&	2.0	\\
$^{136}$Te	&	1.6	&	1.6		&	0.0	&	1.6	&	$^{96}$Rb	&	1.3	&	1.3		&	1.4	&	1.3	&	$^{97}$Rb	&	1.9	&	2.0	&	2.1	&	1.9	\\
$^{143}$Cs	&	1.6	&	1.7		&	1.8	&	1.6	&	$^{86}$As	&	0.7	&	0.8		&	0.8	&	0.7	&	$^{136}$Sb	&	1.7	&	1.1	&	1.3	&	1.1	\\
$^{86}$As	&	0.9	&	1.0		&	1.0	&	0.9	&	$^{135}$Sb	&	0.6	&	0.7		&	0.6	&	0.7	&	$^{140}$I	&	1.7	&	2.1	&	2.2	&	2.0	\\
$^{96}$Rb	&	0.9	&	0.8		&	0.9	&	0.8	&	$^{136}$Te	&	0.6	&	0.6		&	0.0	&	0.6	&	$^{141}$I	&	1.6	&	1.6	&	1.8	&	1.6	\\
$^{137}$Te	&	0.8	&	0.9		&	1.0	&	0.9	&	$^{97}$Rb	&	0.5	&	0.6		&	0.6	&	0.5	&	$^{93}$Rb	&	1.6	&	1.4	&	1.6	&	1.4	\\

\end{tabular}
\end{minipage}

\begin{minipage}{\textwidth}
\begin{tabular}{lcccc|lcccc|lcccc|lcccc} \hline 
 \multicolumn{5}{c}{$^{239}$Pu thermal} & \multicolumn{5}{c}{$^{239}$Pu fast} & \multicolumn{5}{c}{$^{241}$Pu thermal} &  \multicolumn{5}{c}{$^{241}$Pu fast} \\ 
 Nuclide & CRP & ENDF/B & JEFF &  JENDL&Nuclide& CRP & ENDF/B &JEFF& JENDL& Nuclide& CRP & ENDF/B&JEFF & JENDL& Nuclide& CRP & ENDF/B& JEFF& JENDL \\  \hline

$^{137}$I	&	25.8	&	25.0&	24.7&	24.6&	$^{137}$I	&	17.7&	17.4&	16.9	&	17.0&	$^{137}$I	&	24.4	&	23.2&	23.6&	23.1&	$^{137}$I	&	21.9&	21.3&	21.2&	20.9\\
$^{94}$Rb	&10.6	&	11.1&	11.6&	11.0	&$^{94}$Rb	&	13.6&	14.4&	14.8&	14.1&	$^{138}$I	&8.4&	9.0	&	9.6	&	9.0	&	$^{94}$Rb	&	10.4&	10.9&	11.5&	10.7\\
$^{98m}$Y	&	9.4	&	9.7	&	10.6&9.5	&	$^{98m}$Y	&	10.8&	11.2&	12.1	&	10.9	&$^{94}$Rb	&	8.0	&	8.2	&	8.8	&	8.2	&	$^{138}$I&	6.3	&	6.9	&	7.1	&	6.7	\\
$^{89}$Br	&	6.2	&	6.5	&	7.2	&	6.4	&	$^{89}$Br&	9.2	&	9.7	&	10.6&	9.5	&	$^{139}$I&	5.8	&	6.0	&	6.6	&	6.2	&	$^{89}$Br	&	6.2	&	6.4	&	7.2	&	6.3	\\
$^{138}$I	&	5.3	&	5.7	&	5.9	&	5.6	&	$^{88}$Br	&	7.0	&	7.1	&	7.8	&	7.0	&	$^{98m}$Y	&5.2	&	5.3	&	6.0	&	5.3	&	$^{98m}$Y	&	5.9	&	6.0	&	6.6	&	5.9	\\
$^{88}$Br	&	5.0	&	5.0	&	5.6	&	5.0	&	$^{85}$As	&5.2	&	5.2	&	2.1	&	5.1	&	$^{89}$Br&	4.9	&	5.0	&	5.7	&	5.0	&	$^{95}$Rb	&	4.6	&	4.8	&	5.1	&	4.7	\\
$^{93}$Rb	&	4.0	&	3.6	&	3.9	&	3.5	&	$^{90}$Br&	5.1	&	5.2	&	5.5	&	5.1	&	$^{135}$Sb	&	3.9	&	4.4	&	3.5	&	4.4	&	$^{90}$Br&	4.5	&	4.6	&	4.9	&	4.5	\\
$^{99}$Y	&	3.9	&	3.4	&	3.7	&	3.8	&	$^{99}$Y	&	4.6	&	4.2	&	4.5	&	4.6	&	$^{99}$Y&	3.4	&	3.0	&	3.3	&	3.3	&	$^{139}$I	&	3.9	&	4.2	&	4.4	&	4.2	\\
$^{90}$Br	&	3.6	&	3.7	&	3.7	&	3.6	&	$^{93}$Rb	&4.6	&	4.2	&	4.5	&	4.1	&	$^{90}$Br	&	3.3	&	3.3	&	3.6	&	3.3	&	$^{85}$As	&	3.4	&	3.4	&	1.4	&	3.3	\\
$^{95}$Rb	&	3.4	&	3.5	&	3.7	&	3.4	&	$^{95}$Rb	&	4.1	&	4.3		&	4.5	&	4.2	&$^{95}$Rb	&	3.2	&	3.3	&	3.6	&	3.2	&	$^{99}$Y&	3.3	&	2.9	&	3.2	&	3.2	\\
$^{139}$I	&	2.8	&	3.0	&	3.2	&	3.1	&	$^{87}$Br	&	2.8	&	3.0		&	3.1	&	2.9	&	$^{88}$Br	&3.1	&	3.1	&	3.5	&	3.1	&	$^{88}$Br&	3.1	&	3.2	&	3.5	&	3.1	\\
$^{85}$As	&	2.8	&	2.8	&	1.1	&	2.7	&	$^{138}$I	&	2.7	&	3.0	&	3.1	&	2.9	&	$^{85}$As	&	2.5	&	2.4	&	1.0	&	2.4	&	$^{105}$Nb	&	2.4	&	2.5	&	2.7	&	2.5	\\
$^{87}$Br	&	2.5	&	2.6	&	2.8	&	2.6	&$^{91}$Br	&	1.6	&	0.0	&	1.2	&	1.1	&$^{136}$Te	&	2.3	&	2.3	&	0.0	&	2.3	&	$^{93}$Rb	&	2.4	&	2.2	&	2.4	&	2.1	\\
$^{135}$Sb	&	1.5	&	1.7	&	1.4	&	1.7	&$^{105}$Nb	&	1.3	&	1.3	&	1.4	&	1.3	&	$^{93}$Rb&	2.2	&	1.9	&	2.2	&	1.9	&	$^{135}$Sb	&	2.3	&	2.6	&	2.1	&	2.6	\\
$^{143}$Cs	&	1.4	&	1.5	&	1.6	&	1.5	&	$^{139}$I&	1.1	&	1.1	&	1.2	&	1.1	&	$^{143}$Cs&	2.0	&	2.1	&	2.4	&	2.1	&$^{91}$Br	&	2.0	&	0.0	&	1.5	&	1.4		\\
$^{136}$Te	&	1.4	&	1.4	&	0.0	&	1.3	&	$^{96}$Rb&	0.9	&	0.9	&	1.0	&	0.9	&	$^{105}$Nb	&	1.6	&	1.6	&	1.8	&	1.6	&	$^{143}$Cs	&	1.6	&	1.8	&	1.9	&	1.7	\\
$^{105}$Nb	&	1.2	&	1.2	&	1.3	&	1.2	&	$^{143}$Cs	&	0.7	&	0.8	&	0.8	&	0.7	&$^{91}$Br	&	1.2	&	0.0	&	0.9	&	0.8	&	$^{136}$Te	&	1.4	&	1.4	&	0.0	&	1.4	\\
$^{91}$Br	&	1.1	&	0.0	&	0.9	&	0.8	&	$^{86}$As	&	0.6	&	0.7		&	0.7	&	0.6	&	$^{137}$Te	&	1.2	&	1.3	&	1.4	&	1.3	&	$^{96}$Rb	&	1.4	&	1.4	&	1.5	&	1.3	\\
$^{96}$Rb	&	0.8	&	0.7	&	0.8	&	0.7	&	$^{136}$Te	&	0.6	&	0.6		&	0.0	&	0.6	&	$^{87}$Br	&	1.2	&	1.3	&	1.4	&	1.3	&	$^{106}$Nb	&	1.2	&	1.3	&	1.4	&	1.2	\\
$^{137}$Te	&	0.5	&	0.6		&	0.6	&	0.5	& $^{135}$Sb&	0.5	&	0.6		&	0.4	&	0.5	&	$^{145}$Cs	&	1.0	&	1.1	&	1.2	&	1.1	&	$^{87}$Br	&	1.2	&	1.3	&	1.4	&	1.3	\\

\hline 
\end{tabular}
\end{minipage}
\end{sidewaystable*}}

\paragraph{Comparison with recommended values.}

The total  DN  yields  presented  in Table~\ref{tab:TotalYields} are compared with recommended DN yields from ~\cite{DAngelo02} and the recommended DN yield data from JEFF-3.1.1, ENDF/B-VIII.0 and JENDL-4.0~\cite{Shibata2011} libraries in Figs.~\ref{fig:TotalYields1}-\ref{fig:TotalYields3}. It should be noted that the total DN yields from JEFF-3.1.1 are identical to the JEFF-3.3 data therefore the results in Fig.~\ref{fig:TotalYields1} are valid for both libraries. As shown in the figures, the total DN yields obtained with the CRP ($T_{1/2}, P_n$) data for thermal fission of $^{235}$U and $^{239}$Pu are in good agreement with the recommended data ~\cite{DAngelo02} and the data from the other libraries except for the thermal value of $^{239}$Pu from JENDL-4.0. The results for the fast fission of $^{238}$U are substantially closer to the recommended data compared with the previous summation calculations of~\cite{Wilson02}. The differences between the calculated and recommended values of the total DN yields for the fast neutron-induced fission of $^{239}$Pu and $^{235}$U and for both fast and thermal fission of $^{241}$Pu are more than 10$\%$. These differences highlight the need for better treatment of uncertainties and further improvement of the microscopic data for these nuclides, mainly the fission yield data. Another important feature of the CRP results presented in Table~\ref{tab:TotalYields} is the clear evidence of the energy dependence of the total DN yields for fission of $^{235}$U and $^{239}$Pu. This is in agreement with the latest experimental data presented in this paper (see Section~\ref{Sec:Macro-incident-energy}) and supports the need for further investigation of this dependency.

\afterpage{

\begin{figure}[!htb]
\includegraphics[width=1.\linewidth]{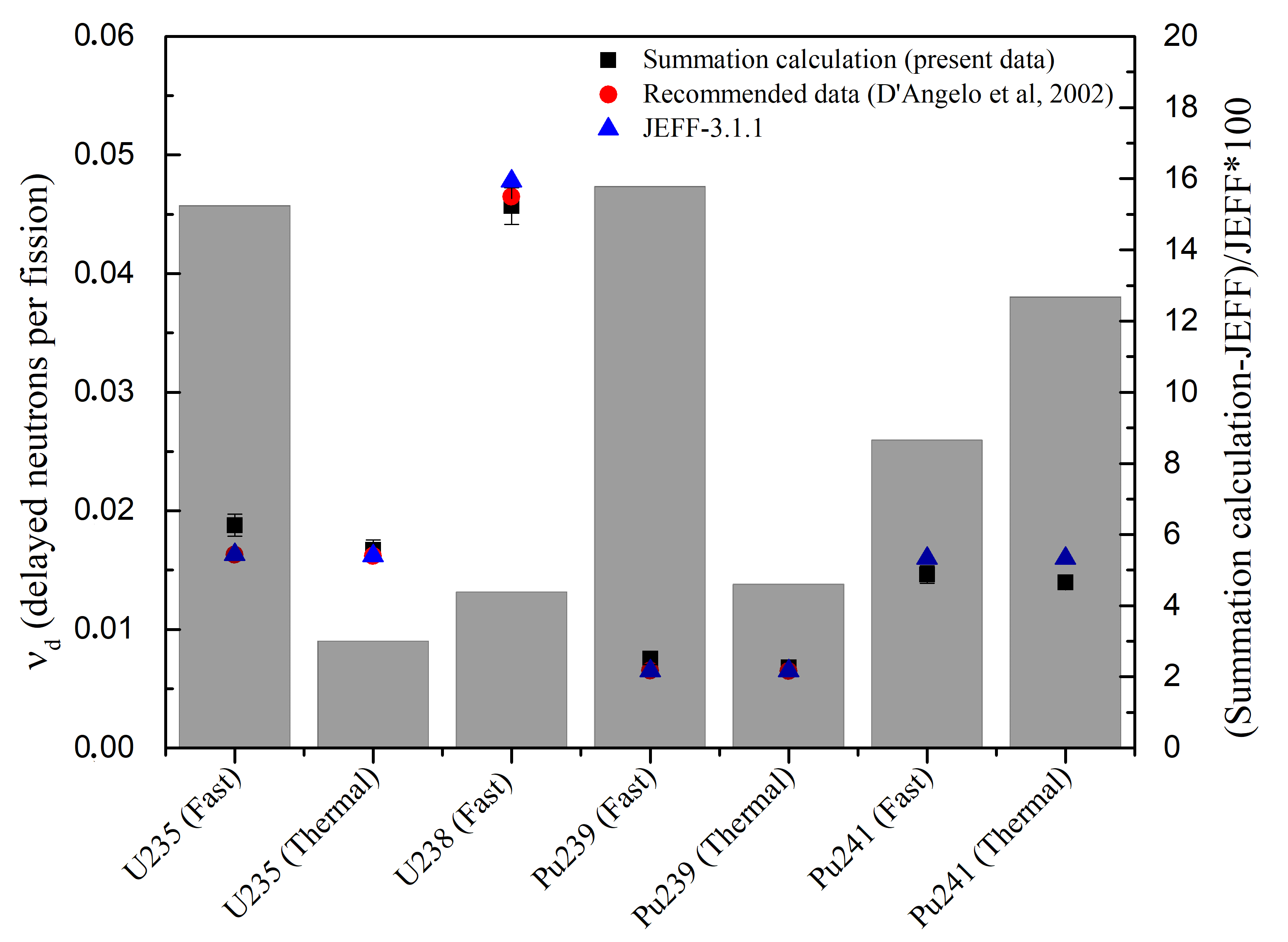}
\caption{Comparison of the total DN yields obtained from the summation method using the CRP ($T_{1/2},P_n$) data and the JEFF-3.1.1 decay-data sub-library, respectively, with the recommended DN yields~\cite{DAngelo02}. The total DN yields for the fission of various nuclides (listed on the abscissa axis) are plotted with symbols. The right ordinate shows the percentage difference between the total DN yield data calculated using the CRP data and the corresponding data from JEFF-3.1.1 (bar chart). \label{fig:TotalYields1}}
\end{figure}

\begin{figure}[!htb]
\includegraphics[width=1.\linewidth]{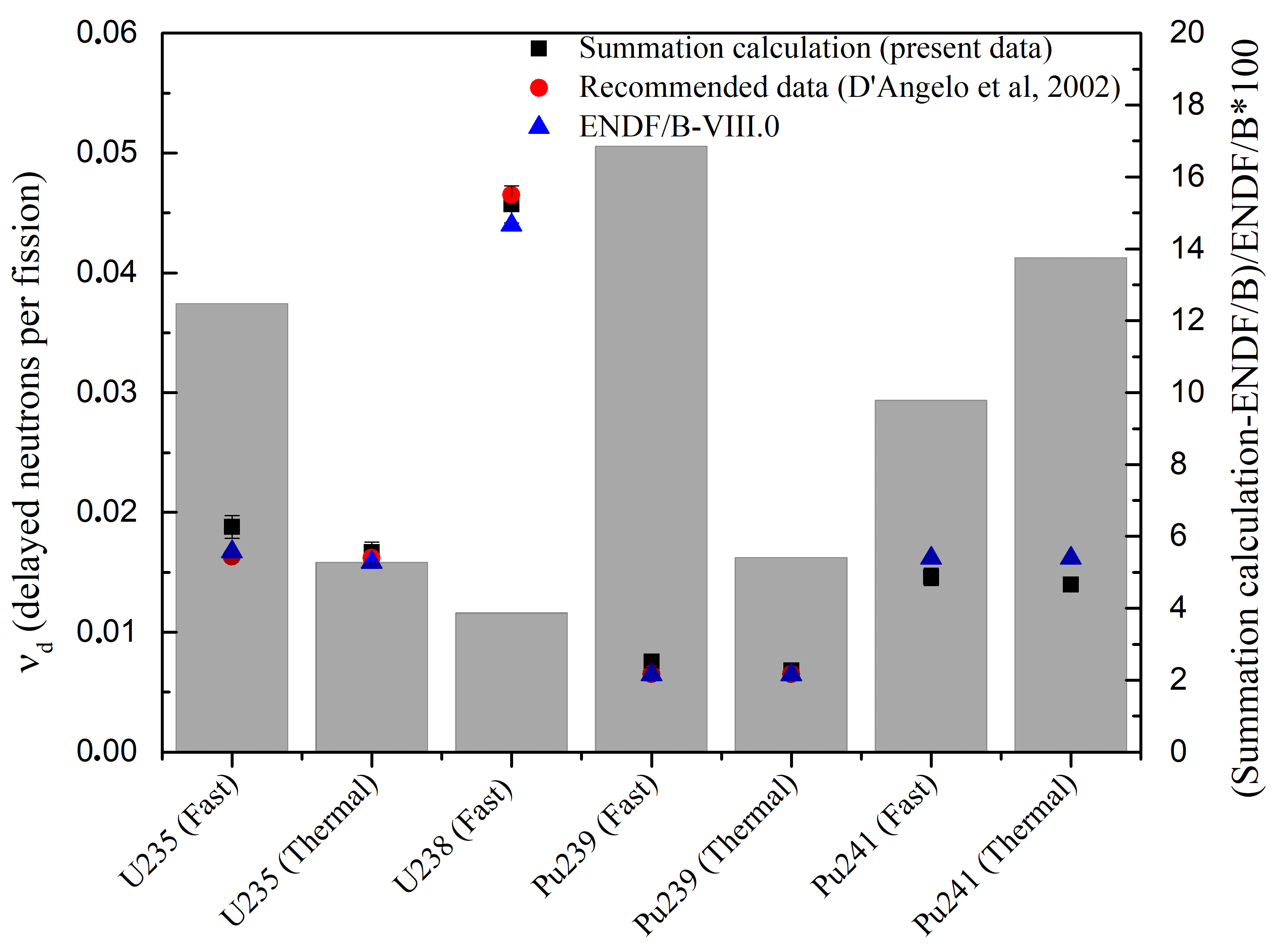}
\caption{Same as in Fig.~\ref{fig:TotalYields1} but for the ENDF/B-VIII.0 decay sub-library. \label{fig:TotalYields2}}
\end{figure}

\begin{figure}[!htb]
\includegraphics[width=1.\linewidth]{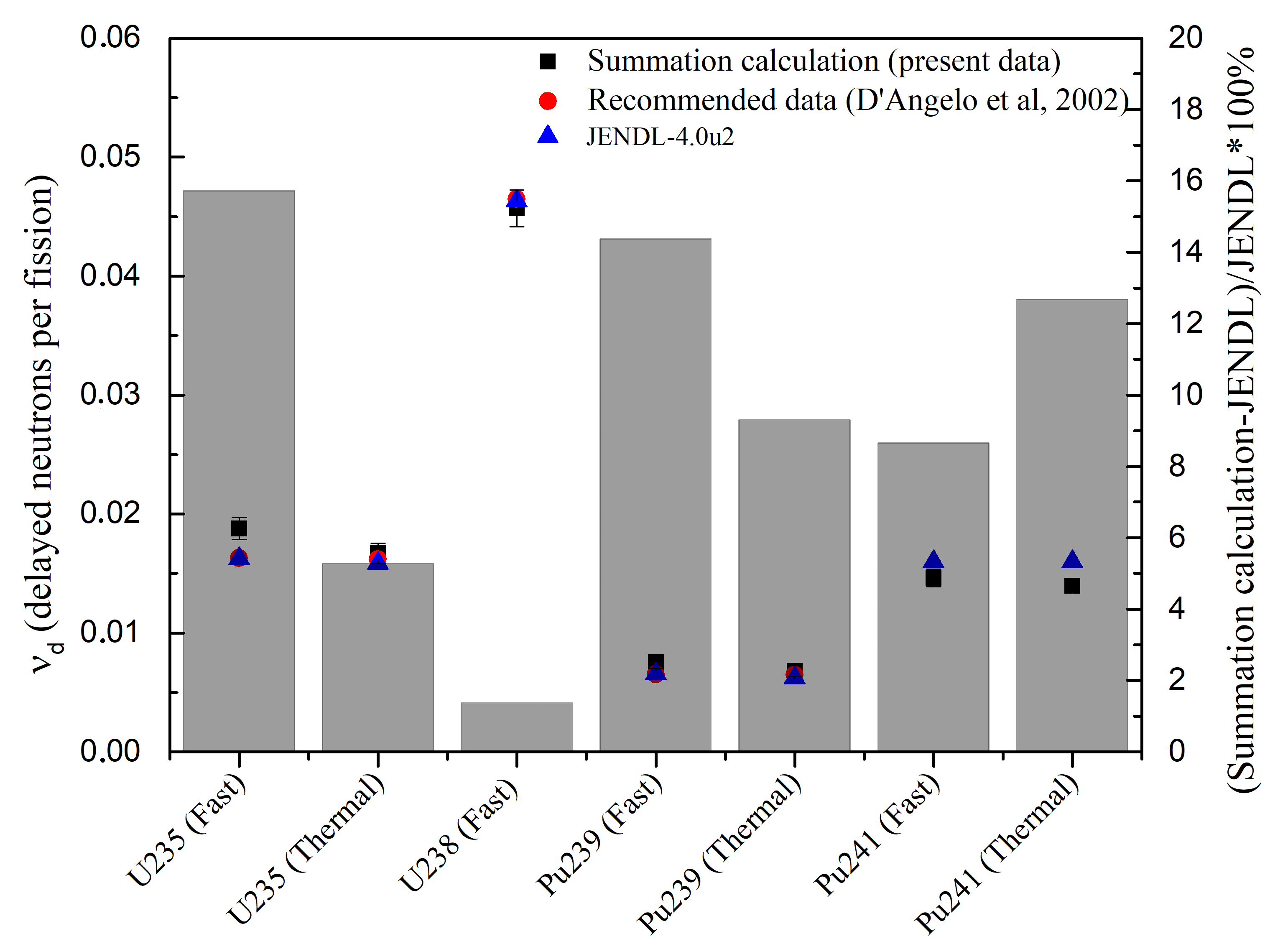}
\caption{Same as in Fig.~\ref{fig:TotalYields1} but for the JENDL-4.0 decay sub-library (bar chart). \label{fig:TotalYields3}}
\end{figure}
}

\subsubsection{Time-dependent calculations }
Apart from summation calculations that use cumulative fission yields to account for the abundances of the fission products in equilibrium, time-dependent calculations that start from independent fission yields and follow the evolution of the system into the decay phase, can also shed light on the impact of the new CRP ($T_{1/2}, P_n$) data on integral quantities. In this section, we describe two different time-dependent calculations of the integral DN activity and total DN yields that were performed with the new CRP data.

\paragraph{Time-dependent calculations during the irradiation and decay phase. }\label{Sec:Macro-summation-yields-timedep-1}

The integral DN activity experiment is simulated by summing up the microscopic contribution of each of the precursors produced during the fission process. The Bateman solver \cite{Foligno19} is a computer program that uses the decay data to reconstruct the family tree of the precursors and then solves the Bateman equations to obtain the precursors' concentration during both the irradiation and the decay phase. The concentrations-in-time are then multiplied by the effective delayed-neutron emission probability of the precursor (Eq.~\ref{eq:ndt}).

\begin{equation}
n_d(t)={ \sum\limits^N_{i=1} C_i(t) \; \overline{P}_{n \; \mathrm{eff}\; i} },
\label{eq:ndt}
\end{equation}

where $\overline{P}_{n \; \mathrm{eff}\; i} = \sum\limits_{x=1}^4 x \; P_{xn}$ is the effective delayed-neutron emission probability of the precursor $i$, which takes into account multiple decay branches leading to the neutron emission.
The following decay-processes are considered in the code~\cite{Foligno19}: $IT$, $\beta^-$, $\beta^-_{n}$, $\beta^-_{2n}$, $\beta^-_{3n}$ and $\beta^-_{4n}$. 
The Bateman solver has been validated through comparisons with the DARWIN code~\cite{Darwin}.

\begin{figure} 
\includegraphics[width=0.95\linewidth]{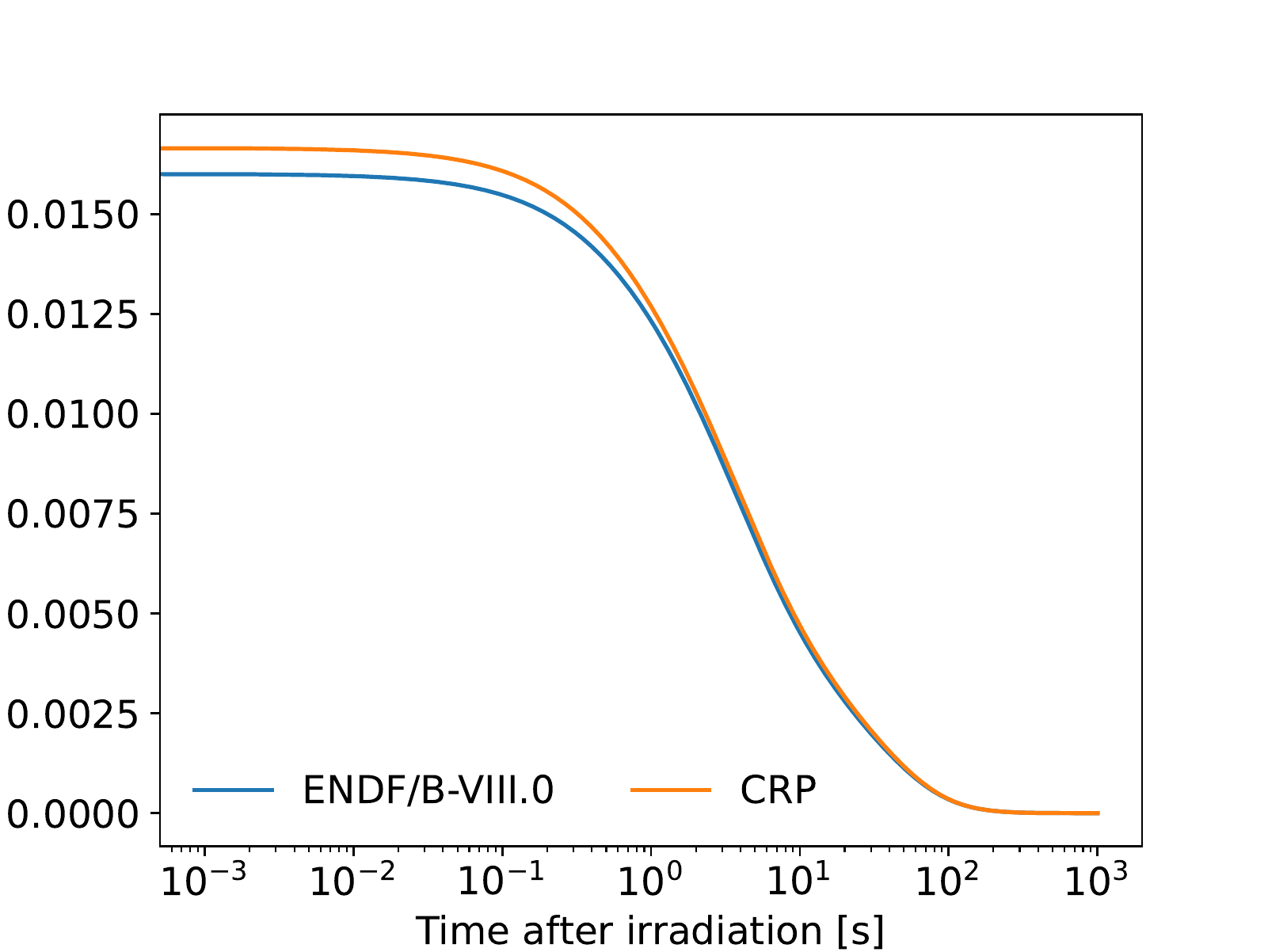}
\includegraphics[width=0.95\linewidth]{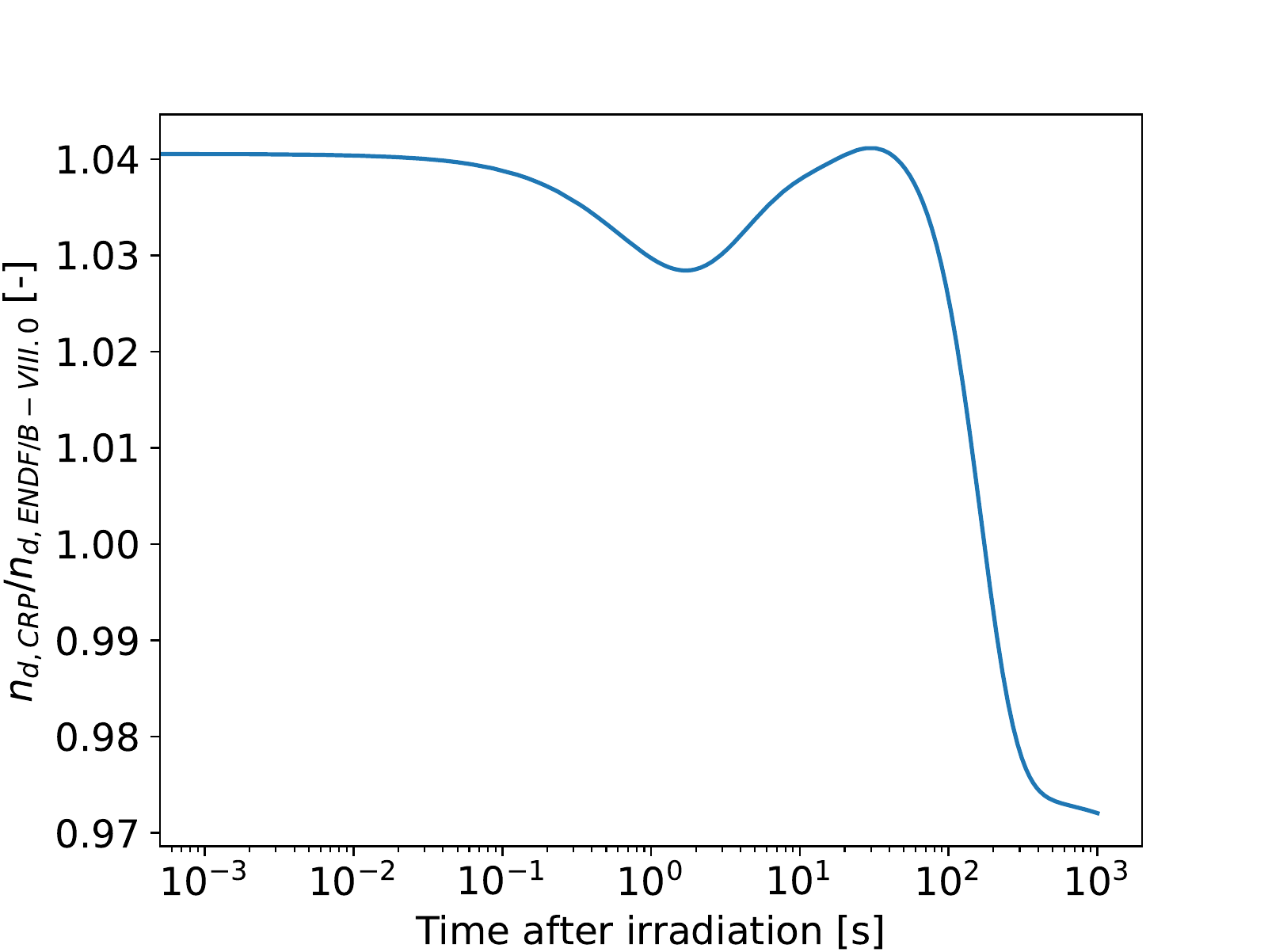}
\caption{Comparison between the CRP+ENDF/B-VIII.0 and ENDF/B-VIII.0 decay data. The two plots represent the two DN activities and their ratio, respectively, following the thermal fission of $^{235}$U.}
\label{fig:DNactivityComparison}
\end{figure}

Figure~\ref{fig:DNactivityComparison} shows the DN activity for thermal neutron-induced fission of $^{235}$U computed by combining the CRP+ENDF/B-VIII.0 and ENDF/B-VIII.0 decay data with JEFF-3.1.1 fission yields. The figure shows the two DN activities as w ell as their ratio, as a function of time after the irradiation.

The ratio does not start from 1 because the $\nu_d$ is different (1.60 neutrons/100 fissions for ENDF/B-VIII.0 and 1.66 neutrons/100 fissions for CRP+ENDF/B-VIII.0 for $^{235}$U). Table \ref{tab:libDiff} displays the 20 most important precursors according to the CRP calculations (see also Table~\ref{tab:nubar-contribution}), together with their partial yield $\nu_{d,i}$, the difference with respect to the partial yields calculated with ENDF/B-VIII.0, and the change in the corresponding precursor $P_n$ emission probability when going from ENDF/B-VIII.0 to CRP+ENDF/BVIII.0.

\begin{table}[ht]
\caption{Detailed comparison of the microscopic data of the 20 most important precursors for thermal neutron-induced fission of $^{235}$U: atomic number ($Z$), mass ($A$), isomeric number (index) ($I$), the partial delayed neutron yield computed by using CRP+ENDF/B-VIII.0 decay data ($\nu_{d,i}$), the difference in the partial yield when using CRP+ENDF/B-VIII.0 rather than ENDF/B-VIII.0 ($\Delta\nu_{d,i}$) and the change in microscopic $P_n$ from ENDF/B-VIII.0 to CRP+ENDF/B-VIII.0 ($P_n$(VIII.0)$\rightarrow$$P_n$(CRP)).}
\label{tab:libDiff}
\begin{tabular}{cccccc} \hline \hline
$Z$ & $A$   & $I$   & $\nu_{d,i}$   &  $\Delta\nu_{d,i}$    & $P_n$(VIII.0)$\rightarrow$$P_n$(CRP) \\ 
\hline
53	&	137	&	0	&	2.72$\times$10$^{-3}$	&	1.75$\times$10$^{-4}$	&	0.071	$\rightarrow$	0.076	\\
35	&	89	&	0	&	1.86$\times$10$^{-3}$	&	--1.365$\times$10$^{-5}$	&	0.138	$\rightarrow$	0.137	\\
37	&	94	&	0	&	1.55$\times$10$^{-3}$	&	--1.64$\times$10$^{-5}$	&	0.105	$\rightarrow$	0.104	\\
35	&	90	&	0	&	1.25$\times$10$^{-3}$	&	1.95$\times$10$^{-5}$	&	0.252	$\rightarrow$	0.256	\\
35	&	88	&	0	&	1.22$\times$10$^{-3}$	&	2.54$\times$10$^{-5}$	&	0.066	$\rightarrow$	0.067	\\
33	&	85	&	0	&	8.93$\times$10$^{-4}$	&	4.43$\times$10$^{-5}$	&	0.594	$\rightarrow$	0.625	\\
53	&	138	&	0	&	7.79$\times$10$^{-4}$	&	--3.82$\times$10$^{-5}$	&	0.056	$\rightarrow$	0.053	\\
39	&	98	&	1	&	6.79$\times$10$^{-4}$	&	7.90$\times$10$^{-6}$	&	0.034	$\rightarrow$	0.034	\\
53	&	139	&	0	&	5.834$\times$10$^{-4}$	&	--1.555$\times$10$^{-5}$	&	0.100	$\rightarrow$	0.097	\\
37	&	95	&	0	&	5.794$\times$10$^{-4}$	&	6.586$\times$10$^{-6}$	&	0.087	$\rightarrow$	0.088	\\
37	&	93	&	0	&	5.67$\times$10$^{-4}$	&	7.44$\times$10$^{-5}$	&	0.014	$\rightarrow$	0.016	\\
35	&	87	&	0	&	5.40$\times$10$^{-4}$	&	--1.50$\times$10$^{-5}$	&	0.026	$\rightarrow$	0.025	\\
35  &   91  &   0   &   4.52$\times$10$^{-4}$    &   4.52$\times$10$^{-4}$    &   0.000   $\rightarrow$   0.298   \\
39	&	99	&	0	&	3.71$\times$10$^{-4}$	&	5.08$\times$10$^{-5}$	&	0.017	$\rightarrow$	0.020	\\
51	&	135	&	0	&	3.57$\times$10$^{-4}$	&	--3.57$\times$10$^{-5}$	&	0.220	$\rightarrow$	0.200	\\
52	&	136	&	0	&	2.76$\times$10$^{-4}$	&	1.21$\times$10$^{-5}$	&	0.013	$\rightarrow$	0.014	\\
55	&	143	&	0	&	2.61$\times$10$^{-4}$	&	--9.55$\times$10$^{-6}$	&	0.016	$\rightarrow$	0.016	\\
33	&	86	&	0	&	1.58$\times$10$^{-4}$	&	--4.45$\times$10$^{-7}$	&	0.355	$\rightarrow$	0.354	\\
37	&	96	&	0	&	1.43$\times$10$^{-4}$	&	8.10$\times$10$^{-6}$	&	0.133	$\rightarrow$	0.141	\\
52	&	137	&	0	&	1.40$\times$10$^{-4}$	&	--3.85$\times$10$^{-6}$	&	0.030	$\rightarrow$	0.029	\\
\hline
\multicolumn{3}{c}{Sum(20)}   & 0.0154  & +7.3$\times$10$^{-4}$  \\
\multicolumn{3}{c}{Sum(all)}  & 0.0167 & +6.4$\times$10$^{-4}$  \\ 
\hline \hline
\end{tabular}
\end{table}

The discrepancies observed in the partial $\nu_{d,i}$ of the first 20 precursors are responsible for a difference in the total $\nu_d$ of +7.3$\times 10^{-4}$. The fact that, when considering all the precursors, such a difference decreases to +6.4$\times 10^{-4}$, means that there is a compensation due to the underestimation of less important precursors in the CRP $P_n$ tables with respect to the ENDF/B-VIII.0 decay library. From a comparison of the individual isotopes, it appears that the largest $\Delta\nu_{d,i}$  belongs to $^{137}$I and $^{91}$Br. The latter precursor in particular, has $P_n=0$ in ENDF/B-VIII.0, whereas in the previous version ENDF/B-VII.1~\cite{Chadwick2011} it had a non-zero $P_n$ value. This is quite a marked difference between the libraries since in the recommended CRP tables this \bdn~precursor alone, accounts for almost 3\% of the $\nu_d$. 

As far as the results for the DN activity shown in Fig.~\ref{fig:DNactivityComparison} are concerned, the two absolute activities seem to agree perfectly, however their ratio shows a discrepancy that varies from +4\% after about 30 s to -3\% after 400 s. The relative uncertainty in these calculations has been estimated to increase with time, due to the model that is used.  At the beginning of the decay (just after an infinite irradiation), the activity corresponds to the $\nu_d$, so the uncertainty associated with the two points is about 5\% per point. This means that the ratio has an uncertainty of about 7\% (not taking into account the correlations induced from using the same fission yields). This is the time after irradiation at which the deviation between the 2 curves is at its maximum (4\%). Therefore, when the deviation between the two curves is at its maximum (4\%), the uncertainty is at its minimum (7\%), and we may conclude that the two DN activity curves agree. 

The variations observed in the ratio of the DN activities are due to the differences in the $P_n$ values of the $^{137}$I and $^{87}$Br precursors in the two libraries.
As can be seen in Table~\ref{tab:libDiff}, the $^{137}$I $P_n$ value from the CRP is increased while the corresponding $^{87}$Br $P_n$ value is decreased. This explains the bump at 30 s - where the importance of $^{137}$I is maximum - as well as the -3\% plateau that extends till the end of the decay when only $^{87}$Br is left. 

\begin{figure}[!tb]
\includegraphics[width=1.0\linewidth]{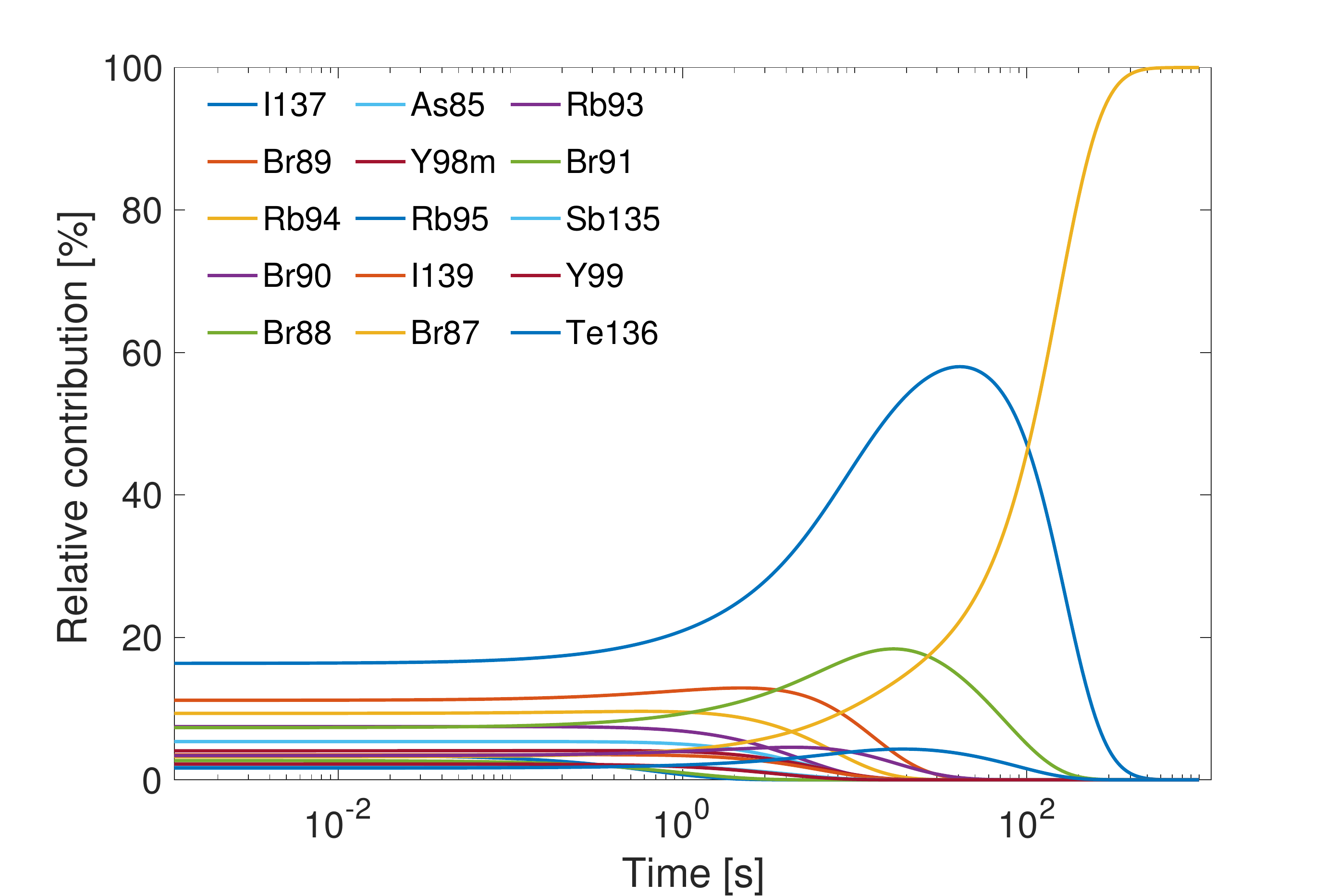}
\caption{Relative contribution of the most important precursors in time, following the thermal fission of $^{235}$U$_{t}$. It is clear that $^{137}$I dominates after about 100 s from the end of the irradiation, while after 300 s the activity is only due to $^{87}$Br}
\label{fig:PrecContributionInTime}
\end{figure}

Figure~\ref{fig:PrecContributionInTime} shows the relative contribution of the most important precursors as a function of time. It is clear that $^{137}$I dominates after about 30 s from the end of the irradiation, while after 300 s the activity is only due to $^{87}$Br. The latter finding for $^{87}$Br is quite interesting, as this precursor did not appear among the most important contributors in the basic summation calculations discussed in the previous section. This confirms that time-dependent calculations allow us to gain deeper insight in the role of each DN precursor in the \bdn process following fission.

Overall, as seen in the previous section, the CRP $P_n$ data differ from those adopted in the ENDF/B-VIII.0 decay data library for several precursor nuclides. In time-dependent calculations, these differences lead to different precursor family trees, both in shape and \textit{intensity}, where intensity is the amount of precursors created by the fathers' decay.
The impact of the new CRP ($T_{1/2}, P_n$) data is evident in the $\nu_d$ and DN activity ratio. In particular, the shape of the curve in the decay phase is strongly affected by the microscopic data of the long-lived precursors, namely $^{137}$I and $^{87}$Br.

\paragraph{Time-evolution calculations of major and minor actinides at thermal, fast and spallation energies. }\label{Sec:Macro-summation-yields-timedep-2}

The CRP data have also been used in time-evolution calculations of the composition of pure spheres filled with major and minor actinides. The calculations have been performed at three incident neutron energies (thermal, fast and spallation) for which fission yield data are available in the evaluated libraries.  To begin with, the JEFF-3.1.1 independent fission yields were combined with the JEFF-3.1 decay data to compute the activities at various time steps using the MURE code~\cite{MURE}. The total DN yields per fission $\nu_{d}$ were then calculated by means of the summation method using the $P_n$ values obtained from the CRP, ENDF/B-VIII.0 and the JEFF-3.1.1 libraries. The results are compared with the recommended values of $\nu_{d}$ found in the JEFF-3.1 evaluated library (Table~\ref{tab:TableSpheres}).

\begin{table*}
\caption{\small{Delayed neutron yields $\nu_d$ obtained from simulations of individual fuels with thermal ("t"), fast ("f") and spallation ("s") independent fission yields from JEFF-3.1.1 coupled to JEFF-3.1 half-lives and $P_n$ values from the CRP (3rd column), ENDF/B-VIII.0 (4th column) and JEFF-3.1.1 (5th column). The $\nu_{d}$ recommended in JEFF-3.1 are displayed in the 2nd column. In cases where the energies of the impinging neutrons differ substantially from the average neutron energies associated with the fission yields, the energies at which the $\nu_d$ were extracted from JEFF-3.1 are given in parentheses.}}
\label{tab:TableSpheres}
\begin{center}
\begin{tabular}{c|c|c|c|c}
\hline
\hline

\multicolumn{1}{c|}{\textbf {Nucleus}}  & \multicolumn{1}{c|}{\textbf {$\nu_{d}$ in JEFF-3.1}}  & \multicolumn{1}{c|}{\textbf {$\nu_{d}$ from CRP $P_n$'s }}& \multicolumn{1}{c|}{\textbf {$\nu_{d}$ from ENDF/B-VIII.0 $P_n$'s }}&\multicolumn{1}{c}{\textbf { $\nu_{d}$ from JEFF-3.1.1 $P_n$'s}} \tabularnewline \hline
\utrois ~t  &6.73E-3 &8.07E-3 &7.82E-3& 7.24E-3\tabularnewline
\utrois ~f  & x & 1.16E-2& 1.12E-2&1.04E-2 \tabularnewline
\utrois ~s & 4.39E-3~(14~MeV) & 6.40E-3&6.19E-3 &5.59E-3 \tabularnewline
\ucinq ~t & 1.62E-2& 1.66E-2&1.60E-2 & 1.47E-2\tabularnewline
\ucinq ~f  & 1.63E-2  ~(200~keV) &1.87E-2 &1.80E-2 & 1.70E-2\tabularnewline
\ucinq ~s  & 8.9E-3~(12~MeV)& 1.04E-2&9.8E-3 &9.3E-3 \tabularnewline
\uhuit ~f  & 4.78E-2~(3.5~MeV)& 4.6E-2& 4.5E-2& 4.1E-2\tabularnewline
\uhuit ~s & 1.88E-2~(20~MeV)&2.75E-2 &2.58E-2 &2.37E-2 \tabularnewline
\thdeux ~f & 5.27E-2~(4~MeV)& 6.35E-2& 6.11E-2 &5.39E-2 \tabularnewline
\thdeux ~s & 3.0E-2~(7-20~MeV)& 3.61E-2&3.35E-2 & 3.05E-2\tabularnewline
\npsept ~t & 1.2E-2& 1.27E-2&1.23E-2 & 1.13E-2\tabularnewline
\npsept ~f & 1.2E-2~(2~MeV)& 1.31E-2&1.26E-2 &1.17E-2 \tabularnewline
\puhuit ~t & 4.71E-3& 3.51E-3& 3.41E-3& 3.19E-3\tabularnewline
\puhuit ~f  & 4.71E-3~(7~MeV)& 5.3E-3& 5.1E-3& 4.8E-3 \tabularnewline
\puneuf ~t & 6.5E-3& 6.8E-3&6.6E-3 & 6.0E-3\tabularnewline
\puneuf ~f  &6.51E-3~(200~keV) & 7.51E-3& 7.23E-3&6.75E-3\tabularnewline
\puzero ~f & 9.0E-3~(9~MeV)& 1.03E-2 &9.9E-3 &9.2E-3 \tabularnewline
\puun ~t & 1.60E-2& 1.39E-2&1.37E-2 & 1.23E-2\tabularnewline
\puun ~f  &1.60E-2~(5~MeV) & 1.46E-2&1.41E-2 &1.29E-2 \tabularnewline
\pudeux ~f  &1.83E-2~(6~MeV) &1.90E-2 &1.85E-2 &1.68E-2 \tabularnewline
\amun ~t &4.27E-3 & 4.22E-3& 4.08E-3& 3.79E-3\tabularnewline
\amun ~f  &4.27E-3~(4~MeV) & 4.58E-3& 4.42E-3& 4.15E-3\tabularnewline
\hline
\hline 
 
\end{tabular}
\end{center}

\end{table*}

A first observation from Table~\ref{tab:TableSpheres} is that the information on delayed neutron yields available in the evaluated database JEFF-3.1 is sparce. For example, there is no value available in the fast range for \utrois, while for several actinides $\nu_{d}$ values are provided at rather different 'fast' energies, i.e. 200~keV in one case and 4~MeV in another case.

Nevertheless, from the comparison of the results obtained at the available incident energies we observe similar trends: the $\nu_{d}$ values obtained with the CRP $P_n$ data (third column in Table~\ref{tab:TableSpheres}) are systematically larger than those obtained with the ENDF/B-VIII.0 (fourth column) and JEFF-3.1.1 $P_n$ data (fifth column). The results obtained with JEFF-3.1.1 decay data are the lowest of the three and deviate from the recommended values the most. This is in agreement with the findings of Sect.~\ref{Sec:Macro-summation-yields-basic}.

For major actinides in the thermal range (fast in the case of \uhuit), the $\nu_{d}$s obtained from the summation calculations are quite close to the recommended values of JEFF-3.1.1 (second column). ENDF/B-VIII.0 is in better agreement with recommended JEFF-3.1.1 values in the case of $^{235}$U and $^{239}$Pu, while for $^{238}$U and $^{241}$Pu the CRP data are closer to the recommended values.  However, once the conditions regarding fuel or impinging neutron energy differ from those of a pressurized water reactor (PWR), the discrepancies among the summation calculations and recommended values increase. Overall the three summation calculations (using different input data) are in better agreement among themselves than with the recommended $\nu_{d}$ values of JEFF-3.1.1. The introduction of the new CRP $P_n$ values does not lead to an improvement with respect to the recommended values, but rather enhances the discrepancies, with the sole exception of $^{238}$U and $^{241}$Pu as already mentioned. This is due to the enhanced $P_n$ values of the most important contributors which was also observed in the previous sections~\ref{Sec:Macro-summation-yields-basic} and ~\ref{Sec:Macro-summation-yields-timedep-1}.

The results show that additional work is needed to improve the other microscopic data used in the summation calculations besides the \bdn~emission data, i.e. the fission yield data. It is of high importance to improve the evaluated fission yield data for a range of incident neutron energies needed in various applications.  Furthermore, considering the inconsistencies observed in some of the recommended data in JEFF-3.1.1 (identical values for thermal and fast fission), and the availability of integral data at incident neutron energies that are not directly comparable with those of the fission yields, it is clear that new integral measurements of $\nu_d$ is merited.

\subsection{Time-dependent delayed-neutron integral spectra }\label{Sec:Macro-Integralspec}

In this section we use the new CRP ($T_{1/2}, P_n$) data in calculations of time-dependent DN integral spectra for thermal neutron-induced fission of $^{235}$U and compare them with measured, as well as recommended spectra in the existing evaluated libraries.

\begin{figure*}[!htb]
	\centering
\includegraphics[width=\linewidth]{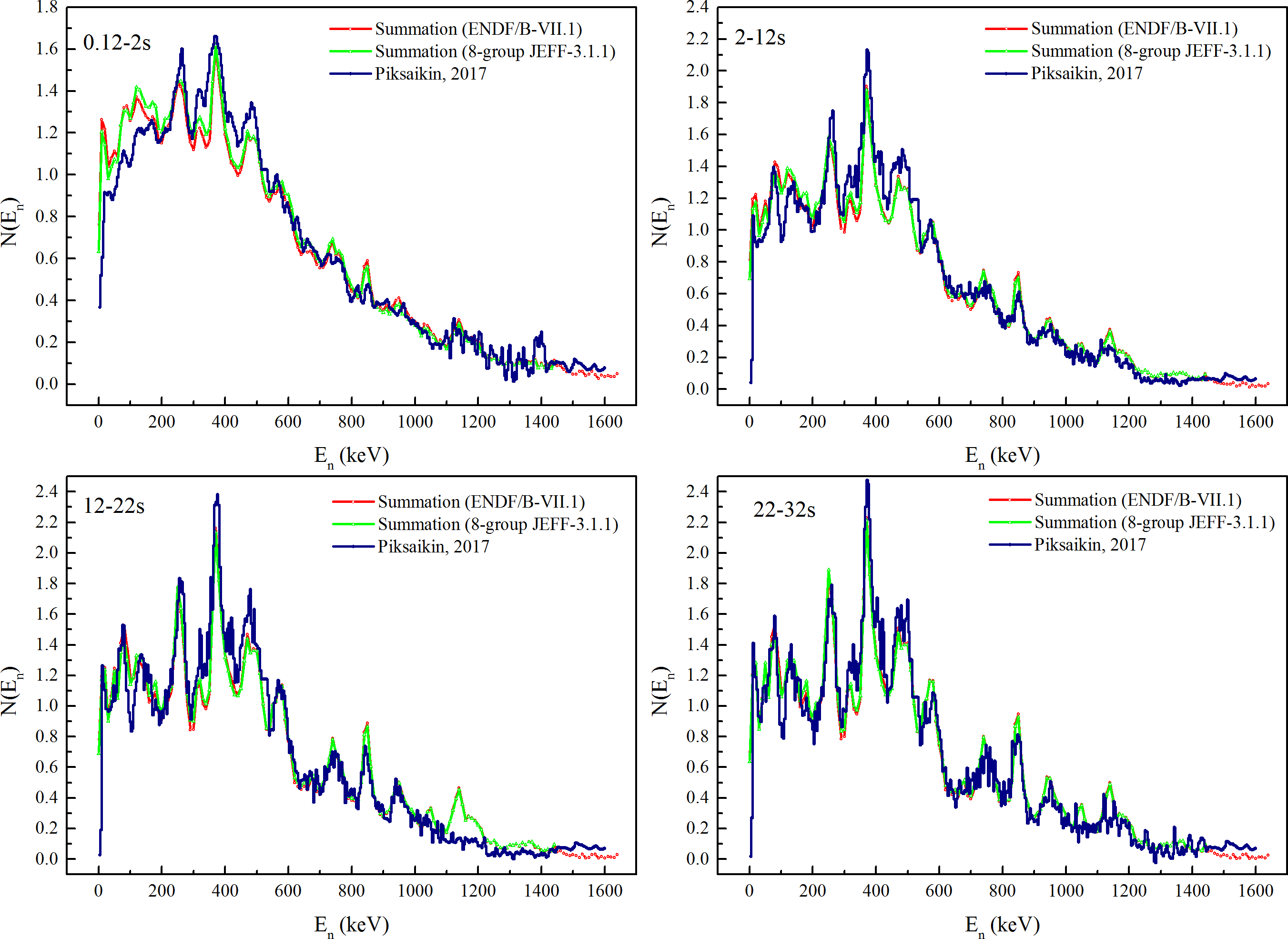}
\caption{Comparison of the integral delayed spectra obtained from thermal neutron-induced fission of $^{235}$U in the time intervals 0.12-2 s, 2-12 s, 12-22 s, 22-32 s \cite{Piksaikin17} with corresponding spectra calculated by means of the summation method using different sets of microscopic delayed neutron data (irradiation time of experiments \mbox{-} 120 s).}
\label{fig:5_8}
\end{figure*}

\begin{figure*}[!htb]
	\centering
\includegraphics[width=\linewidth]{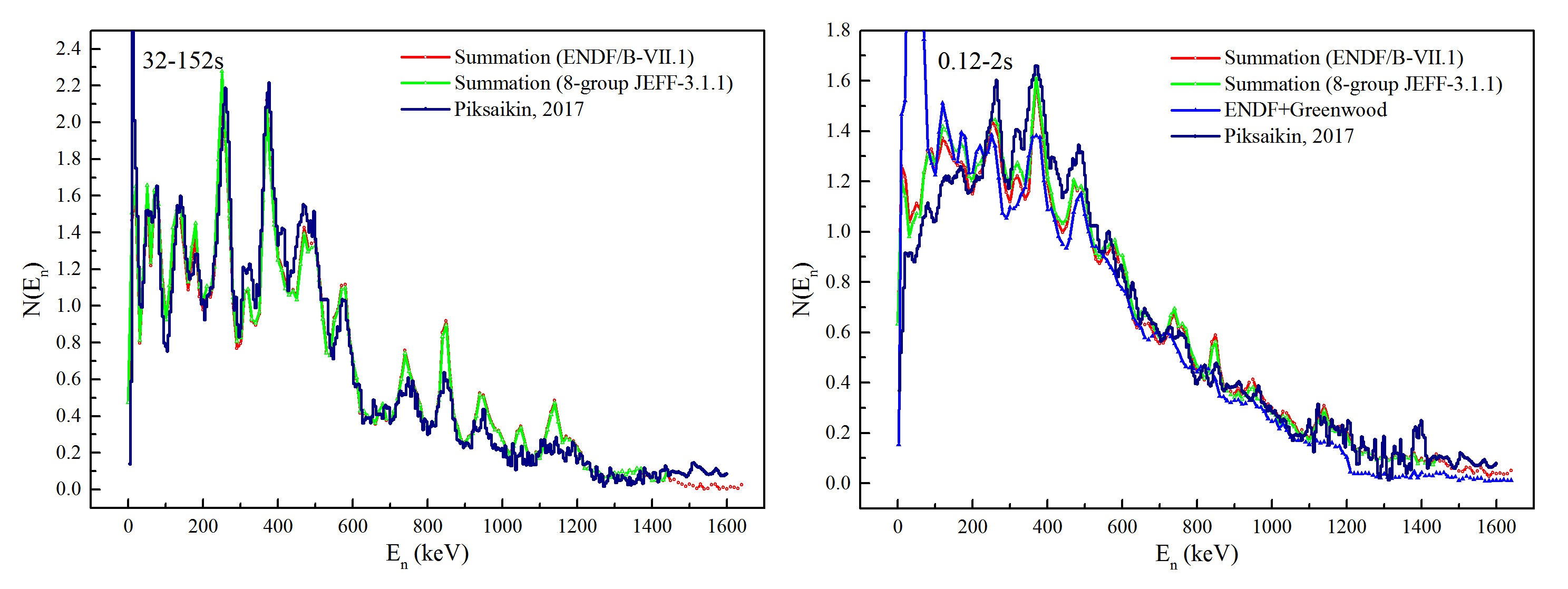}
\caption{Thermal neutron-induced fission of $^{235}$U. Left side: comparison of the integral delayed spectra in the time interval 32-152 s \cite{Piksaikin17} with corresponding spectra calculated by means of the summation method using different sets of microscopic delayed neutron data. Right side: the integral delayed spectra in the time interval 0.12-2 s. The Greenwood \textit{et al.} \cite{Greenwood1985,Greenwood1997} spectra are added to the ENDF/B-VII.1 data base~\cite{Chadwick2011} (irradiation time of experiments \mbox{-} 120 s).}
\label{fig:5_9}
\end{figure*}

\begin{figure*}[!htb]
	\centering
\includegraphics[width=\linewidth]{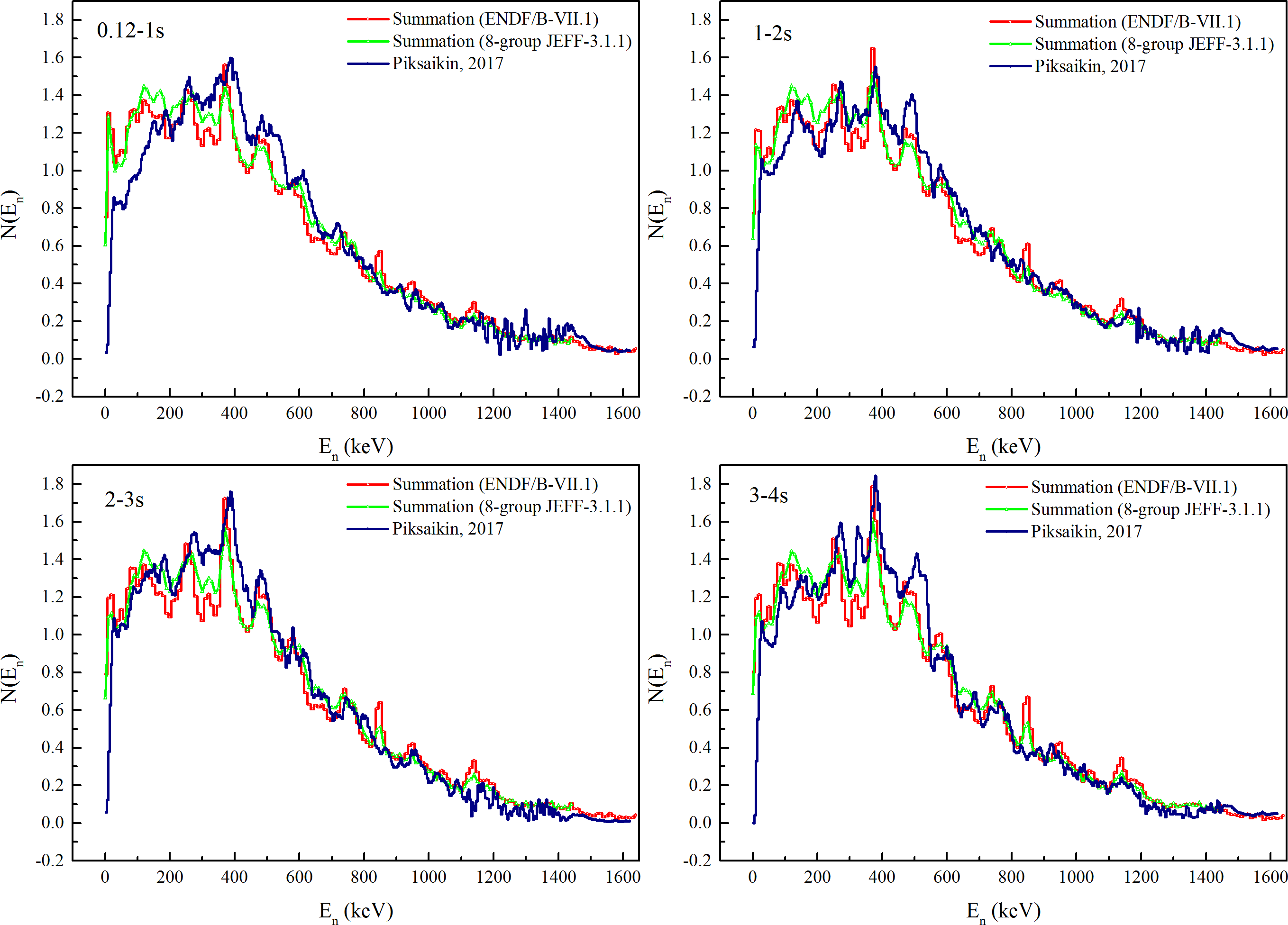}
\caption{Comparison of the integral delayed spectra obtained for thermal neutron-induced fission of $^{235}$U in the time intervals  0.12-1 s, 1-2 s, 2-3 s and 3-4 s \cite{Piksaikin17} with corresponding spectra calculated by means of the summation method using different sets of microscopic delayed neutron data (irradiation time of experiments \mbox{-} 20 s).}
\label{fig:5_10}
\end{figure*}

\begin{figure*}[!htb]
	\centering
\includegraphics[width=\linewidth]{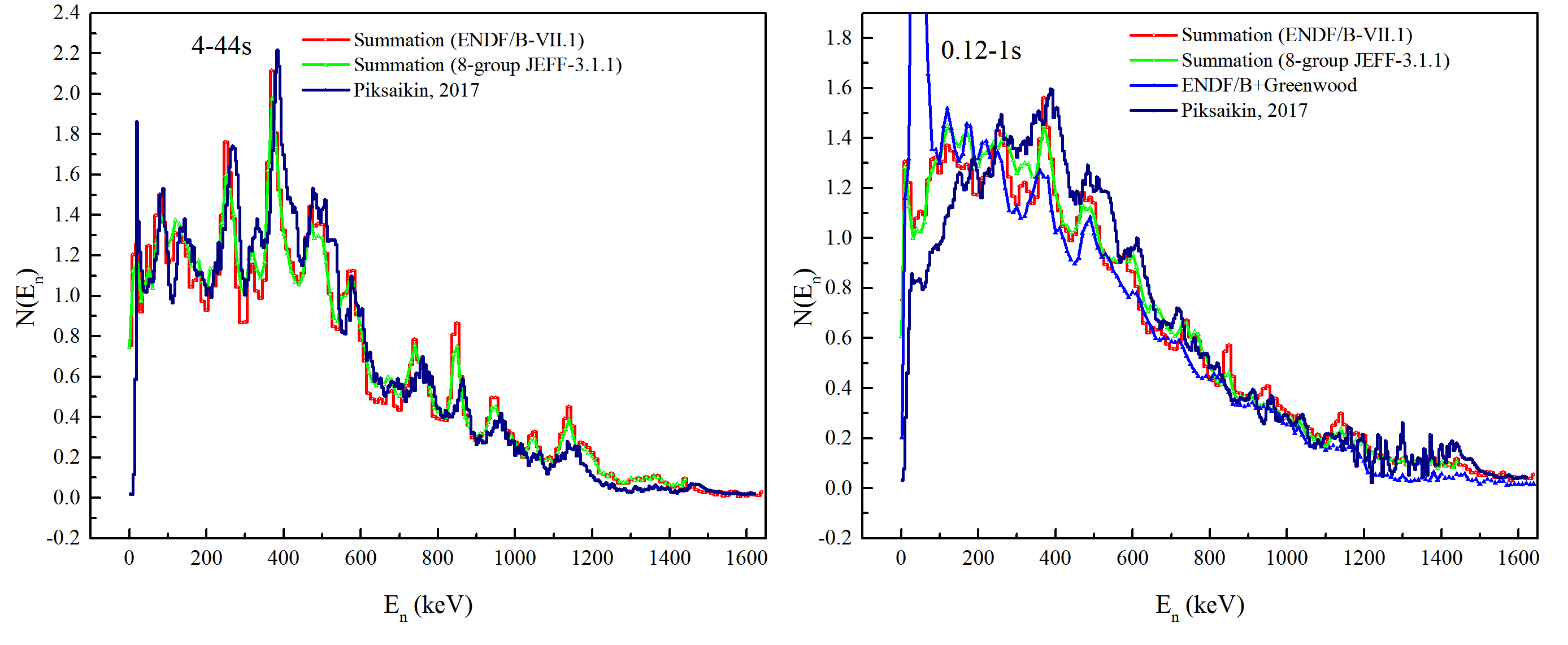}
\caption{Thermal neutron-induced fission of $^{235}$U. Left side: comparison of the integral delayed spectra in the time interval 4-44 s \cite{Piksaikin17} with corresponding spectra calculated by means of the summation method using different sets of microscopic delayed neutron data. Right side: the integral delayed spectra in the time interval 0.12-1 s. The Greenwood \textit{et al.} \cite{Greenwood1985,Greenwood1997} spectra are added to the ENDF/B-VII.1 data base~\cite{Chadwick2011} (irradiation time of experiments \mbox{-} 20 s).}
\label{fig:5_11}
\end{figure*}

In Figs.~\ref{fig:5_8}-\ref{fig:5_11}, the spectra measured for thermal neutron-induced fission of $^{235}$U at different time intervals by \cite{Piksaikin17} as described in Section~\ref{Sec:Macro-meas-new.2} are compared with the corresponding spectra calculated on the basis of the microscopic delayed neutron data, namely, the delayed neutron spectra from individual precursors~\cite{IAEA0683}, the emission probability of delayed neutrons $P_n$ and the half-life of delayed neutron precursors $T_{1/2}$ taken from the CRP tables ($T_{1/2}, P_n$). Three sets of delayed neutron spectra were used in the calculations. The first one is the delayed neutron spectra from the ENDF/B-VII.1~\cite{Chadwick2011} decay data file which has been developed by Brady and England \cite{Brady89}. These data comprise thirty four experimental spectra combined and adjusted on the basis of the raw spectra data measured in Mainz \cite{Kratz79}, Studsvik \cite{Rudstam74,Rudstam77} and INEL \cite{Greenwood1985,Greenwood1997} and additionally, the spectra for 235 nuclides calculated with the help of the evaporation model. The second set of energy spectra represents the data from the ENDF/B-VII.1 file in which the spectra for $^{89-90}$Br, $^{93-97}$Rb, $^{136}$Te, $^{138-139}$I, $^{143-145}$Cs precursors were replaced by the corresponding data of Greenwood \textit{et al.}~\cite{Greenwood1985,Greenwood1997}. The third set is the 8-group spectra from the JEFF-3.1.1 file~\cite{jeff3.1.1}. The cumulative fission yield data used in all these calculations were taken from the JEFF-3.1.1 FY file~\cite{jeff3.1.1}. 

The energy spectra in definite time-windows were calculated with the help of the time-dependent formula written for $N$ delayed neutron precursors \cite{Piksaikin17}
\begin{align}
\chi(E_n,t) dE_n = & A\cdot\sum_{i=1}^{N} \left( \frac{P_{ni}\cdot CY_i}{\lambda_i}\right)\left(1-e^{-\lambda_i \cdot t_{irr}}\right) \notag\\ & \times \left(e^{-\lambda_i\cdot t_d}\right)\left(1-e^{-\lambda_i\cdot\Delta t_c}\right) T_i\chi_i (E_n) dE_n\, ,
\end{align}
\begin{align*}
T_i=\left[ \frac{n}{1-e^{-\lambda_i\cdot T}}-e^{-\lambda_i \cdot T}\cdot\frac{1-e^{-n\lambda_i T}}{\left(1-e^{-\lambda_i\cdot T}\right)^2} \right]\,,
\end{align*}
where $T_{i}$ term describes the dependence of delayed neutron activity on the number of cycles; A the saturation activity; $P_{ni}$ the emission probability of the i-th precursor; $CY_i$ the cumulative yield of the $i$-th precursor taken from JEFF-3.1.1 file~\cite{jeff3.1.1}; $\chi_i(E_n)$ the delayed neutron energy spectrum associated with the i-th precursor; $\lambda_i$ the decay constant of the $i$-th precursor; $t_{irr}$ the irradiation time, s; $t_{d}$ the delay time (cooling), s; $\Delta t_c$ the neutron counting time (time window); $n$ the number of cycles; $T$ the period of one cycle (irradiation-cooling-counting-delay).

As can be seen from Figs.~\ref{fig:5_8} and ~\ref{fig:5_9} both the general form and peak structure of the summation spectra in time intervals 2-12 s, 12-22 s, 22-32 s, 32-152 s and in the whole energy range are in excellent agreement with the corresponding integral spectra measured with the irradiation time 120 s. In the first time interval 0.12-2 s both summation data overestimate the low-energy part of the energy spectrum as compared with the experimental data but nevertheless, they show the same peak structure at 10, 50, 80, 120 and 170 keV. Upon inclusion in the ENDF/B-VII.1 decay data file~\cite{Chadwick2011} of the Greenwood \textit{et al.} spectra for precursors $^{89-90}$Br, $^{93-97}$Rb, $^{136}$Te, $^{138-139}$I, $^{143-145}$Cs \cite{Greenwood1985,Greenwood1997}, the agreement between the ENDF/B-VII.1 based summation spectra and the experimental data severely deteriorates (see Figs.~\ref{fig:5_8} and ~\ref{fig:5_10}). The main difference between these spectra is that the low-energy part of the spectrum (below 100 keV) is heavily overestimated when the Greenwood precursor spectra are included. The summation energy spectra calculated on the basis of the 8-group data are very close to the ENDF/B-VII.1 based data~\cite{Chadwick2011}. This observation can be considered as a clear indication that the procedure used for the generation of the 8-group spectra from microscopic data is reliable and that the 8-group spectra reproduce the time evolution of the composite spectra.  
In the case of the short irradiation data (see Fig.~\ref{fig:5_11}), we observe good agreement between experimental and calculated spectra data in the 4-44 s time interval except for some deviations at the high energy peaks at 740, 850, 950, 1050, and 1140 keV, which are overestimated in the summation spectra. The main difference between the summation spectra and experimental short irradiation data in the other time intervals is the low intensity of the 317 and 352 keV peaks and the unresolved peaks at about 420 and 513 keV in the summation spectra (see Fig.~\ref{fig:5_10}).

Thus, the agreement between the summation and composite experimental spectra observed in the long irradiation data validates the approach taken by Brady and England \cite{Brady89} in developing the ENDF/B-VII.1 precursor spectra. Furthermore, this agreement justifies our confidence in the summation method using the current microscopic CRP \bdn database. The discrepancies observed in the short irradiation data show the necessity to continue efforts to improve the microscopic data on the short-lived precursor spectra on the basis of the up-to-date model  developed by \cite{Kawano2008}.  The IPPE integral energy spectra obtained from thermal neutron-induced fission of $^{235}$U in different time intervals \cite{Piksaikin17} that were successfully used in the comparison studies in this section, should be used further as a benchmark for both testing individual precursor spectra and improving the short-lived group spectra.

\subsection{Time-dependent delayed-neutron parameters}\label{subsec:time-dependence}
The CRP data are also used to calculate time-dependent parameters in the 8-group model. For this purpose, the DN activity curve is obtained from summation calculations over the microscopic \bdn~CRP data ($T_{1/2}, P_n$). The 8-group parameters ($a_i, \lambda_i$) are extracted from the activity curve and compared with recommended 8-group parameters~\cite{NEA-WPEC6}. This is another way of assessing the impact of the CRP ($T_{1/2}, P_n$) data on integral quantities.

\paragraph{Time-dependent parameters from microscopic calculations}


The delayed neutron-activity is simulated through microscopic summation calculations, as explained in Subsection~\ref{subsec:time-dependence}. Fission Yields (FY) are taken from JEFF-3.1.1 and Radioactive Decay Data (RDD) from CRP+ENDF/B-VIII.0 and ENDF/B-VIII as described in Section~\ref{Sec:Macro-summation-yields-basic}. Where needed, the excitation state of the daughter nucleus after the decay ($RFS$) is taken from ENDF/B-VIII.0~\cite{Brown2018}.
 
The DN activity obtained from the summation calculations is then fitted by the CONRAD code~\cite{Conrad}, which is developed at CEA (France) with the purpose of performing high-quality uncertainty analysis. 
Instead of fitting the delayed neutron activity directly, we have chosen to fit the difference between the two models: microscopic and macroscopic (Eq.~\ref{eq:fit}).

\begin{equation}
{\frac{n_d(t)}{n_d(0)}} - \sum\limits_{i=1}^8 a_i(1-e^{-\lambda_i \; t_{\mathrm{irr}}})e^{-\lambda_i \; (t-t_{\mathrm{irr}})} = 0,
\label{eq:fit}
\end{equation}

where $t_{\mathrm{irr}}$ is the irradiation time and $n_d(0)$ is the DN emission rate at the end of the irradiation phase, which corresponds to the average DN yield in case of an infinite irradiation.

\begin{figure} 
\includegraphics[width=1.1\linewidth]{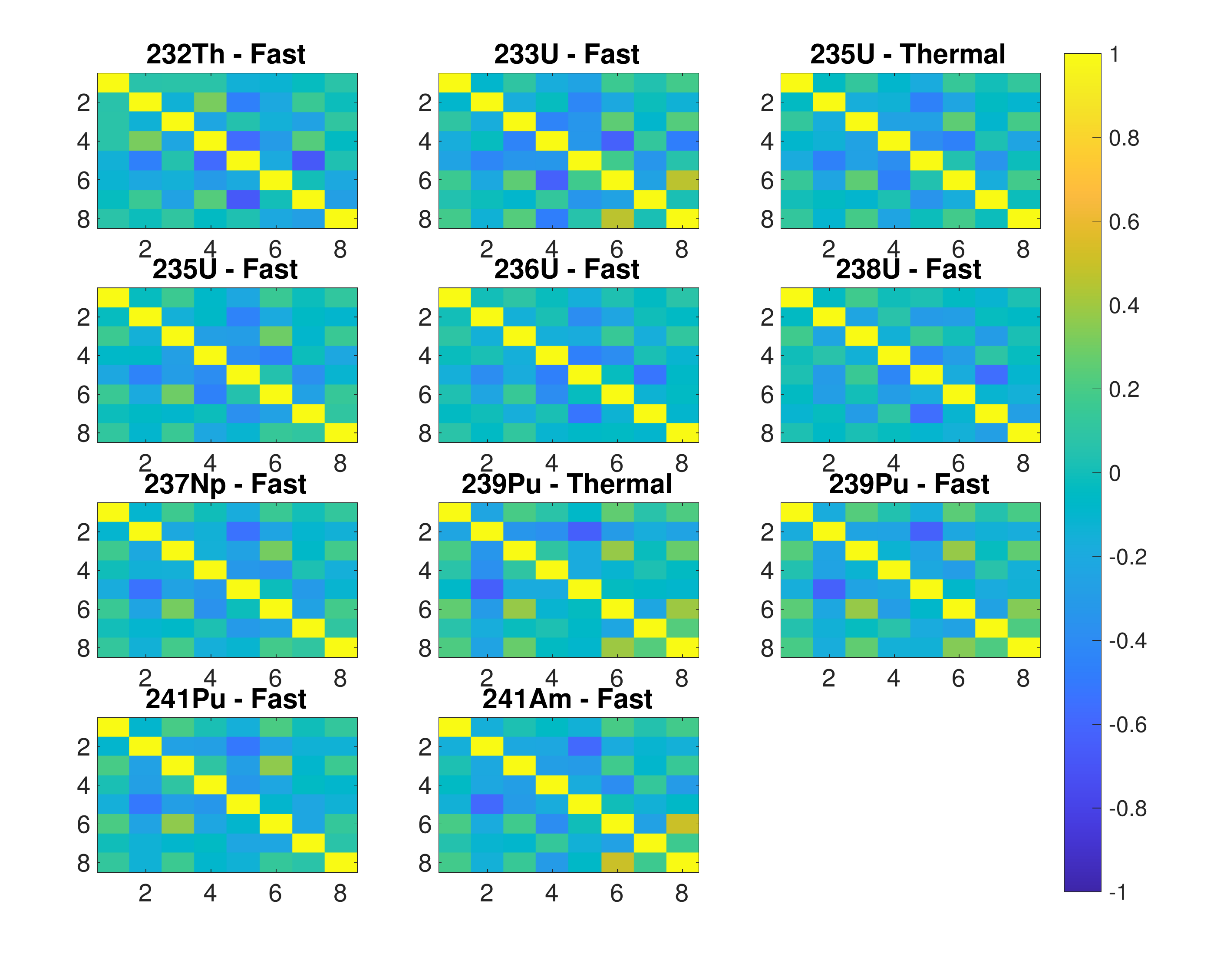}
\caption{Group abundances correlation matrix for several fissioning systems and energies. The correlations are produced by the CONRAD code through an analytical marginalization technique.}
\label{fig:corr}
\end{figure}

The CONRAD code implements Baye's theorem, which allows adjusting the model parameters - in this case, the group abundances - according to the available experimental information. 
In this work, however, the experimental data are replaced by the calculated DN emission rates. The parameters of the macroscopic model are therefore adjusted to fit the microscopic model. This procedure, coupled with an analytical marginalization technique, allows one to propagate the uncertainty in hundreds of microscopic data to just the eight abundances $a_i$. At the same time, CONRAD provides the correlations among the uncertainties of the fitted parameters (see Fig.~\ref{fig:corr}), which compensate for the apparently higher uncertainty in the single abundances. 

The results of the simulation of the experiment, i.e. of the DN activity curve, and the subsequent extraction of the group parameters are presented in Tab.~\ref{tab:ai}. The associated uncertainties given in brackets reflect the uncertainties arising from the hundreds of parameters marginalized in the CONRAD fit. Tables ~\ref{tab:corr_Th232f} to~\ref{tab:corr_Pu241f} display the correlation matrices of the abundances obtained with the analytical marginalization technique. The small correlations among the abundances derive from the fact that each group is responsible for a part of the decay curve and is supposed to represent only those precursors having a similar half-life. A \textit{physical} correlation among abundances could only arise from the family tree, for example, a short-lived precursor being created by the decay of a long-lived precursor. This would create a small correlation between the two groups containing the mentioned precursors. The average precursors' half-lives, computed using the new set of abundances, are listed in Tab.~\ref{tab:meanHL}, together with the uncertainty calculated with and without correlations. 

\begin{table*}[!htb]
\caption{Temporal parameters obtained from summation calculations [per 100 fissions], computed using JEFF-3.1.1 fission yields and CRP+ENDF/B-VIII.0 decay data. Relative uncertainties are given in brackets.}
\label{tab:ai}
\begin{tabular}{c|cccccccc} \hline \hline
Sample & a$_1$ & a$_2$ & a$_3$ & a$_4$ & a$_5$ & a$_6$ & a$_7$ & a$_8$  \\ \hline
$^{232}$Th$_{f}$ &   2.69 (14.4\%) &   7.27 (14.7\%) &   7.17 (14.0\%) &  11.98 (14.2\%) &  34.94 (8.8\%) &  20.33 (8.2\%) &  11.65 (16.7\%) &   3.98 (17.9\%)   \\
$^{233}$U$_{f}$ &   7.27 (16.2\%) &   9.22 (28.9\%) &  13.23 (21.7\%) &  18.53 (19.9\%) &  39.45 (9.5\%) &   5.74 (27.0\%) &   5.74 (24.4\%) &   0.82 (35.3\%)  \\
$^{235}$U$_{t}$ &   3.37 (14.9\%) &  16.93 (14.4\%) &   9.13 (17.3\%) &  17.59 (14.7\%) &  33.92 (9.1\%) &  11.07 (12.0\%) &   6.01 (24.8\%) &   1.97 (16.9\%)  \\
$^{235}$U$_{f}$ &   3.17 (13.4\%) &  13.28 (16.1\%) &   8.90 (16.4\%) &  17.50 (13.9\%) &  38.80 (7.8\%) &   8.54 (14.9\%) &   7.49 (22.7\%) &   2.32 (18.9\%)   \\
$^{236}$U$_{f}$ &   2.24 (17.2\%) &  11.49 (14.2\%) &   6.76 (16.9\%) &  14.17 (14.6\%) &  37.29 (7.9\%) &  14.58 (9.3\%) &   9.41 (20.6\%) &   4.05 (15.1\%)  \\
$^{238}$U$_{f}$ &   0.93 (14.9\%) &   9.92 (12.5\%) &   4.24 (14.1\%) &  12.40 (10.0\%) &  31.33 (7.0\%) &  20.30 (7.3\%) &  13.56 (13.3\%) &   7.32 (14.2\%)   \\
$^{237}$Np$_{f}$ &   3.02 (14.2\%) &  16.05 (18.1\%) &   7.66 (17.1\%) &  15.53 (13.8\%) &  38.64 (8.3\%) &  10.54 (12.9\%) &   5.98 (27.1\%) &   2.57 (15.4\%)  \\
$^{239}$Pu$_{t}$ &   2.54 (16.8\%) &  26.53 (17.8\%) &   6.26 (22.6\%) &  16.18 (15.3\%) &  32.80 (12.1\%) &  10.15 (13.5\%) &   3.34 (38.0\%) &   2.19 (13.2\%)   \\
$^{239}$Pu$_{f}$ &   2.86 (14.7\%) &  18.17 (21.8\%) &   7.49 (19.5\%) &  15.97 (14.2\%) &  40.30 (9.5\%) &   8.04 (17.9\%) &   5.38 (27.6\%) &   1.76 (18.4\%)   \\
$^{241}$Pu$_{f}$ &   1.25 (15.1\%) &  22.72 (12.3\%) &   4.48 (19.4\%) &  15.95 (12.4\%) &  32.66 (8.7\%) &  12.92 (10.2\%) &   6.26 (23.2\%) &   3.75 (13.3\%)   \\
$^{241}$Am$_{f}$ &   5.06 (21.9\%) &  17.54 (28.8\%) &   8.33 (39.1\%) &  16.65 (21.3\%) &  39.94 (12.4\%) &   7.19 (24.7\%) &   4.13 (35.9\%) &   1.17 (26.3\%)  \\
 \hline\hline 
 \end{tabular}
\end{table*}

\begin{table*}[h]
\begin{minipage}{0.5\textwidth}

\caption{$^{232}$Th-fast fission. $a_i$ correlation matrix. }
\label{tab:corr_Th232f}
\begin{tabular}{cccccccc} \hline \hline
1  & 0.068   & 0.076    & 0.071    & -0.146    & -0.124    & -0.028    & 0.076    \\ 
0.068    & 1 & -0.138    & 0.315    & -0.465    & -0.206    & 0.149    & -0.012    \\ 
0.076    & -0.138   & 1  & -0.224    & 0.052    & -0.149    & -0.251    & 0.108    \\ 
0.071    & 0.315   & -0.224    & 1  & -0.572    & -0.299    & 0.228    & -0.042    \\ 
-0.146    & -0.465   & 0.052    & -0.572    & 1  & -0.198    & -0.665    & 0.035    \\ 
-0.124    & -0.206   & -0.149    & -0.299    & -0.198    & 1  & 0.004    & -0.209    \\ 
-0.028    & 0.149   & -0.251    & 0.228    & -0.665    & 0.004    & 1  & -0.269    \\ 
0.076    & -0.012   & 0.108    & -0.042    & 0.035    & -0.209    & -0.269    & 1 \\ 

\hline\hline 
\end{tabular}
\end{minipage}\hfill
\begin{minipage}{0.5\textwidth}
\caption{$^{233}$U-fast fission. $a_i$ correlation matrix.}
\label{tab:corr_U233f}
\begin{tabular}{cccccccc} \hline \hline
1  & -0.088   & 0.099    & -0.178    & -0.247    & 0.157    & 0.044    & 0.180    \\ 
-0.088    & 1 & -0.173    & -0.025    & -0.424    & -0.208    & -0.013    & -0.140    \\ 
0.099    & -0.173   & 1  & -0.438    & -0.333    & 0.250    & -0.086    & 0.231    \\ 
-0.178    & -0.025   & -0.438    & 1  & -0.322    & -0.631    & 0.123    & -0.474    \\ 
-0.247    & -0.424   & -0.333    & -0.322    & 1  & 0.173    & -0.331    & 0.055    \\ 
0.157    & -0.208   & 0.250    & -0.631    & 0.173    & 1  & -0.252    & 0.464    \\ 
0.044    & -0.013   & -0.086    & 0.123    & -0.331    & -0.252    & 1  & 0.021    \\ 
0.180    & -0.140   & 0.231    & -0.474    & 0.055    & 0.464    & 0.021    & 1 \\ 
         \hline\hline 
\end{tabular}
\end{minipage}\hfill
\begin{minipage}{0.5\textwidth}
\caption{$^{235}$U-fast fission. $a_i$ correlation matrix.}
\label{tab:corr_U235f}
\begin{tabular}{cccccccc} \hline \hline
 1  & -0.026   & 0.161    & -0.076    & -0.216    & 0.151    & -0.008    & 0.111    \\ 
-0.026    & 1 & -0.150    & -0.067    & -0.446    & -0.199    & -0.060    & -0.098    \\ 
0.161    & -0.150   & 1  & -0.271    & -0.282    & 0.292    & -0.087    & 0.155    \\ 
-0.076    & -0.067   & -0.271    & 1  & -0.385    & -0.458    & -0.012    & -0.212    \\ 
-0.216    & -0.446   & -0.282    & -0.385    & 1  & 0.053    & -0.374    & -0.113    \\ 
0.151    & -0.199   & 0.292    & -0.458    & 0.053    & 1  & -0.264    & 0.139    \\ 
-0.008    & -0.060   & -0.087    & -0.012    & -0.374    & -0.264    & 1  & 0.102    \\ 
0.111    & -0.098   & 0.155    & -0.212    & -0.113    & 0.139    & 0.102    & 1 \\
 \hline\hline 
\end{tabular}
\end{minipage}\hfill
\begin{minipage}{0.5\textwidth}

\caption{$^{235}$U-thermal fission. $a_i$ correlation matrix.}
\label{tab:corr_U235t}
\begin{tabular}{cccccccc} \hline \hline
1  & -0.041   & 0.122    & -0.087    & -0.193    & 0.133    & 0.007    & 0.118    \\ 
-0.041    & 1 & -0.171    & -0.134    & -0.456    & -0.228    & -0.042    & -0.107    \\ 
0.122    & -0.171   & 1  & -0.256    & -0.264    & 0.253    & -0.098    & 0.188    \\ 
-0.087    & -0.134   & -0.256    & 1  & -0.380    & -0.454    & 0.024    & -0.240    \\ 
-0.193    & -0.456   & -0.264    & -0.380    & 1  & 0.044    & -0.346    & -0.013    \\ 
0.133    & -0.228   & 0.253    & -0.454    & 0.044    & 1  & -0.173    & 0.187    \\ 
0.007    & -0.042   & -0.098    & 0.024    & -0.346    & -0.173    & 1  & -0.009    \\ 
0.118    & -0.107   & 0.188    & -0.240    & -0.013    & 0.187    & -0.009    & 1 \\ 
 \hline\hline 
\end{tabular}
\end{minipage}\hfill
\begin{minipage}{0.5\textwidth}
\caption{$^{236}$U-fast fission. $a_i$ correlation matrix.}
\label{tab:corr_U236f}
\begin{tabular}{cccccccc} \hline \hline
1     & 0.005	  & 0.086     & -0.013     & -0.156	& 0.027     & -0.047	 & 0.078  \\
0.005	  & 1	  & -0.158     & 0.025     & -0.385	& -0.233     & -0.012	  & -0.046  \\
0.086	  & -0.158     & 1     & -0.163     & -0.173	 & 0.145     & -0.175	  & 0.117  \\
-0.013     & 0.025     & -0.163     & 1     & -0.461	 & -0.382     & 0.016	  & -0.109  \\
-0.156     & -0.385	& -0.173     & -0.461	  & 1	  & -0.026     & -0.521     & -0.07  \\
0.027	  & -0.233     & 0.145     & -0.382	& -0.026     & 1     & -0.131	  & -0.062  \\
-0.047     & -0.012	& -0.175     & 0.016	 & -0.521     & -0.131     & 1     & -0.096  \\
0.078	  & -0.046     & 0.117     & -0.109	& -0.07     & -0.062	 & -0.096     & 1  \\ 
     \hline\hline 
\end{tabular}
\end{minipage}\hfill
\begin{minipage}{0.5\textwidth}
\caption{$^{237}$Np-fast fission. $a_i$ correlation matrix.}
\label{tab:corr_Np237f}
\begin{tabular}{cccccccc} \hline \hline
 1     & -0.109     & 0.165	& 0.002     & -0.18	& 0.141     & -0.006	 & 0.132  \\
 -0.109     & 1     & -0.211	 & -0.143     & -0.542     & -0.236	& -0.106     & -0.148  \\
 0.165     & -0.211	& 1	& -0.162     & -0.244	  & 0.298     & -0.081     & 0.191  \\
 0.002     & -0.143	& -0.162     & 1     & -0.303	  & -0.36     & 0.019	  & -0.159  \\
 -0.18     & -0.542	& -0.244     & -0.303	  & 1	  & -0.007     & -0.311     & -0.111  \\
 0.141     & -0.236	& 0.298     & -0.36	& -0.007     & 1     & -0.25	 & 0.169  \\
 -0.006     & -0.106	 & -0.081     & 0.019	  & -0.311     & -0.25     & 1     & 0.108  \\
 0.132     & -0.148	& 0.191     & -0.159	 & -0.111     & 0.169	  & 0.108     & 1  \\
     	  \hline\hline 
\end{tabular}
\end{minipage}\hfill
\begin{minipage}{0.5\textwidth}
\caption{$^{238}$U-fast fission. $a_i$ correlation matrix.}
\label{tab:corr_U238f}
\begin{tabular}{cccccccc} \hline \hline
1     & -0.051     & 0.179     & 0     & 0.021     & -0.033	& -0.119     & 0.032  \\
-0.051     & 1     & -0.243	& 0.077     & -0.302	 & -0.288     & -0.025     & -0.057  \\
0.179	  & -0.243     & 1     & -0.149     & 0.161	& 0.029     & -0.305	 & 0.019  \\
0     & 0.077	  & -0.149     & 1     & -0.448     & -0.284	 & 0.098     & -0.017  \\
0.021	  & -0.302     & 0.161     & -0.448	& 1	& -0.166     & -0.572	  & -0.114  \\
-0.033     & -0.288	& 0.029     & -0.284	 & -0.166     & 1     & -0.113     & -0.15  \\
-0.119     & -0.025	& -0.305     & 0.098	 & -0.572     & -0.113     & 1     & -0.27  \\
0.032	  & -0.057     & 0.019     & -0.017	& -0.114     & -0.15	 & -0.27     & 1  \\
        \hline\hline 
\end{tabular}
\end{minipage}\hfill
\begin{minipage}{0.5\textwidth}
\caption{$^{239}$Pu-fast fission. $a_i$ correlation matrix.}
\label{tab:corr_Pu239f}
\begin{tabular}{cccccccc} \hline \hline
 1     & -0.181     & 0.235     & 0.048     & -0.161	& 0.231     & 0.034	& 0.216  \\
-0.181     & 1     & -0.237	& -0.241     & -0.634	  & -0.209     & -0.145     & -0.184  \\
0.235	  & -0.237     & 1     & -0.127     & -0.234	 & 0.362     & -0.036	  & 0.272  \\
0.048	  & -0.241     & -0.127     & 1     & -0.193	 & -0.297     & 0.062	  & -0.152  \\
-0.161     & -0.634	& -0.234     & -0.193	  & 1	  & -0.078     & -0.196     & -0.142  \\
0.231	  & -0.209     & 0.362     & -0.297	& -0.078     & 1     & -0.252	  & 0.324  \\
0.034	  & -0.145     & -0.036     & 0.062	& -0.196     & -0.252	  & 1	  & 0.217  \\
0.216	  & -0.184     & 0.272     & -0.152	& -0.142     & 0.324	 & 0.217     & 1  \\
   	 \hline\hline 
\end{tabular}
\end{minipage}\hfill
\begin{minipage}{0.5\textwidth}
\caption{$^{239}$Pu-thermal fission. $a_i$ correlation matrix.}
\label{tab:corr_Pu239t}
\begin{tabular}{cccccccc} \hline \hline
1     & -0.234     & 0.201     & 0.076     & -0.069	& 0.25     & 0.053     & 0.229  \\
-0.234     & 1     & -0.332	& -0.364     & -0.648	  & -0.295     & -0.168     & -0.258  \\
0.201	  & -0.332     & 1     & 0.093     & -0.178	& 0.371     & -0.032	 & 0.286  \\
0.076	  & -0.364     & 0.093     & 1     & -0.186	& -0.125     & 0.024	 & -0.039  \\
-0.069     & -0.648	& -0.178     & -0.186	  & 1	  & -0.032     & -0.076     & -0.076  \\
0.25	 & -0.295     & 0.371	  & -0.125     & -0.032     & 1     & -0.222	 & 0.388  \\
0.053	  & -0.168     & -0.032     & 0.024	& -0.076     & -0.222	  & 1	  & 0.224  \\
0.229	  & -0.258     & 0.286     & -0.039	& -0.076     & 0.388	 & 0.224     & 1  \\ 
	   \hline\hline 
\end{tabular}
\end{minipage}\hfill
\begin{minipage}{0.5\textwidth}
\caption{$^{241}$Am-fast fission. $a_i$ correlation matrix.}
\label{tab:corr_Am241f}
\begin{tabular}{cccccccc} \hline \hline
  1     & -0.162     & 0.025	& -0.045     & -0.139	  & 0.191     & 0.04	 & 0.196  \\
 -0.162     & 1     & -0.211	 & -0.214     & -0.588     & -0.187	& -0.11     & -0.157  \\
 0.025     & -0.211	& 1	& -0.279     & -0.296	  & 0.176     & -0.088     & 0.147  \\
 -0.045     & -0.214	 & -0.279     & 1     & -0.173     & -0.398	& 0.126     & -0.302  \\
 -0.139     & -0.588	 & -0.296     & -0.173     & 1     & 0.006     & -0.15     & -0.061  \\
 0.191     & -0.187	& 0.176     & -0.398	 & 0.006     & 1     & -0.263	  & 0.499  \\
 0.04	  & -0.11     & -0.088     & 0.126     & -0.15     & -0.263	& 1	& 0.173  \\
 0.196     & -0.157	& 0.147     & -0.302	 & -0.061     & 0.499	  & 0.173     & 1  \\
   \hline\hline 
\end{tabular}
\end{minipage}\hfill
\begin{minipage}{0.5\textwidth}
\caption{$^{241}$Pu-fast fission. $a_i$ correlation matrix.}
\label{tab:corr_Pu241f}
\begin{tabular}{cccccccc} \hline \hline
 1     & -0.1	  & 0.205     & 0.021	  & -0.148     & 0.19	  & -0.02     & 0.129  \\
 -0.1	  & 1	  & -0.262     & -0.264     & -0.504	 & -0.242     & -0.139     & -0.142  \\
 0.205     & -0.262	& 1	& 0.091     & -0.267	 & 0.346     & -0.098	  & 0.174  \\
 0.021     & -0.264	& 0.091     & 1     & -0.316	 & -0.221     & -0.058     & -0.088  \\
 -0.148     & -0.504	 & -0.267     & -0.316     & 1     & -0.103	& -0.231     & -0.125  \\
 0.19	  & -0.242     & 0.346     & -0.221	& -0.103     & 1     & -0.209	  & 0.109  \\
 -0.02     & -0.139	& -0.098     & -0.058	  & -0.231     & -0.209     & 1     & 0.068  \\
 0.129     & -0.142	& 0.174     & -0.088	 & -0.125     & 0.109	  & 0.068     & 1  \\
  \hline\hline 
\end{tabular}
\end{minipage}\hfill
\end{table*}

\begin{table}[h]
\caption{Average half-lives $\langle T  \rangle$ and uncertainties for fast (f) and thermal (t) neutron-induced fission of minor and major actinides.}
\label{tab:meanHL}
\begin{tabular}{c|c|c|c} \hline \hline
Nuclide &   $\langle T  \rangle$ [s]& $\sigma$ (no corr.) [\%] &  $\sigma$  (corr.) [\%] \\
\hline
$^{232}$Th$_{f}$  &     6.16 &   5.64 &   5.91   \\
$^{233}$U$_{f}$   &    10.44 &   7.55 &   8.30  \\
$^{235}$U$_{t}$   &      9.38 &   5.75 &   6.42  \\
$^{235}$U$_{f}$   &     8.42 &   5.80 &   6.42  \\
$^{236}$U$_{f}$   &     6.98 &   5.94 &   6.30  \\
$^{238}$U$_{f}$   &    5.31 &   5.49 &   5.58  \\
$^{237}$Np$_{f}$  &     8.72 &   6.64 &   7.50  \\
$^{239}$Pu$_{t}$  &      10.68 &   7.44 &   9.01  \\
$^{239}$Pu$_{f}$  &    9.16 &   7.90 &   9.20  \\
$^{241}$Pu$_{f}$  &     8.77 &   5.92 &   6.81 \\
$^{241}$Am$_{f}$  &    10.38 &   9.83 &  11.14   \\ \hline\hline 
\end{tabular}
\end{table}

\paragraph{Comparison with recommended values.}
The kinetic parameters obtained by summation calculations are compared with those recommended by the CRP in the 8-group model (see Section~\ref{Sec:Macro-Recommended}). The CRP recommended values consist of those recommended by WPEC-SG6 (2002)~\cite{NEA-WPEC6}, except for the sets of abundances measured by IPPE after 2002 (\cite{Piksaikin02a,Piksaikin02c,Roshchenko06,Piksaikin13}) for which special recommendations are made in Section~\ref{Sec:Macro-Recommended}. The abundances ($a_i$) and decay-constants ($\lambda_i$) recommended by WPEC-SG6 (2002) were obtained from the expansion of the 6-group parameter set measured by Keepin in 1957, to an 8-group parameter set~\cite{NEA-WPEC6}. 
In the framework of WPEC-SG6~\cite{NEA-WPEC6}, it also appeared convenient to fix the decay-constants for all fissioning systems in order to simplify the kinetic calculations involving more than one fissioning system.

\begin{figure}[!htb]
\includegraphics[width=0.95\linewidth]{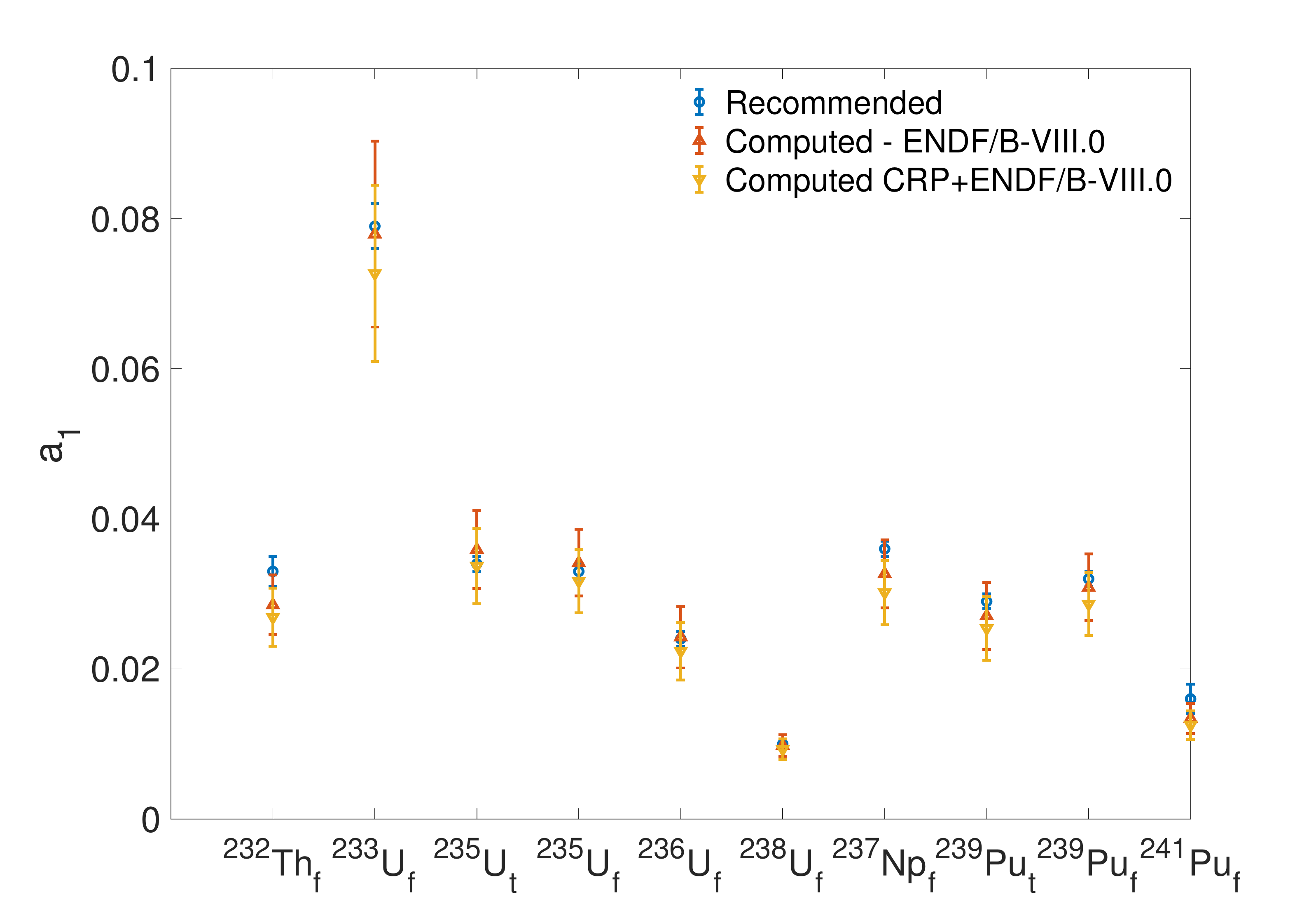}
\includegraphics[width=0.95\linewidth]{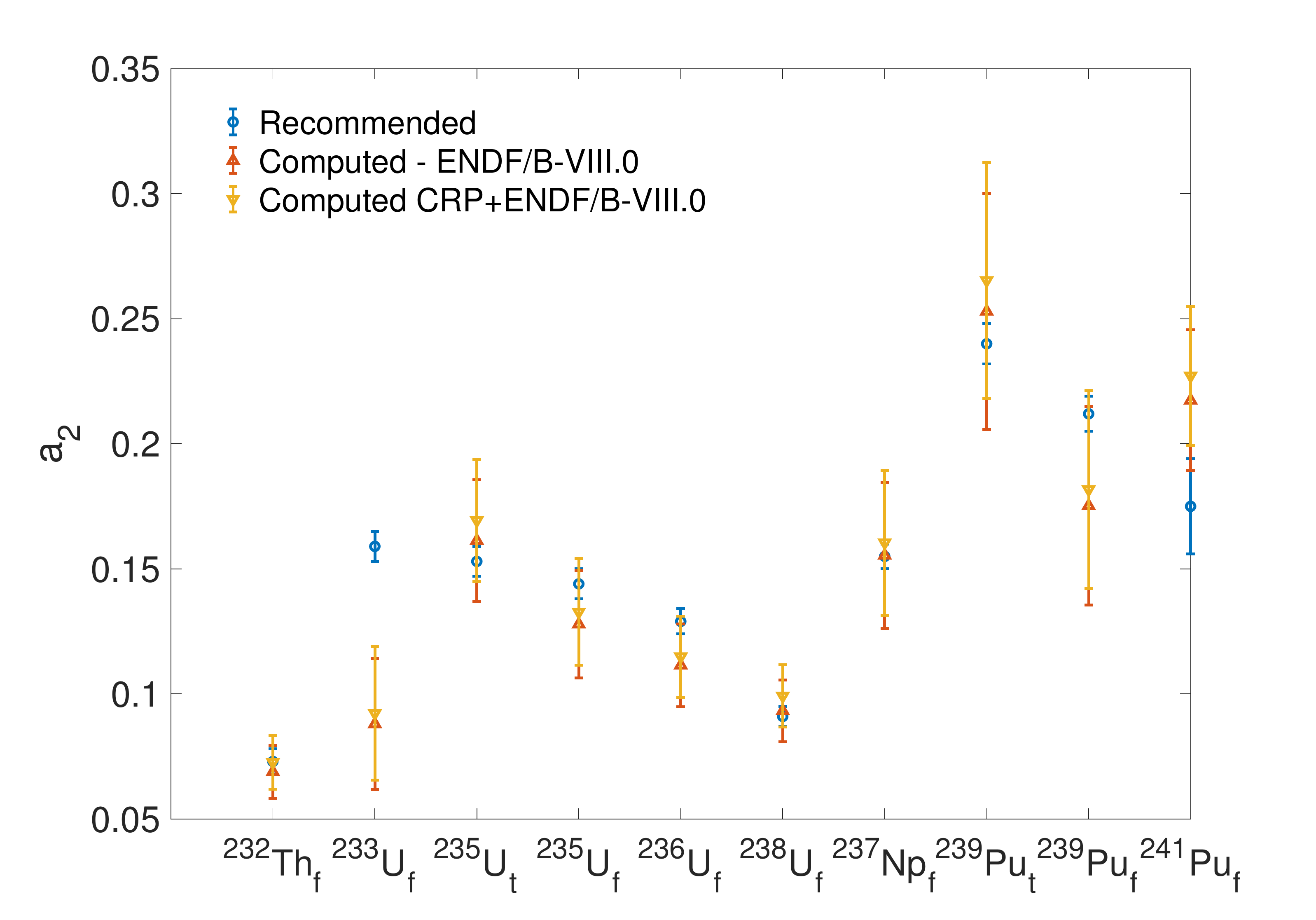}
\caption{First and Second group abundances for several fissioning systems and energies. Comparison between summation calculation results and recommended values. The uncertainty in the calculated abundances is obtained through an analytical marginalization technique.}
\label{fig:a12}
\end{figure}

\begin{figure}[!htb]
\includegraphics[width=0.95\linewidth]{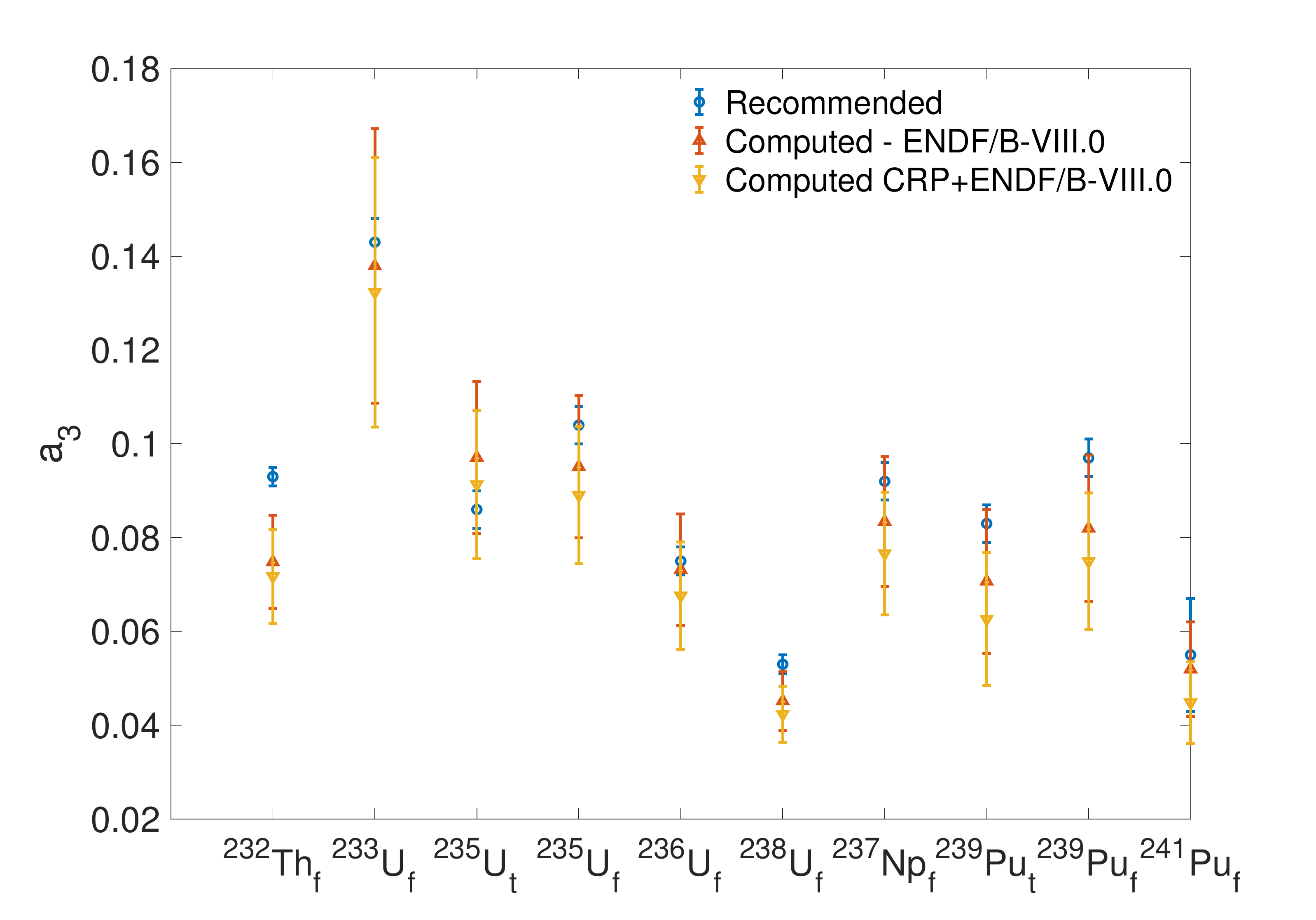}
\includegraphics[width=0.95\linewidth]{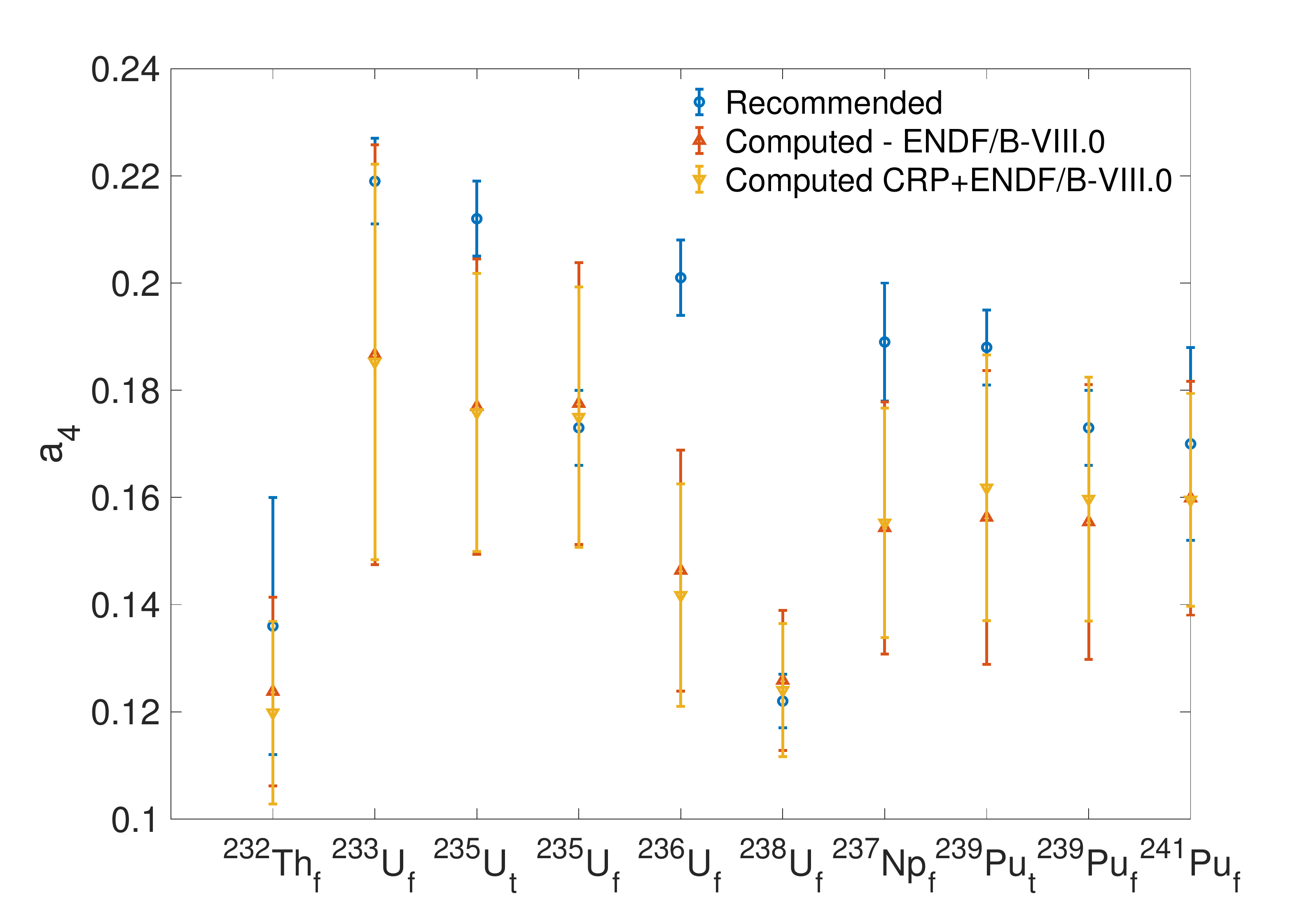}
\caption{Third and Fourth group abundance for several fissioning systems and energies. Comparison between summation calculation results and recommended values. The uncertainty in the calculated abundances is obtained through an analytical marginalization technique.}
\label{fig:a34}
\end{figure}

\begin{figure}[!htb]
\includegraphics[width=0.95\linewidth]{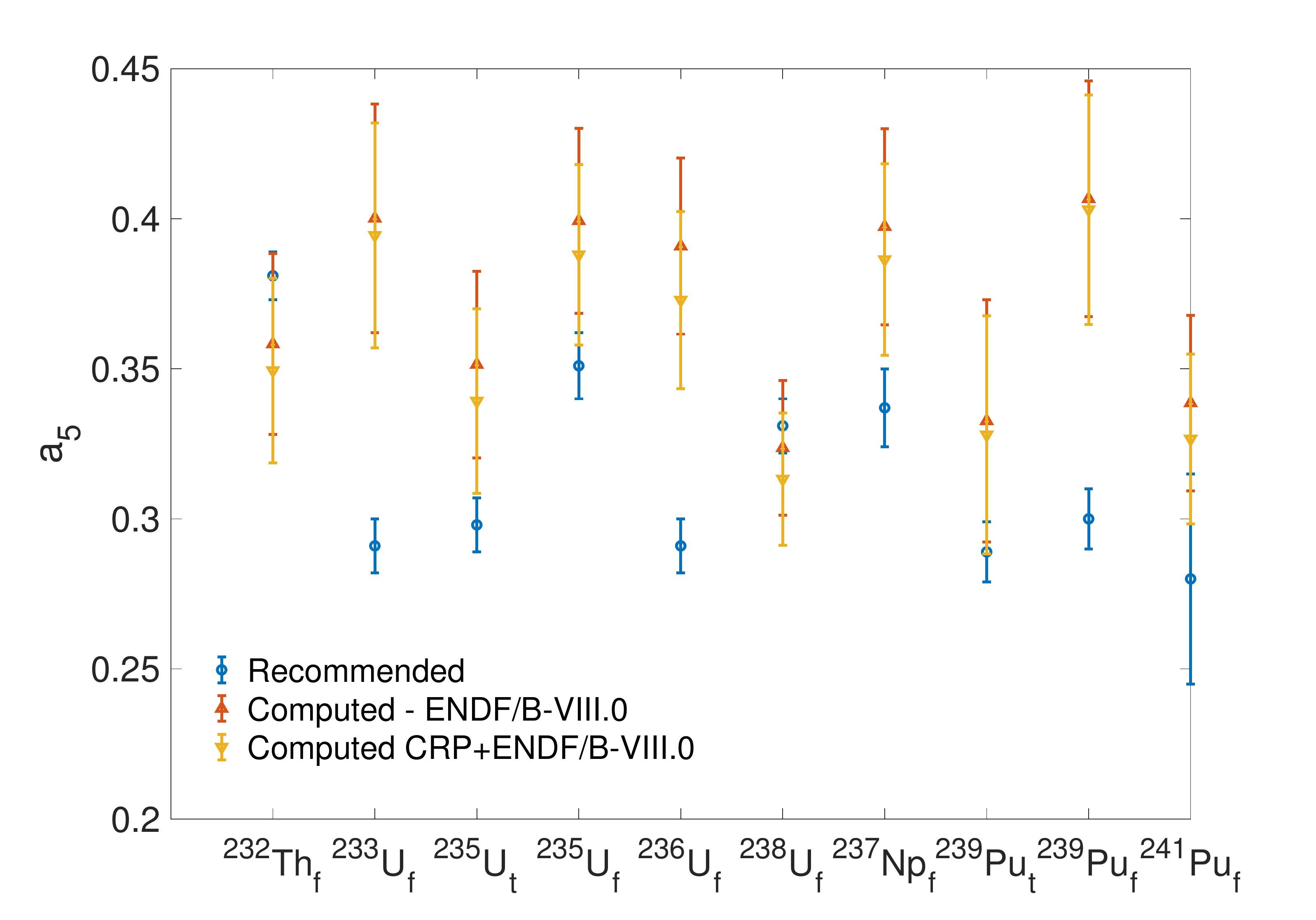}
\includegraphics[width=0.95\linewidth]{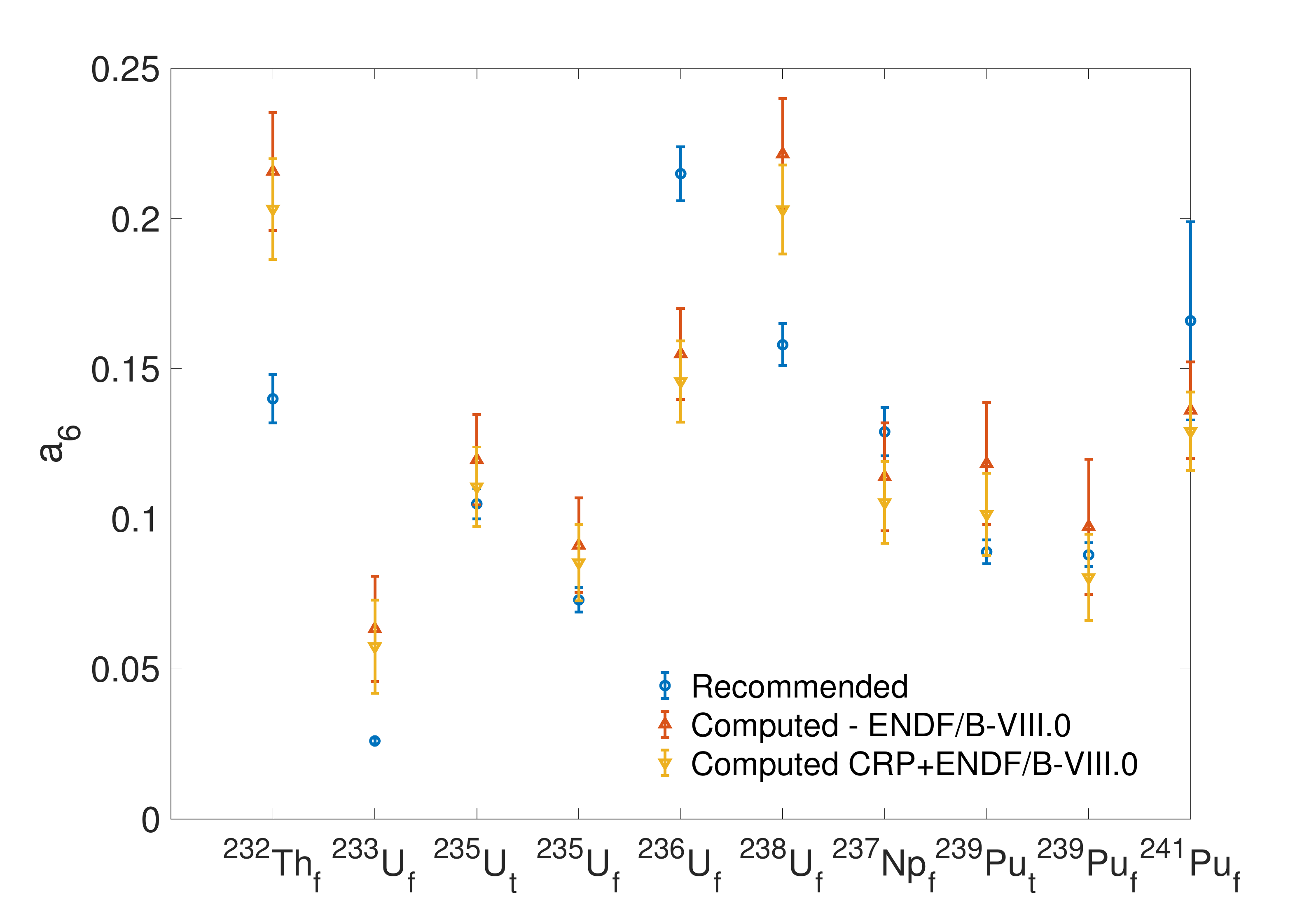}
\caption{Fifth and Sixth group abundance for several fissioning systems and energies. Comparison between summation calculation results and recommended values. The uncertainty in the calculated abundances is obtained through an analytical marginalization technique.}
\label{fig:a56}
\end{figure}

\begin{figure*}[!htb]
\begin{minipage}{0.47\linewidth}
\includegraphics[width=0.95\linewidth]{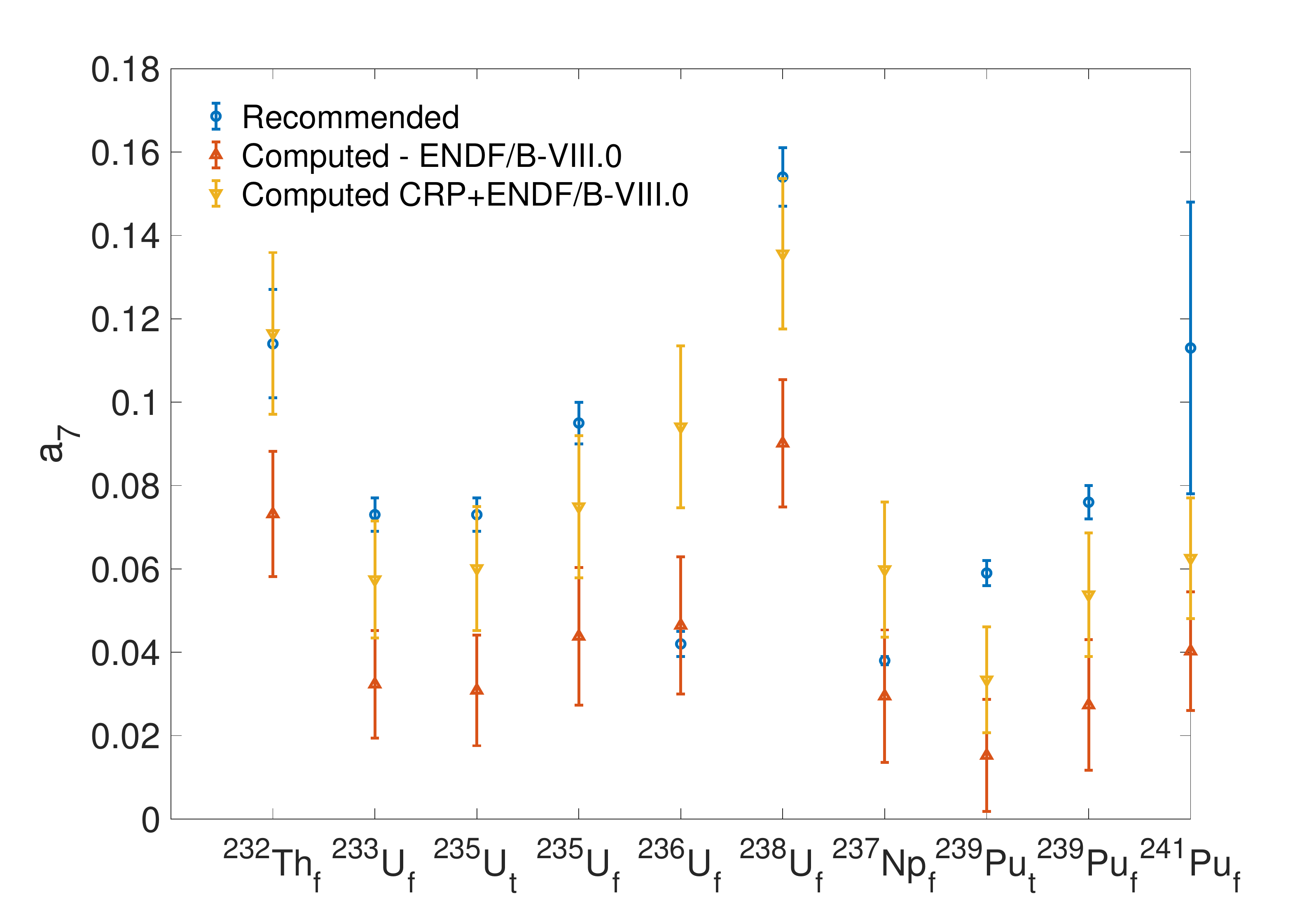}
\end{minipage}
\begin{minipage}{0.47\linewidth}
\includegraphics[width=0.95\linewidth]{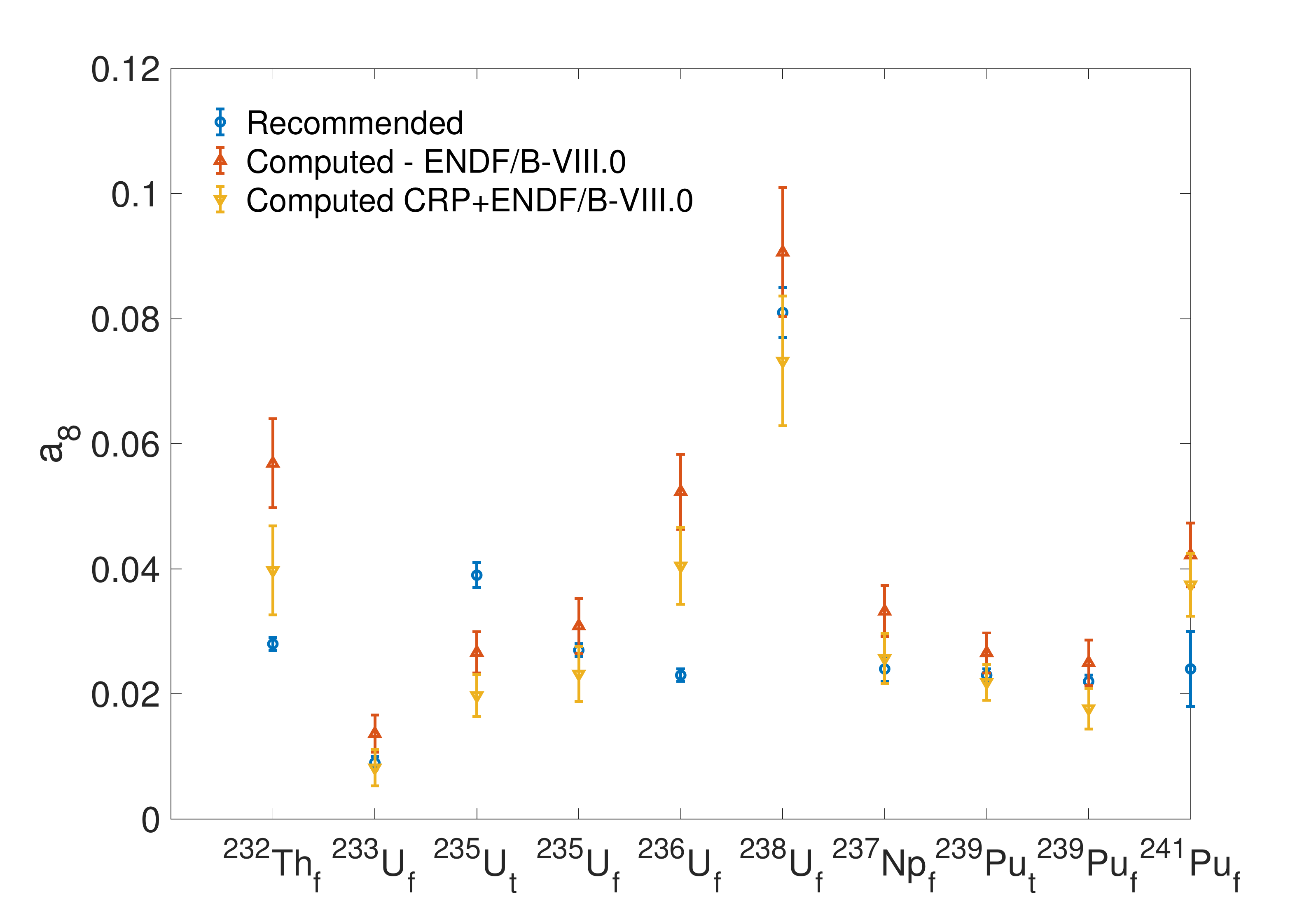}
\end{minipage}
\caption{Seventh and Eight group abundance for several fissioning systems and energies. Comparison between summation calculation results and recommended values. The uncertainty in the calculated abundances is obtained through an analytical marginalization technique.}
\label{fig:a78}
\end{figure*}

Figures~\ref{fig:a12} to~\ref{fig:a78} show the comparison between measured and calculated kinetic parameters. Each figure refers to one group abundance and compares several fissioning systems and energies.

From the results it is evident that the calculated abundances are affected by larger uncertainties since they reflect the uncertainties in hundreds of microscopic parameters. The integral experiment does not consider the uncertainties in the microscopic data explicitly, but rather captures the global behavior, which leads to smaller uncertainties on the fitted abundances. Furthermore, the summation calculations have been performed starting from the independent fission yields, which - due to the complexity of the measurements - are affected by larger uncertainties than the cumulative fission yields. The comparison with the experimental values unveils some patterns between calculated and measured parameters. Depending on the group, the calculated $a_i$ is either overestimated or underestimated, and this is true independently of the fissioning system or of the energy. The first abundance is fairly well estimated. The values are very close with the recommended data and the uncertainty ranges overlap (see Fig.~\ref{fig:a12}). This is true for all fissioning system and all energies, suggesting that the long-lived precursors, such as $^{87}$Br and $^{91}$Rb, are well identified and their data are well known.
The CRP+ENDF/B-VIII.0 database gives a slightly smaller $a_1$ than ENDF/B-VIII.0, because the $P_n$ of $^{91}$Rb has been set to zero. The reason is that its $Q_{\beta n}$ happens to be very small and experimental values are still preliminary (see Table~\ref{tab::iso_a1}). On the other hand, the ENDF/B-VIII.0 value is adopted from systematics~\cite{McCutchan2012}.

\begin{table}
\centering
\caption{Precursors contributing to the first delayed neutron group}
\label{tab::iso_a1}
\begin{tabular}{c|c|c|c} \hline \hline
Isotope & Quantity & ENDF/B-VIII.0 & CRP+ENDF/B-VIII.0 \\ \hline
\multirow{2}{*}{$^{91}$Rb} & $P_n$ [-]& 2.0319E-07 & 0 \\
                      & T$_{1/2}$ [s]& 58.40 & 57.90 \\ \hline
\multirow{2}{*}{$^{87}$Br} & $P_n$ [-]& 0.026 & 0.0253 \\
                      & T$_{1/2}$ [s]& 55.65 & 55.64 \\                       
\hline\hline  
\end{tabular}
\end{table}

The second abundance, which should only include $^{137}$I and $^{141}$Cs, is very well estimated for the fast fission of $^{232}$Th, while it is badly estimated for the fast fission of $^{233}$U. For the rest of the fissioning systems and energies, the calculated values always agree with the experimental $a_2$ within the range of uncertainty. In some cases, ENDF/B-VIII.0 is closer to the recommended value than CRP+ENDF/B-VIII.0 ($^{235}$U$_{t}$, $^{238}$U$_{f}$, $^{237}$Np$_{f}$, $^{239}$Pu$_{t}$, $^{241}$Pu$_{f}$). In other cases, however, it is the other way around ($^{235}$U$_{f}$, $^{236}$U$_{f}$, $^{239}$Pu$_{f}$). The CRP+ENDF/B-VIII.0 $a_2$ is always slightly larger than the ENDF/B-VIII.0 value, due to the fact that the $P_n$ value for $^{137}$I has been increased from 0.0714 to 0.0765 (see Table~\ref{tab::iso_a2}). It is evident though, that both libraries slightly underestimate $a_2$. This may indicate that a revision of the microscopic data of $^{141}$Cs and $^{137}$I is needed or that the fission yields need to be investigated.    

\begin{table}
\centering
\caption{Precursors contributing to the second delayed neutron group}
\label{tab::iso_a2}
\begin{tabular}{c|c|c|c} \hline \hline
Isotope & Quantity & ENDF/B-VIII.0 & CRP+ENDF/B-VIII.0 \\ \hline
\multirow{2}{*}{$^{141}$Cs} & $P_n$ [-]& 0.00035 & 0.00034\\
                      & T$_{1/2}$ [s]& 24.84 & 24.91 \\ \hline
\multirow{2}{*}{$^{137}$I} & $P_n$ [-]& 0.0714 & 0.0763 \\
                      & T$_{1/2}$ [s]& 24.50 & 24.59 \\                    
\hline\hline  
\end{tabular}
\end{table}

The third abundance, which should only include $^{136}$Te and $^{88}$Br, is very well estimated for the thermal fission of $^{235}$U and the fast fission of $^{241}$Pu, while it is badly estimated for the fast fission of $^{232}$Th. For the rest of the fissioning systems, it seems that the ENDF/B-VIII.0 $a_3$ values are always closer to the experimental values than the CRP. The differences between the two libraries are reported in Tab.~\ref{tab::iso_a3} and concern the $^{136}$Te half-life and $^{88}$Br branching ratio $P_n$. However, for both libraries, the general trend is a slight underestimation of $a_3$ with respect to experiments. 

\begin{table}
\centering
\caption{Precursors contributing to the third delayed neutron group}
\label{tab::iso_a3}
\begin{tabular}{c|c|c|c} \hline \hline
Isotope & Quantity & ENDF/B-VIII.0 & CRP+ENDF/B-VIII.0 \\ \hline
\multirow{2}{*}{$^{136}$Te} & $P_n$ [-]& 0.0131 & 0.0137\\
                      & T$_{1/2}$ [s]& 17.50 & 17.67 \\ \hline
\multirow{2}{*}{$^{88}$Br} &$P_n$ [-]& 0.0658 & 0.0672 \\
                      & T$_{1/2}$ [s] & 16.29 & 16.29 \\        
\hline\hline  
\end{tabular}
\end{table}

Both ENDF/B-VIII.0 and CRP+ENDF/B-VIII.0 converge towards the same largely underestimated $a_4$. The only actinides for which the parameter is fairly well estimated are $^{232}$Th$_{f}$, $^{235}$U$_{f}$, $^{238}$U$_{f}$ and $^{241}$Pu$_{f}$. On the other hand, $^{235}$U$_{t}$, $^{236}$U$_{f}$, and $^{237}$Np$_{f}$ are very far from the recommended values. 

Figures ~\ref{fig:a56} and ~\ref{fig:a78} show that the estimation of the short-lived precursors' abundances needs improvement. The fifth abundance which includes precursors $^{94}$Rb, $^{139}$I, $^{85}$As and $^{98m}$Y, is systematically overestimated with the exception of some isolated cases ($^{238}$U$_{f}$ and $^{232}$Th$_{f}$). On the other hand, the seventh abundance, which involves $^{91}$Br and $^{95}$Rb, is systematically underestimated, with the exception of $^{236}$U$_{f}$ and $^{237}$Np$_{f}$. 
The sixth and eight abundances do not follow any specific pattern. 
It should be stressed that these results and conclusions are dependent upon the FY dat used to calculate the DN activity, and could be different if a different FY library was adopted.


\paragraph{Effect of the abundances set on the quantities of interest.}
In this paragraph,the effect of the abundances sets on quantities of interest, such as the mean delayed neutron half-life and the delayed neutron activity, is discussed. 
Table~\ref{tab:meanHL_comparison} shows the average half-life values $\langle T  \rangle$ computed using different sets of DN abundances. In particular,   $\langle T  \rangle$ values obtained using ENDF/B-VIII.0 and CRP+ENDF/B-VIII.0 decay data are compared with recommended values for the fast neutron-induced fission of $^{233}$U, $^{235}$U, $^{239}$Pu and the thermal-induced fission of $^{235}$U, $^{239}$Pu. The average half-life of DN precursors obtained by summation calculations has larger uncertainties than the recommended ones, due to the marginalization of hundreds of microscopic input data. Correlations have negligible effect on the uncertainty since they account for approximately 1\% of the total uncertainty. The mean half-lives for $^{239}$Pu (thermal), $^{235}$U (thermal) and $^{238}$U (fast) are well estimated. On the other hand, the results for $^{233}$U (fast) and $^{232}$Th (fast) deviate from the measured values therefore the relevant microscopic nuclear data require significant improvement. For the other isotopes and energies, the results are rather satisfactory, even though further work is needed to reduce the observed disagreement with recommended data.   

\begin{table}[h]
\caption{Average half-lives $\langle T  \rangle$ and uncertainties compared with measured values for fast (f) and thermal (t) neutron-induced fission of minor and major actinides presented in Tables~\ref{tab:9.5}-\ref{tab:9.6}.}
\label{tab:meanHL_comparison}
\begin{tabular}{c|c|c|c} \hline \hline
Nuclide & CRP+ENDF/B-VIII.0 & ENDF/B-VIII.0 &  Measured \\
$^{233}$U$_{f}$      &  10.44 $\pm$ 0.87 &    10.74 $\pm$  0.89&     12.50  $\pm$ 0.20 \\
$^{235}$U$_{t}$      &  9.38 $\pm$ 0.60 &     9.43  $\pm$  0.60&     8.98  $\pm$ 0.11 \\
$^{235}$U$_{f}$      &  8.42 $\pm$ 0.54 &     8.57  $\pm$  0.55&     8.96  $\pm$  0.06 \\
$^{239}$Pu$_{t}$     &   10.68 $\pm$ 0.96&    10.59 $\pm$  0.96&     10.59  $\pm$  0.17 \\
$^{239}$Pu$_{f}$     &   9.16 $\pm$ 0.84&     9.24  $\pm$  0.85&     10.27  $\pm$  0.13 \\ \hline\hline 
\end{tabular}
\end{table}

\begin{figure*}[!htb]
\includegraphics[width=1\linewidth]{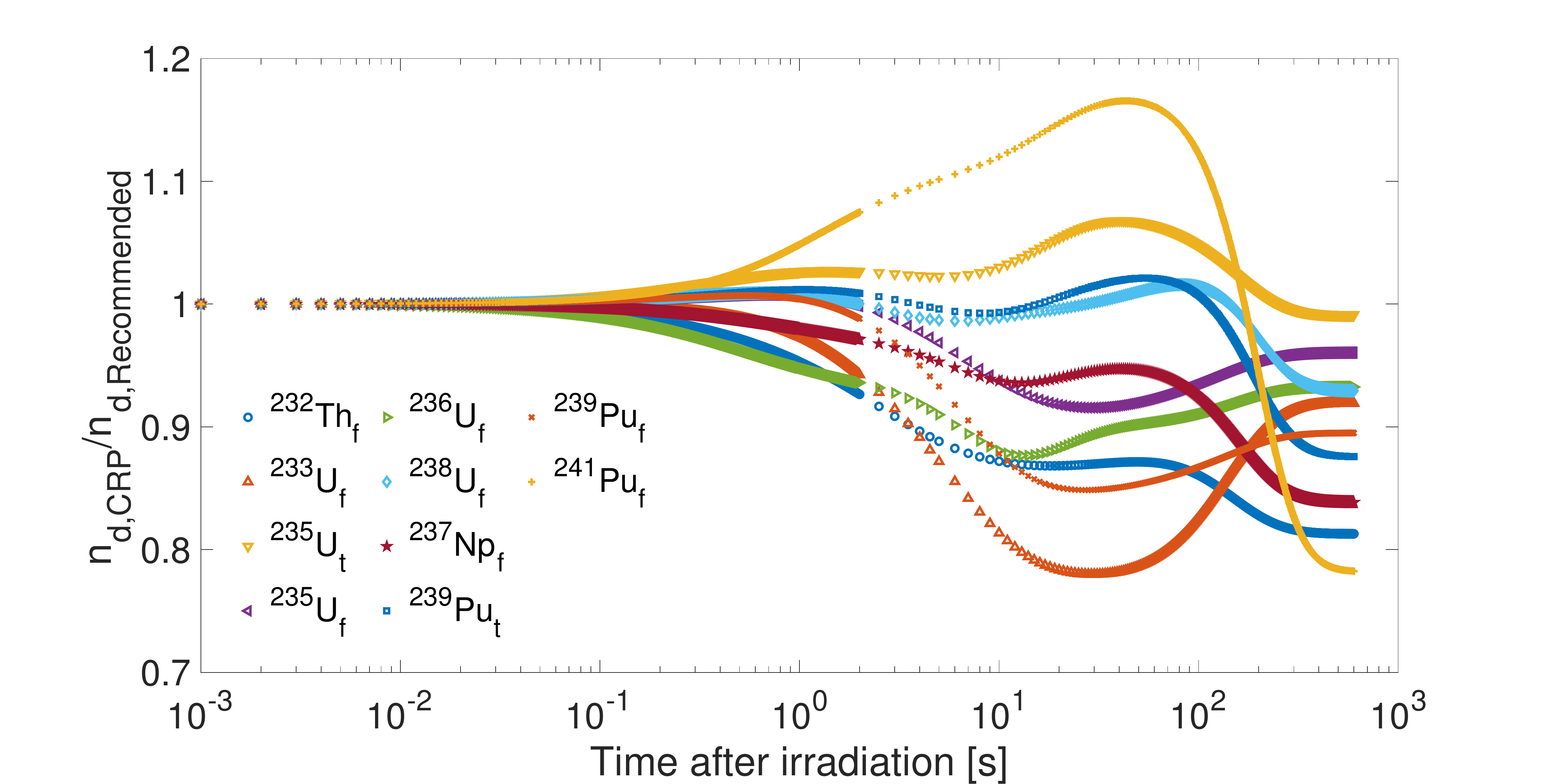}
\caption{Delayed neutron activity ratio for several fissioning systems and energies. The CRP activity has been computed using the abundances derived from the summation calculation using CRP decay data ($T_{1/2},P_{n}$). The Recommended activity has been obtained by using the recommended sets of abundances.}
\label{fig:comparisonActivity}
\end{figure*}

Figure~\ref{fig:comparisonActivity} shows the DN activity obtained using the calculated set of $a_i$ versus the delayed neutron activity obtained using the measured set of $a_i$. The plotted quantity is the ratio of the two activities. The graph is composed of several curves, each representing a different fissioning system or a different incident neutron energy. The behavior of the ratio reflects the differences in the group abundances. The recommended set of abundances is expected to reproduce the measured DN activity fairly well, so the comparison with the calculated one reflects the quality of the nuclear data. 
From the figure, it appears that, with the current microscopic data, it is possible to perfectly estimate the DN activity for the thermal fission of $^{235}$U. The same is not true for other fissioning systems or energies, where the discrepancies can reach up to 20\% (see $^{232}$Th$_{f}$ or $^{241}$Pu$_f$).
Since the activity is represented by a sum of exponentials of the abundances, it is expected to be highly sensitive to small variations in the abundances. The asymptotic value of the ratio between the activities is directly linked to $a_1$, which is, in principle, fairly well estimated. As far as $^{241}$Pu$_{f}$ is concerned, the calculated $a_1$ is 0.013, while the recommended one is 0.016. Such a small difference is responsible for the 20\% difference in the DN activity. 

To conclude, in this section we attempted to verify the quality of the new microscopic CRP data ($T_{1/2},P_{n}$) by extracting 8-group parameters ($a_i, \lambda_i$) from the DN activity obtained from summation calculations over these data.
The first abundance is fairly well estimated for all fissioning systems and energies, while the others depend strongly on the case under investigation. A pattern seems to emerge in the systematic overestimation of $a_5$ and underestimation of $a_7$, implying that precursors belonging to the fifth ($^{94}$Rb, $^{139}$I, $^{85}$As and $^{98m}$Y) and seventh group ($^{91}$Br and $^{95}$Rb) may need to be revisited. Abundance correlation matrices have been obtained and used in the estimation of the uncertainty of quantities of interest, like $\langle T  \rangle$. Although the correlations appear to be weak, their consideration in the estimation of the total uncertainty of $\langle T  \rangle$ is important. In spite of the discrepancies observed between calculated and measured $a_i$, the average half-lives are comparable while the large underestimation obtained for $^{233}$U (fast) and $^{239}$Pu (fast) suggests that too much weight is assigned to short-lived precursors in the fast neutron-induced fission. As far as the DN activity is concerned, the current microscopic data are not able to reproduce the experimental curve for all fissioning systems and energies. Figure~\ref{fig:comparisonActivity} shows that the discrepancy can reach $\pm$ 20\% for $^{241}$Pu$_f$ and $^{233}$U$_f$, while $^{235}$U$_t$ is perfectly reproduced. One should bear in mind though, that the reactivity, which is the quantity of interest for reactor physicists, is less sensitive to the abundances than the activity. Therefore, a 20\% discrepancy in the DN activity does not necessarily translate into a similar discrepancy in the reactivity. Rather, 300 s after irradiation, the latter has a tendency to depend on the mean half-life of the precursors.

On the other hand, one should bear in mind that the results for the time-dependent parameters depend not only on the microscopic decay data but also on the FY library that was used. Therefore, apart from revisiting certain decay data ($T_{1/2}, P_n$) as suggested from the comparison of abundances, it is also crucial to review and improve the existing evaluated FY libraries before any final conclusion can be drawn on the CRP \bdn~emission data.

\section{ INTEGRAL CALCULATIONS}
\label{Sec:Macro-Integral}

In Sect.~\ref{Sec:Macro-Summation}, summation calculations were used to determine the total delayed neutron yields from fission as a function of the incident neutron energy $\nu_{d}$(E). Moreover, they were used to check the consistency of the new ($T_{1/2}, P_n$) data through systematic comparisons with recommended values. In this section, we attempt to validate the new CRP ($T_{1/2}, P_n$) data by studying their impact on specific reactor designs and comparing their results against well-measured benchmark experiments.

\subsection{Comparison with integral experiments }

\begin{figure}[!htb]
\includegraphics[width=\linewidth]{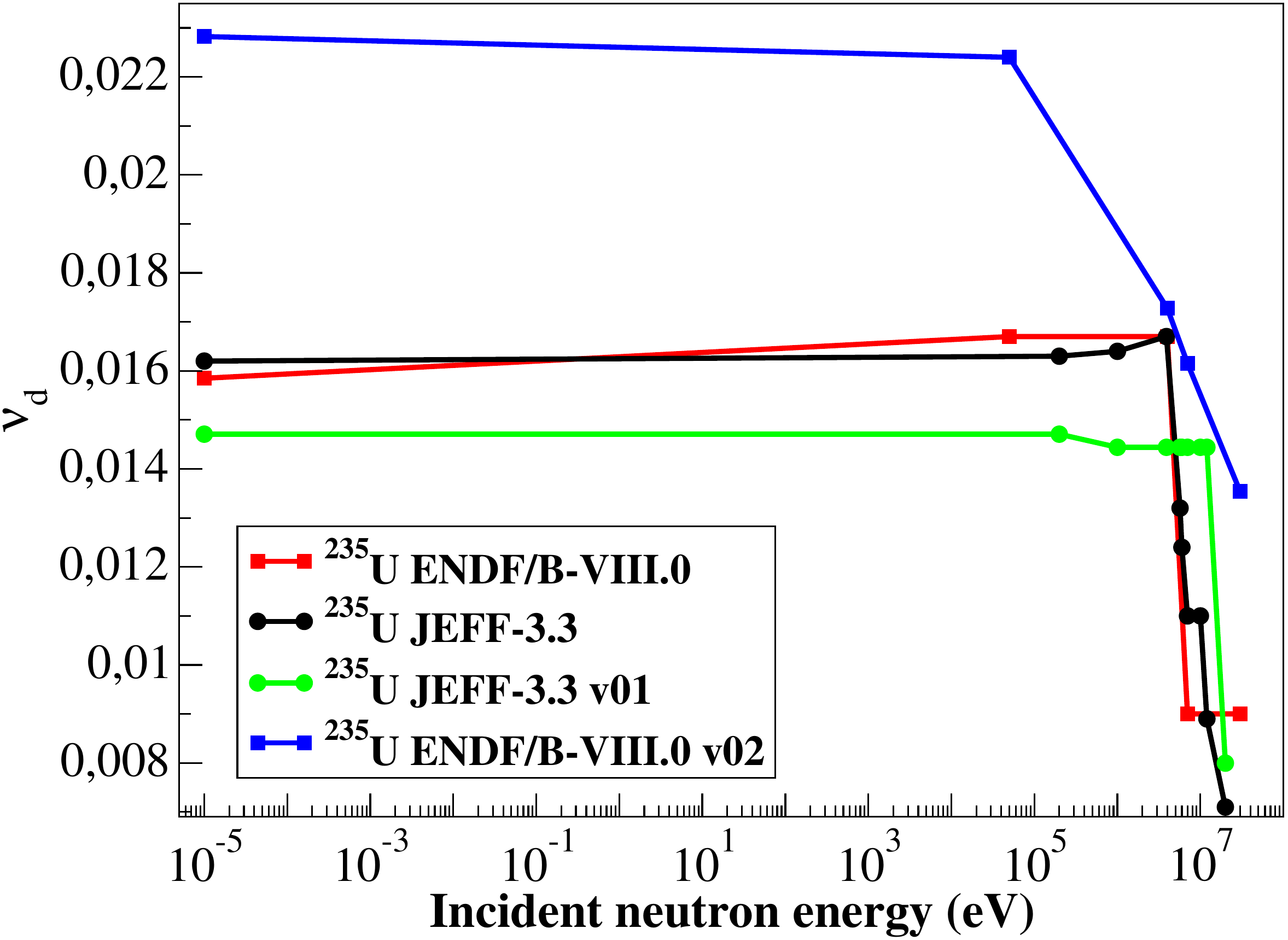}
\caption{Comparison between recommended and modified delayed neutron yields for $^{235}$U.}
\label{fig:neutron-prod-comp}
\end{figure}

The first step in the comparison with integral experiments was to perform summation calculations of $\nu_{d}$(E) using the following combinations of input data from ENDF/B-VIII.0~\cite{Brown2018} and JEFF-3.3~\cite{Plompen2020} evaluated nuclear data libraries, and the new CRP $P_{n}$ tables: i) fission yields and decay data from the same nuclear data library (v01); ii) fission yields from nuclear data libraries and CRP $P_{n}$ values complemented with decay data from the same library (v02); iii) JEFF-3.3 fission yields and decay data form ENDF/B-VIII.0 (v03); and iv) JEFF-3.3 fission yields and CRP $P_{n}$ values complemented with data from ENDF/B-VIII.0 decay library (v04) which is identical to CRP+ENDF/B-VIII.0 from the previous sections. In the following, these calculated $\nu_{d}$ values will be referred to as \emph{modified} $\nu_{d}$, while the recommended $\nu_{d}$ contained in the evaluated nuclear data libraries will be referred to as \emph{recommended}. Differences between 5\% and 10\% can be found between recommended JEFF-3.3 and ENDF/B-VIII.0 $\nu_{d}$ values. Furthermore, recommended $\nu_{d}$ and modified $\nu_{d}$ values differ by up to 70\% in some cases. An example of this behaviour can be seen in Figure~\ref{fig:neutron-prod-comp} for $^{235}$U.

To validate the CRP ($T_{1/2}, P_n$) data, the $\nu_{d}$ values obtained from the above-mentioned summation calculations have been used in simulations of integral experiments from the International Criticality Safety Benchmark Evaluation Project (ICSBEP) Handbook~\cite{ICSBEP} using the MCNP 6.1.1 code~\cite{MCNP611b}. Integral benchmarks consist of extensively peer-reviewed data that can be used by the international nuclear data community for testing and improvement of nuclear data files. They are also used by the international reactor physics, criticality safety, and math $\&$ computation communities for validation of analytical methodologies used for reactor physics, fuel cycle, nuclear facility safety analysis and design, and advanced modeling and simulation efforts~\cite{Briggs2014}. 

For our purposes, we have chosen 9 integral experiments with available measured effective delayed neutron fraction $\beta_{eff}$: POPSY, TOPSY, JEZEBEL, SKIDOO and FLATTOP-23, which all have a simple spherical geometry; and in addition, BIG TEN, SNEAK-7A, SNEAK-7B and IPEN, all of which have a reactor-type configuration. The IPEN integral experiment is the only one based on a thermal spectrum, whereas the rest of them are fast spectrum systems.

According to Keepin~\cite{Keepin1965}, the effective delayed neutron fraction is defined as the ratio between the adjoint weighted delayed ($P_{d,eff}$) and total ($P_{eff}$) neutron production:

\begin{align}
\beta_{eff}=\frac{P_{d,eff}}{P_{eff}}
\end{align}

with:

\begin{align}
\begin{split}
P_{eff}=\int{\phi^*(\vec{r},E',\Omega')}\chi(E')\nu(E)\Sigma_{f}(\vec{r},E)\\\phi(\vec{r},E,\Omega)\delta{E}\delta{\Omega}\delta{E'}\delta{\Omega'}\delta{\vec{r}}
\end{split}
\end{align}
\begin{align}
\begin{split}
P_{d,eff}=\int{\phi^*(\vec{r},E',\Omega')}\chi_{d}(E')\nu_{d}(E)\Sigma_{f}(\vec{r},E)\\\phi(\vec{r},E,\Omega)\delta{E}\delta{\Omega}\delta{E'}\delta{\Omega'}\delta{\vec{r}}
\end{split}
\end{align}

where $\phi(\vec{r},E,\Omega)$ and $\phi^*(\vec{r},E',\Omega')$ are the direct and adjoint angular fluxes, $\chi(E')$ is the energy spectrum of the generated fission neutrons, $\nu(E)$ is the average neutron multiplicity per fission. Correspondingly, $\chi_{d}(E')$ and $\nu_{d}(E)$ are the same quantities but for delayed neutrons and $\Sigma_{f}(\vec{r},E)$ is the macroscopic fission cross section of the material. $\beta_{eff}$ has been selected as reference integral parameter since the sensitivity of $\beta_{eff}$ to changes in delayed neutron data is significantly greater than to other data~\cite{DAngelo1993}.

A sensitivity analysis was carried out first with the SUMMON~\cite{Romojaro2019} code to identify the main fissioning systems contributing to $\beta_{eff}$ in each integral experiment. The results are presented in Table~\ref{tab:beta-contributions}. Although some experiments are characterized by a major contribution to $\beta_{eff}$ coming from a single isotope, such as JEZEBEL, in others such as BIG TEN, we observe nearly equal contributions coming from two isotopes. In the latter case, it is difficult to assess the impact of $\nu_{d}$ changes in $\beta_{eff}$ due to compensation effects.

\begin{table*}[!htb]
\caption{Relative contributions of main isotopes to the $\beta_{eff}$ for every integral experiment.}
\label{tab:beta-contributions}
\centering

\begin{tabular}{c|c|c}
\hline \hline
\multicolumn{1}{c|}{Integral exp.} & \multicolumn{1}{c|}{Relative contribution to $\beta_{eff}$} & Brief description\tabularnewline
\hline
Popsy & 52.5\% $^{239}$Pu, 40.9\% $^{238}$U & Pu core, U-nat reflector\tabularnewline
Topsy & 72.7\% $^{235}$U, 25.4\% $^{238}$U & HEU core, U-nat reflector\tabularnewline
Big Ten & 52.8\% $^{238}$U, 46.6\% $^{235}$U & U cylindrical shape core\tabularnewline
Jezebel & 90.9\% $^{239}$Pu, 7.1\% $^{240}$Pu & Pu bare sphere\tabularnewline
Sneak 7A & 46.1\% $^{239}$Pu, 45.2\% $^{238}$U & MOX core\tabularnewline
Sneak 7B & 51.9\% $^{238}$U, 38.5\% $^{239}$Pu & MOX core\tabularnewline
Skidoo & 97.2\% $^{233}$U, 1.3\% $^{238}$U & $^{233}$U bare sphere\tabularnewline
Flattop-23 & 65.7\% $^{233}$U, 30.3\% $^{238}$U & $^{233}$U core, U-nat reflector\tabularnewline
Ipen & 88.8\% $^{235}$U, 11.1\% $^{238}$U & UO$_{2}$ core (thermal)\tabularnewline
\hline \hline
\end{tabular}
\end{table*}

\begin{figure} 
\centering
\includegraphics[width=\linewidth]{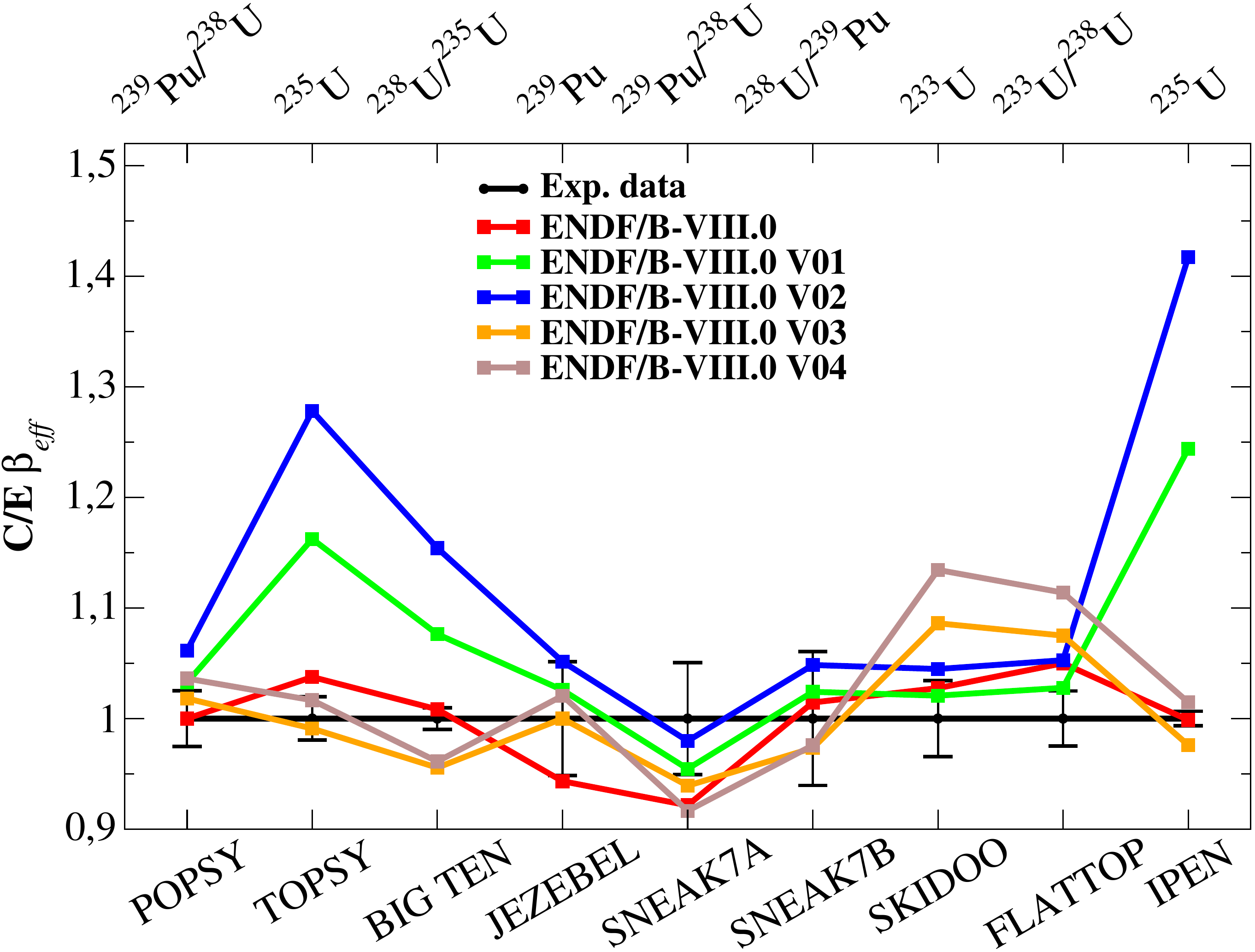}
\includegraphics[width=\linewidth]{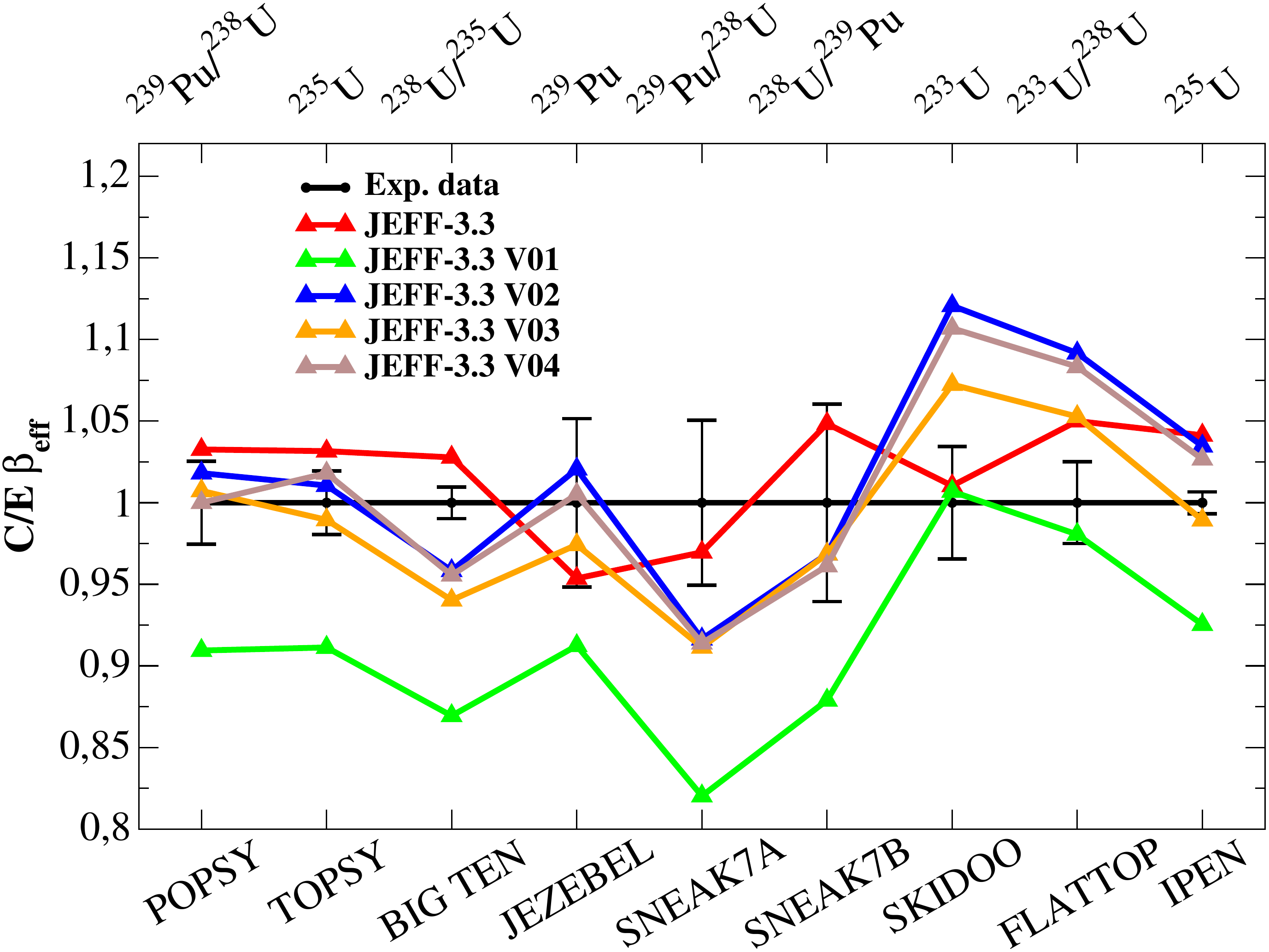}
\caption{C/E comparison with the recommended and modified $\nu_{d}$ values from ENDF/B-VIII.0 (top) and JEFF-3.3 (bottom) for the 9 integral experiments.}
\label{fig:beta-comparison}
\end{figure}

The next step was to perform the simulations using an adequate number of particle histories to achieve negligible statistical uncertainties ($<2$ pcm) in $\beta_{eff}$. The results of the simulations are shown in Fig.~\ref{fig:beta-comparison}, where the top x-axis shows the isotopes which are the main contributors to $\beta_{eff}$ for each integral experiment (see Table~\ref{tab:beta-contributions}).

The $\beta_{eff}$ results obtained with the recommended ENDF/B-VIII.0 $\nu_{d}$ values are in reasonable agreement with the experiments. Good agreement is also observed for the recommended JEFF-3.3 $\nu_{d}$ values. Regarding the DN yields obtained from the summation calculations, the v01 values obtained from JEFF-3.3 severely underestimate $\beta_{eff}$ in most of the cases. 
Since this is the only combination of JEFF-3.3 FYs that gives such large deviations, it is very likely that the source of these deviations is the JEFF-3.3 decay data library. 
On the contrary, the v01 values derived from ENDF/B-VIII.0 agree with six out of the nine experimental results. In the remaining three cases, which are the ones where the main contributor to $\beta_{eff}$ is $^{235}$U, an overestimation is observed. 
Of all the summation combinations using ENDF/B-III.0 libraries, v02 and v03 which use ENDF/B-VIII.0 FYs are the most discrepant, so it is most likely that the latter ENDF/B-VIII.0 FYs do not perform that well.
The combination of fission yields from JEFF-3.3 and decay data from ENDF/B-VIII.0 (v03) gives overall good agreement with the experimental results which is comparable to the recommended data. Finally, when the ENDF/B-VIII.0 decay $P_n$ data in v03 are replaced by the new CRP $P_{n}$ data to form v04, the agreement is not improved instead it is slightly worsened as we observe an increase in the respective $\beta_{eff}$. Overall, we find that the CRP $P_n$ data lead to an increase in $\beta_{eff}$ with respect to the $\beta_{eff}$ obtained when using the two other decay data libraries. Note that the modified decay-data libraries (v03) and (v04) give different results when used with ENDF/B-VIII.0 and JEFF-3.3 cross-section libraries, as seen in the upper and lower panels of Fig.~\ref{fig:beta-comparison}, respectively.

We have shown that the summation technique is a useful tool for calculating $\nu_{d}$ values that can reproduce reactor parameters with acceptable accuracy. This is particularly relevant for the minor actinides, for which no experimental fission pulse data are available, but for which fission yields and decay data are available. On the other hand, the use of the new CRP data in integral calculations results in increased values of $\beta_{eff}$, at least for the 9 benchmark experiments studied herein. These results agree with the findings of the previous section (Sect.~\ref{Sec:Macro-summation}) and confirm the need to investigate the evaluated fission yield libraries before conclusions are drawn with respect to the \bdn~data.

\subsection{Impact on reactor calculations }
The impact of the new CRP data on reactor calculations has also been assessed by performing two different analyses described in this section. 

\paragraph{Analysis I.} A pressurized water reactor (PWR) and two GEN-IV fast reactor concepts have been selected in this first analysis: a pin design (with reflective boundary surfaces) representative of the Obrigheim PWR~\cite{ICE}, the sodium cooled ASTRID (Advanced Sodium Technological Reactor for Industrial Demonstration) reactor~\cite{ASTRID} and the lead-bismuth cooled MYRRHA (Multi-purpose hYbrid Research Reactor for High-tech Applications) reactor~\cite{MYRRHA}.

Calculations of $\beta_{eff}$ were performed using the modified libraries (v01-v04), as well as recommended libraries, described in Sect.~\ref{Sec:Macro-Integral}. Results from these calculations are presented in Table~\ref{tab:beta-reactors}, where statistical uncertainties of the order of 4\% in $\beta_{eff}$ have been neglected. The $\beta_{eff}$ values obtained with the modified libraries deviate from those obtained using the recommended $\nu_d$ data by up to $\sim$10\% in ASTRID and MYRRHA for both ENDF/B-VIII.0 and JEFF-3.3 libraries. This is in agreement with the differences observed in the integral experiment calculations presented in the previous Sect.~\ref{Sec:Macro-Integral}. Regarding the PWR model, for JEFF-3.3 the largest difference observed is again $\sim$10\%, however in the ENDF/B-VIII.0 cases the deviations are larger by up to $\sim$44\% (v02). This result agrees with the findings for the Ipen experiment (the only one with thermal spectrum) discussed in the previous section which showed larger deviations with respect to the recommended values when using the ENDF/B-VIII.0 library combinations.

\begin{table}[!htb]
\caption{Calculated effective delayed neutron fraction values for PWR, ASTRID
and MYRRHA reactors using recommended and modified libraries.}
\label{tab:beta-reactors}

\center
\begin{tabular}{c|c|c|c|c}
\hline \hline
\multicolumn{5}{c}{PWR}\tabularnewline
\hline 
& \multicolumn{2}{c|}{ENDF/B-VIII.0} & \multicolumn{2}{c}{JEFF-3.3}\tabularnewline
\hline 
& $\beta_{\text{eff}}$ & Relative to rec. & $\beta_{\text{eff}}$ & Relative to rec.\tabularnewline
\hline \hline 
rec. & 657 & - & 695 & -\tabularnewline
v01 & 843 & 1.28 & 620 & 0.89\tabularnewline
v02 & 945 & 1.44 & 728 & 1.05\tabularnewline
v03 & 667 & 1.02 & 666 & 0.96\tabularnewline
v04 & 704 & 1.07 & 685 & 0.99\tabularnewline
\hline
\end{tabular}

\begin{tabular}{c|c|c|c|c}
\hline
\multicolumn{5}{c}{ASTRID}\tabularnewline
\hline 
& \multicolumn{2}{c|}{ENDF/B-VIII.0} & \multicolumn{2}{c}{JEFF-3.3}\tabularnewline
\hline 
& $\beta_{\text{eff}}$ & Relative to rec. & $\beta_{\text{eff}}$ & Relative to rec.\tabularnewline
\hline \hline 
rec. & 344 & - & 338 & -\tabularnewline
v01 & 348 & 1.01 & 316 & 0.93\tabularnewline
v02 & 361 & 1.05 & 342 & 1.01\tabularnewline
v03 & 339 & 0.99 & 343 & 1.01\tabularnewline
v04 & 305 & 0.89 & 367 & 1.09\tabularnewline
\hline
\end{tabular}

\begin{tabular}{c|c|c|c|c}
\hline
\multicolumn{5}{c}{MYRRHA}\tabularnewline
\hline 
& \multicolumn{2}{c|}{ENDF/B-VIII.0} & \multicolumn{2}{c}{JEFF-3.3}\tabularnewline
\hline 
& $\beta_{\text{eff}}$ & Relative to rec. & $\beta_{\text{eff}}$ & Relative to rec.\tabularnewline
\hline \hline 
rec. & 319 & - & 321 & -\tabularnewline
v01 & 337 & 1.06 & 299 & 0.93\tabularnewline
v02 & 355 & 1.11 & 332 & 1.03\tabularnewline
v03 & 321 & 1.01 & 321 & 1.00\tabularnewline
v04 & 326 & 1.02 & 317 & 0.99\tabularnewline
\hline \hline 
\end{tabular}
\end{table}

To conclude, similar to the results of $\beta_{eff}$ obtained in the previous section, we find that the best combination of input data is v03 with JEFF-3.1.1 FYs and ENDF/B-VIII.0 decay data. When the ENDF/B-VIII.0 $P_n$ data are replaced with the CRP values (v04), the agreement is slightly worsened except for one case, the PWR JEFF-3.1.1 recommended value. The deviations with respect to recommended data increase when the ENDF/B-VIII.0 FY library or the JEFF-3.1.1 decay data library is used in the combination.

\paragraph{Analysis II.} In the second analysis, a Sodium Fast Breeder reactor (FBR) model has been used to compute the $\beta_{eff}$ with MCNPX2.6.f coupled to an evolution code within the MCNP Utility for Reactor Evolution (MURE) software \cite{MURE}. The model is extracted from \cite{GlaserRamana} and the simulation has been originally developed for the study of the anti-neutrino emission from a sodium FBR (Na-FBR). Several versions of the model associated with various fuel compositions were used. The detailed description of the model can be found in \cite{Cormon2012},  here we present only the information necessary for understanding the present results. 

 In the FBR model, a central zone corresponds to the internal core, which is enriched at 21$\%$ in plutonium, while the external core is enriched at 28$\%$ in plutonium, while axial and radial blankets surround the core. A set of assemblies made of stainless steel surrounds the whole configuration, acting as a reflector.  

The core composition in its internal and external parts is given in Table \ref{tab:TableFBRCompo}. The isotopic plutonium vector corresponds to that of the Monju reactor \cite{Cormon2012}. This Mixed Oxide fuel (MOX) has a density of 11\,$g/cm^3$.
\begin{table}[!htb]
\caption{\small{Fuel composition in the internal and external cores of the sodium Fast Breeder model reactor (FBR) which are enriched by 21$\%$ and 28$\%$ in plutonium, respectively.}}
\label{tab:TableFBRCompo}
\begin{tabular}{c|c|c}
\hline
\hline
\multicolumn{1}{c|}{Isotope} &\multicolumn{1}{c|} {Internal Core}  & External core 
\tabularnewline \hline
\uhuit{} & 0.79 & 0.72\tabularnewline
\puneuf{} & 0.1218 &0.1624\tabularnewline
\puzero{} &  0.0504 &0.0672\tabularnewline
\puun{} & 0.0294  &0.0392  \tabularnewline
\pudeux{} & 0.0084  &0.0112\tabularnewline
\hline
\hline
\end{tabular}

\end{table}

\begin{table}[!htb]
\caption{\small{Composition in Minor Actinides (MA) of the transmutation blanket surrounding the cores of the FBR model reactor used in the calculations.}}
\label{tab:TableMACompo}
\begin{center}
\begin{tabular}{c|c}
\hline
\hline
\multicolumn{1}{c|}{Nuclei }& \multicolumn{1}{c}{\% weight of Minor Actinides} 
\tabularnewline \hline
$^{242}$Cm & 0.001 \tabularnewline
$^{243}$Cm & 0.01 \tabularnewline
$^{244}$Cm & 0.4 \tabularnewline
$^{245}$Cm & 0.17 \tabularnewline
$^{246}$Cm & 0.06 \tabularnewline
$^{247}$Cm & 0.01 \tabularnewline
$^{248}$Cm & 0.002 \tabularnewline
\hline
$^{241}$Am & 10.4 \tabularnewline
$^{242m}$Am & 0.4 \tabularnewline
$^{243}$Am & 2.9 \tabularnewline
\hline
$^{237}$Np & 3.5 \tabularnewline
\hline
other & 0.001\\
\hline
\hline 
 
\end{tabular}
\end{center}

\end{table}

The composition of the radial and axial blankets is typical for transmutation. $79\%$ of depleted uranium extracted from CANDU used fuel and 21\% of actinides are mixed according to the table \ref{tab:TableMACompo}.

With regards to the depleted uranium composition, the isotopic composition of a depleted plutonium/uranium vector extracted from a CANDU reactor at 6.7GWd/t was used as simulated. This composition does not account for the decay of $^{241}$Pu of half-life 14.33\,yrs \cite{jeff3.1.1}. 
The fuel temperature is 1500\,K. 

A  Protected-Plutonium-Production version of the fuel composition has been used as well (called "Fuel 2"), in which a significant fraction of $^{238}$Pu is added to the plutonium vector in order to prevent fuel diversion. The fission rates associated with the two fuel compositions are displayed in table~\ref{tab:TableFuelsAndBetaEff}.

The $\beta_{eff}$ was extracted at the beginning of the cycle. Two methods were used. In the first method, $\beta_{eff}$ was extracted using the fission rate of each actinide obtained from the reactor model combined with the recommended delayed and total neutron fractions $\nu_{t,d}$ from the JEFF3.1 database \cite{jeff3.1.1}. In the second method, the summation method was used, whereby the fission product activities obtained in the various simulations using JEFF-3.1.1 fission yields and JEFF-3.1 decay data were coupled to the new CRP $P_{n}$ data. 

Table~\ref{tab:TableFuelsAndBetaEff} shows the results of the computed $\beta_{eff}$ as a function of the loaded fuel. 

\begin{table} 
\caption{{\small{Fission rates at the beginning of the cycle corresponding to the two fuels loaded in the FBR model, and corresponding $\beta_{eff}$ (pcm) obtained with the recommended JEFF-3.1 decay data (JEFF) and the CRP $P_n$ data.}}}
\label{tab:TableFuelsAndBetaEff}

\begin{center}
\begin{tabular}{c|c|c}
\hline
\hline
\multicolumn{1}{c|}{Nuclides} & \multicolumn{1}{c|}{\% fission rate} &\multicolumn{1}{c}{\% fission rate} 
\tabularnewline
  & Fuel 1 & Fuel 2 
\tabularnewline \hline
$^{238}$U & 12.29& 12.25 \tabularnewline 
$^{239}$Pu& 59.15 & 54.28 \tabularnewline 
$^{240}$Pu& 6.68& 3.48 \tabularnewline
$^{241}$Pu& 18.73& 12.23 \tabularnewline
$^{242}$Pu& 0.83& 0.72 \tabularnewline
$^{235}$U& 0.30& 0.30 \tabularnewline
$^{236}$U& 0.00& 0.00 \tabularnewline
$^{238}$Pu& - & 14.75 \tabularnewline
$^{242}$Cm& 0.02& 0.02 \tabularnewline
$^{243}$Cm& 0.03& 0.03 \tabularnewline
$^{244}$Cm& 0.06& 0.06 \tabularnewline
$^{245}$Cm& 0.43& 0.43 \tabularnewline
$^{246}$Cm& 0.01& 0.01 \tabularnewline
$^{247}$Cm& 0.02& 0.02 \tabularnewline
$^{248}$Cm& 0.00& 0.00 \tabularnewline
$^{241}$Am& 0.91& 0.91 \tabularnewline
$^{242}$Am& 0.02& 0.02 \tabularnewline
$^{243}$Am& 0.19& 0.16 \tabularnewline
$^{237}$Np& 0.34& 0.34 \tabularnewline
\hline
$\beta_{eff}$~(JEFF) & 471 &437
\tabularnewline
\hline
$\beta_{eff}$~(CRP)& 478 & 447\tabularnewline
\hline
\hline 
 
\end{tabular}
\end{center}

\end{table}

The $\beta_{eff}$ values obtained with both methods are displayed in the two bottom lines of Table~\ref{tab:TableFuelsAndBetaEff}.  One can see that the CRP results are in very good quantitative agreement with the results using recommended JEFF-3.1.1 $\nu_d$ values for both fuels, with a difference of less than 10~pcm. The $\beta_{eff}$ value obtained for fuel 1 is larger than for fuel 2 because of the larger fission rate of $^{239}Pu$ in the former fuel compared to the latter.This is due to the added fraction of $^{238}Pu$ in fuel 2, which has a very low $\nu_d$. 

The results confirm the findings of Analysis I showing that the combination of JEFF-3.1.1 FYs with the CRP $P_n$ data yields overall good results for $\beta_{eff}$ which are comparable to the recommended values.
However, there is an obvious trend that has been observed in all the previous integral calculations, and that is that the CRP $P_n$ data yield larger $\nu_d$ values and subsequently, larger $\beta_{eff}$ values. Possible culprits for this increase have been suggested to be $^{137}$I and $^{91}$Br based on the comparison of the corresponding $P_n$ data in Sect.~\ref{Sec:Macro-summation-yields-basic}. However, due to compensation effects from the variations of $P_n$ values of several important contributors and the dependence of these results on the fission yield library used in the summation calculations, it is not straightforward to draw a definitive conclusion on the impact of the new CRP data.

\section{ SYSTEMATICS OF MACROSCOPIC DATA: Time-dependent parameters}\label{Sec:Macro-Systematics}


\subsection{New approach for estimation of temporary parameters for unmeasured nuclides}\label{Sec:Macro-Timedependent}
The systematics of the delayed neutron (DN) characteristics and their correlation properties gives valuable information for developing a reliable database of DN data. The total delayed neutron yields $\nu_d$ are represented by an exponential function of the parameters involving the mass $A_c$ and the atomic numbers $Z$ of the compound fissioning nucleus ($A_c$, $Z$) \cite{Piksaikin02b}. Moscati and Goldemberg \cite{Moscati62} proposed the $exp\left[-K\cdot(2\cdot Z-N)\right]$ dependence which was modified to $exp\left[-K\cdot(3\cdot Z-A_c)\right]$ by Caldwell and Dowdy \cite{Caldwell75} and then investigated in detail by Pai \cite{Pai76}, Waldo and co-workers \cite{Waldo81} and Ronen \cite{Ronen96}. The parameterization $-(A_c-3\cdot Z)\cdot(A_c/Z)$ has been used by Tuttle \cite{Tuttle75} along with the experimental data in the evaluation of the total DN yields for the isotopes $^{231}$Pa, $^{234}$U, $^{236}$U, $^{238}$Pu, $^{240}$Pu, $^{241}$Pu and $^{242}$Pu.
The systematics of the temporal DN parameters became possible after introducing in practice the value of the average half-life $\langle T  \rangle$~of DN precursors \cite{Piksaikin02b}. The $\langle T  \rangle$~value can be calculated using expression 
\begin{align}
\langle T \rangle = \sum_{i=1}^{N} a_i\cdot T_i\,, \sum_{i=1}^{N} a_i = 1\,,
\label{eq.viii.1}
\end{align}

where $a_i$ and $T_i$ is the relative abundance and the half-life of the $i$-th DN group, respectively; $N$ the number of DN groups. When calculating this value on the basis of the microscopic DN data and fission yield data the following expression is used 
\begin{align}
\langle T \rangle = \frac{\sum_{i=1}^{N} CY_i \cdot P_{ni} \cdot T_{1/2i}}{\sum_{i=1}^{N} CY_i\cdot P_{ni}}\,,
\label{eq.viii.2}
\end{align}

where $CY_i$ is the cumulative yield of the i-th DN precursor; $P_{ni}$ and $T_{1/2i}$ the probability of DN emission and the half-life of the i-th DN precursor, respectively. 

It turns out that the   $\langle   T\rangle$~data for the isotopes of one element can be  systematized with the help of the exponential dependence using the Tuttle's parameter – $-(A_c-3\cdot Z)/(A_c/Z)$, where $A_c$ and $Z$ is the mass and the atomic numbers of the fissioning compound nucleus, respectively \cite{Piksaikin02b}. Furthermore it was found that the total DN yields $  \nu_d$ and the average half-life values   $\langle   T\rangle$ ~for isotopes of one element are related to each other by the dependence $  \nu_d =a \cdot\langle T\rangle^b$, $a$ and $b$ being constants. These findings made it possible to show that the exponential dependence of the $  \nu_d$ value on the parameter $-(A_c-3\cdot Z)\cdot (A_c/Z)$ has essentially an isotopic character \cite{Piksaikin02b}. This means that isotopes of each fissionable element have their own dependence of $  \nu_d$ on the parameter $-(A_c-3\cdot Z)\cdot(A_c/Z)$ that must be taken into account in the evaluation process especially for the nuclei lying far from the valley of $\beta$-stability. The properties of the systematic $-(A_c-3\cdot Z)\cdot(A_c/Z)$ both for the total DN yields and the average half-life of DN precursors are discussed in detail in \cite{Piksaikin02b}. 
The systematics of the average half-life of DN precursors is useful for the validation of the DN group parameters for the isotopes of Th, U, Pu and Am and the prediction of the average half life of unmeasured isotopes of these elements. But unlike the $-(A_c-3\cdot Z)\cdot (A_c/Z)$ systematics of the total DN yield, this systematics does not allow us to predict the average half-life of DN precursors   $\langle   T\rangle$~for isotopes of such elements as Pa, Cm, Cf and others. Below a new approach for the prediction of the average half-life of DN precursors and the appropriate set of relative abundances and periods of individual DN groups for unmeasured nuclides is considered.

\subsection{The $(A_c/Z)\cdot 92$ systematics of the average half-life}
\begin{figure} 
	\centering
\includegraphics[width=\linewidth]{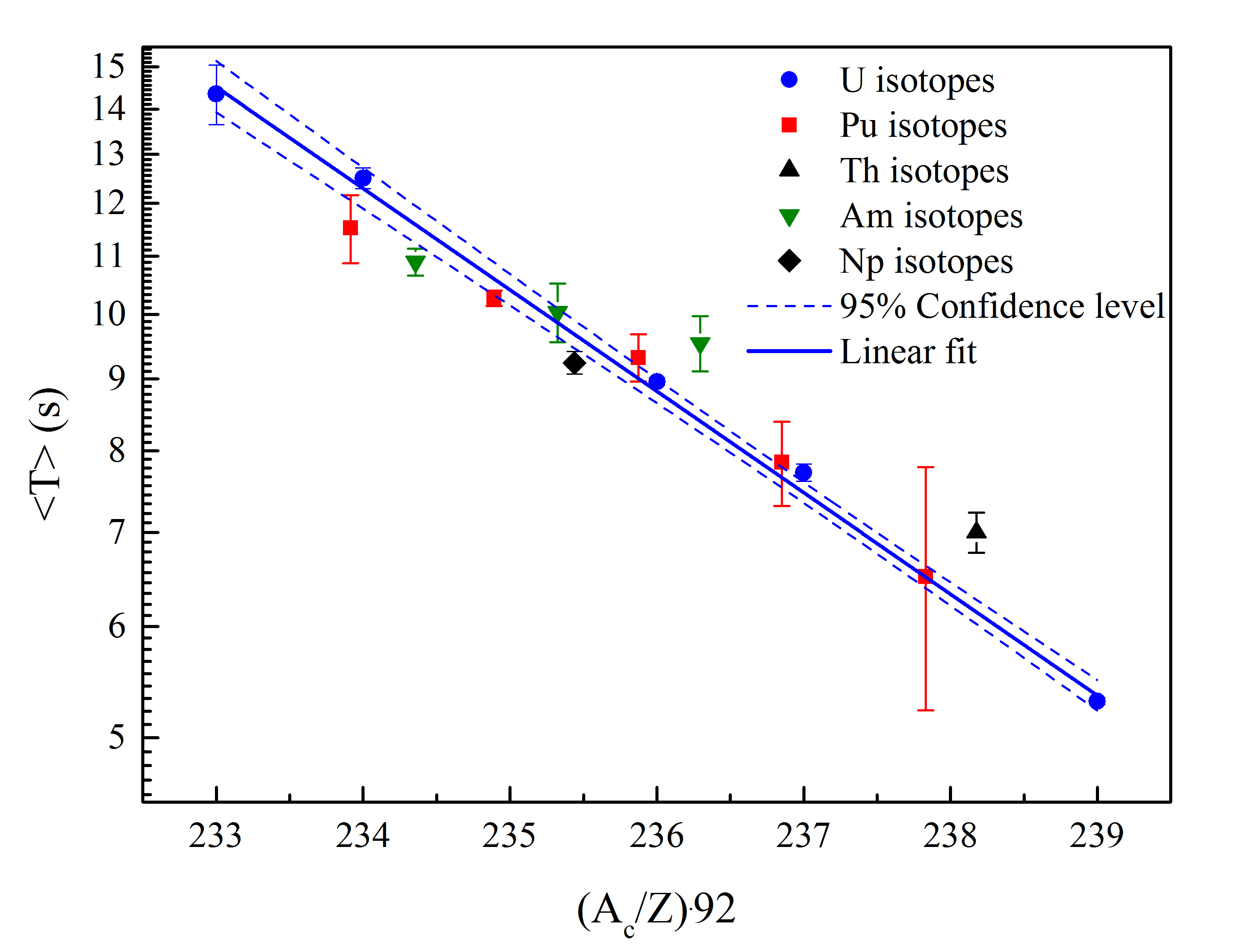}
\caption{The systematics $(A_c/Z) \cdot 92$ of the average half-life of DN precursors. The solid line is the approximation of the average half-life by expression $\langle T\rangle = exp \left[ a+b\cdot (A_c/Z)\cdot 92\right]$. The average half-life values   $\langle   T\rangle$   ~were taken from the recommended data sets $(a_i,\,T_i)$ \cite{Spriggs02}.}
\label{fig:8_1}
\end{figure}

It has been shown in \cite{Roshchenko10} that the average half-life of the DN precursors of all isotopes of heavy nuclides measured in the fast neutron induced fission can be approximated by the expression $\langle T\rangle = exp \left[a+b \cdot (A_c/Z)\cdot 92\right]$, where $A_c$ and $Z$ is the mass and the atomic numbers of the fissioning compound nucleus, respectively. As compared with the systematics $-(A_c-3\cdot Z)\cdot A_c/Z$ \cite{Piksaikin02b}, which is valid for isotopes of one element, the systematics $(A_c/Z)\cdot 92$ allows to predict the average half-life for isotopes of all elements with the help of the only one set of constants $a$ and $b$. In Fig.~\ref{fig:8_1} the values of the average half-life calculated on the basis of the recommended experimental data $(a_i,T_i)$ for the fast neutron induced fission of U, Pu, Am, Np and Th isotopes \cite{Spriggs02} are shown by separate points. The solid line is the approximation of these data by the dependence $\langle T\rangle = exp\left[a+b \cdot (A_c/Z)\cdot 92\right]$, where $a = 41.371 \pm 1.042$, $b = -0.166 \pm 0.004$. This dependence can be used for the prediction of the average half-life of Pa, Cm, Bk, Cf, etc., isotopes. It should be noted that the observed   $\langle   T\rangle$~dependence has a fine structure that is due to the differences in the dependences observed for U, Pu, and Am elements. So for example the coefficients $a$ and $b$ for the $\langle T\rangle = exp\left[ a+b\cdot(A_c/Z) \cdot 92\right]$ dependence for uranium isotopes are  41.219 $\pm$ 1.072 and -0.165 $\pm$ 0.005, respectively.

\subsection{Systematics $(A_c-2\cdot Z)$ of relative abundances in the 8-group model}

\begin{figure*} 
	\centering
\includegraphics[width=0.95\textwidth]{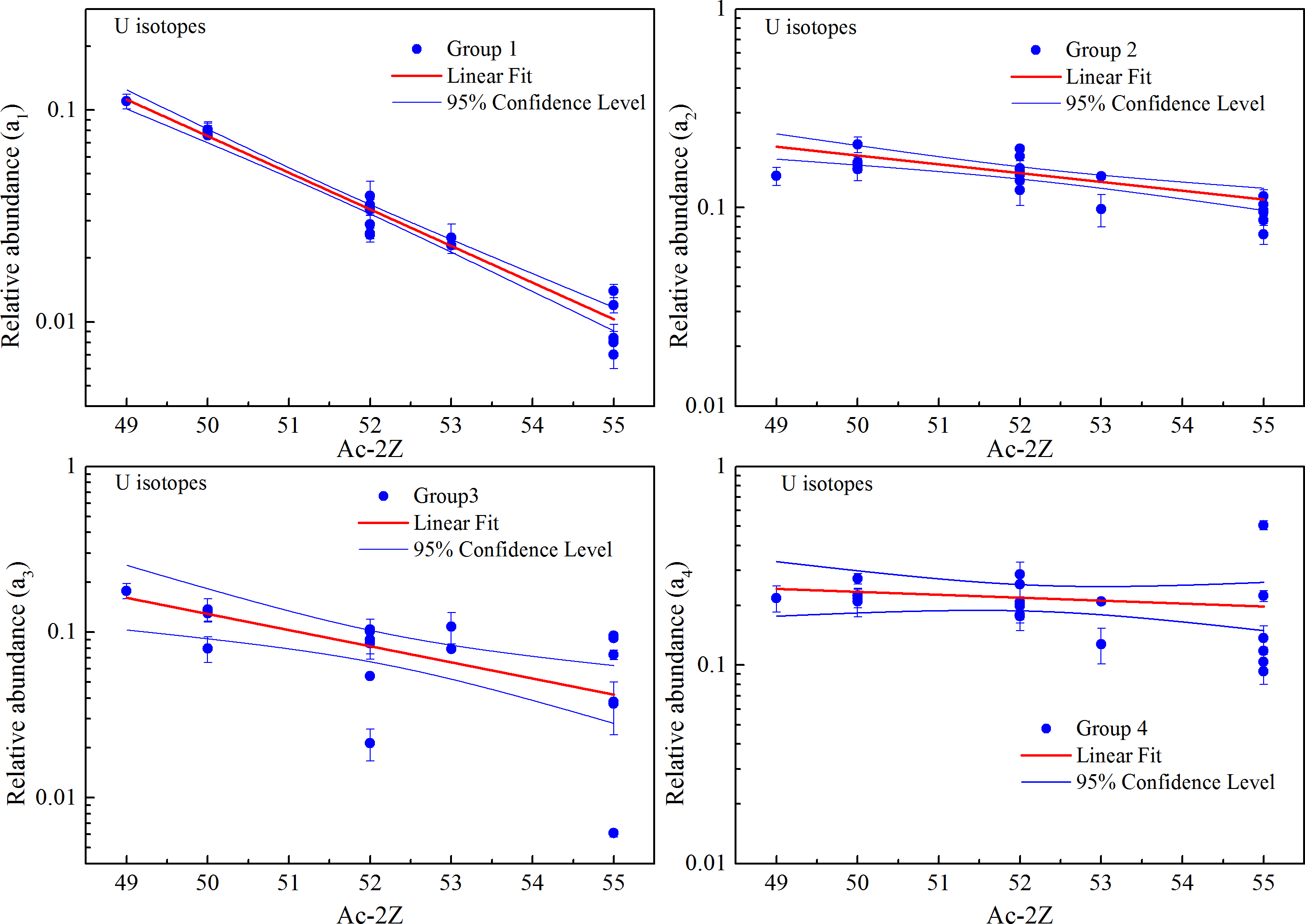}
\caption{The $(A_c-2\cdot Z)$ systematics of the relative abundances of delayed neutrons from fission of uranium isotopes (1-4 groups). The values of the relative abundances were taken from \cite{Spriggs99}.}
\label{fig:8_2}
\end{figure*}

\begin{figure*}[!htb]
	\centering
\includegraphics[width=0.95\textwidth]{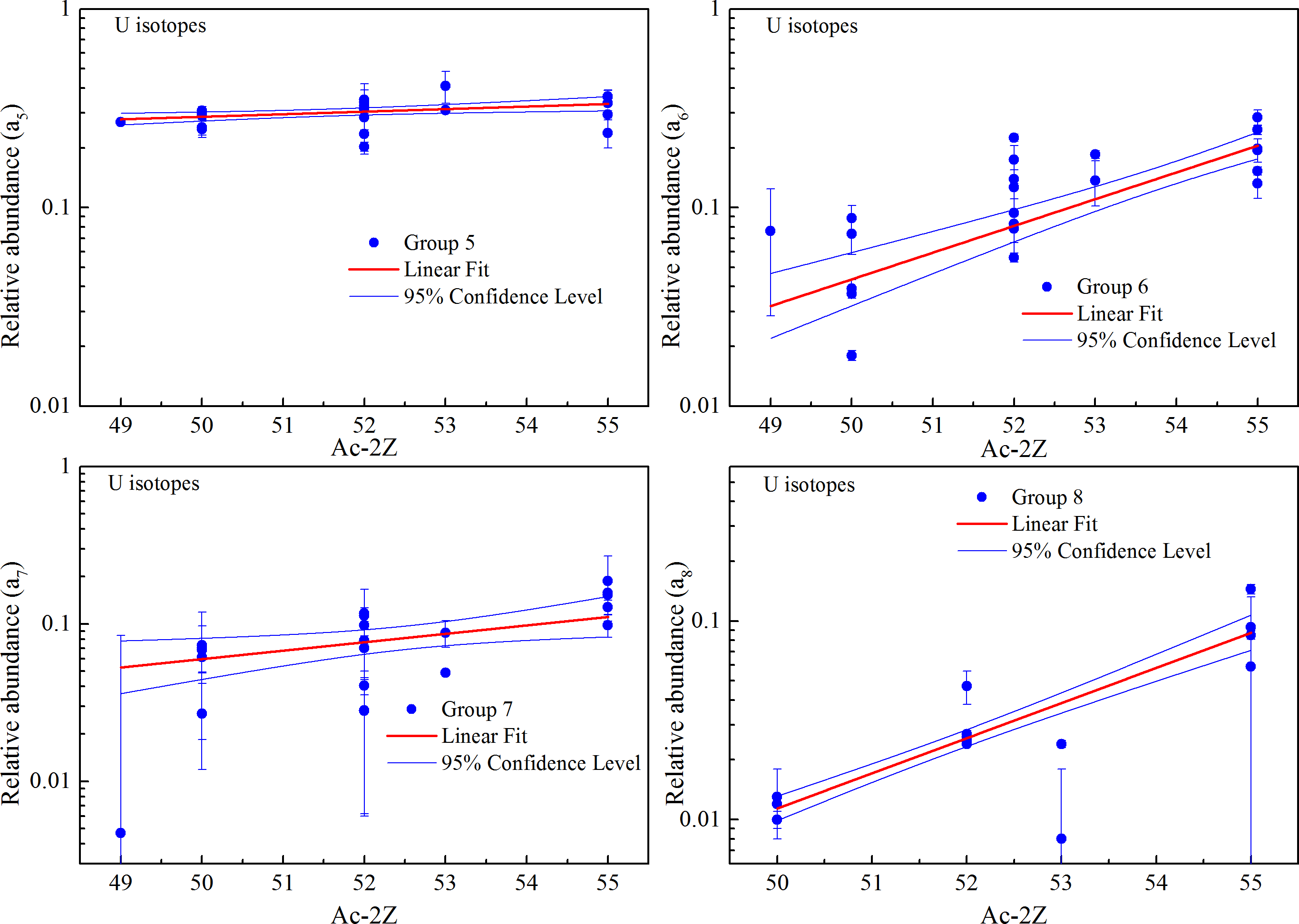}
\caption{The $(A_c-2\cdot Z)$ systematics of the relative abundances of delayed neutrons from fission of uranium isotopes (5-8 groups). The values of the relative abundances were taken from \cite{Spriggs99}.}
\label{fig:8_3}
\end{figure*}

In practice, the time parameters of DN are represented by such characteristics as the relative yields and the periods of individual DN groups. Therefore, the development of the systematics of the relative yields and periods of DN for individual groups is required. This task cannot be performed using the 6-group model, because of the strong correlation of the periods and the relative abundances of DN. Until now the only method for prediction of the temporary characteristics $(a_i,T_i)$ of DN was based on the summation calculations \cite{Wilson02}. The 8-group model should be used for the systematics of the relative abundances because in this representation of the group parameters there is no correlation between the relative abundances and periods of DN, and the values of DN periods are universal for the isotopes of all the elements. \\
To estimate the relative abundances of delayed neutrons from the fast neutron induced fission of the uranium isotopes the parameter $(A_c-2\cdot Z)$ was used. In Figs.~\ref{fig:8_2}  and \ref{fig:8_3} the 8-group relative abundances of uranium isotopes $a_i\cdot(i = 1,\ldots,8)$ are approximated by expression  $a_i= exp\left[c_i+b_i\cdot (A_c-2\cdot Z)\right]$, where $c_i$, $b_i$ are constants. These constants are estimated for each DN group and then the obtained dependences $a_i$ have been used for the calculation of the relative abundances of all uranium isotopes including unmeasured uranium isotopes $^{234}$U and $^{237}$U. The results of this estimate are presented in Table~\ref{tab:8.1}.

\begin{table*}
\caption{Relative abundances of uranium isotopes estimated by the $(A_c-2\cdot Z)$ systematics.}
\label{tab:8.1}
\begin{tabular}{c|c|c|c|c|c|c|c} \hline \hline
\T Uranium isotopes &	$^{232}$U&	$^{233}$U&	$^{234}$U&	$^{235}$U&	$^{236}$U&	$^{237}$U&	$^{238}$U \\ \hline
\T Half-life, s&	\multicolumn{7}{c}{Relative abundances, $a_i$} \\ \hline
\T 55.6 (Group 1) &	0.1002&	0.0726&	0.0514&	0.0354&	0.0238&	0.0155&	0.0098  \\  
24.5 (Group 2) &	0.1811&	0.1766&	0.1680&	0.1558&	0.1406&	0.1232&	0.1045 \\
16.3 (Group 3) &	0.1433&	0.1236&	0.1040&	0.0854&	0.0681&	0.0528&	0.0397 \\
5.21 (Group 4) &	0.2454&	0.2414&	0.2316&	0.2167&	0.1972&	0.1743&	0.1492 \\
2.37 (Group 5) &	0.2478&	0.2757&	0.2992&	0.3165&	0.3257&	0.3256&	0.3152 \\
1.04 (Group 6) &	0.0285&	0.0419&	0.0602&	0.0843&	0.1149&	0.1520&	0.1948 \\
0.424 (Group 7) &	0.0469&	0.0573&	0.0683&	0.0793&	0.0897&	0.0984&	0.1046 \\
0.195 (Group 8) &	0.0067&	0.0109&	0.0172&	0.0266&	0.0399&	0.0582&	0.0823 \\
Average half-life   $\langle   T\rangle$~(s)& 14.26&	12.36&	10.68&	9.18&	7.84&	6.63&	5.54 \\
Systematics $(A_c-2\cdot Z) $   &	   &         &       &      &      &       &         \\

Average half-life   $\langle   T\rangle$~(s)&14.60$\pm$0.74&	12.38$\pm$0.50&	10.49$\pm$0.34&	8.90$\pm$0.28&	7.54$\pm$0.27&	6.38$\pm$0.19&5.42$\pm$0.30 \\
Systematics $(A_c/Z)\cdot 92$&   &         &       &      &      &       &         \\
Average half-life   $\langle   T\rangle$~(s)&14.35$\pm$0.71 &	12.38$\pm$0.37&	-&	9.10$\pm$0.09&	7.37$\pm$0.66&	-&	5.30$\pm$0.16 \\
 (recommended data~\cite{Spriggs02})&    &         &       &      &      &       &         \\
Average half-life   $\langle   T\rangle$~(s)&  	14.44&	12.36&	10.49&	7.54&	7.77&	6.90&	5.05 \\
summation~\cite{Wilson02}&		&    &         &       &      &      &           \\  \hline \hline
\end{tabular}
\end{table*}

\begin{figure}[!htb]
	\centering
\includegraphics[width=\linewidth]{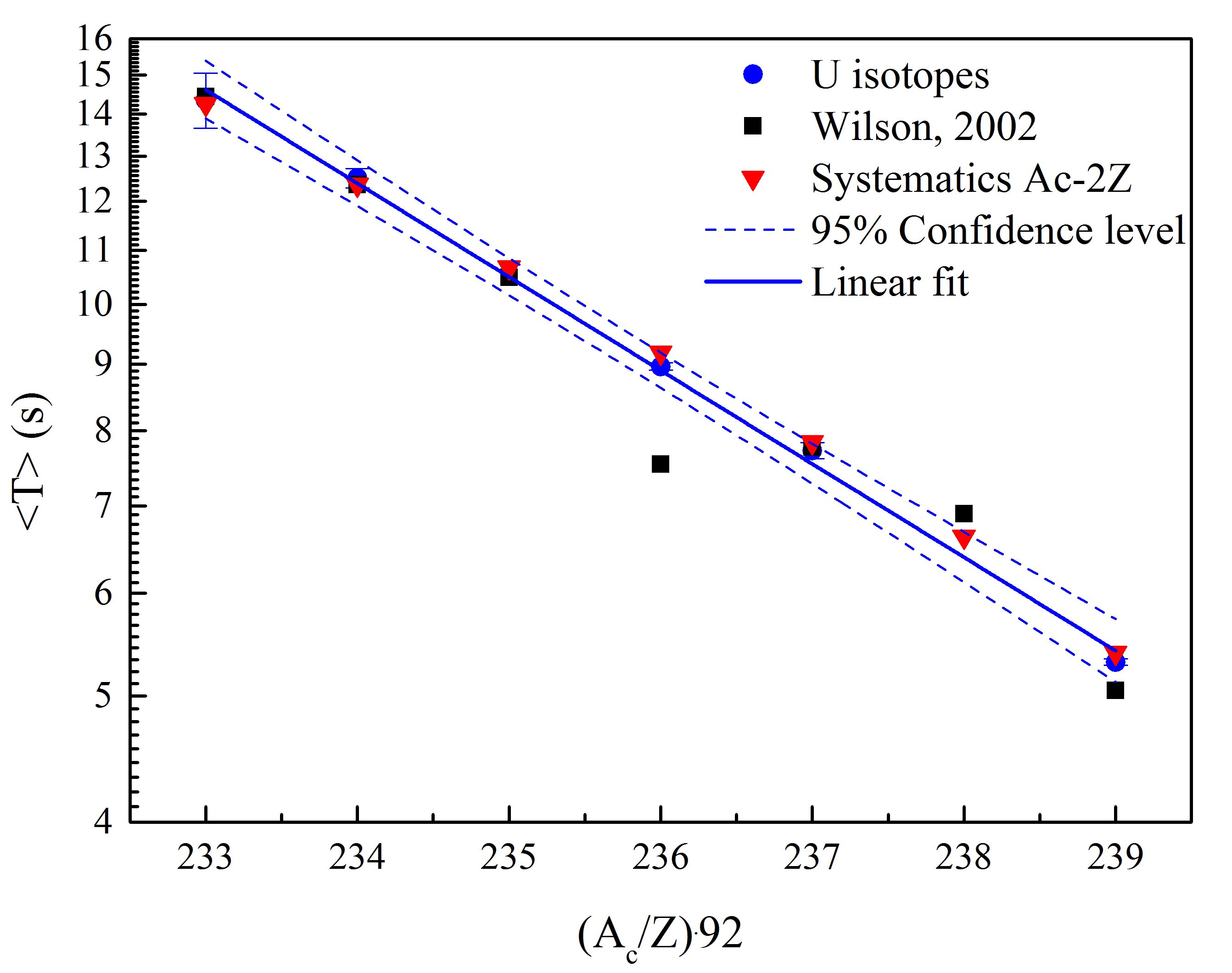}
\caption{The $(A_c/Z)\cdot 92$ systematics of the average half-life of DN precursors for uranium isotopes. Full circles - the average half-life of uranium isotopes (recommended data), triangles - the average half-life estimated with the help of $(A-2\cdot Z)$ systematic, squares - the average half-life calculated by the summation method \cite{Wilson02}, solid line - the approximation of the average half-life by expression $\langle T \rangle =exp \left[ a+b \cdot (A_c/Z) \cdot 92 \right]$.}
\label{fig:8_4}
\end{figure}

It can be seen from Table~\ref{tab:8.1} that the average half-lives   $\langle   T\rangle$~obtained on the basis of the $(A_c-2\cdot Z)$ systematics agree both with the recommended data and the data calculated by the summation techniques with the exception of the value for $^{235}$U calculated with the help of the summation method \cite{Wilson02}. These data are also presented in Fig.~\ref{fig:8_4} from which one can see that the predicted values for $^{234}$U and $^{237}$U agree with the systematics $(A_c/Z)\cdot 92$.

From the present studies it may be concluded that both the $(A_c/Z)\cdot 92$ systematics of the average half-life of DN precursors and the $(A_c-2\cdot Z)$ systematics of the relative  abundances $a_i$ have a good predictive power and can be used for the prediction of the temporary DN parameters of unmeasured nuclides.


\section{ RECOMMENDED DATA IN 6- and 8-GROUP MODELS }
\label{Sec:Macro-Recommended}


In modern nuclear power development programs, priority is given to the development of new concepts of nuclear reactors characterized by a more stringent energy spectrum of neutrons, a complex composition of nuclear fuel, and the possibility of their use for transmutation of nuclear waste. Issues such as safety and efficient operation of  reactors have increased the need for a reliable and complete set of nuclear physical constants used in reactor practice, including the database on delayed neutrons (DN).

Since the latest recommendations by WPEC-SG6, i.e. the DN group parameters based on the experimental data presented in \cite{Spriggs99,Spriggs02}, a number of measurements of the DN group parameters has been performed at IPPE (Obninsk) for neutron induced fission of uranium, plutonium, thorium, americium, and neptunium isotopes for neutron energies ranging from thermal to 18 MeV. The short review of these data can be found in Section~\ref{Sec:Macro-meas} of this report. The compilation of these data is included in the database created within the IAEA Coordinated Research Project \cite{IAEA0683}. There is an increasing need for reliable DN data uncertainties, including information on the covariance matrix of the DN group parameters. The task of estimating the DN group parameters (the relative abundances $a_i$ and the half-life $T_i$ of their precursor) has become more pressing due to the availability of new experimental data and current trends in nuclear technologies.

In reactor physics practice two models of the DN group parameters are used: the unconstrained 6- and constrained 8-group model \cite{Spriggs99,Spriggs02}. In the former, the relative abundances $a_i$ and the half-life $T_i$ of the individual DN groups are obtained from analysis of the decay curves of the DN activity by the least-squares method (LSM) with 12 free parameters ($a_i$,$T_i$) \cite{Keepin57,Piksaikin02a}. In the 8- group model the set of the group half-lives $T_i$ is universal for all nuclides and primary neutron energies.  The $T_i$ values of this data set are presented in Table~\ref{tab:9.1}.

\begin{table}[hb]
\caption{Consistent set of half-lives $T_i$ in the 8-group model.}
\label{tab:9.1}
\begin{tabular}{c|c|c|c|c|c|c|c|c} \hline \hline
\T Group number &	1&	2&	3&	4&	5&	6&	7&	8 \\  \hline
\T Half-life $T_i$ (s)&	55.6&	24.50&	16.30&	5.21&	2.37&	1.039&	0.424&	0.195 \\ \hline \hline
\end{tabular}
\end{table}

The motivation for the development of the 8-group model and the resulting advantages, as compared with the 6-group model, are discussed in detail in \cite{Spriggs99,Spriggs02} and include: a significant reduction in the cross correlation of the DN group parameters, a more simple dynamic model for generating DN in a multi-component mixture of fissile nuclides, an improved procedure for the estimation of the DN group energy spectra and a better description of the relationship between reactivity and periods for negative values of reactivity. In the present evaluation the recommended sets of DN group parameters are given in 6- and 8-group representation for all nuclides under consideration. 

\begin{figure*} 
\centering
\includegraphics[width=\textwidth]{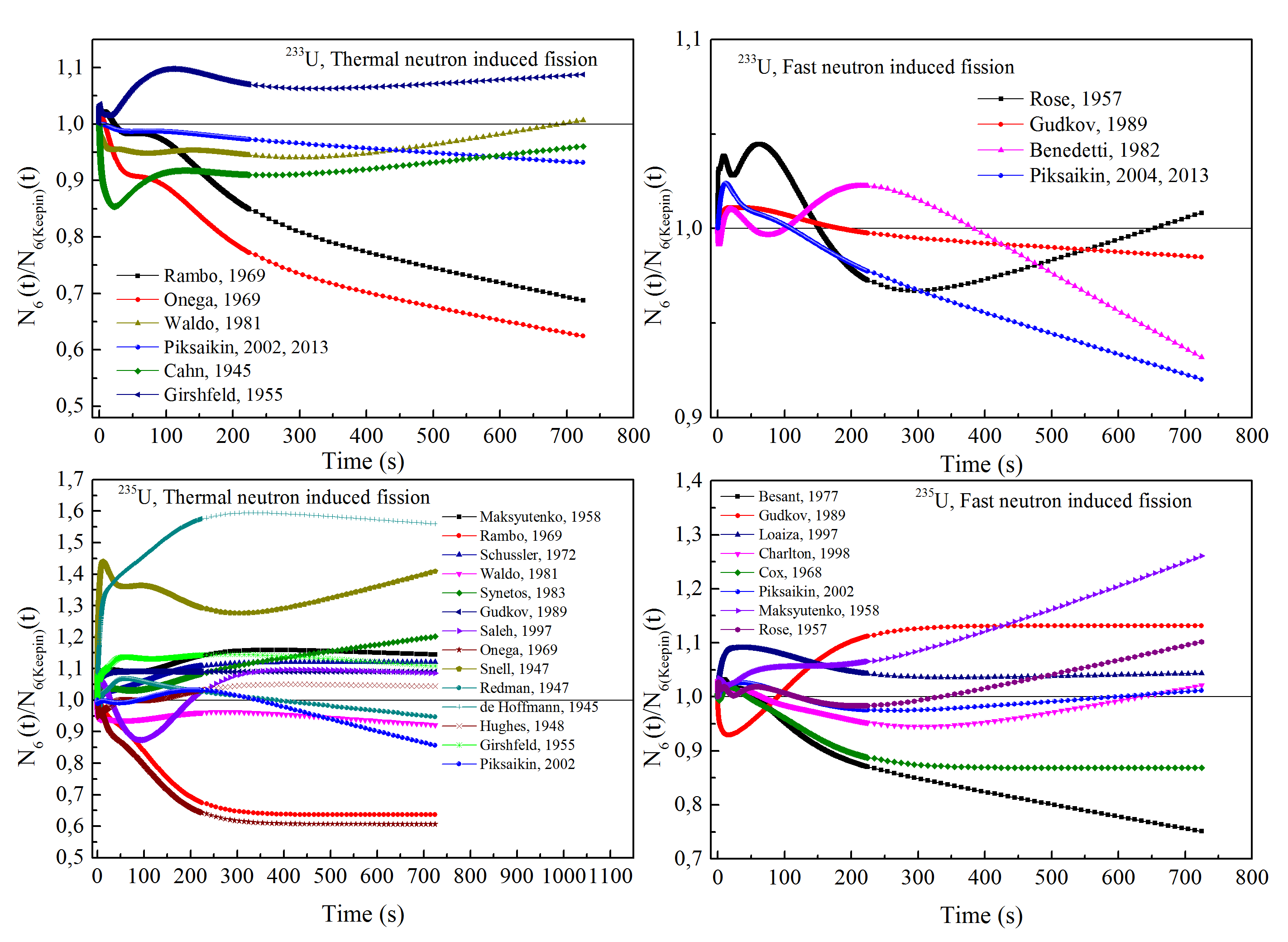}
\caption{The ratio of the DN decay curves in the 6-group representation measured from fission of $^{233}$U and $^{235}$U by thermal and fast neutrons to the corresponding DN decay curve calculated using the ($a_i$,$T_i$) data sets from Keepin's data \cite{Keepin57}. The data references can be taken from the compilation by Spriggs and Campbell \cite{Spriggs02a} except for the data by Piksaikin \textit{et al.} \cite{Piksaikin02a,Piksaikin13,Piksaikin02e,Piksaikin04}}
\label{fig:9_1}
\end{figure*}

\begin{figure*} 
\centering
\includegraphics[width=\textwidth]{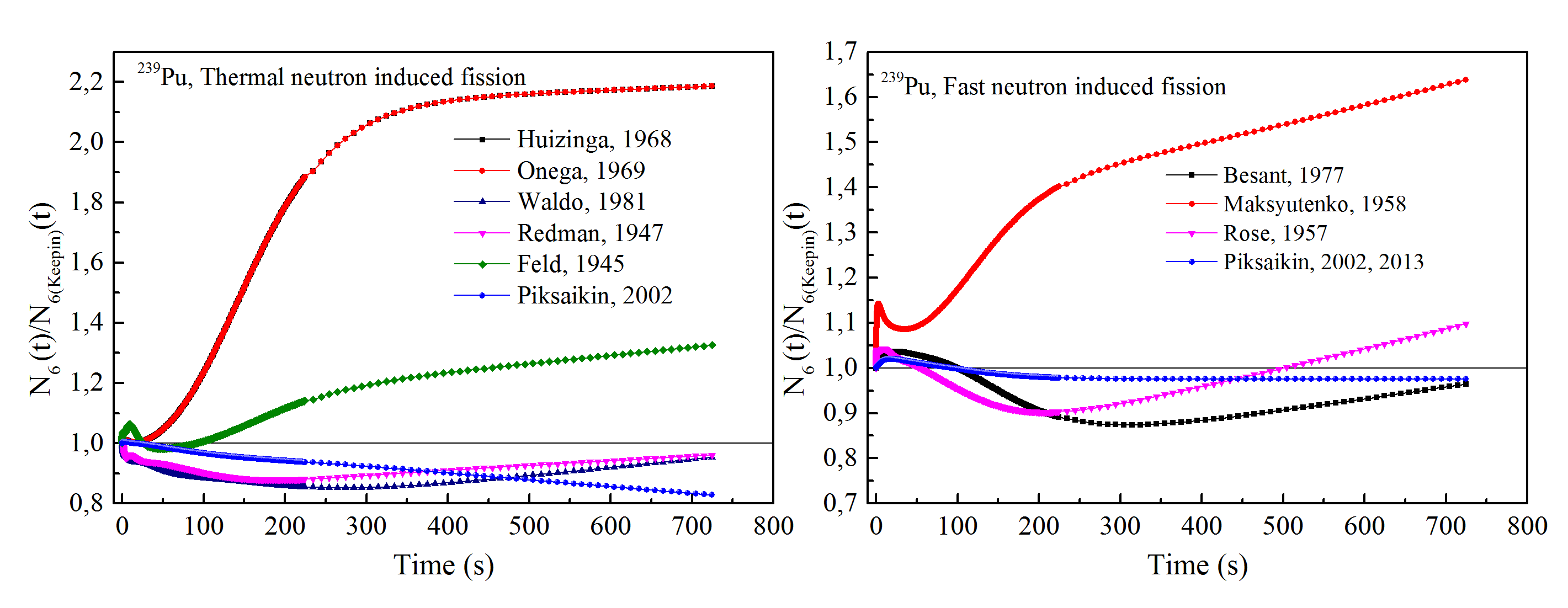}
\caption{The ratio of the DN decay curves in the 6-group representation measured from the fission of $^{239}$Pu by thermal and fast neutrons to the corresponding DN decay curve calculated using the ($a_i,T_i$) data sets from Keepin's data \cite{Keepin57}. The data references can be taken from the compilation by Spriggs and Campbell \cite{Spriggs02a} except for the data by Piksaikin \textit{et al.} \cite{Piksaikin02a,Piksaikin13,Piksaikin02d,Piksaikin02e}}
\label{fig:9_2}
\end{figure*}

\paragraph{Analysis of the relative abundances and half-lives of delayed neutrons.} 
The selection of the DN decay curve and the corresponding set of the temporal DN parameters has been carried out on the basis of the time of the sample transfer from an irradiation position to the neutron detector, the time of recording the decay curve of DN neutron activity, the number of DN groups resolved in the experiment, the method for estimating the DN parameters (graphical or LSM), the uncertainties of the DN parameters, the consistency of the DN group parameters with the systematics of the average half-life of the DN precursors, and the energy dependence of the temporal DN parameters expressed in the average half-life of the DN precursors \cite{Spriggs99,Spriggs02}.
In the present evaluation additional criteria were considered that allow making a wider analysis of the compared DN data: the ratio of the decay curves in the 6- and 8-group model to the corresponding recommended data and the quality of the expansion of the 6-group data into the 8-group model. When carrying out this analysis, it was assumed that the 6-group parameters reproduce the experimental DN decay curve more accurately in comparison to the 8-group representation. However, it should be borne in mind that this statement is only valid for the time interval in which these data were measured. In the overwhelming number of experiments this interval did not exceed 500 s \cite{Spriggs02a}.
The aggregate DN curve for each DN data set ($a_i$,$T_i$) was modeled in the range 0-724 s using expression

\begin{equation}
N(t)=A\cdot\sum_{i=1}^m (1-e^{-\lambda_i\cdot t_{irr}})\cdot a_i\cdot e^{-\lambda\cdot t}\,,
\label{eq.ix.1}
\end{equation}
where $A$ is the saturation activity, $t_{irr}$ the irradiation time (300 s), ($a_i$, $\lambda_i$)  the relative abundance and the decay constant of the $i$-th DN group, $m$  the number of DN groups. The DN data sets ($a_i$, $T_{i}=ln2/\lambda_i$) were taken from the compilation by Spriggs and Campbell~\cite{Spriggs02a}. The 8-group data sets were taken from the evaluation by Spriggs \textit{et al.} \cite{Spriggs99,Spriggs02} with the exception of the 8-group data sets from the IPPE data obtained after 2002 which are not included in the evaluation by Spriggs \textit{et al.} \cite{Spriggs02}. A procedure used for the selection of the DN group data is shown in the example of the DN data sets measured from the thermal and fast neutron induced fission of $^{233}$U, $^{235}$U and $^{239}$Pu. The DN decay curves of different authors $N_{6}(t)$ are presented as the ratio to the corresponding DN decay curves calculated using the DN sets ($a_i$, $T_i$) from the Keepin data $N_{6(K)}(t)$ \cite{Keepin57}. The obtained data are shown in Figs.~\ref{fig:9_1} and ~\ref{fig:9_2}.

In the interval 0-500 s, the IPPE data~\cite{Piksaikin02a,Piksaikin13,Piksaikin02d,Piksaikin02e,Piksaikin04} show the smallest discrepancy with Keepin's data \cite{Keepin57}. The difference between these data does not exceed 5$\%$ for all fissioning systems with the exception of the thermal neutron induced fission of $^{239}$Pu.  At the end of the considered interval ($\Delta t$ = 0-724 s) this difference is increased and reaches 18$\%$ for the thermal neutron induced fission of $^{239}$Pu. The observed difference between decay curves at the end of the considered interval is mainly caused by the differences between the half-life values of the 1-st DN groups. The half-life values of the 1-st DN group $T_1$ and the average half-life of DN precursors   $\langle   T\rangle$~for fission of $^{233}$U, $^{235}$U and $^{239}$Pu are presented in Table~\ref{tab:9.2} for both the IPPE and the Keepin data \cite{Keepin57}. Table~\ref{tab:9.2} shows that for all fissioning systems the $T_1$ values in the Keepin data are systematically lower  than the corresponding values of the IPPE data sets. One of the reasons for this may be related to the averaging procedure used for the estimation of DN parameters on the basis of different experimental runs which does not take into account the correlation between DN parameters. The second reason for the difference in the $T_1$ values is related to the different time intervals in which the DN curves were measured. The IPPE temporary DN parameters have been estimated on the basis of the experimental data measured in the extended time interval (0.12-724 s) \cite{Piksaikin06} while in Keepin's experiment this interval was 500 s. In addition a new averaging procedure for data sets ($a_i$,$T_i$) measured in different experimental runs has been developed at the IPPE that accounts for a correlation property of DN group parameters \cite{Piksaikin02a,Piksaikin02e}.

\begin{table}
\caption{Average half-life of DN precursors   $\langle   T\rangle$~and half-life of the first group of DN $T_1$ in [s] from fission of $^{233}$U, $^{235}$U and $^{239}$Pu. F is for fast incident neutrons and T for thermal incident neutrons.}
\label{tab:9.2}
\begin{tabular}{c|c|c|c|c|c} \hline \hline
Target &	&	$T_1$ &	  $\langle   T\rangle$  &	$T_1$ &   $\langle   T\rangle$ \\ \hline
       &   & \multicolumn{2}{c|}{~\cite{Keepin57}}  &\multicolumn{2}{c}{~\cite{Piksaikin02a,Piksaikin13}} \\ \hline
\T $^{233}$U &	F&	55.11$\pm$1.86	&12.38$\pm$0.40	&54.55$\pm$0.38	&12.50$\pm$0.20 \\
	      &T&	55.00$\pm$0.54&	12.80$\pm$0.38&	54.64$\pm$0.19&	12.69$\pm$0.11  \\
$^{235}$U &	F&54.51$\pm$0.94	&8.82$\pm$0.25	&54.90$\pm$0.15	&8.96$\pm$0.06 \\
	      &T	&54.72$\pm$1.28	&9.03$\pm$0.34	&53.95$\pm$0.28	&8.98$\pm$0.11 \\
$^{239}$Pu	&F	&53.75$\pm$0.95	&10.16$\pm$0.24	&53.81$\pm$0.41	&10.27$\pm$0.13 \\
	        &T	&54.28$\pm$2.34	&10.70$\pm$1.11	&53.1$\pm$0.46	&10.59$\pm$0.17 \\ \hline \hline
\end{tabular}
\end{table}

As mentioned above the measurements of the DN decay curves in \cite{Keepin57} have been carried out during 500 s after the end of irradiation while in \cite{Piksaikin06,Piksaikin02a} this interval is much wider - 724 s. Consequently, the comparison of DN data in the region $t\, \textgreater$ 500 s is not entirely correct since the Keepin data and data of other authors (with a few exceptions) in this region have been \textit{extrapolated} because the shape of the DN decay curve here is mainly determined by the half-life of the 1-st DN group obtained from data in the range of  $t\, \textless$ 500 s. In addition, the comparison of the uncertainty of the DN parameters of the 1-st DN group obtained in \cite{Keepin57} and \cite{Piksaikin02a,Piksaikin13} shows that the IPPE data have better statistical accuracy and, most likely, better background conditions. 
To study the influence of interval width used for DN counting on the estimate of the half-life value of the 1-st DN group, the experimental DN decay curves from the IPPE data \cite{Piksaikin02a,Piksaikin13,Roshchenko10} measured in separate measurement runs from fission of $^{235}$U, $^{238}$U and $^{232}$Th by fast neutrons have been analyzed in the time intervals 224-714 s, 300-714 s and 400-714 s in the frame of a single-group approximation. The results are shown in Table~\ref{tab:9.3}.

\begin{table*}
\caption{Half-lives of DN precursors obtained in different time intervals of DN decay curves for the fast neutron induced fission of $^{232}$Th, $^{235}$U, $^{238}$U.}
\label{tab:9.3}
\begin{tabular}{c|c|c|c|c|c} \hline \hline
Target nuclides	&\multicolumn{3}{c|}{Single-group approximation}&6-group approximation~\cite{Keepin57}& 6-group approximation~\cite{Piksaikin02a,Piksaikin13,Roshchenko10}         \\
&\multicolumn{3}{c|}{Time interval (s)} & half-life of  & half-life of  \\ 
   &\multicolumn{3}{c|}{Half-life $T_{1/2}$ (s)} &1st DN group $T_1$ &1st DN group $T_1$  \\ \hline \hline

\T	  &224-724&	300-724&	400-724&	Interval 0-500 s&	Interval 0-724 s \\
$^{238}$U&	50.72$\pm$0.59&	55.41$\pm$1.65&	57.51$\pm$5.37&	52.38$\pm$1.29&	52.9$\pm$0.17\\
$^{232}$Th&	53.09$\pm$0.96&	54.17$\pm$2.58&	54.3$\pm$10.32&	56.03$\pm$0.095&	55.42$\pm$0.27\\
$^{235}$U&	51.24$\pm$0.68&	54.02$\pm$1.09&	55.27$\pm$3.22&	54.51$\pm$0.94&	54.9$\pm$0.16\\ \hline \hline
\end{tabular}
\end{table*}

\begin{figure*} 
\centering
\includegraphics[width=\textwidth]{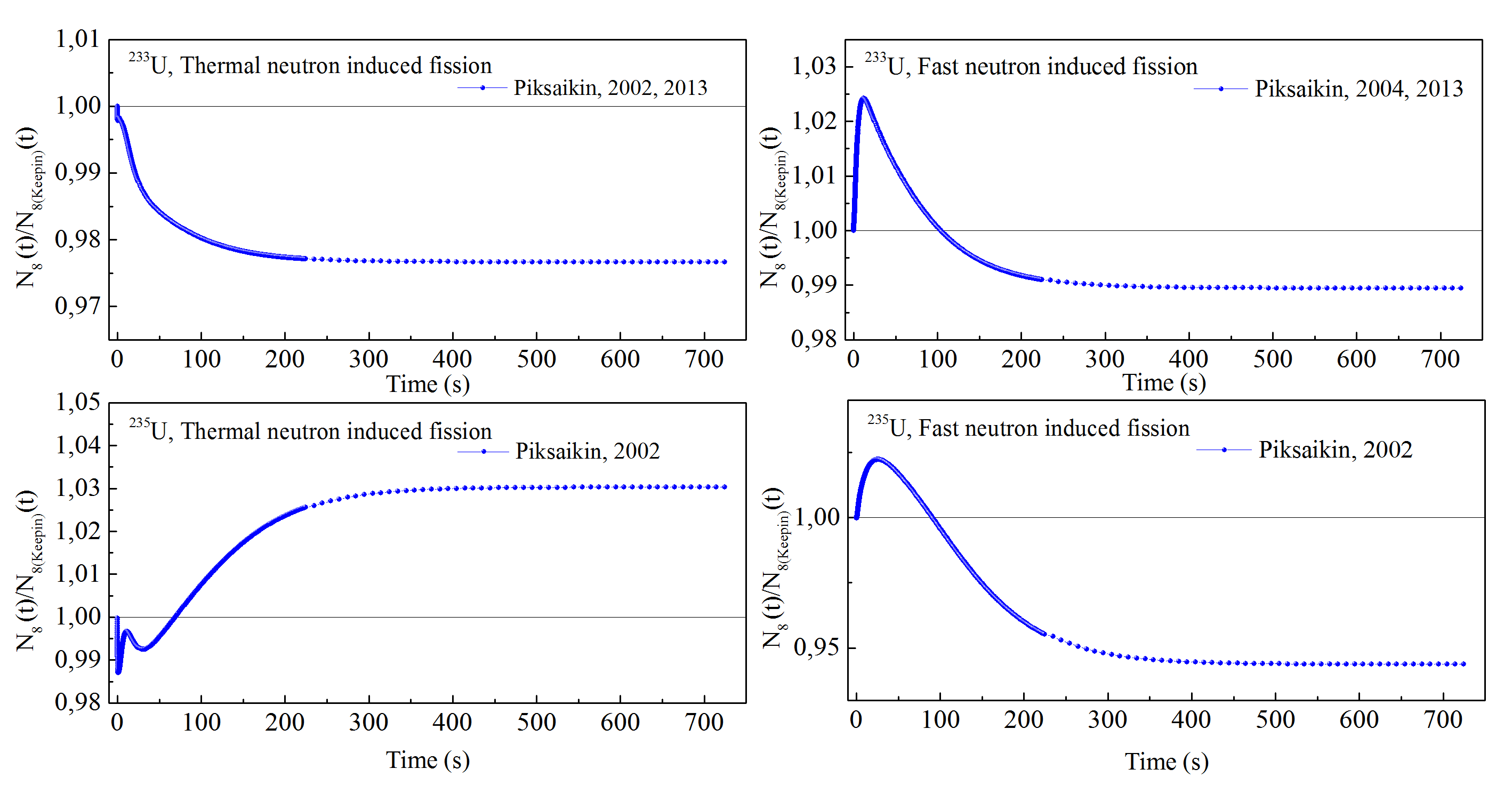}
\caption{The ratio of the DN decay curves in the 8-group representation from the IPPE data \cite{Piksaikin02a,Piksaikin13,Piksaikin02e,Piksaikin04} to the corresponding DN decay curve calculated by using the data sets ($a_i$,$T_i$) from the Keepin data \cite{Keepin57} measured in the fission of $^{233}$U, $^{235}$U by thermal and fast neutrons.}
\label{fig:9_3}
\end{figure*}

\begin{figure*} 
\centering
\includegraphics[width=\textwidth]{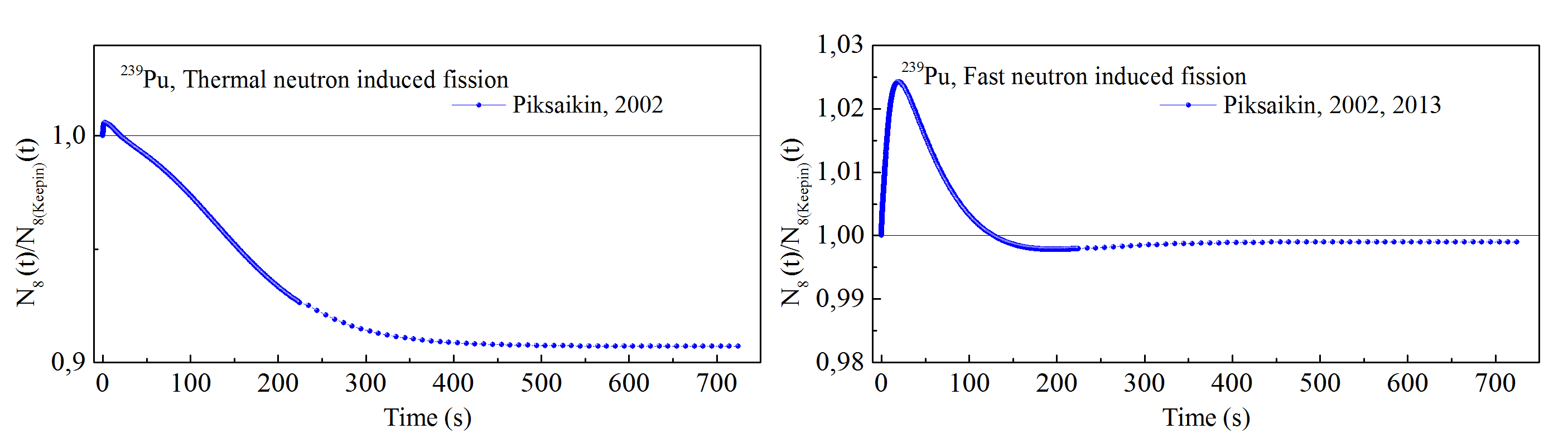}
\caption{The ratio of the DN decay curves in the 8-group representation from IPPE data \cite{Piksaikin02a,Piksaikin13,Piksaikin02d,Piksaikin02e} to the corresponding DN decay curve calculated by using Keepin's data ($a_i$,$T_i$) \cite{Keepin57} measured in the fission of $^{239}$Pu by thermal and fast neutrons.}
\label{fig:9_4}
\end{figure*}

It can be seen from Table~\ref{tab:9.3} that the value for all nuclides in the interval $224-724$ s is lower than the  half-life of $^{87}$Br ($T_{1/2} = 55.6$ s) which can be explained by the contribution of $^{137}$I ($T_{1/2} = 24.5$ s) to the aggregate DN decay curve. In the interval 300-724 s the values of the half-life are comparable with the half-life of $^{87}$Br which implies a small contribution of $^{137}$I activity which in turn means that the $T_{1/2}$ value reaches its asymptotic value. The $T_{1/2}$ value in the interval 400-724 s has large uncertainties because of the low statistics at the end of the decay curve. The results show that the half-life value of the 1-st DN group obtained by the LSM using the 6-group model, can differ significantly from the $^{87}$Br half-life (see the data for $^{238}$U in Table~\ref{tab:9.3}) in spite of the fact that single-group processing of the experimental curve in the interval 300-724 s gives a value close to the half-life of $^{87}$Br - $55.41\pm 1.65$ s. \textit{Thus, the value of the parameters of the 1-st DN group obtained in the measurements of the DN decay curves is determined not only by the time interval in which the measurements are performed, but also by the peculiarities of the data processing procedure (LSM) and strong cross-correlation between group parameters as it has already been concluded by Spriggs \textit{et al.}} \cite{Spriggs99,Spriggs02}.
In order to eliminate the disagreement in the DN decay curves in the range $t\,\textgreater$ 400 s arising from the uncertainties in the parameters of the 1-st DN group observed in the 6-th group model one has to transfer 6-group data to the 8-group model. Figs.~\ref{fig:9_3} and \ref{fig:9_4} show the ratio of the DN decay curves in the 8-th group representation from the IPPE data $N_{8}(t)$ to the corresponding Keepin 8-group data $N_{8(K)}(t)$ \cite{Keepin57} for thermal and fast neutron induced fission of $^{233}$U, $^{235}$U and $^{239}$Pu. 

It can be seen from  Figs.~\ref{fig:9_3} and \ref{fig:9_4} that the discrepancy between the Keepin and the IPPE data for all considered fissioning systems with the exception of the thermal neutron induced fission of $^{239}$Pu significantly decreased and does not exceed 5$\%$ in the whole time interval 0-724 s. The residual difference in the shape of the DN decay curves is due to a distinction in the values of the relative abundances of the 1-st DN group $a_1$. 

This observation as well as the resulting similar values of the average half-life   $\langle T\rangle$  (see Table~\ref{tab:9.2}) can be considered as the confirmation of the agreement between the IPPE and Keepin data for the fission of $^{233}$U and $^{235}$U by thermal and fast neutrons and for the fission of $^{239}$Pu by fast neutrons, if they are presented in the 8-th group model. \textit{Thus the data presented in Figs.\ref{fig:9_3}-\ref{fig:9_4} demonstrate the advantage of the 8-th group model, since it eliminates the discrepancies in the time range $t$ \textgreater 400 s associated with uncertainties in determining the parameters of the 1-st DN group within the framework of the 6-model.} The divergence in case of the thermal neutron-induced fission of $^{239}$Pu is caused by a large difference in the relative abundances of the 1-st DN group ($a_1=0.038\pm 0.003$ and $0.035\pm 0.001$ in the Keepin \cite{Keepin57} and the IPPE data \cite{Piksaikin02a} sets, respectively). This may be connected to the poor statistics in Keepin's experiment which can be seen seen from the large uncertainties of the average half-life of DN precursors observed for the thermal neutron induced fission of $^{239}$Pu: $\langle T\rangle = 10.70 \pm 1.11$ s (see Table~\ref{tab:9.2}).

\paragraph{Quality of the expansion method.}
Validation of the expansion methods \cite{Spriggs02,Piksaikin02a} used for the transformation of 6-group data to 8-group model data has been made on the basis of the ratio of the DN decay curve in the 8-group representation to the corresponding DN decay curve in the 6-group model. As a result of this comparison it has been shown that in the recommended DN data set \cite{Spriggs02} the 8-group expanded data and the corresponding original 6-group data set for the fast neutron induced fission of $^{238}$U are not consistent. The 8-group data set for $^{238}$U obtained on the basis of the IPPE data has been proposed as the recommended one.

\paragraph{High energy data.}
The IPPE measurements in the high energy range (14-18 MeV) have revealed the possible reasons for the large discrepancy observed in the delayed neutron temporary parameters: a deterioration of the counting property of the neutron detector in a high intensity neutron flux coming from the T(d,n)$^4$He reaction \cite{Piksaikin07} and the influence of a concomitant D(d,n)$^3$He neutron source generated at an accelerator target in measurements using the T(d,n)$^4$He reaction \cite{Piksaikin06a,Piksaikin07}. At present, these data are only available in the 6-group representation. The expansion of these data to the 8-group model will be undertaken in the near future.

\paragraph{Conclusion.}
The efforts undertaken at IPPE for improving the experimental techniques and data processing procedures included the shortening of the transportation time, the extension of delayed neutron counting time, the exact determination of the incident neutron energy as well as a new approach for averaging the temporary DN parameters obtained in different experimental runs. As a result, the accuracy of the DN group parameters ($a_i$,$T_i$) was greatly improved. In general, these studies can be considered as the next step following the last evaluation by Spriggs \textit{et al.} \cite{Spriggs02} in 2002 leading to the improvement of the macroscopic database of the relative abundances of delayed neutrons and half-lives of their precursors. The revisions introduced to the WPEC-SG6 recommended DN temporary data sets \cite{Spriggs02} are presented in Table~\ref{tab:9.4}. A complete set of the present recommendations for thermal, fast and high energy fission are given in Tables~\ref{tab:9.5} and ~\ref{tab:9.6}. The relative abundances and half-lives of delayed neutrons are presented in the 6- and 8-group model representation with their uncertainties and the values of the average half-life. Information on correlations and covariance matrices can be found in the IAEA compilation database created under the CRP project \cite{IAEA0683}. The references for the original 6-group data sets and the 8-group data sets can be found in the compilation by Spriggs and Campbell \cite{Spriggs02a} and Spriggs \textit{et al.} \cite{Spriggs02,Spriggs99} with the exception of the IPPE data, the references for which are included in the present report. 

\begin{table}
\caption{The revisions made to the WPEC-SG6 recommended sets of temporary DN parameters presented in the evaluation by Spriggs \textit{et al.}~\cite{Spriggs02}. 6 and 8 stand for 6-group and 8-group DN parameters, respectively.}
\label{tab:9.4}
\begin{tabular}{c|c|c|c} \hline \hline
&	\multicolumn{3}{c}{Neutron energy} \\ \hline
Target& Thermal & Fast & High  \\ \hline \hline
\T $^{232}$Th &	-	& No revision	& New (6) \\
$^{233}$U &	New (6 $\&$ 8) &	New (6 $\&$ 8)&	No revision \\
$^{235}$U &	New (6 $\&$ 8) &	New (6 $\&$ 8)&	No revision \\
$^{236}$U &	-	& New (6 $\&$ 8) &	- \\
$^{238}$U &	-	& New (6 $\&$ 8)& New (6) \\
$^{237}$Np &	-	& New (6 $\&$ 8)&	New (6) \\
$^{239}$Pu &	New (6 $\&$ 8)&	New (6 $\&$ 8)&	New (6) \\
$^{241}$Am &	-	&New (6 $\&$ 8)&	New (6) \\ \hline \hline
\end{tabular}
\end{table}

\afterpage{

\begin{sidewaystable*}
\caption{Recommended 6-Group Delayed Neutron Parameters, where $T_i$ and   $\langle   T\rangle$  in (s); (F)  for fast neutrons, (T) for thermal neutrons, (SF) for spontaneous fission. All the references for the data are included in the review of Spriggs \textit{et al.}~\cite{Spriggs02}.}\label{tab:9.5}
\centering
\begin{tabular}{l|c|c|c|c|c|c|c|c|c|c}
 \hline \hline
Target& $E_n$ [MeV]	&Author	& &	gr 1 &	gr 2	&	gr 3 &	gr 4 &	gr 5 &	gr 6 &	  $\langle   T\rangle$ \\ \hline  \hline 
   
\T $^{232}$Th	&(F)&Keepin &	$T_i$&56.0$\pm$0.095	&20.8$\pm$0.66&5.74$\pm$0.24&2.16$\pm$0.08&0.57$\pm$0.042&0.211$\pm$0.019&	6.98$\pm$0.23 \\
	&	&('57)	&	$a_i$	&0.034$\pm$0.002&0.15$\pm$0.005&0.155$\pm$0.021&0.446$\pm$0.015&0.172$\pm$0.013	&0.043$\pm$0.006&	\\ \hline
	&14.23&~\cite{Roshchenko10} &$T_i$&56.28$\pm$0.411	&21.25$\pm$0.197&5.45$\pm$0.082&2.27$\pm$0.03&0.578$\pm$0.012&0.214$\pm$ 0.005&7.43$\pm$0.08 \\
	&	&	&$a_i$&	0.039$\pm$0.001	&0.149$\pm$	0.003&0.161$\pm$0.003&0.482$\pm$0.006&0.13$\pm$	0.003&0.038$\pm$0.001&	\\ \hline
\T $^{231}$Pa	&(F)&Anoussis&$T_i$&	57.0$\pm$0.000&24.6$\pm$0.000&15.4$\pm$0.000&6.10$\pm$0.000&4.10$\pm$0.000&	1.90$\pm$0.000	&14.33$\pm$0.19 \\
	&	& ('73)&$a_i$&	0.104$\pm$0.001	&0.155$\pm$0.003&0.178$\pm$0.005&0.053$\pm$0.017&0.251$\pm$0.026&0.259$\pm$0.027& \\ \hline
	&14.8 &Brown&$T_i$&53.4$\pm$0.000&19.2$\pm$0.000&5.50$\pm$0.000&2.00$\pm$0.000&-	&- 	&14.33$\pm$0.91 \\
	&	& ('71)&$a_i$&0.136$\pm$0.014&0.249$\pm$0.025&0.299$\pm$0.030&0.316$\pm$0.032&	-&-& \\ \hline	
\T $^{232}$U&(T)&Waldo&$T_i$&54.3$\pm$0.170&19.8$\pm$0.160&4.820$\pm$0.200&1.75$\pm$0.200&0.513$\pm$0.150&-	&14.37$\pm$0.70 \\
	&	& ('81)&$a_i$&0.120$\pm$0.009&0.300$\pm$0.023&0.306$\pm$0.032&0.258$\pm$0.027&0.016$\pm$0.089&	- &	\\ \hline
\T $^{233}$U&(T)&~\cite{Piksaikin02e} &$T_i$&54.64$\pm$0.19&20.77$\pm$0.13&5.25$\pm$0.05&2.143$\pm$0.03&0.611$\pm$0.013&0.276$\pm$0.006&12.69$\pm$0.11 \\
	&	& &$a_i$&	0.085$\pm$0.001&0.291$\pm$0.004&0.254$\pm$0.004&0.285$\pm$0.003&0.051$\pm$0.001&0.034$\pm$0.001& \\ \hline	
	&(F)&~\cite{Piksaikin04} &$T_i$&54.55$\pm$0.39&20.92$\pm$0.23&5.663$\pm$0.104&2.308$\pm$0.058&0.539$\pm$0.019&0.221$\pm$0.008&12.50$\pm$0.20 \\
	& 0.59 & &$a_i$&0.086$\pm$0.002&0.275$\pm$0.006&0.233$\pm$0.006&0.310$\pm$0.007&0.074$\pm$0.003&0.023$\pm$0.001&	\\ \hline	
	&14.7 &East &$T_i$&55.6$\pm$0.400&19.3$\pm$0.200&5.040$\pm$0.200&2.18$\pm$0.080&0.570$\pm$0.030&0.221$\pm$0.000&11.29$\pm$0.17 \\
	&	&('70)	&$a_i$&	0.095$\pm$0.002&0.208$\pm$0.002&0.242$\pm$0.016&0.327$\pm$0.014&0.087$\pm$0.004&0.041$\pm$0.003& \\	\hline
\T $^{235}$U&(T)&~\cite{Piksaikin02a,Piksaikin02e} &$T_i$&53.95$\pm$0.28&22.34$\pm$0.13&6.400$\pm$0.083&2.258$\pm$0.033&0.494$\pm$0.017&0.179$\pm$0.006&8.98$\pm$0.11 \\ 
	&	& &$a_i$&0.038$\pm$0.001&0.211$\pm$0.004&0.197$\pm$0.004&0.396$\pm$0.005&0.132$\pm$0.004&0.026$\pm$0.001& \\ \hline	
	&(F) &~\cite{Piksaikin02a} &$T_i$&54.90$\pm$0.16&21.86$\pm$0.081&5.88$\pm$0.05&2.232$\pm$0.020&0.488$\pm$0.009&	0.180$\pm$0.003	&8.96$\pm$0.06 \\
	&0.81 	& &$a_i$&0.0363$\pm$0.0004&0.224$\pm$0.002&0.188$\pm$0.003&0.398$\pm$0.003&0.129$\pm$0.002&0.025$\pm$0.0004& \\ \hline	
	&	14.7 &East&$T_i$&54.6$\pm$0.500&20.2$\pm$0.200&5.36$\pm$0.300&2.38$\pm$0.100&0.770$\pm$0.070&	0.240$\pm$0.010&8.97$\pm$0.15 \\
    &	& ('70) &$a_i$&0.057$\pm$0.001&0.192$\pm$0.002&0.190$\pm$0.020&0.357$\pm$0.013&0.120$\pm$0.009&0.084$\pm$0.007&	\\ \hline 
\T $^{236}$U	&	(F) 	&~\cite{Piksaikin13,Isaev98} &	$T_i$	&	55.81	$\pm$	0.38	&	21.92	$\pm$	0.14	&	4.50	$\pm$	0.05	&	1.495	$\pm$	0.036	&	0.484	$\pm$	0.017	&	0.214	$\pm$	0.008	&	7.72	$\pm$	0.11 \\
	& 3.75 & 	&	$a_i$	&	0.026	$\pm$	0.001	&	0.195	$\pm$	0.004	&	0.322	$\pm$	0.006	&	0.347	$\pm$	0.008	&	0.099	$\pm$	0.003	&	0.013	$\pm$	0.0005	&	\\ \hline
\T $^{238}$U	&	(F) 	&~\cite{Piksaikin02a} &	$T_i$	&	52.91	$\pm$	0.17	&	21.83	$\pm$	0.06	&	5.12	$\pm$	0.03	&	1.973	$\pm$	0.013	&	0.496	$\pm$	0.007	&	0.173	$\pm$	0.003	&	5.31	$\pm$	0.03 \\
	&3.75 & 	&	$a_i$	&	0.0130	$\pm$	0.0001	&	0.133	$\pm$	0.001	&	0.164	$\pm$	0.002	&	0.389	$\pm$	0.003	&	0.227	$\pm$	0.003	&	0.074	$\pm$	0.001	&	\\ \hline	
	&	14.23 	&~\cite{Piksaikin07} &	$T_i$	&	53.4	$\pm$	0.400	&	20.9	$\pm$	0.200	&	4.64	$\pm$	0.080	&	1.92	$\pm$	0.040	&	0.500	$\pm$	0.010	&	0.176	$\pm$	0.005	&	5.90	$\pm$	0.20 \\
	&		& &	$a_i$	&	0.022	$\pm$	0.001	&	0.144	$\pm$	0.003	&	0.188	$\pm$	0.005	&	0.412	$\pm$	0.008	&	0.180	$\pm$	0.005	&	0.054	$\pm$	0.002	& \\ \hline	
\T $^{237}$Np	&	(F) 	&~\cite{Piksaikin98} &	$T_i$	&	54.7	$\pm$	0.550	&	21.8	$\pm$	0.260	&	5.74	$\pm$	0.190	&	2.26	$\pm$	0.056	&	0.719	$\pm$	0.034	&	0.199	$\pm$	0.010	&	9.24	$\pm$	0.17 \\
	&3.75 &		&	$a_i$	&	0.040	$\pm$	0.001	&	0.228	$\pm$	0.006	&	0.191	$\pm$	0.009	&	0.395	$\pm$	0.016	&	0.116	$\pm$	0.005	&	0.030	$\pm$	0.002	&	\\ \hline		
	&	14.23	&~\cite{Gremyachkin17}  &	$T_i$	&	58.20	$\pm$	0.50	&	20.20	$\pm$	0.20	&	4.58	$\pm$	0.11	&	2.12	$\pm$	0.04	&	0.460	$\pm$	0.500	&	0.196	$\pm$	0.500	&	8.24	$\pm$	0.23 \\
	&		&	&	$a_i$	&	0.045	$\pm$	0.001	&	0.181	$\pm$	0.003	&	0.211	$\pm$	0.006	&	0.451	$\pm$	0.010	&	0.100	$\pm$	0.003	&	0.013	$\pm$	0.004	&	\\ \hline	
\T $^{238}$Pu	&	(T)	&	Waldo	&	$T_i$	&	54.9	$\pm$	0.570	&	22.9	$\pm$	0.260	&	8.15	$\pm$	1.100	&	3.52	$\pm$	0.410	&	1.95	$\pm$	0.280	&	0.513	$\pm$	0.150	&	11.58	$\pm$	1.31 \\
	&		& ('81)	&	$a_i$	&	0.043	$\pm$	0.007	&	0.307	$\pm$	0.048	&	0.114	$\pm$	0.067	&	0.176	$\pm$	0.028	&	0.327	$\pm$	0.052	&	0.033	$\pm$	0.190	&	\\ \hline		
	&	(F)	&	Benedetti &	$T_i$	&	49.5	$\pm$	1.800	&	21.9	$\pm$	1.100	&	4.95	$\pm$	0.210	&	1.87	$\pm$	0.110	&	0.578	$\pm$	0.290	&	-  	&	11.52	$\pm$	0.64 \\
	&		&('82)	&	$a_i$	&	0.068	$\pm$	0.007	&	0.280	$\pm$	0.018	&	0.300	$\pm$	0.020	&	0.285	$\pm$	0.030	&	0.069	$\pm$	0.023	&		- 	&	\\ \hline
    \multicolumn{11}{r}{--Continued to next page} \\ 
\end{tabular}
\end{sidewaystable*}
\pagebreak

\begin{sidewaystable*}
\addtocounter{table}{-1}
\caption{ -- Continued from previous page} 
\begin{tabular}{l|c|c|c|c|c|c|c|c|c|c} \hline
 Target&	$E_n$ [MeV]	&Author&	 &	gr 1 &	gr 2	&	gr 3 &	gr 4 &	gr 5 &	gr 6 &	  $\langle   T\rangle$	\\ \hline	
\T $^{239}$Pu	&	(T)	&	~\cite{Piksaikin02a,Piksaikin02e} &	$T_i$	&	53.19	$\pm$	0.46	&	22.61	$\pm$	0.11	&	5.59	$\pm$	0.08	&	2.17	$\pm$	0.04	&	0.621	$\pm$	0.022	&	0.256	$\pm$	0.009	&	10.59	$\pm$	0.17 \\
	&		&	&	$a_i$	&	0.035	$\pm$	0.001	&	0.302	$\pm$	0.007	&	0.204	$\pm$	0.005	&	0.332	$\pm$	0.006	&	0.084	$\pm$	0.003	&	0.043	$\pm$	0.002	&	\\ \hline
	&	0.86	&~\cite{Piksaikin02a,Piksaikin02d} &	$T_i$	&	53.81	$\pm$	0.41	&	22.19	$\pm$	0.10	&	5.25	$\pm$	0.07	&	2.06	$\pm$	0.03	&	0.547	$\pm$	0.016	&	0.217	$\pm$	0.006	&	10.27	$\pm$	0.13 \\
	&  & &	$a_i$	&	0.037	$\pm$	0.001	&	0.288	$\pm$	0.005	&	0.216	$\pm$	0.005	&	0.321	$\pm$	0.005	&	0.102	$\pm$	0.003	&	0.034	$\pm$	0.001	&	 \\ \hline		
    &	15.8 	&~\cite{Roshchenko06a}&	$T_i$	&	55.76	$\pm$	0.69	&	20.97	$\pm$	0.27	&	4.75	$\pm$	0.09	&	2.13	$\pm$	0.05	&	0.560	$\pm$	0.020	&	0.215	$\pm$	0.006	&	8.92	$\pm$	0.34 \\
	&		& &	$a_i$	&	0.054	$\pm$	0.001	&	0.186	$\pm$	0.004	&	0.241	$\pm$	0.006	&	0.362	$\pm$	0.008	&	0.118	$\pm$	0.003	&	0.039	$\pm$	0.001	&	\\ \hline 
\T $^{240}$Pu	&	(F)	&	Keepin 	&	$T_i$	&	53.6	$\pm$	1.200	&	22.1	$\pm$	0.380	&	5.14	$\pm$	0.420	&	2.08	$\pm$	0.190	&	0.511	$\pm$	0.077	&	0.172	$\pm$	0.033	&	9.32	$\pm$	0.36 \\
	&		&	('57)	&	$a_i$	&	0.028	$\pm$	0.003	&	0.273	$\pm$	0.004	&	0.192	$\pm$	0.053	&	0.350	$\pm$	0.020	&	0.128	$\pm$	0.018	&	0.029	$\pm$	0.006	&		\\ \hline	
\T $^{241}$Pu	&	(T)	&	Cox 	&	$T_i$	&	54	$\pm$	1.000	&	23.2	$\pm$	0.500	&	5.6	$\pm$	0.600	&	1.97	$\pm$	0.100	&	0.43	$\pm$	0.040	&	- 		&	7.77	$\pm$	0.32 \\
	&		&	('61)	&	$a_i$	&	0.010	$\pm$	0.003	&	0.233	$\pm$	0.006	&	0.176	$\pm$	0.026	&	0.396	$\pm$	0.051	&	0.185	$\pm$	0.019	&	- 		&	\\ \hline		
	&	(F)	&	Gudkov 	&	$T_i$	&	55.9	$\pm$	0.000	&	23.0	$\pm$	0.000	&	5.64	$\pm$	0.000	&	2.10	$\pm$	0.000	&	0.578	$\pm$	0.000	&	0.201	$\pm$	0.000	&	7.85	$\pm$	0.54 \\ 
	&		&	('89)	&	$a_i$	&	0.017	$\pm$	0.003	&	0.218	$\pm$	0.022	&	0.179	$\pm$	0.019	&	0.362	$\pm$	0.040	&	0.189	$\pm$	0.042	&	0.035	$\pm$	0.014	&		\\ \hline	
\T $^{242}$Pu	&	(F)	&	Waldo 	&	$T_i$	&	51.7	$\pm$	1.000	&	23.5	$\pm$	1.200	&	16.9	$\pm$	5.800	&	5.46	$\pm$	0.240	&	1.75	$\pm$	0.150	&	0.312	$\pm$	0.120	&	6.51	$\pm$	1.28 \\ 
	&		&	('81)	&	$a_i$	&	0.011	$\pm$	0.001	&	0.161	$\pm$	0.053	&	0.031	$\pm$	0.005	&	0.164	$\pm$	0.015	&	0.367	$\pm$	0.036	&	0.266	$\pm$	0.086	&	\\	\hline
	&	14.7 	&	East 	&	$T_i$	&	55.4	$\pm$	4.500	&	21.8	$\pm$	0.500	&	5.50	$\pm$	0.300	&	2.20	$\pm$	0.100	&	0.760	$\pm$	0.060	&	0.250	$\pm$	0.010	&	6.67	$\pm$	0.33 \\
	&		&	('70)	&	$a_i$	&	0.022	$\pm$	0.004	&	0.168	$\pm$	0.002	&	0.150	$\pm$	0.015	&	0.366	$\pm$	0.009	&	0.159	$\pm$	0.010	&	0.135	$\pm$	0.009	&	\\ \hline		
\T $^{241}$Am	&	1.06 	&~\cite{Piksaikin11} 	&	$T_i$	&	54.0	$\pm$	1.1	&	23.92	$\pm$	0.45	&	6.13	$\pm$	0.12	&	2.31	$\pm$	0.05	&	0.49	$\pm$	0.01	&	0.179	$\pm$	0.004	&	10.9	$\pm$	0.2 \\
	&	&	&	$a_i$	&	0.044	$\pm$	0.001	&	0.276	$\pm$	0.006	&	0.202	$\pm$	0.004	&	0.308	$\pm$	0.006	&	0.148	$\pm$	0.003	&	0.023	$\pm$	0.001	&	\\ \hline	
	&	15.83 	&	~\cite{Gremyachkin17} &	$T_i$	&	55.10	$\pm$	0.50	&	24.40	$\pm$	0.20	&	5.26	$\pm$	0.04	&	3.23	$\pm$	0.03	&	0.490	$\pm$	0.004	&	0.175	$\pm$	0.002	&10.81	$\pm$	0.08 \\
	&		&		&	$a_i$	&	0.0420	$\pm$	0.0003	&	0.229	$\pm$	0.002	&	0.427	$\pm$	0.003	&	0.215	$\pm$	0.002	&	0.067	$\pm$	0.001	&	0.0191	$\pm$	0.0002	&		\\ \hline	
\T $^{242m}$Am	&	(T)	&	Waldo 	&	$T_i$	&	54.4	$\pm$	0.210	&	23.1	$\pm$	0.085	&	7.45	$\pm$	0.430	&	2.82	$\pm$	0.077	&	1.06	$\pm$	0.130	&	0.513	$\pm$	0.150	&	10.04	$\pm$	0.48 \\
	&		&	('81)	&	$a_i$	&	0.026	$\pm$	0.002	&	0.284	$\pm$	0.019	&	0.120	$\pm$	0.013	&	0.355	$\pm$	0.038	&	0.173	$\pm$	0.019	&	0.044	$\pm$	0.065	&		\\ \hline	
\T $^{245}$Cm	&	(T)	&	Waldo 	&	$T_i$	&	51.9	$\pm$	0.350	&	22.9	$\pm$	0.860	&	6.66	$\pm$	0.090	&	3.29	$\pm$	0.170	&	1.29	$\pm$	0.180	&	0.513	$\pm$	0.150	&	10.06	$\pm$	0.60 \\
	&		&	('81)	&	$a_i$	&	0.024	$\pm$	0.002	&	0.303	$\pm$	0.020	&	0.091	$\pm$	0.029	&	0.294	$\pm$	0.052	&	0.230	$\pm$	0.027	&	0.059	$\pm$	0.100	&		\\ \hline	
\T $^{249}$Cf	&	(T)	&	Waldo 	&	$T_i$	&	53.9	$\pm$	0.084	&	22.8	$\pm$	0.008	&	4.13	$\pm$	0.091	&	1.28	$\pm$	0.150	&	- 			&	- 			&	11.47	$\pm$	0.62  \\
	&		&	('81)	&	$a_i$	&	0.029	$\pm$	0.002	&	0.353	$\pm$	0.026	&	0.383	$\pm$	0.032	&	0.236	$\pm$	0.026	&	- 			&	- 			&			\\ \hline
\T $^{252}$Cf	&	(SF)	&	Chulick 	&	$T_i$	&	26.8	$\pm$	1.100	&	6.10	$\pm$	1.400	&	2.00	$\pm$	0.300	&	0.500	$\pm$	0.100	&	- 			&	- 			&	10.34	$\pm$	0.55 \\
	&		&	('69)	&	$a_i$	&	0.310	$\pm$	0.010	&	0.220	$\pm$	0.020	&	0.300	$\pm$	0.030	&	0.170	$\pm$	0.050	&	 -			&	 -			&		\\ \hline
     \multicolumn{11}{c}{Recommended 7-Group Delayed Neutron Parameters for $^{243}$Am. Parameters and symbols as above} \\ \hline
    $E_n$	&Author	& &	gr 1 &	gr 2	&	gr 3 &	gr 4 &	gr 5 &	gr 6 &	gr 7 &   $\langle   T\rangle$ \\ \hline  

\T (F)	&	Charlton 	&	$T_i$	&	55.9	$\pm$	1.800	&	24.5	$\pm$	0.780	&	16.7	$\pm$	0.360	&	4.59	$\pm$	0.061	&	1.77	$\pm$	0.027	&	0.774	$\pm$	0.035	&	0.283	$\pm$	0.005	&	9.54	$\pm$	0.43 \\
 		&	('98)	&	$a_i$	&	0.017	$\pm$	0.005	&	0.226	$\pm$	0.011	&	0.088	$\pm$	0.006	&	0.206	$\pm$	0.011	&	0.335	$\pm$	0.011	&	0.088	$\pm$	0.009	&	0.041	$\pm$	0.008	&	\\ \hline \hline
\end{tabular}
\end{sidewaystable*}
}

\afterpage{
\begin{sidewaystable*}
\caption{Recommended 8-Group Delayed Neutron Parameters. Parameters and symbols as in Table~\ref{tab:9.5}.}\label{tab:9.6}
\centering
\begin{tabular}{l|c|c|c|c|c|c|c|c|c|c|c|c} 
\hline \hline 
Target& $E_n$	&Author	& &	gr 1 &	gr 2	&	gr 3 &	gr 4 &	gr 5 &	gr 6 &	gr 7 & gr 8&  $\langle   T\rangle$ \\ \hline 
\T	&	[MeV]	&		&	$T_i$	&	55.6		 	&	24.5		 	&	16.3		 	&	5.21		 	&	2.37		 	&	1.04		 	&	0.424		 	&	0.195		 	&	 \\ \hline	\hline  
\T    $^{232}$Th	&	(F)	&	Keepin ('57)	&	$a_i$	&	0.033	$\pm$	0.002	&	0.073	$\pm$	0.005	&	0.093	$\pm$	0.002	&	0.136	$\pm$	0.024	&	0.381	$\pm$	0.008	&	0.14	$\pm$	0.008	&	0.114	$\pm$	0.013	&	0.028	$\pm$	0.001	&	6.98	$\pm$	0.23 \\
\T $^{231}$Pa	&	(F)	&	Anoussis ('73)	&	$a_i$	&	0.115	$\pm$	0.001	&	0.099	$\pm$	0.002	&	0.228	$\pm$	0.006	&	0.181	$\pm$	0.026	&	0.353	$\pm$	0.030	&	0.024	$\pm$	0.010	&	 -	&	- 	&	14.33	$\pm$	0.19 \\
	&	14.8 	&	Brown ('71)	&	$a_i$	&	0.126	$\pm$	0.013	&	0.068	$\pm$	0.016	&	0.232	$\pm$	0.021	&	0.205	$\pm$	0.028	&	0.341	$\pm$	0.031	&	0.028	$\pm$	0.012	&	- 	&- 	&	14.37	$\pm$	0.91 \\
\T $^{232}$U	&	(T)	&	Waldo ('81)	&	$a_i$	&	0.110	$\pm$	0.009	&	0.144	$\pm$	0.015	&	0.178	$\pm$	0.019	&	0.218	$\pm$	0.033	&	0.270	$\pm$	0.005	&	0.076	$\pm$	0.048	&	0.005	$\pm$	0.080	&	- 	&	14.38	$\pm$	0.71 \\
\T $^{233}$U	&	(T)	&~\cite{Piksaikin13,Piksaikin02e} &	$a_i$	&	0.078	$\pm$	0.003	&	0.164	$\pm$	0.006	&	0.150	$\pm$	0.005	&	0.209	$\pm$	0.008	&	0.287	$\pm$	0.009	&	0.046	$\pm$	0.002	&	0.048	$\pm$	0.002	&	0.017	$\pm$	0.001	&	12.65	$\pm$	0.23 \\
	&	0.59	&~\cite{Piksaikin13}&	$a_i$	&	0.079	$\pm$	0.003	&	0.159	$\pm$	0.006	&	0.143	$\pm$	0.005	&	0.219	$\pm$	0.008	&	0.291	$\pm$	0.009	&	0.026	$\pm$	0.001	&	0.073	$\pm$	0.004	&	0.009	$\pm$	0.001	&	12.52	$\pm$	0.23 \\
    & 14.7 & East ('70) & $a_i$ & 0.093 $\pm$ 0.002 & 0.078 $\pm$ 0.002 & 0.140 $\pm$ 0.003 & 0.204 $\pm$ 0.018 & 0.330 $\pm$ 0.008 & 0.058 $\pm$ 0.009 & 0.072 $\pm$ 0.001 & 0.026 $\pm$ 0.002 & 11.29 $\pm $0.17   \\
\T $^{235}$U	&	(T)	&~\cite{Piksaikin02a} &	$a_i$	&	0.034	$\pm$	0.001	&	0.153	$\pm$	0.006	&	0.086	$\pm$	0.004	&	0.212	$\pm$	0.007	&	0.298	$\pm$	0.009	&	0.105	$\pm$	0.005	&	0.073	$\pm$	0.004	&	0.039	$\pm$	0.002	&	9.00	$\pm$	0.18 \\
	&	0.81 	&~\cite{Piksaikin02a} &	$a_i$	&	0.033	$\pm$	0.001	&	0.144	$\pm$	0.006	&	0.104	$\pm$	0.004	&	0.173	$\pm$	0.007	&	0.351	$\pm$	0.011	&	0.073	$\pm$	0.004	&	0.095	$\pm$	0.005	&	0.027	$\pm$	0.001	&	8.91	$\pm$	0.18 \\
	&	14.7 	&	East ('70)	&	$a_i$	&	0.052	$\pm$	0.001	&	0.099	$\pm$	0.002	&	0.107	$\pm$	0.004	&	0.185	$\pm$	0.022	&	0.346	$\pm$	0.007	&	0.079	$\pm$	0.009	&	0.087	$\pm$	0.002	&	0.045	$\pm$	0.008	&	8.97	$\pm$	0.15 \\

\T $^{236}$U	&	(F)	&~\cite{Piksaikin13} &	$a_i$	&	0.024	$\pm$	0.001	&	0.129	$\pm$	0.005	&	0.075	$\pm$	0.003	&	0.201	$\pm$	0.007	&	0.291	$\pm$	0.009	&	0.215	$\pm$	0.009	&	0.042	$\pm$	0.003	&	0.023	$\pm$	0.001	&	7.72	$\pm$	0.15 \\
\T $^{238}$U	&	3.75 	&	~\cite{Piksaikin02a}&	$a_i$	&	0.0100	$\pm$	0.0004	&	0.091	$\pm$	0.004	&	0.053	$\pm$	0.002	&	0.122	$\pm$	0.005	&	0.331	$\pm$	0.009	&	0.158	$\pm$	0.007	&	0.154	$\pm$	0.007	&	0.081	$\pm$	0.004	&	5.31	$\pm$	0.11 \\
\T $^{237}$Np	&	3.75 &~\cite{Piksaikin98}	&	$a_i$	&	0.036	$\pm$	0.001	&	0.155	$\pm$	0.005	&	0.092	$\pm$	0.004	&	0.189	$\pm$	0.011	&	0.337	$\pm$	0.013	&	0.129	$\pm$	0.008	&	0.038	$\pm$	0.001	&	0.024	$\pm$	0.002	&	9.23	$\pm$	0.17 \\
\T $^{238}$Pu	&	(T)	&	Waldo ('81)	&	$a_i$	&	0.042	$\pm$	0.009	&	0.219	$\pm$	0.027	&	0.137	$\pm$	0.058	&	0.134	$\pm$	0.065	&	0.386	$\pm$	0.008	&	0.066	$\pm$	0.100	&	0.017	$\pm$	0.170	&	 - 	&	11.59	$\pm$	1.31 \\
	&	(F)	&	Benedetti ('82)	&	$a_i$	&	0.045	$\pm$	0.008	&	0.250	$\pm$	0.018	&	0.052	$\pm$	0.001	&	0.256	$\pm$	0.014	&	0.251	$\pm$	0.035	&	0.120	$\pm$	0.012	&	0.027	$\pm$	0.016	&	-  	&	11.52	$\pm$	0.64 \\
\T $^{239}$Pu	&	(T)	&~\cite{Piksaikin02a,Piksaikin13}&	$a_i$	&	0.029	$\pm$	0.001	&	0.240	$\pm$	0.008	&	0.083	$\pm$	0.004	&	0.188	$\pm$	0.007	&	0.289	$\pm$	0.010	&	0.089	$\pm$	0.004	&	0.059	$\pm$	0.003	&	0.023	$\pm$	0.001	&	10.63	$\pm$	0.22 \\
	&	0.86 	&~\cite{Piksaikin02a,Piksaikin13}&	$a_i$	&	0.032	$\pm$	0.001	&	0.212	$\pm$	0.007	&	0.097	$\pm$	0.004	&	0.173	$\pm$	0.007	&	0.300	$\pm$	0.010	&	0.088	$\pm$	0.004	&	0.076	$\pm$	0.004	&	0.022	$\pm$	0.001	&	10.29	$\pm$	0.20 \\
\T $^{240}$Pu	&	(F)	&	Keepin ('57)	&	$a_i$	&	0.022	$\pm$	0.003	&	0.207	$\pm$	0.005	&	0.080	$\pm$	0.002	&	0.161	$\pm$	0.055	&	0.314	$\pm$	0.009	&	0.105	$\pm$	0.010	&	0.079	$\pm$	0.017	&	0.032	$\pm$	0.003	&	9.32	$\pm$	0.36 \\
\T $^{241}$Pu	&	(T)	&	Cox ('61)	&	$a_i$	&	0.011	$\pm$	0.003	&	0.166	$\pm$	0.003	&	0.095	$\pm$	0.011	&	0.100	$\pm$	0.025	&	0.382	$\pm$	0.043	&	0.073	$\pm$	0.030	&	0.174	$\pm$	0.012	&	  -	&	7.78	$\pm$	0.32 \\
	&	(F)	&	Gudkov ('89)	&	$a_i$	&	0.016	$\pm$	0.002	&	0.175	$\pm$	0.019	&	0.055	$\pm$	0.012	&	0.170	$\pm$	0.018	&	0.280	$\pm$	0.035	&	0.166	$\pm$	0.033	&	0.113	$\pm$	0.035	&	0.024	$\pm$	0.006	&	7.85	$\pm$	0.54 \\
\T $^{242}$Pu	&	(F)	&	Waldo ('81)	&	$a_i$	&	0.014	$\pm$	0.0003	&	0.095	$\pm$	0.051	&	0.134	$\pm$	0.015	&	0.033	$\pm$	0.020	&	0.404	$\pm$	0.008	&	0.001	$\pm$	0.060	&	0.258	$\pm$	0.046	&	0.062	$\pm$	0.052	&	6.52	$\pm$	1.28 \\
	&	14.7	&	East ('70)	&	$a_i$	&	0.022	$\pm$	0.005	&	0.097	$\pm$	0.002	&	0.090	$\pm$	0.002	&	0.108	$\pm$	0.018	&	0.366	$\pm$	0.004	&	0.111	$\pm$	0.002	&	0.143	$\pm$	0.010	&	0.064	$\pm$	0.006	&	6.67	$\pm$	0.28 \\
\T $^{241}$Am  &  1.06 & \citep{Piksaikin13}   & $a_i$ & 0.040$\pm$0.001 &	0.261$\pm$0.008	&0.031$\pm$0.002&	0.231$\pm$0.009&	0.220$\pm$0.010&	0.091$\pm$0.005&	0.086$\pm$0.004&	0.042$\pm$0.002&	10.96$\pm$0.23 \\
\T $^{242m}$Am	&	(T)	&	Waldo ('81)	&	$a_i$	&	0.021	$\pm$	0.0004	&	0.245	$\pm$	0.018	&	0.060	$\pm$	0.008	&	0.205	$\pm$	0.025	&	0.261	$\pm$	0.029	&	0.179	$\pm$	0.040	&	0.030	$\pm$	0.056	&	- 	&	10.03	$\pm$	0.48 \\
\T $^{243}$Am	&	(F)	&	Charlton ('98)	&	$a_i$	&	0.018	$\pm$	0.006	&	0.220	$\pm$	0.012	&	0.098	$\pm$	0.002	&	0.121	$\pm$	0.009	&	0.316	$\pm$	0.013	&	0.170	$\pm$	0.003	&	0.043	$\pm$	0.011	&	0.015	$\pm$	0.002	&	9.54	$\pm$	0.44 \\
\T $^{245}$Cm	&	(T)	&	Waldo ('81)	&	$a_i$	&	0.016	$\pm$	0.004	&	0.269	$\pm$	0.020	&	0.045	$\pm$	0.001	&	0.204	$\pm$	0.046	&	0.255	$\pm$	0.040	&	0.179	$\pm$	0.050	&	0.033	$\pm$	0.084	&	- 	&	10.06	$\pm$	0.60 \\
\T $^{249}$Cf	&	(T)	&	Waldo ('81)	&	$a_i$	&	0.024	$\pm$	0.0005	&	0.293	$\pm$	0.023	&	0.064	$\pm$	0.011	&	0.228	$\pm$	0.028	&	0.265	$\pm$	0.026	&	0.128	$\pm$	0.017	&	0.033	$\pm$	0.084	&	- 	&	11.48	$\pm$	0.61 \\
$^{252}$Cf	&	(SF)	&	Chulick ('69)	&	$a_i$	&	0.014	$\pm$	0.006	&	0.318	$\pm$	0.006	&	0.001	$\pm$	0.024	&	0.209	$\pm$	0.018	&	0.200	$\pm$	0.004	&	0.144	$\pm$	0.031	&	0.112	$\pm$	0.044	&	-  	&	10.38	$\pm$	0.55 \\ \hline \hline 
\end{tabular}
\end{sidewaystable*}
}


\section{ REFERENCE DATABASE }\label{Sec:Database}

All the compilations, evaluations, systematics as well as theoretical results and recommendations produced by the CRP are available on the online Reference Database for $\beta$-delayed neutron emission~\cite{database}. 

The database is split into two parts, the microscopic data and macroscopic data section. Each section contains a search engine that allows the user to search per nuclide, energy, and in the case of the microscopic database, also per ranges of half-lives $T_{1/2}$ and $P_n$ values.

All the microscopic data, including compilations, comments, and evaluations can be downloaded as Excel spreadsheets while the theoretical results can be downloaded as '.csv' files. The final numerical file of microscopic ($T_{1/2}, P_{n}$) values which is a computer readable file that has been especially prepared for use in summation calculations (Sect.~\ref{Sec:Macro-Summation}) and other applications can also be downloaded as a '.csv' file.

In the following, some special features of the two distinct microscopic and macroscopic sections are described.
\subsection{Microscopic data}
This database contains all the compiled and evaluated values of $T_{1/2}$ and $P_n$ produced by the CRP evaluators and published in ~\cite{Birch2015} ($Z < 29$) and ~\cite{Liang2020} ($Z > 28$). The data are displayed in tabular form following the same nomenclature adopted in the respective published tables. In addition, the user can retrieve all the comments on the data and evaluation process which are not included in the published articles. The references of the compiled articles are given in terms of their NSR key-numbers which are also hyperlinked to the NSR database. 

Beta-delayed neutron spectra are also included in the database. In cases where these data were published in graphical form only, the graphs were digitized and the numerical values were included in the database in simple ASCII files. This is a major development as prior to this CRP there was no database dedicated to a compilation of experimental delayed neutron spectra for individual precursors. The database includes all the graphical spectra published in the thesis of M. Brady in 1989~\cite{Brady1989a}. 

The complete microscopic data tables (compilation and evaluation separately) can be downloaded in the format they were published (\cite{Birch2015,Liang2020}), but also as pure numerical files which can be readily used in calculations. The spectra files can also be downloaded in a zipped file.

The systematics that were developed on the basis of the evaluated data of ~\cite{Birch2015} and ~\cite{Liang2020} (Sect.~\ref{Sec:Eval-Syst}) are also available from the database: these use the empirical formula or approaches described in McCutchan \textit{et al.} \cite{McCutchan2012} and \cite{Miernik2013} but were adapted to the new evaluated data.

In addition to the compiled experimental and evaluated data, the online database contains theoretical $T_{1/2}$ and $P_n$ values, for comparison. At present, the global calculations of Ref.~\cite{Moller2003} and ~\cite{Marketin2016} (see Sect.~\ref{Sec:Micro-theory}) have been included, while more models are planned to be made available in the near future.

The online interface provides a search tool which allows the user to search in terms of $Z,\, N,\, A$ and ranges of half-lives and $P_n$ values. The result of the search can also be plotted. Different options for plotting are available (in terms of $T_{1/2}$, $Q_{\beta}$, \Qbn, $P_n$). 



\subsection{Macroscopic data}

The macroscopic database includes three types of aggregate data:

\paragraph{Total delayed neutron yields.}

This part contains previous compilations of measured total delayed neutron yields published in ~\cite{Tuttle79, Cox68, Cox70, Cox74, Zemyatnin76}, as well as new measurements performed by the IPPE group~\cite{Piksaikin13,Piksaikin97,Piksaikin12} (see Sect.~\ref{Sec:Macro-methods}). The data are accompanied by recommendations and comments on necessary adjustments to the data. The whole table including comments can be downloaded as a .csv file.

In addition to the experimental total delayed neutron yields, the user can also retrieve the respective recommended yields included in the various evaluated libraries such as ENDF/B-VII.1~\cite{Chadwick2011}, JEFF-3.1.1~\cite{jeff3.1.1} and JENDL-4.0~\cite{Shibata2011}. 


\paragraph{Time-dependent group parameters ($a_i,T_i$).}

It includes a compilation of all the time-dependent group parameters measured over the past decades using 4-, 5-, 6-, 7- and 8-group models including the new measurements performed at IPPE~\cite{Piksaikin02a,Piksaikin07,Piksaikin17,Piksaikin17} (see Sect.~\ref{Sec:Macro-meas} and \ref{Sec:Macro-Recommended}). In the case of the latter data, experimental covariance matrices revealing the correlations between the members $i$ of the groups are also provided. The recommended group parameters of ~\cite{Spriggs02} are included as well as the recommended group-parameters adopted  by the evaluated libraries (ENDF/B-VII.1~\cite{Chadwick2011}, JEFF-3.1.1~\cite{jeff3.1.1} and JENDL-4.0~\cite{Shibata2011}).

\paragraph{Composite delayed neutron spectra.}

A complete set of measured delayed neutron spectra obtained from  Ref.~\cite{Piksaikin17} (see Sect.~\ref{Sec:Macro-meas}) are contained in this part. The spectra can be downloaded in numerical tables. Plots comparing the measured spectra with results of the summation method using the newly evaluated $P_n$ values are also available.

\section{  SUMMARY AND CONCLUSIONS }
\label{Sec:Conclusions}

In 2019 we celebrated the 80th anniversary of the discovery of \bdns by Roberts \etal \cite{Roberts1939a,Roberts1939b}. The next generation of radioactive beam facilities that will become operational in the next years will allow the access of even more neutron-rich nuclei in the experimental "Terra Incognita". This will be a busy time for \bdn measurements since the vast majority of these ~3000--4000 still-to-be-discovered nuclides will decay by \bdn emission. However, the main driver for these new measurements are no longer fission reactors but the question how about half of the elements heavier than iron are created in explosive stellar events like the "rapid neutron capture ($r$) process" in core collapse supernova explosions and binary neutron star mergers \cite{Horowitz2018}.




The last decade has seen an increased interest in the measurement of individual nuclear decay properties of the most neutron-rich nuclei that can be produced. The need for more reliable data for $r$-process calculations is strongly intertwined with the better understanding of the evolution of single-particle and collective levels in neutron-rich matter.

Detection techniques have been developed and improved for operation at the next generation of radioactive beam facilities, and thoroughly used to map the existing \bne landscape. Worldwide collaborative efforts have lead to the discovery of many new neutron-rich isotopes, and the measurement of their half-lives and neutron-branching ratios. The majority of these recent results has yet to be published, and then evaluated and included in the database. 


The collaborative work performed under the auspices of the International Atomic Energy Agency (IAEA) was timely as it will not only guide future experiments on \bn~emitters but will also lead to more reliable calculations for $r$-process nucleosynthesis and the new generation of fission reactors.

\subsection{Microscopic experiments}

The members of the CRP assessed all methods for the measurement of half-lives, neutron-branching ratios, and neutron energy spectra, taking into account recent technological developments and the requirements for future measurements. The "$\beta$-n" method using the coincident detection of $\beta$'s and neutrons (and optional with subsequent detection of $\gamma$'s) was identified as the most reliable method for the measurement of neutron-branching ratios (see Sect.~\ref{Sec:micro-bncoin}). If the decay scheme of a nucleus is well-known (including absolute decay intensities and isomeric states), also detection methods without neutrons can provide very reliable results, for example the coincident "Double $\gamma$ counting method" ("$\gamma,\gamma$", Sect.~\ref{Sec:Micro:gammagamma}). 

In the past years also new, pure ion-counting methods have been developed that circumvent the necessity to detect neutrons, $\beta$'s or $\gamma$'s. These experiments have been grouped in the "Double ion counting" method ("ion,ion", Sect.~\ref{Sec:methods-ionion}) and comprise techniques using various types of traps, time-projection chambers, or even storage rings (see Sect.~\ref{Sec:Micro-newmethods})

Some of these new ion-counting techniques, namely the ion-recoil detection methods, can also be used to extract the respective neutron energy spectra. This has been shown recently for the Beta-decay Paul trap (BPT) where a good agreement with the measured spectrum for $^{137}$I was found, although with limited resolution due to the trap geometry. 

Besides traditional techniques the neutron time-of-flight (TOF) method with fast plastic or liquid scintillators has shown to be the best method to measure the neutron energy spectra. Thanks to their relatively high intrinsic efficiency they can be also employed for the measurement of rare isotopes but their limitation is the achievable resolution at higher neutron energies ($>$3~ MeV).

Total Absorption $\gamma$-ray Spectrometers (TAGS) have been used since many years to measure $\beta$- and $\gamma$-strength functions, as well as level densities. More recently it has been shown that they can also be used to determine $P_{1n}$ values and delayed neutron spectra (see Sect.~\ref{Sec:TAGS}). This offers new exciting opportunities for the investigation of important \bne at RIB facilities worldwide with complementary detection techniques.

Recent experimental studies included a revisiting of the properties of \bne abundantly produced during a nuclear fuel cycle in power and research reactors. Re-measurements of half-lives and neutron-branching ratios for nuclei on priority lists for the most important contributors to the delayed neutron fraction $ \nu_d$ (see e.g. p. 15 in Ref.~\cite{IAEA0643}) have been carried out, as well as complete studies of the $\beta$-strength distribution below and above the neutron emission threshold and resulting \bn-energy spectra. These measurements helped to refine the analysis of the processes occurring in nuclear fuel during the operation of nuclear reactors and provided data for a better theoretical analysis of the $\beta$-strength distribution.

One important focus for \bne studies in the following years will be the role of competitions between different decay channels. This includes the competition between one- and multi-neutron emission with the de-excitation via $\gamma$ transitions from highly excited states above the neutron separation energy. In earlier theoretical models these competitions have not been included (in the case of $\gamma$ competition) or only with a simplified cut-off picture (as in the case of competition between the 1n and 2n channel). However, as has been shown in the past years, the correct treatment of these competing decay channels is crucial since it can lead to systematically wrong predictions. This can have an influence, for example, on the amount of neutrons available in late phases of the astrophysical $r$ process and thus lead to different calculated abundance distribution patterns.

\subsection{Compilation and evaluation of microscopic data}
The compilation and evaluation of the half-lives and neutron-emission probabilities of all 653 presently identified \bn~emitters between $^{8}$He ($Z$=2) and $^{233}$Fr ($Z$=87) has been carried out \cite{Birch2015, Liang2020}. 

In total, 507 of these \bne have measured $\beta$-decay half-lives and 306 have a measured $P_{1n}$ value. However, only for 32 nuclides the $\beta$-delayed two-neutron emission ratio ($P_{2n}$) has been measured. The $P_{3n}$ value is only known for four nuclei so far ($^{11}$Li, $^{17,19}$B, and $^{23}$N), and the result of the $P_{4n}$ of $^{17}$B is only tentative.


Nine standards for $P_{1n}$ measurements with a precision of better than 5\% and four with a precision between 5-10\% have been recommended by the CRP: $^{9}$Li, $^{16}$C, $^{17}$N, $^{49}$K, $^{82}$Ga, $^{87,88}$Br, $^{94,95}$Rb, $^{137}$I, and $^{145,146,147}$Cs (see Table~\ref{tab:standards}). 

\begin{figure*}[!htb]
	\centering
\includegraphics[width=0.9\textwidth]{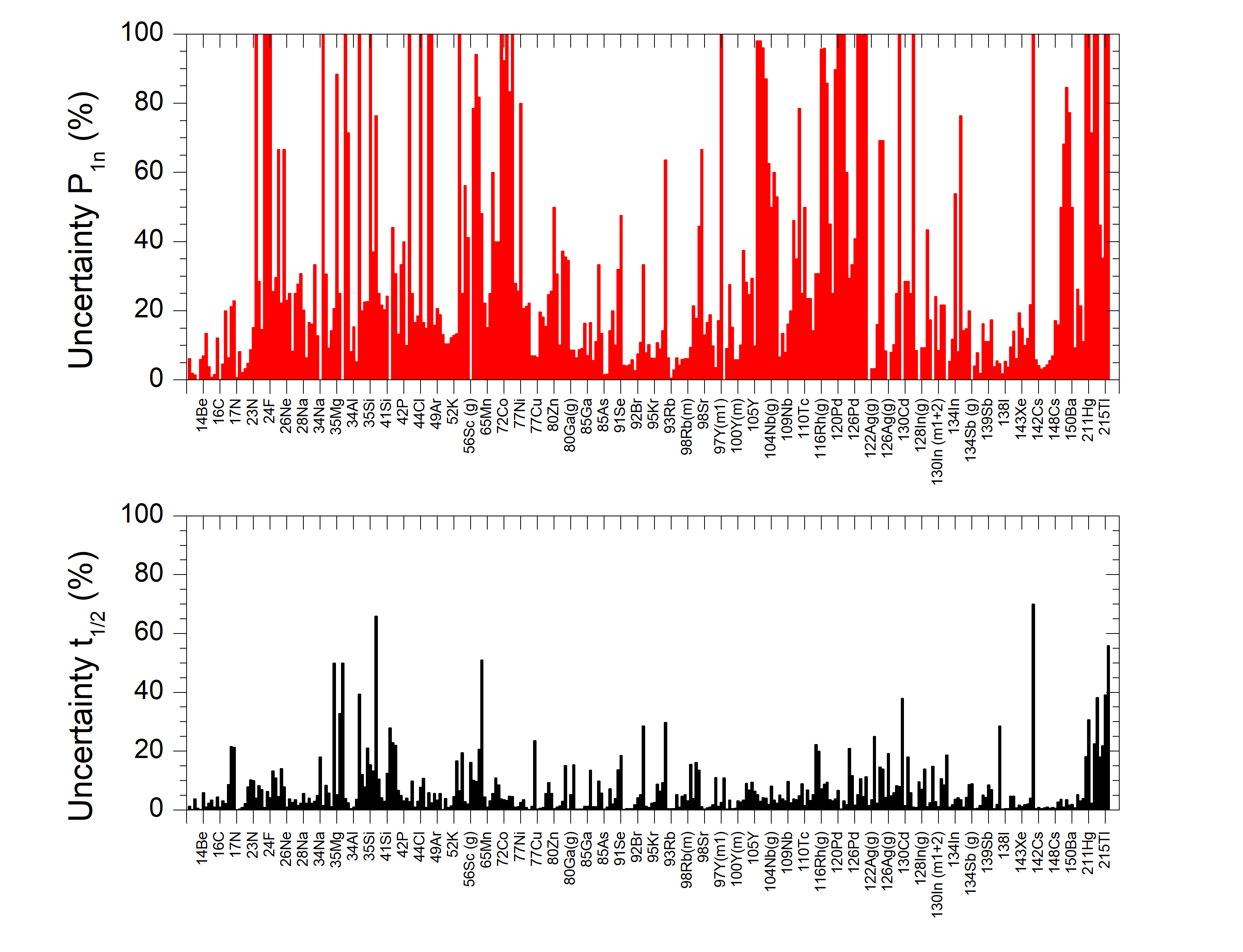}
\caption{Schematic comparison of uncertainties for $P_{1n}$ values (top) and half-lives (bottom) for evaluated \bn~emitters \cite{Birch2015, Liang2020}. Upper limits are shown with 100\% uncertainty.}
\label{fig:eval-comp}
\end{figure*}

In general, the precision of the measured neutron-branching ratios is much lower than for $\beta$-decay half-lives (see Fig.~\ref{fig:eval-comp}), and the neutron branching ratios of many nuclei are only available as upper (shown with 100\% uncertainty) or lower limits. 


With the expected vast amount of new data in the next years, an annual re-evaluation of \bn~emitters is planned so that this new data can be included in the IAEA Reference Database~\cite{database} in a timely manner. The evaluation of the light mass region ($Z$=2--28) has been published more than 5 years ago \cite{Birch2015}, and work on the update has started recently by the Canadian nuclear data community.


\subsection{Microscopic theory}

Two global self-consistent microscopic models have been used in comparisons with the CRP evaluated microscopic data ($T_{1/2}, P_n$): the spherical continuum pnQRPA~\cite{Borzov2003} (DF+CQRPA) based on the Fayans functional ~\cite{Fayans2000} in its new version DF3a that is suitable for (quasi-) spherical nuclei with low or zero deformation, and the spherical relativistic pnQRPA model ~\cite{Marketin2016} based on the D3C* functional ~\cite{Marketin2007} (RHB+RQRPA) that has been applied to all the nuclei across the nuclear chart including the deformed ones. 

The comparisons have focused on spherical and deformed nuclei both in the light and heavy fission fragment regions, and on isotopes around the doubly closed shell nuclei $^{78}$Ni and  $^{132}$Sn which are important for the astrophysical $r$-process nucleosynthesis. 

The models perform equally well or even better than the widely used restricted (Q)RPA-like scheme which is based on the microscopic-macroscopic Finite Range Liquid Drop Model (FRDM12+(Q)RPA+HF ~\cite{Moller2019}). The global self-consistent models can thus be recommended for planning RIB experiments and for predictions in mass regions where measurements are not possible and which are important for astrophysical modeling. 

Both models, DF+CQRPA and RHB+QRPA, treat the contributions of the first-forbidden (FF) and the allowed Gamow-Teller (GT) transitions fully microscopically and on equal footing. However, the comparison with the evaluated ($T_{1/2}, P_n$) data shows that the $\beta$-decay observables are extremely sensitive to the competition between the GT and FF strengths, the isoscalar pn-pairing-like NN-interaction, and the ground-state spin inversion caused by the neutron-proton tensor interaction which may cause a \emph{stabilization effect}. This leads to a slowing down of the rate at which the half-life decreases with increasing mass number, as well as other irregularities in the $P_{n}$ values. 

These findings underscore the need to further improve the nuclear energy-density functional, in particular the pairing and tensor parts of the functional. The recently proposed hybrid density functional~\cite{Rei17} that uses the Skyrme functional with a pairing part depending on the gradient of density, as in the Fayans functional~\cite{Fayans2000}, has the potential to describe both the odd-even staggering effects in the $Q_{\beta}$ values and charge radii~\cite{Saperstein2016}, and the two-neutron emission probabilities~\cite{Borzov2017,Borzov2018}. Also promising are the newly developed ``beyond the QRPA" self-consistent models that take into account the impact of the phonon-phonon ~\cite{Severyukhin2017,Sushenok2018} or quasiparticle-phonon~\cite{Niu2015,Robin2016,Niu2018, Litvinova2014} couplings on the $\beta$-decay properties.

\subsection{Summation and time-dependent calculations}
The evaluated ($T_{1/2},P_n$) values were used in summation and time-dependent calculations of total delayed neutron yields $\nu_d$, DN activity, and DN spectra in combination with the cumulative fission yields from the JEFF-3.1.1 library. Overall, the CRP microscopic data lead to enhanced $\nu_d$ values, partly due to the increased $P_n$ value of $^{137}$I which is the main contributor to nearly all the fission systems studied. The results of the present summation calculations, nevertheless, are in better agreement with the $\nu_d$~values recommended by WPEC-SG6~\cite{NEA-WPEC6} compared to those of Wilson~\cite{Wilson02} which were based on previous compilations of $P_n$ data. The best agreement with recommended $\nu_d$ values is obtained for thermal fission of $^{235}$U and $^{239}$Pu, for which we have the most reliable fission yield data, while for $^{238}$U fast fission there is significant improvement. In the case of fast fission of $^{235}$U and $^{239}$Pu, as well as fission of $^{241}$Pu, however, the recommended values are overestimated.

Time-dependent kinetic parameters were calculated starting from the microscopic recommended CRP data ($T_{1/2},P_n$) and the JEFF-3.1.1 independent fission yields, taking into account correlations in the decay path. The estimated group abundances are affected by larger uncertainties than the measured ones indicating that a more comprehensive approach to treating correlations is needed that would include contributions from fission yield data.
Both summation and time-dependent calculations provide clear evidence of the energy dependence of the total DN yields for the fission of $^{235}$U and $^{239}$Pu. These results confirm the need for a more detailed investigation of the energy dependence of fission yields and a subsequent updating of the fission yield libraries.

Though the new microscopic CRP ($T_{1/2}, P_n$) database is significantly improved with respect to previous dated compilations, the impact on integral DN properties also depends strongly on the fission yield library that was used. In addition, the uncertainties in the integral quantities do not take into account correlations affecting the fission yield data, therefore they could be underestimated. We strongly recommend, as a next step, that the evaluated fission yield libraries are updated with due consideration of correlation effects and the dependence on incident energy.

\subsection{Integral experiments and reactor calculations}
The impact of the new evaluated microscopic data on integral experiments and specific reactor designs has been assessed. Integral calculations were performed using different evaluated libraries and the CRP ($T_{1/2}, P_n$) data. The results show that $\beta_{eff}$ values obtained from the combination of JEFF-3.1.1 fission yields with the CRP data and supplementary ENDF/B-VIII.0 decay data give slightly increased $\beta_{eff}$ values with respect to the other libraries for the benchmark experiments POPSY, TOPSY, JEZEBEL, SKI-DOO, FLATTOP-23, BIG TEN, SNEAK-7A,SNEAK-7B and IPEN. 

Similar results were obtained for three different reactor designs, a pressurized water reactor (PWR), the sodium cooled ASTRID reactor and the lead-bismuth cooled MYRRHA reactor. The combination of the CRP data with JEFF-3.1.1 fission yields leads to $\beta_{eff}$ values that deviate from the reference values by 1-2\% for PWR and MYRRHA and 11\% for ASTRID.

These results depend strongly on the fission yield library that was used, therefore, as has been stressed above, further studies using updated fission yield libraries would be required before a definitive conclusion is drawn.

\subsection{Macroscopic measurements and systematics}
The total delayed neutron yields have been measured in the energy range from 0.3 to 5 MeV for nuclides of importance to reactor technology and applications. In contrast to the results obtained from $\beta_{eff}$ methods, these measurements show an energy dependence that does not agree with the recommended data currently available in the evaluated libraries. 

The time-dependent integral DN energy spectra have been measured for thermal neutron-induced fission of $^{235}$U. For the first time, the consistency of macroscopic DN spectra with the spectra obtained from the summation calculations over microscopic data ($T_{1/2},P_n$) for individual precursors has been demonstrated.

The energy dependence of DN temporary group parameters ($a_i,T_i$) have been studied in the energy range from thermal to 18 MeV. A procedure for averaging DN parameters obtained in different experimental runs has been developed. DN group parameters have been determined experimentally in an extended time of DN counting compared to previous measurements by Keepin~\cite{Keepin57}. The correlation and covariance matrices for the DN temporary group parameters have been extracted. 

These studies have revealed the reasons behind the large discrepancies observed in the  temporary parameters in the energy range 14-18 MeV. They have also led to the development of new systematics for the DN temporary parameters which have enhanced predictive power.

\subsection{New recommendations of group constants}

Finally, new recommendations have been made for the DN temporary group parameters ($a_i,T_i$) in 6- and 8-group models based on the evaluation of the new experimental macroscopic data. As the validation of these group models was beyond the scope of this CRP, it remains for the user community to use these new recommended data in verification and validation studies. The new recommended ($a_i,T_i$) data are available in this report and from the online CRP database~\cite{database}.

\section*{ Acknowledgments}
This work has been performed in the framework of a Coordinated Research Project of the International Atomic Energy Agency (IAEA) on "Development of a Reference Database for $\beta$-delayed neutron emission data" (F41030).

I.D.'s work was partly funded by the Natural Sciences and Engineering Research Council (NSERC) and the Natural Research Council (NRC) Canada. B. S.'s work was partly funded by the Nuclear Data Section of the IAEA. I.N.B. was supported by the Russian Science Foundation under Grant No. 16-12-10161. 

K.R. was partially supported by the US DOE Office of Science, Nuclear Physics) under the contract number DE-AC05-00OR22725 with UT Battelle, LLC. Work at Brookhaven National Laboratory was sponsored by the Office of Nuclear Physics, Office of Science of the U.S. Department of Energy under Contract No. DE-AC02-98CH10886 with Brookhaven Science Associates, LLC. Work at Lawrence Livermore National Laboratory was performed under the auspices of the U.S. Department of Energy under Contract DE-AC52-07NA27344.

Work at IFIC (A.A., J.L.T.) was supported by the Spanish Ministerio de Econom\'ia y Competitividad under Grants No. FPA2011-24553, No. FPA2014-52823-C2-1-P, FPA2017-83946-C2-1-P, and the program Severo Ochoa (SEV-2014-0398). 

Work at CIEMAT was supported by the Spanish national company for radioactive waste management ENRESA, through the CIEMAT-ENRESA agreements on ``Transmutaci\'{o}n de radionucleidos de vida larga como soporte a la gesti\'{o}n de residuos radioactivos de alta actividad'' and by the Spanish Ministerio de Econom\'ia y Competitividad under Grants No. FPA2014-53290-C2-1-P and FPA2016-76765-P.

Work at SUBATECH was supported by the CNRS challenge NEEDS and the associated NACRE project, and by the CNRS/in2p3 Master Project OPALE.
\clearpage
\bibliography{NDS}

\begin{thebibliography}{100}

\bibitem{Roberts1939a}
{R.B. Roberts}, {R.C. Meyer}, and P.~Wang, ``Further observations on the
  splitting of {U}ranium and {T}horium,'' {\sc \protect\JournalTitle{Phys.
  Rev.}} \textbf{55},  510 (1939).

\bibitem{Roberts1939b}
{R.B. Roberts}, {L.R. Hafstad}, {R.C. Meyer}, and P.~Wang, ``The delayed
  neutron emission which accompanies fission of {U}ranium and {T}horium,'' {\sc
  \protect\JournalTitle{Phys. Rev.}} \textbf{55},  664 (1939).

\bibitem{Meitner1938}
L.~Meitner, F.~Strassmann, and O.~Hahn, ``Kuenstliche {U}mwandlungsprozesse bei
  {B}estrahlung des {T}horiums mit {N}eutronen; {A}uftreten isomerer {R}eihen
  durch {A}bspaltung von $\alpha$-{S}trahlen,'' {\sc \protect\JournalTitle{Z.
  Physik}} \textbf{109},  538 (1938).

\bibitem{Bohr1939}
{N. Bohr and J.A. Wheeler}, ``The mechanism of nuclear fission,'' {\sc
  \protect\JournalTitle{Phys. Rev.}} \textbf{56},  426 (1939).

\bibitem{Azuma1979}
{R.E. Azuma}, {L.C. Carraz}, {P.G. Hansen},\textit{  {et~al.}}, ``First
  {O}bservation of {B}eta-{D}elayed {T}wo-{N}eutron {R}adioactivity:
  $^{11}${L}i,'' {\sc \protect\JournalTitle{Phys. Rev. Lett.}} \textbf{43},
  1652 (1979).

\bibitem{Azuma80}
{R.E. Azuma, \textit{et al.}}, ``{Beta-delayed three-neutron radioactivity of
  $^{11}\mathrm{Li}$},'' {\sc \protect\JournalTitle{Physics Letters B}}
  \textbf{96},  31  (1980).

\bibitem{Dufour1988}
{J.P. Dufour}, {R. Del Moral}, F.~Hubert,\textit{  {et~al.}}, ``{Beta delayed
  multi-neutron radioactivity of $^{17}$B, $^{14}$Be, $^{19}$C},'' {\sc
  \protect\JournalTitle{Phys. Lett. B}} \textbf{206},  195 (1988).

\bibitem{Booth1939}
{E.T. Booth, J.R. Dunning and F.G. Slack}, ``Delayed neutron emission from
  {U}ranium,'' {\sc \protect\JournalTitle{Phys. Rev.}} \textbf{55},  876
  (1939).

\bibitem{Brostrom1939}
{K.J. Brostrom, J. Koch and T. Lauritsen}, ``Delayed neutron emission
  accompanying {U}ranium fission,'' {\sc \protect\JournalTitle{Nature}}
  \textbf{144},  830 (1939).

\bibitem{Snell1947}
{A.H. Snell}, {V.A. Nedzel}, {H.W. Ibsen}, {J.S. Levinger}, {R.G. Wilkinson},
  and {M.B. Sampson}, ``Studies of the delayed neutrons. {I}. {T}he decay curve
  and the intensity of the delayed neutrons,'' {\sc \protect\JournalTitle{Phys.
  Rev.}} \textbf{72},  541 (1947).

\bibitem{DeHoffmann1948}
F.~{De Hoffmann}, {B.T. Feld}, and {P.R. Stein}, ``Delayed neutrons from
  $^{235}${U},'' {\sc \protect\JournalTitle{Phys. Rev.}} \textbf{74},  1330
  (1948).

\bibitem{Hughes1948}
{D.J. Hughes}, J.~Dabbs, A.~Cahn, and D.~Hall, ``Delayed neutrons from fission
  of {U}$^{235}$,'' {\sc \protect\JournalTitle{Phys. Rev.}} \textbf{73},  111
  (1948).

\bibitem{Keepin57}
{G.R. Keepin, T.F. Wimett and R.K. Zeigler}, ``{Delayed Neutrons from
  Fissionable Isotopes of Uranium, Plutonium and Thorium},'' {\sc
  \protect\JournalTitle{Phys. Rev.}} \textbf{107},  1044 (1957).

\bibitem{NEA-WPEC6}
{G. Rudstam, Ph. Finck, A. Filip, A. D'Angelo and R.D. McKnight}, ``{Delayed
  Neutron Data for the Major Actinides},'' Tech. Rep. NEA/WPEC-6, NEA/OECD,
  Paris, France (2002).
\newblock Available at
  \url{https://www.oecd-nea.org/science/wpec/volume6/volume6.pdf}.

\bibitem{Rudstam74}
{G. Rudstam and S. Shalev}, ``Energy spectra of delayed neutrons from separated
  fission products,'' {\sc \protect\JournalTitle{Nucl. Phys. A}} \textbf{235},
  397 (1974).

\bibitem{Rudstam77}
{G. Rudstam and E. Lund}, ``{Energy Spectra of Delayed Neutrons from the
  Precursors $^{79}$(Zn, Ga), $^{80}$Ga, $^{81}$Ga, $^{94}$Rb, $^{95}$Rb,
  $^{129}$In, and $^{130}$In},'' {\sc \protect\JournalTitle{Nucl. Sci. and
  Eng.}} \textbf{64},  749 (1977).

\bibitem{Greenwood1985}
{R.C. Greenwood and A.J. Caffrey}, ``Delayed neutron spectra of $^{93-97}${Rb}
  and $^{143-145}${Cs},'' {\sc \protect\JournalTitle{Nucl. Sci. Eng.}}
  \textbf{91},  305 (1985).

\bibitem{Greenwood1997}
{R.C. Greenwood and K.D. Watts}, ``Delayed neutron energy spectra of
  $^{87}${Br}, $^{88}${Br}, $^{89}${Br}, $^{90}${Br}, $^{137}${I}, $^{138}${I},
  $^{139}${I} and $^{136}${Te},'' {\sc \protect\JournalTitle{Nucl. Sci. Eng.}}
  \textbf{126},  324 (1997).

\bibitem{Kratz79}
{K.-L. Kratz}, ``Review of delayed neutron energy spectra,'' tech. rep., {in
  Report INDC(NDS)-0107, IAEA, Vienna, Austria} (1979).

\bibitem{Rudstam1980}
G.~Rudstam, ``The uncertainty of neutron energy spectra deduced from measured
  pulse spectra in a $^{3}${He}-spectrometer,'' {\sc
  \protect\JournalTitle{Nuclear Instruments and Methods}} \textbf{177},  529
  (1980).

\bibitem{Rudstam1993}
G.~Rudstam, K.~Aleklett, and L.~Sihver, ``Delayed neutron branching ratios of
  precursors in the fission product region,'' {\sc \protect\JournalTitle{At.
  Data Nucl. Data Tables}} \textbf{53},  1 (1993).

\bibitem{Pfeiffer2002}
B.~Pfeiffer, K.-L. Kratz, and P.~M{\"o}ller, ``Status of delayed neutron
  precursor data: half-lives and neutron emission probabilities,'' {\sc
  \protect\JournalTitle{Prog. Nucl. Energy}} \textbf{41},  39 (2002).

\bibitem{ENSDF}
``{Evaluated Nuclear Structure Data File (ENSDF)}.''
\newblock Online at \url{https://www.nndc.bnl.gov/ensdf/}.

\bibitem{Chadwick2011}
{M.B. Chadwick}, M.~Herman, P.~Oblo\v{z}insky,\textit{  {et~al.}},
  ``{{ENDF/B-VII.1} Nuclear Data for Science and Technology: Cross Sections,
  Covariances, Fission Product Yields and Decay Data},'' {\sc
  \protect\JournalTitle{Nuclear Data Sheets}} \textbf{112},  2887  (2011).
\newblock Special Issue on ENDF/B-VII.1 Library.

\bibitem{Brown2018}
{D.A. Brown, M.B. Chadwick, R. Capote, \textit{et al.}}, ``{{ENDF/B-VIII.0:}
  The 8th Major Release of the Nuclear Reaction Data Library with
  {CIELO}-project Cross Sections, New Standards and Thermal Scattering Data},''
  {\sc \protect\JournalTitle{Nucl. Data Sheets}} \textbf{148},  1 (2018).
\newblock Special Issue on ENDF/B-VIII.0 Library.

\bibitem{Katakura11}
{J. Katakura}, ``{JENDL FP Decay Data File 2011 and Fission Yields Data File
  2011},'' (2011).
\newblock JAEA-Data/Code 2011-025, Tokai, Japan, 2011.

\bibitem{Katakura2015}
J.~Katakura and F.~Minato, ``{JENDL Decay Data File 2015},'' (2015).
\newblock {JAEA}-Data/Code 2015-030.

\bibitem{jeff3.1.1}
{M.A. Kellett, O. Bersillon, R.W. Mills}, ``The {JEFF}-3.1/-3.1.1 radioactive
  decay data and fission yields sub-libraries,'' (OECD 2009).
\newblock {ISBN 978-92-64-99087-6}.

\bibitem{Plompen2020}
{Plompen, A.J.M.}, {Cabellos, O.}, {De Saint Jean, C.},\textit{  {et~al.}},
  ``{The joint evaluated fission and fusion nuclear data library, JEFF-3.3},''
  {\sc \protect\JournalTitle{Eur. Phys. J. A}} \textbf{56},  181 (2020).

\bibitem{Nichols1998}
{A.L. Nichols}, ``{Recommended Decay Data for Short-lived Fission Products:
  Status and Needs},'' in \textit{ Proc. Second International Workshop on
  Nuclear Fission and Fission-product Spectroscopy, Seyssins, France}, AIP
  Conference Proceedings 447, 100-115 (1998), American Institute of Physics,
  Woodbury, New York, USA (1998).

\bibitem{Brady1989a}
{M.C. Brady}, \textit{ {Evaluation and Application of Delayed Neutron Precursor
  Data}}.
\newblock PhD thesis, Texas A \& M University, Los Alamos National Laboratory
  (1989).

\bibitem{Kawano2008}
{T. Kawano, P. M\"oller, and W.B. Wilson}, ``{Calculation of delayed-neutron
  energy spectra in a quasiparticle random-phase approximation--Hauser-Feshbach
  model},'' {\sc \protect\JournalTitle{Phys. Rev. C}} \textbf{78},  054601
  (2008).

\bibitem{AME16}
{M. Wang}, {G. Audi}, {F.G. Kondev}, {W.J. Huang}, {S. Naimi}, and {X. Xu},
  ``{The AME2016 atomic mass evaluation (II). Tables, graphs and references},''
  {\sc \protect\JournalTitle{Chin. Phys. C}} \textbf{41},  030003 (2017).

\bibitem{Stetter1962}
G.~Stetter, ``Investigation of the decay of {T}l-210 {(RaC'')},'' {\sc
  \protect\JournalTitle{Nucl. Sci. Abstr.}} \textbf{16},  1409 (1962).
\newblock Abstract 10963.

\bibitem{Birch2015}
M.~Birch, B.~Singh, I.~Dillmann, D.~Abriola, {T.D. Johnson}, {E.A. McCutchan},
  and {A.A. Sonzogni}, ``{Evaluation of Beta-Delayed Neutron Emission
  Probabilities and Half-Lives for {Z = 2--28}},'' {\sc
  \protect\JournalTitle{Nucl. Data Sheets}} \textbf{128},  131 (2015).

\bibitem{Liang2020}
J.~Liang, B.~Singh, {E.A. McCutchan},\textit{  {et~al.}}, ``{Compilation and
  Evaluation of Beta-Delayed Neutron Emission Probabilities and Half-Lives for
  {Z$>$28} Precursors},'' {\sc \protect\JournalTitle{Nucl. Data Sheets}}
  \textbf{168},  1 (2020).

\bibitem{BRIKEN2017}
A.~{Tarife{\~n}o-Saldivia}, J.~L. {Tain}, C.~{Domingo-Pardo},\textit{
  {et~al.}}, ``{Conceptual design of a hybrid neutron-gamma detector for study
  of {$\beta$}-delayed neutrons at the RIB facility of RIKEN},'' {\sc
  \protect\JournalTitle{Journal of Instrumentation}} \textbf{12},  P04006
  (2017).

\bibitem{NuDat27}
``{NuDat 2.7},'' National Nuclear Data Center, Brookhaven National Laboratory,
  Brookhaven/ USA (2018), [Online] Available at
  \url{http://www.nndc.bnl.gov/nudat2/}.

\bibitem{IAEA0599}
D.~Abriola, B.~Singh, and I.~Dillmann, ``{Summary Report of Consultants’
  Meeting about Beta-delayed neutron emission evaluation},'' Tech. Rep.
  INDC(NDS)-0599, International Atomic Energy Agency, Vienna, Austria (2011).
\newblock Available at
  \url{https://www-nds.iaea.org/publications/indc/indc-nds-0599.pdf}.

\bibitem{IAEA0643}
I.~Dillmann, P.~Dimitriou, and B.~Singh, ``{Summary Report of 1st Research
  Coordination Meeting on Development of Reference Database for Beta-delayed
  neutron emission},'' Tech. Rep. INDC(NDS)-0643, International Atomic Energy
  Agency, Vienna, Austria (2013).
\newblock Available at
  \url{https://www-nds.iaea.org/publications/indc/indc-nds-0643.pdf}.

\bibitem{IAEA0683}
I.~Dillmann, P.~Dimitriou, and B.~Singh, ``{Summary Report of 2nd Research
  Coordination Meeting on Development of Reference Database for Beta-delayed
  neutron emission},'' Tech. Rep. INDC(NDS)-0683, International Atomic Energy
  Agency, Vienna, Austria (2015).
\newblock Available at
  \url{https://www-nds.iaea.org/publications/indc/indc-nds-0683.pdf}.

\bibitem{IAEA0735}
I.~Dillmann, P.~Dimitriou, and B.~Singh, ``{Summary Report of 3rd Research
  Coordination Meeting on Development of Reference Database for Beta-delayed
  neutron emission},'' Tech. Rep. INDC(NDS)-0683, International Atomic Energy
  Agency, Vienna, Austria (2017).
\newblock Available at
  \url{https://www-nds.iaea.org/publications/indc/indc-nds-0735.pdf}.

\bibitem{database}
``{Reference Database for beta-delayed neutron emission}.''
\newblock Online at
  \url{https://www-nds.iaea.org/beta-delayed-neutron/database.html}.

\bibitem{Crane1991}
{T.W. Crane and M.P. Baker}, ``Neutron detectors,'' in \textit{ Passive
  Nondestructive Assay of Nuclear Materials} ({O. Reilly, \textit{et al.}},
  ed.),  379 (1991).
\newblock NUREG/CR-5550.

\bibitem{Hanson1947}
{A.O. Hanson and J.L. McKibben}, ``{A Neutron Detector Having Uniform
  Sensitivity from 10 keV to 3 MeV},'' {\sc \protect\JournalTitle{Phys. Rev.}}
  \textbf{72},  673 (1947).

\bibitem{Pereira2010}
{J. Pereira, \textit{et al.}}, ``{The neutron long counter {NERO} for studies
  of $\beta$-delayed neutron emission in the r-process},'' {\sc
  \protect\JournalTitle{Nuclear Instruments and Methods in Physics Research
  Section A: Accelerators, Spectrometers, Detectors and Associated Equipment}}
  \textbf{618},  275 (2010).

\bibitem{Mathieu2012}
{L. Mathieu, \textit{et al.}}, ``{New neutron long-counter for delayed neutron
  investigations with the LOHENGRIN fission fragment separator},'' {\sc
  \protect\JournalTitle{Journal of Instrumentation}} \textbf{7},  P08029
  (2012).

\bibitem{Grzywacz2014}
{R.K. Grzywacz, \textit{et al.}}, ``{Hybrid $^{3}$Hen - new detector for gammas
  and neutrons},'' {\sc \protect\JournalTitle{Acta Physica Polonica B}}
  \textbf{45},  217 (2014).

\bibitem{Agramunt16}
{J. Agramunt, \textit{et al.}}, ``Characterization of a neutron-beta counting
  system with beta-delayed neutron emitters,'' {\sc
  \protect\JournalTitle{Nuclear Instruments and Methods in Physics Research
  Section A: Accelerators, Spectrometers, Detectors and Associated Equipment}}
  \textbf{807},  69  (2016).

\bibitem{Testov2016}
{D. Testov, \textit{et al.}}, ``{The $^{3}$He long-counter TETRA at the ALTO
  ISOL facility},'' {\sc \protect\JournalTitle{Nuclear Instruments and Methods
  in Physics Research Section A: Accelerators, Spectrometers, Detectors and
  Associated Equipment}} \textbf{815},  96 (2016).

\bibitem{Tolosa2018}
A.~Tolosa-Delgado, J.~Agramunt, J.~L. Tain,\textit{  {et~al.}},
  ``{Commissioning of the BRIKEN detector for the measurement of very exotic
  $\beta$-delayed neutron emitters},'' {\sc \protect\JournalTitle{Nucl. Instr.
  Meth. A}} \textbf{925},  133 (2019).

\bibitem{Madurga2016}
{M. Madurga, \textit{et al.}}, ``{Evidence for Gamow-Teller Decay of $^{78}$Ni
  Core from Beta-Delayed Neutron Emission Studies},'' {\sc
  \protect\JournalTitle{Phys. Rev. Lett.}} \textbf{117},  092502 (2016).

\bibitem{Reeder1977}
{P.L. Reeder, \textit{et al.}}, ``Average neutron energies from separated
  delayed-neutron precursors,'' {\sc \protect\JournalTitle{Phys. Rev. C}}
  \textbf{15},  2098 (1977).

\bibitem{Caballero18}
R.~Caballero, I.~Dillmann, J.~Agramunt,\textit{  {et~al.}}, ``{First
  determination of $\beta$-delayed multiple neutron emission beyond A $=$ 100
  through direct neutron measurement: The P2n value of 136Sb},'' {\sc
  \protect\JournalTitle{Phys. Rev. C}} \textbf{98},  034310 (2018).

\bibitem{Rasco2018}
{B.C. Rasco, \textit{et al.}}, ``{The ORNL Analysis Technique for Extracting
  $\beta$-Delayed Multi-Neutron Branching Ratios with BRIKEN},'' {\sc
  \protect\JournalTitle{Nuclear Instruments and Methods in Physics Research
  Section A: Accelerators, Spectrometers, Detectors and Associated Equipment}}
  \textbf{911},  79 (2018).

\bibitem{Stehney1953}
A.~Stehney and N.~Sugarman, ``{Characteristics of $^{87}\mathrm{Br}$, a Delayed
  Neutron Activity},'' {\sc \protect\JournalTitle{Phys. Rev.}} \textbf{89},
  194 (1953).

\bibitem{Wahl88}
{A.C. Wahl}, ``Nuclear-charge distribution and delayed neutron yields for
  thermal-neutron-induced fission of $^{235}${U}, $^{233}${U}, and $^{239}${P}u
  and and for spontaneous fission of $^{252}${C}f,'' {\sc
  \protect\JournalTitle{At. Data Nucl. Data Tables}} \textbf{39},  1 (1988).

\bibitem{Talbert1969}
{W.L. Talbert, \textit{et al.}}, ``Delayed neutron emission in the decays of
  short-lived separated isotopes of gaseous fission products,'' {\sc
  \protect\JournalTitle{Phys. Rev.}} \textbf{177},  1805 (1969).

\bibitem{Okano1986}
{K. Okano, \textit{et al.}}, ``Delayed neutron emission in the decays of
  short-lived separated isotopes of gaseous fission products,'' {\sc
  \protect\JournalTitle{Ann. Nucl. Energy}} \textbf{13},  467 (1986).

\bibitem{Shearman2018}
R.~Shearman, \textit{ {Development of a novel gamma-ray detection system for
  decay data measurement and beta-delayed spectroscopy of primary fission
  fragments at N$\approx$82}}.
\newblock PhD thesis, University of Surrey, UK (2018).

\bibitem{Reeder1980b}
{P.L. Reeder and R.A. Warner}, ``{Delayed neutron emission probabilities of Rb
  and Cs precursos measured by both ion and beta counting techniques},'' Tech.
  Rep. PNL-SA-8766, Pacific Northwest Laboratory (1980).

\bibitem{Amarel1969}
{I. Amarel, \textit{et al.}}, ``{Delayed neutron emission probabilities of Rb
  and Cs precursors. The half-life of $^{97}\mathrm{Rb}$},'' {\sc
  \protect\JournalTitle{J. Inorg. Nucl. Chem.}} \textbf{31},  577 (1969).

\bibitem{Winger2009}
J.~A. Winger, S.~V. Ilyushkin, K.~P. Rykaczewski,\textit{  {et~al.}}, ``{Large
  $\ensuremath{\beta}$-Delayed Neutron Emission Probabilities in the
  $^{78}\mathrm{Ni}$ Region},'' {\sc \protect\JournalTitle{Phys. Rev. Lett.}}
  \textbf{102},  142502 (2009).

\bibitem{Yee13}
{R.M. Yee}, {N.D. Scielzo}, {P.F. Bertone},\textit{  {et~al.}},
  ``$\beta$-delayed neutron spectroscopy using trapped radioactive ions,'' {\sc
  \protect\JournalTitle{Phys. Rev. Lett.}} \textbf{110},  092501 (2013).

\bibitem{Mianowski10}
S.~Mianowski, H.~Czyrkowski, R.~Dabrowski,\textit{  {et~al.}}, ``Imaging the
  decay of $^8${H}e,'' {\sc \protect\JournalTitle{Acta Phys. Polon. B}}
  \textbf{41},  449 (2010).

\bibitem{Evdokimov13}
A.~Evdokimov, I.~Dillmann, M.~Marta,\textit{  {et~al.}}, ``{An alternative
  approach to measure $\beta$-delayed neutron emission},'' in \textit{ XII
  International Symposium on Nuclei in the Cosmos, August 5-12, 2012, Cairns,
  Australia},  PoS(NIC XII)115, Proceedings of Science (2012).

\bibitem{Mardor2017}
I.~Mardor, ``{A Novel Method for Measuring $\beta$-Delayed Neutron Emission},''
  tech. rep., GSI Helmholtz Center for Heavy Ion Research Darmstadt (2017).
\newblock Available at
  \url{https://www.gsi.de/en/work/research/library_documentation/gsi_scientific_reports.htm}.

\bibitem{Shalev1973}
{S. Shalev and J.M. Cuttler}, ``The energy distribution of delayed fission
  neutrons,'' {\sc \protect\JournalTitle{Nuclear Science and Engineering}}
  \textbf{51},  52 (1973).

\bibitem{Franz1977}
H.~Franz, W.~Rudolph, H.~Ohm, K.-L. Kratz, G.~Herrmann, F.~Nuh, D.~Slaughter,
  and S.~Prussin, ``Delayed-neutron spectroscopy with $^{3}${He}
  spectrometers,'' {\sc \protect\JournalTitle{Nuclear Instruments and Methods}}
  \textbf{144},  253  (1977).

\bibitem{Owen1981}
{J.G. Owen, D.R. Weaver and J. Walker}, ``The calibration of a $^{3}${He}
  spectrometer and its use to measure the neutron spectrum from an {Am/Li}
  source,'' {\sc \protect\JournalTitle{Nuclear Instruments and Methods in
  Physics Research}} \textbf{188},  579  (1981).

\bibitem{Chichester2012}
{D.L. Chichester, J.T. Johnson and E.H. Seabury}, ``Fast-neutron spectrometry
  using a $^{3}${He} ionization chamber and digital pulse shape analysis,''
  {\sc \protect\JournalTitle{Applied Radiation and Isotopes}} \textbf{70},
  1457  (2012).

\bibitem{Beimer1986}
K.-H. Beimer, G.~Nyman, and O.~Tengblad, ``Response function for $^{3}${He}
  neutron spectrometers,'' {\sc \protect\JournalTitle{Nuclear Instruments and
  Methods in Physics Research Section A: Accelerators, Spectrometers, Detectors
  and Associated Equipment}} \textbf{245},  402  (1986).

\bibitem{Reeder1980}
{P.L. Reeder}, {L.J. Alquist}, {R.L. Kiefer}, {F.H. Ruddy}, and {R.A. Warner},
  ``{Energy spectra of delayed neutrons from the separated precursors
  Rubidium-93, -94, -95, and Cesium-143},'' {\sc \protect\JournalTitle{Nuclear
  Science and Engineering}} \textbf{75},  140 (1980).

\bibitem{Bennett1972}
{E.F. Bennett and T.J. Yule}, ``Response functions for proton-recoil
  proportional counter spectrometer,'' {\sc \protect\JournalTitle{Nuclear
  Instruments and Methods}} \textbf{98},  393  (1972).

\bibitem{Gold1968}
{R. Gold and E.F. Bennett}, ``Effects of finite size in 4$\pi$-recoil
  proportional counters,'' {\sc \protect\JournalTitle{Nuclear Instruments and
  Methods}} \textbf{63},  285  (1968).

\bibitem{Sloan1978}
{W.R. Sloan and G.L. Woodruff}, ``Frequency filtering of proton-recoil data,''
  {\sc \protect\JournalTitle{Nuclear Instruments and Methods}} \textbf{150},
  253  (1978).

\bibitem{Eccleston1977}
{G.W. Eccleston and G.L. Woodruff}, ``Measured near-equilibrium delayed neutron
  spectra produced by fast-neutron-induced fission of $^{232}${Th},
  $^{233}${U}, $^{235}${U}, $^{238}${U}, and $^{239}${Pu},'' {\sc
  \protect\JournalTitle{Nucl. Sci. Eng.}} \textbf{62},  636 (1977).

\bibitem{ANL7763}
{E.F. Bennett and T.J. Yule}, ``Techniques and analyses of fast-reactor neutron
  spectroscopy with proton-recoil counters,'' Tech. Rep. ANL-7763, Argonne
  National Laboratory, Argonne, Illinois, USA (1971).
\newblock Available at
  \url{https://inldigitallibrary.inl.gov/Reports/ANL-7763.pdf}.

\bibitem{Kor09}
{N.V. Kornilov, I. Fabry, S. Oberstedt and F.-J. Hambsch}, ``{Total
  characterization of neutron detectors with a $^{252}$Cf source and a new
  light output determination},'' {\sc \protect\JournalTitle{Nuclear Instruments
  and Methods in Physics Research Section A: Accelerators, Spectrometers,
  Detectors and Associated Equipment}} \textbf{599},  226  (2009).

\bibitem{Zai12}
N.~Zaitseva, B.~L. Rupert, I.~PaweŁczak, A.~Glenn, H.~P. Martinez, L.~Carman,
  M.~Faust, N.~Cherepy, and S.~Payne, ``{Plastic scintillators with efficient
  neutron/gamma pulse shape discrimination},'' {\sc
  \protect\JournalTitle{Nuclear Instruments and Methods in Physics Research
  Section A: Accelerators, Spectrometers, Detectors and Associated Equipment}}
  \textbf{668},  88  (2012).

\bibitem{Buta2000}
A.~Buta, T.~Martin, C.~Timis,\textit{  {et~al.}}, ``{TONNERRE}: an array for
  delayed-neutron decay spectroscopy,'' {\sc \protect\JournalTitle{Nucl. Instr.
  Meth. A}} \textbf{455},  412 (2000).

\bibitem{Har91}
R.~Harkewicz, D.~J. Morrissey, B.~A. Brown, J.~A. Nolen, N.~A. Orr, B.~M.
  Sherrill, J.~S. Winfield, and J.~A. Winger, ``\ensuremath{\beta}-decay
  branching ratios of the neutron-rich nucleus $^{15}\mathrm{B}$,'' {\sc
  \protect\JournalTitle{Phys. Rev. C}} \textbf{44},  2365 (1991).

\bibitem{Morrissey1997}
{D.J. Morrissey}, {K.N. McDonald}, D.~Bazin,\textit{  {et~al.}}, ``Single
  neutron emission following $^{11}${L}i $\beta$-decay,'' {\sc
  \protect\JournalTitle{Nucl. Phys. A}} \textbf{627},  222 (1997).

\bibitem{Sum10}
C.~S. Sumithrarachchi, D.~J. Morrissey, A.~D. Davies,\textit{  {et~al.}},
  ``States in $^{22}\mathrm{O}$ via $\ensuremath{\beta}$ decay of
  $^{22}\mathrm{N}$,'' {\sc \protect\JournalTitle{Phys. Rev. C}} \textbf{81},
  014302 (2010).

\bibitem{Tim05}
C.~Timis, J.~C. Angélique, A.~Buta,\textit{  {et~al.}}, ``{Spectroscopy around
  N = 20 shell closure: $\beta$–n decay study of $^{35}$Al},'' {\sc
  \protect\JournalTitle{Journal of Physics G: Nuclear and Particle Physics}}
  \textbf{31},  S1965 (2005).

\bibitem{Per06}
F.~Perrot, F.~Mar\'echal, C.~Jollet,\textit{  {et~al.}},
  ``\ensuremath{\beta}-decay studies of neutron-rich {K} isotopes,'' {\sc
  \protect\JournalTitle{Phys. Rev. C}} \textbf{74},  014313 (2006).

\bibitem{Mat08}
C.~Matei, {D.W. Bardayan}, {J.C. Blackmon}, {J.A. Cizewski}, {R.K. Grzywacz},
  {S.N. Liddick}, {W.A. Peters}, and F.~Sarazin, ``{The Versatile Array of
  Neutron Detectors at Low Energy (VANDLE)},'' in \textit{ 10th Symposium on
  Nuclei in the Cosmos (NIC X)} (2008).

\bibitem{Pet15}
{W.A. Peters}, S.~Ilyushkin, M.~Madurga,\textit{  {et~al.}}, ``{Performance of
  the Versatile Array of Neutron Detectors at Low Energy (VANDLE)},'' {\sc
  \protect\JournalTitle{Nuclear Instruments and Methods in Physics Research
  Section A: Accelerators, Spectrometers, Detectors and Associated Equipment}}
  \textbf{836},  122  (2016).

\bibitem{Bil13}
V.~Bildstein, {P.E. Garrett}, J.~Wong,\textit{  {et~al.}}, ``Comparison of
  deuterated and normal liquid scintillators for fast-neutron detection,'' {\sc
  \protect\JournalTitle{Nuclear Instruments and Methods in Physics Research
  Section A: Accelerators, Spectrometers, Detectors and Associated Equipment}}
  \textbf{729},  188  (2013).

\bibitem{Garrett2014}
{P.E. Garrett}, ``{DESCANT – the deuterated scintillator array for neutron
  tagging},'' {\sc \protect\JournalTitle{Hyp. Int.}} \textbf{225},  137 (2014).

\bibitem{Mar14}
T.~Mart\'{\i}nez, D.~Cano-Ott, J.~Castilla,\textit{  {et~al.}}, ``{MONSTER: a
  TOF Spectrometer for $\beta$-delayed Neutron Spectroscopy},'' {\sc
  \protect\JournalTitle{Nuclear Data Sheets}} \textbf{120},  78  (2014).

\bibitem{Pau14}
{S.V. Paulauskas}, M.~Madurga, R.~Grzywacz, D.~Miller, S.~Padgett, and H.~Tan,
  ``{A digital data acquisition framework for the Versatile Array of Neutron
  Detectors at Low Energy (VANDLE)},'' {\sc \protect\JournalTitle{Nuclear
  Instruments and Methods in Physics Research Section A: Accelerators,
  Spectrometers, Detectors and Associated Equipment}} \textbf{737},  22
  (2014).

\bibitem{Cut71}
{J.M. Cuttler and S. Shalev}, ``{ Annual. Meeting of Israel Phys. Soc.,
  Haifa},'' (1971).

\bibitem{Mei73}
{R.J. De Meijer}, C.~Delaune, D.~McShan, {J.W. Nelson}, and {H.A. Van
  Rinsvelt}, ``{$\beta$-Decay of $^{17}$N to unbound states of 17O},'' {\sc
  \protect\JournalTitle{Nuclear Physics A}} \textbf{209},  424  (1973).

\bibitem{Pol73}
{A.R. Poletti and J.G. Pronko}, ``{Beta Decay of $^{17}\mathrm{N}$},'' {\sc
  \protect\JournalTitle{Phys. Rev. C}} \textbf{8},  1285 (1973).

\bibitem{Alb76}
{D.E. Alburger and D.H. Wilkinson}, ``{Beta decay of $^{16}\mathrm{C}$ and
  $^{17}\mathrm{N}$},'' {\sc \protect\JournalTitle{Phys. Rev. C}} \textbf{13},
  835 (1976).

\bibitem{Ohm76}
H.~Ohm, W.~Rudolph, and K.-L. Kratz, ``{Beta-delayed neutron emission following
  the decay of $^{17}$N},'' {\sc \protect\JournalTitle{Nuclear Physics A}}
  \textbf{274},  45  (1976).

\bibitem{Miyatake2003}
H.~Miyatake, H.~Ueno, Y.~Yamamoto,\textit{  {et~al.}}, ``{Spin-parity
  assignments in $^{15}\mathrm{C}$ by a new method: $\beta$ -delayed
  spectroscopy for a spin-polarized nucleus},'' {\sc
  \protect\JournalTitle{Phys. Rev. C}} \textbf{67},  014306 (2003).

\bibitem{Rac84}
J.~Rachidi, \textit{ PhD thesis}.
\newblock PhD thesis, Université Louis Pasteur de Strasbourg (1984).

\bibitem{Men14}
E.~Mendoza, D.~Cano-Ott, T.~Koi, and C.~Guerrero, ``{New Standard Evaluated
  Neutron Cross Section Libraries for the GEANT4 Code and First
  Verification},'' {\sc \protect\JournalTitle{IEEE Transactions on Nuclear
  Science}} \textbf{61},  2357  (2014).

\bibitem{Gar17}
{A.R. Garcia}, E.~Mendoza, D.~Cano-Ott, R.~Nolte, T.~Martinez, A.Algora, {J.L.
  Tain}, K.~Banerjee, and C.Bhattacharya, ``{New physics model in GEANT4 for
  the simulation of neutron interactions with organic scintillation
  detectors},'' {\sc \protect\JournalTitle{Nuclear Instruments and Methods in
  Physics Research Section A: Accelerators, Spectrometers, Detectors and
  Associated Equipment}} \textbf{868},  73  (2017).

\bibitem{Duke1970}
{C.L. Duke, \textit{et al.}}, ``Strength-function phenomena in electron-capture
  beta decay,'' {\sc \protect\JournalTitle{Nuclear Physics A}} \textbf{151},
  609 (1970).

\bibitem{Algora2010}
{A. Algora, D. Jordan, J.L. Tain, \textit{et al.}}, ``Reactor decay heat in
  $^{239}${P}u: Solving the $\gamma$ discrepancy in the 4-3000-s cooling
  period,'' {\sc \protect\JournalTitle{Physical Review Letters}} \textbf{105},
  202501 (2010).

\bibitem{Spyrou2014}
{ A. Spyrou, \textit{et al.}}, ``{Novel technique for constraining r-process
  (n,$\gamma$) reaction rates},'' {\sc \protect\JournalTitle{Phys. Rev. Lett.}}
  \textbf{113},  232502 (2014).

\bibitem{Rasco2016}
B.~C. Rasco, M.~Woli\'{n}ska-Cichocka, A.~Fija\l{}kowska,\textit{  {et~al.}},
  ``{Decays of the Three Top Contributors to the Reactor
  ${\overline{\ensuremath{\nu}}}_{e}$ High-Energy Spectrum, $^{92}\mathrm{Rb}$,
  $^{96\mathrm{gs}}\mathrm{Y}$, and $^{142}\mathrm{Cs}$, Studied with Total
  Absorption Spectroscopy},'' {\sc \protect\JournalTitle{Phys. Rev. Lett.}}
  \textbf{117},  092501 (2016).

\bibitem{Rasco2017}
B.~C. Rasco, K.~P. Rykaczewski, A.~Fija\l{}kowska,\textit{  {et~al.}},
  ``Complete $\ensuremath{\beta}$-decay pattern for the high-priority
  decay-heat isotopes $^{137}\mathrm{I}$ and $^{137}\mathbf{Xe}$ determined
  using total absorption spectroscopy,'' {\sc \protect\JournalTitle{Phys. Rev.
  C}} \textbf{95},  054328 (2017).

\bibitem{Tain2015}
J.~L. Tain, E.~Valencia, A.~Algora,\textit{  {et~al.}}, ``{Enhanced
  $\ensuremath{\gamma}$-Ray Emission from Neutron Unbound States Populated in
  $\ensuremath{\beta}$ Decay},'' {\sc \protect\JournalTitle{Phys. Rev. Lett.}}
  \textbf{115},  062502 (2015).

\bibitem{guadilla2018}
{V. Guadilla, \textit{et al.}}, ``{Characterization and performance of the DTAS
  detector},'' {\sc \protect\JournalTitle{Nuclear Instruments and Methods in
  Physics Research Section A: Accelerators, Spectrometers, Detectors and
  Associated Equipment}} \textbf{910},  79 (2018).

\bibitem{Tain2015b}
{J.L. Tain, \textit{et al.}}, ``{The sensitivity of LaBr$_{3}$:Ce scintillation
  detectors to low energy neutrons: Measurement and Monte Carlo simulation},''
  {\sc \protect\JournalTitle{Nuclear Instruments and Methods in Physics
  Research Section A: Accelerators, Spectrometers, Detectors and Associated
  Equipment}} \textbf{774},  17 (2015).

\bibitem{Scielzo2012}
{N.D. Scielzo, \textit{et al.}}, ``The $\beta$-decay {P}aul trap: A
  radiofrequency-quadrupole ion trap for precision $\beta$-decay studies,''
  {\sc \protect\JournalTitle{Nucl. Instrum. Methods Phys. Res. A}}
  \textbf{681},  94 (2012).

\bibitem{Czeszumska20}
{A. Czeszumska, \textit{et al.}} {\sc \protect\JournalTitle{Phys. Rev. C}}
  \textbf{101},  024312 (2020).

\bibitem{Wang20}
{B.S. Wang, \textit{et al.}} {\sc \protect\JournalTitle{Phys. Rev. C}}
  \textbf{101},  025806 (2020).

\bibitem{Hirsh2016}
{T.Y. Hirsh, \textit{et al.}}, ``{First operation and mass separation with the
  CARIBU MR-TOF},'' {\sc \protect\JournalTitle{Nucl. Instrum. Methods Phys.
  Res. B}} \textbf{376},  229 (2016).

\bibitem{Savard2016}
{G. Savard, \textit{et al.}}, ``The {CARIBU} gas catcher,'' {\sc
  \protect\JournalTitle{Nucl. Instrum. Methods Phys. Res. B}} \textbf{376},
  246 (2016).

\bibitem{Mia18}
S.~Mianowski, \textit{ {Study of beta decay of $^{8}$He with charged particle
  emission}}.
\newblock PhD thesis, Faculty of Physics, University of Warzawa, Poland (2018).

\bibitem{Litvinov2004}
{Yu.A. Litvinov}, H.~Geissel, {Yu.N. Novikov},\textit{  {et~al.}}, ``Precision
  experiments with time-resolved {S}chottky mass spectrometry,'' {\sc
  \protect\JournalTitle{Nucl. Phys. A}} \textbf{434},  473 (2004).

\bibitem{Najafi16}
{M.A. Najafi}, I.~Dillmann, F.~Bosch,\textit{  {et~al.}}, ``{CsI-}{Silicon}
  {Particle} detector for {Heavy} ions {Orbiting} in {Storage} rings
  ({CsISiPHOS}),'' {\sc \protect\JournalTitle{Nucl. Instr. Meth. A}}
  \textbf{836},  1 (2016).

\bibitem{ILIMA2017}
R.~Gernh{\"a}user, T.~Faestermann, and I.~Dillmann, ``The {ILIMA} ring detector
  for particle identification, life-time measurement and beam diagnostics --
  {T}echnical {R}eport for the {D}esign, {C}onstruction and {C}ommissioning of
  the {H}eavy {I}on {D}etector,'' tech. rep., ILIMA Collaboration, GSI
  Helmholtz Center for Heavy Ion Research Darmstadt (2017).
\newblock Available at
  \url{https://fair-center.eu/for-users/experiments/nustar/documents/technical-design-reports.html}.

\bibitem{Miyatake2018}
H.~Miyatake, M.~Wada, {X.Y. Watanabe},\textit{  {et~al.}}, ``{Present status of
  the KISS project},'' {\sc \protect\JournalTitle{AIP Conf. Proceedings}}
  \textbf{1947},  020018 (2018).

\bibitem{Plass2013}
W.~R. Plass, T.~Dickel, S.~Purushothaman,\textit{  {et~al.}}, ``The {FRS} ion
  catcher – {A} facility for high-precision experiments with stopped
  projectile and fission fragments,'' {\sc \protect\JournalTitle{Nucl. Instr.
  Meth. B}} \textbf{317},  457 (2013).

\bibitem{Horowitz2018}
{C.J. Horowitz}, A.~Arcones, B.~Cote,\textit{  {et~al.}}, ``{r-Process
  Nucleosynthesis: Connecting Rare-Isotope Beam Facilities with the Cosmos},''
  {\sc \protect\JournalTitle{Journal Phys. G}} (2019).
\newblock accepted.

\bibitem{IAEA0687}
E.~Ricard-McCutchan, P.~Dimitriou, and A.~Nichols, ``{International Network of
  Nuclear Structure and Decay Data (NSDD) Evaluators},'' Tech. Rep.
  INDC(NDS)-0687, International Atomic Energy Agency, Vienna, Austria (2015).
\newblock Available at
  \url{https://www-nds.iaea.org/publications/indc/indc-nds-0687.pdf}.

\bibitem{JWu2019}
J.~Wu, S.~Nishimura, P.~M\"oller,\textit{  {et~al.}},
  ``$\ensuremath{\beta}$-decay half-lives of 55 neutron-rich isotopes beyond
  the {N=82} shell gap,'' {\sc \protect\JournalTitle{Phys. Rev. C}}
  \textbf{101},  042801 (2020).

\bibitem{AME16a}
{W.J. Huang}, {G. Audi}, {M. Wang}, {F.G. Kondev}, {S. Naimi}, and {X. Xu},
  ``{The {AME2016} atomic mass evaluation (I). Evaluation of input data; and
  adjustment procedures},'' {\sc \protect\JournalTitle{Chin. Phys. C}}
  \textbf{41},  030002 (2017).

\bibitem{Caballero2016}
R.~Caballero-Folch, C.~Domingo-Pardo, J.~Agramunt,\textit{  {et~al.}}, ``{First
  measurement of several $\beta$-delayed neutron emitting isotopes beyond
  N=126},'' {\sc \protect\JournalTitle{Phys. Rev. Lett.}} \textbf{115},  012501
  (2016).

\bibitem{Caballero2017a}
R.~Caballero-Folch, C.~Domingo-Pardo, J.~Agramunt,\textit{  {et~al.}},
  ``{$\beta$-decay half-lives and $\beta$-delayed neutron emission
  probabilities for several isotopes of {A}u, {H}g, {T}l, {P}b and {B}i, beyond
  N=126},'' {\sc \protect\JournalTitle{Phys. Rev. C}} \textbf{95},  064322
  (2017).

\bibitem{Amiel1970}
S.~Amiel and H.~Feldstein, ``A semi-empirical treatment of neutron emission
  probabilities from delayed neutron precursors,'' {\sc
  \protect\JournalTitle{Phys. Lett.}} \textbf{31B},  59 (1970).

\bibitem{KHF1973}
K.-L. Kratz and G.~Herrmann, ``Systematics of neutron emission probabilities
  from delayed neutron precursors,'' {\sc \protect\JournalTitle{Z. Phys. A}}
  \textbf{263},  435 (1973).

\bibitem{McCutchan2012}
{E.A. McCutchan, A.A. Sonzogni, T.D. Johnson, D. Abriola, M. Birch, and B.
  Singh}, ``Improving systematic predictions of $\beta$-delayed
  neutron-emission probabilities,'' {\sc \protect\JournalTitle{Phys. Rev. C}}
  \textbf{86},  041305(R) (2012).

\bibitem{Miernik2013}
K.~Miernik, ``Phenomenological model of beta-delayed neutron-emission
  probability,'' {\sc \protect\JournalTitle{Phys. Rev. C}} \textbf{88},
  041301(R) (2013).

\bibitem{Miernik2014}
K.~Miernik, ``$\ensuremath{\beta}$-delayed multiple-neutron emission in the
  effective density model,'' {\sc \protect\JournalTitle{Phys. Rev. C}}
  \textbf{90},  054306 (2014).

\bibitem{Brady89}
{M.C. Brady and T. England}, ``Delayed neutron data and group parameters for 43
  fissioning systems,'' {\sc \protect\JournalTitle{Nucl. Sci. Eng.}}
  \textbf{103},  129 (1989).

\bibitem{GSYS2.4}
``Gsys2.4, graph suchi yomitori system,'' (2014).
\newblock Available at \url{https://www.jcprg.org/gsys/ver24/gsys-e.html}.

\bibitem{Otuka2014}
{N. Otuka, E. Dupont, V. Semkova, B. Pritychenko, A.I. Blokhin, M. Aikawa, S.
  Babykina, M. Bossant, G. Cheng, S. Dunaeva, R.A. Forrest, T.Fukahori, N.
  Furutachi, S. Ganesan, Z. Ge, O. Gritzay, M. Herman, S. Hlava\v{c}, Y.
  Zhuang}, ``{Towards a more complete and accurate experimental nuclear
  reaction data library {(EXFOR)}: International Collaboration Between Nuclear
  Reaction Data Centres {(NRDC)}},'' {\sc \protect\JournalTitle{Nucl. Data
  Sheets}} \textbf{120},  272 (2014).

\bibitem{Bjornstad1981}
T.~Bjornstad, {H.A. Gustafsson}, B.~Jonson, {P.O. Larsson}, V.~Lindfors,
  S.~Mattsson, G.~Nyman, A.~Poskanzer, H.~Ravn, and D.Schardt, ``The decay of
  $^8${H}e,'' {\sc \protect\JournalTitle{Nucl. Phys. A}} \textbf{366},  461
  (1981).

\bibitem{Nyman1990}
G.~Nyman, {R.E. Azuma}, {P.G. Hansen},\textit{  {et~al.}}, ``The beta decay of
  $^9${L}i to levels in $^9${B}e: {A} new look,'' {\sc
  \protect\JournalTitle{Nucl. Phys. A}} \textbf{510},  189 (1990).

\bibitem{Hirayama2015}
Y.~Hirayama, T.~Shimoda, H.~Miyatake, H.~Izumi, A.~Hatakeyama, K.~Jackson,
  C.~Levy, M.~Pearson, M.~Yagi, and H.~Yano, ``Unexpected spin-parity
  assignments of the excited states in $^9${B}e,'' {\sc
  \protect\JournalTitle{Phys. Rev. C}} \textbf{91},  024308 (2015).

\bibitem{Aoi1997}
N.~Aoi, K.~Yoneda, H.~Miyatake,\textit{  {et~al.}}, ``$\beta$-delayed neutron
  decay of drip line nuclei $^{11}${L}i and $^{14}${B}e,'' {\sc
  \protect\JournalTitle{Zeit. Physik}} \textbf{358},  253 (1997).

\bibitem{Hirayama2004}
Y.~Hirayama, T.~Shimoda, H.~Izumi, H.~Yano, M.~Yagi, A.~Hatakeyama, C.~Levy,
  K.~Jackson, and H.~Miyatake, ``Structure of $^{11}${B}e studied in
  $\beta$-delayed neutron- and $\gamma$- decay from polarized $^{11}${L}i,''
  {\sc \protect\JournalTitle{Nucl. Phys. A}} \textbf{746},  71c (2004).

\bibitem{Grevy2001}
S.~Grevy, {N.L. Achouri}, {J.C. Angelique},\textit{  {et~al.}}, ``Observation
  of a new transition in the $\beta$-delayed neutron decay of $^{16}${C},''
  {\sc \protect\JournalTitle{Phys. Rev. C}} \textbf{63},  037302 (2001).

\bibitem{Yamamoto1997}
Y.~Yamamoto, S.~Tanimoto, H.~Miyatake,\textit{  {et~al.}} Tech. Rep. Annual
  Report 1996, p. 25, Osaka University Lab. Nuclear Studies, Osaka, Japan
  (1997).

\bibitem{Lou2007}
{J.-L. Lou}, {Z.H. Li}, {Y.L. Ye},\textit{  {et~al.}}, ``Observation of a new
  transition in the $\beta$-delayed neutron decay of $^{18}${N},'' {\sc
  \protect\JournalTitle{Phys. Rev. C}} \textbf{75},  057302 (2007).

\bibitem{Lou2008}
{J.-L. Lou}, {Z.-H. Li}, {Y.-L. Ye},\textit{  {et~al.}}, ``The decay of
  $^{21}${N},'' {\sc \protect\JournalTitle{Chin. Phys. Lett.}} \textbf{25},
  1992 (2008).

\bibitem{Li2009}
{Z.H. Li}, {J.-L. Lou}, {Y.L. Ye},\textit{  {et~al.}}, ``Experimental study of
  the $\beta$ -delayed neutron decay of $^{21}${N},'' {\sc
  \protect\JournalTitle{Phys. Rev. C}} \textbf{80},  054315 (2009).

\bibitem{Ziegert1981}
W.~Ziegert, {L.C. Carraz}, {P.G. Hansen}, B.~Jonson, K.-L. Kratz, G.~Nyman,
  H.Ohm, H.~Ravn, and A.~Schroder, ``{Investigation of the beta strength
  function at high energy: gamma-ray spectroscopy of the decay of 5.3-s
  $^{84}$As to $^{84}$Se},'' in \textit{ Proc. 4th International Conference on
  Nuclei Far from Stability, Helsingor/ Denmark}, CERN Yellow Reports:
  Conference Proceedings (CERN-81-09) (1981).

\bibitem{Tengblad1987}
O.~Tengblad, K.-H. Beimer, and G.~Nyman, ``Influence of neutron scattering in
  the walls of a {3He} spectrometer,'' {\sc \protect\JournalTitle{Nuclear
  Instruments and Methods in Physics Research Section A: Accelerators,
  Spectrometers, Detectors and Associated Equipment}} \textbf{258},  230
  (1987).

\bibitem{Moller2003}
P.~M{\"o}ller, B.~Pfeiffer, and K.-L. Kratz, ``New calculations of gross
  b-decay properties for astrophysical applications: Speeding-up the classical
  r process,'' {\sc \protect\JournalTitle{Phys. Rev. C}} \textbf{67},  055802
  (2003).

\bibitem{Moller2019}
{ P. M\"oller, M.R. Mumpower, T. Kawano, W.D. Myers}, ``Nuclear properties for
  astrophysical and radioactive-ion-beam applications (ii),'' {\sc
  \protect\JournalTitle{At. Data N. Data Tables}} \textbf{125},  1 (2019).

\bibitem{Fayans2000}
{S.A. Fayans, S.V. Tolokonnikov, E.L. Trykov and D. Zawischa}, ``Nuclear
  isotope shifts within the local energy-density functional approach,'' {\sc
  \protect\JournalTitle{Nuclear Physics A}} \textbf{676},  49 (2000).

\bibitem{Borzov1996}
{I.N. Borzov, S.A. Fayans, E. Kromer, D. Zawischa}, ``Ground state properties
  and $\beta$-decay half-lives near $^{132}${S}n in a self-consistent theory,''
  {\sc \protect\JournalTitle{Z. Phys. A}} \textbf{355},  177 (1996).

\bibitem{Borzov2003}
{I.N. Borzov}, ``{Gamow-Teller and first-forbidden decays near the r-process
  paths at N=50, 82, and 126},'' {\sc \protect\JournalTitle{Physical Review C}}
  \textbf{67},  025802 (2003).

\bibitem{Borzov2005}
{I.N. Borzov}, ``{$\beta$-delayed neutron emission in the 78Ni region},'' {\sc
  \protect\JournalTitle{Physical Review C}} \textbf{71},  065801 (2005).

\bibitem{Marketin2016}
T.~Marketin, L.~Huther, and G.~Mart\'{\i}nez-Pinedo, ``Large-scale evaluation
  of $\ensuremath{\beta}$-decay rates of $r$-process nuclei with the inclusion
  of first-forbidden transitions,'' {\sc \protect\JournalTitle{Phys. Rev. C}}
  \textbf{93},  025805 (2016).

\bibitem{Marketin2007}
T.~Marketin, D.~Vretenar, and P.~Ring, ``Calculation of $\beta$-decay rates in
  a relativistic model with momentum-dependent self-energies,'' {\sc
  \protect\JournalTitle{Physical Review C}} \textbf{75},  024304 (2007).

\bibitem{Moeller2016}
P.~M{\"o}ller, A.~J. Sierk, T.~Ichikawa, and H.~Sagawa, ``Nuclear ground-state
  masses and deformations: {FRDM}(2012),'' {\sc \protect\JournalTitle{Atomic
  Data and Nuclear Data Tables}} \textbf{109-110},  1 (2016).

\bibitem{Nakatsukasa2007}
T.~Nakatsukasa, T.~Inakura, and K.~Yabana, ``Finite amplitude method for the
  solution of the random-phase approximation,'' {\sc
  \protect\JournalTitle{Physical Review C}} \textbf{76},  024318 (2007).

\bibitem{Mustonen2014}
{M.T. Mustonen, T. Shafer, Z. Zenginerler, and J. Engel}, ``Finite-amplitude
  method for charge-changing transitions in axially deformed nuclei,'' {\sc
  \protect\JournalTitle{Physical Review C}} \textbf{90},  024308 (2014).

\bibitem{Mustonen2016}
{M.T. Mustonen and J. Engel}, ``Global description of
  ${\ensuremath{\beta}}^{\ensuremath{-}}$ decay in even-even nuclei with the
  axially-deformed skyrme finite-amplitude method,'' {\sc
  \protect\JournalTitle{Physical Review C}} \textbf{93},  014304 (2016).

\bibitem{Shafer2016}
{T. Shafer, J. Engel, C. Fr\"ohlich, G.C. McLaughlin, M.R. Mumpower, and R.
  Surman}, ``{$\ensuremath{\beta}$ decay of deformed $r$-process nuclei near
  $A=80$ and $A=160$, including odd-$A$ and odd-odd nuclei, with the Skyrme
  finite-amplitude method},'' {\sc \protect\JournalTitle{Phys. Rev. C}}
  \textbf{94},  055802 (2016).

\bibitem{Bender2001}
M.~Bender, J.~Dobaczewski, J.~Engel, and W.~Nazarewicz, ``{Gamow-Teller}
  strength and the spin-isospin coupling constants of the {S}kyrme energy
  functional,'' {\sc \protect\JournalTitle{Phys. Rev. C}} \textbf{65},  054322
  (2001).

\bibitem{Zhi2013}
{Q. Zhi, E. Caurier, J.J. Cuenca-Garc\'{i}a, K. Langanke, G.
  Mart\'{i}nez-Pinedo, and K. Sieja}, ``Shell-model half-lives including
  first-forbidden contributions for $r$-process waiting-point nuclei,'' {\sc
  \protect\JournalTitle{Physical Review C}} \textbf{87},  025803 (2013).

\bibitem{Yoshida2018}
S.~Yoshida, ``Systematic shell-model {A = }40,'' {\sc
  \protect\JournalTitle{Phys. Rev. C}} \textbf{97},  054321 (2018).

\bibitem{Severyukhin2014}
{A.P. Severyukhin, V.V. Voronov, I.N. Borzov, N.N. Arsenyev, and N. Van Giai},
  ``Influence of 2p-2h configurations on $\beta$-decay rates,'' {\sc
  \protect\JournalTitle{Physical Review C}} \textbf{90},  044320 (2014).

\bibitem{Severyukhin2017}
{A.P. Severyukhin, N.N. Arsenyev, I.N. Borzov and E.O. Sushenok},
  ``{Multi-neutron emission of Cd isotopes},'' {\sc \protect\JournalTitle{Phys.
  Rev. C}} \textbf{95},  034314 (2017).

\bibitem{Sushenok2018}
{E.O. Sushenok, A.P. Severyukhin, N.N. Arsenyev, I.N. Borzov}, ``{Impact of
  Tensor Interaction on Beta-Delayed Neutron Emission from Neutron-Rich Nickel
  Isotopes},'' {\sc \protect\JournalTitle{Physics of Atomic Nuclei}}
  \textbf{81},  24 (2018).

\bibitem{Niu2015}
{Y.F. Niu, Z.M. Niu, G. Col\`{o}, and E. Vigezzi}, ``{Particle-Vibration
  Coupling Effect on the $\beta$ Decay of Magic Nuclei},'' {\sc
  \protect\JournalTitle{Physical Review Letters}} \textbf{114},  142501 (2015).

\bibitem{Niu2018}
{Y.F. Niu, Z.M. Niu, G. Col\`{o}, E. Vigezzi}, ``Interplay of
  quasiparticle-vibration coupling and pairing correlations on $\beta$-decay
  half-lives,'' {\sc \protect\JournalTitle{Physics Letters B}} \textbf{780},
  325 (2018).

\bibitem{Robin2016}
C.~Robin and E.~Litvinova, ``{Nuclear response theory for spin-isospin
  excitations in a relativistic quasiparticle-phonon coupling framework},''
  {\sc \protect\JournalTitle{European Physical Journal A}} \textbf{52},  205
  (2016).

\bibitem{Borzov2000}
{I.N. Borzov and S. Goriely}, ``{Weak interaction rates of neutron rich nuclei
  and the r process nucleosynthesis},'' {\sc \protect\JournalTitle{Phys. Rev.}}
  \textbf{C62},  035501 (2000).

\bibitem{Minato2016}
F.~Minato and O.~Iwamoto, ``Calculation of beta decay half-lives and delayed
  neutron branching ratio of fission fragments with {S}kyrme-{QRPA},'' {\sc
  \protect\JournalTitle{EPJ Web of Conferences}} \textbf{122},  10001 (2016).

\bibitem{Yoshida2013}
K.~Yoshida, ``{Spin-isospin response of deformed neutron-rich nuclei in a
  self-consistent Skyrme energy-density-functional approach},'' {\sc
  \protect\JournalTitle{PTEP}} \textbf{2013},  113D02 (2013).

\bibitem{Yoshida2019}
K.~Yoshida, ``Suddenly shortened half-lives beyond $^{78}\mathrm{Ni}:{N=50}$
  magic number and high-energy nonunique first-forbidden transitions,'' {\sc
  \protect\JournalTitle{Phys. Rev. C}} \textbf{100},  024316 (2019).

\bibitem{Sarriguren2014}
P.~Sarriguren, A.~Algora, and J.~Pereira, ``{Gamow-Teller response in deformed
  even and odd neutron-rich Zr and Mo isotopes},'' {\sc
  \protect\JournalTitle{Phys. Rev. C}} \textbf{89},  034311 (2014).

\bibitem{Martini2014}
M.~Martini, S.~P\'eru, and S.~Goriely, ``Gamow-{T}eller strength in deformed
  nuclei within the self-consistent charge-exchange quasiparticle random-phase
  approximation with the {G}ogny force,'' {\sc \protect\JournalTitle{Physical
  Review C}} \textbf{89},  044306 (2014).

\bibitem{Borzov2020}
{I.N. Borzov}, ``{Global calculations of the beta decay properties with the
  Fayans functional},'' {\sc \protect\JournalTitle{Phys. Atom. Nucl.}}
  \textbf{83},  24 (2020).
\newblock [Yad. Fiz.83,no.5,(2020)].

\bibitem{Borzov2006}
{I.N. Borzov}, ``{Beta-decay rates},'' {\sc \protect\JournalTitle{Nuclear
  physics A}} \textbf{777},  645 (2006).

\bibitem{Borzov2017}
{I.N. Borzov}, ``Self-consistent approach to beta decay and delayed neutron
  emission,'' {\sc \protect\JournalTitle{Physics of Atomic Nuclei}}
  \textbf{79},  910 (2017).

\bibitem{Migdal1967}
{A.B. Migdal}, \textit{ Theory of Finite Fermi Systems: And Applications to
  Atomic Nuclei}.
\newblock Interscience Publishers (1967).

\bibitem{Tolokonnikov2010}
{S.V. Tolokonnikov and E.E. Saperstein}, ``{Description of superheavy nuclei
  with a modified version of the energy density functional DF3},'' {\sc
  \protect\JournalTitle{Physics of Atomic Nuclei}} \textbf{73},  1684 (2010).

\bibitem{Madurga2012}
M.~Madurga, R.~Surman, I.~N. Borzov,\textit{  {et~al.}}, ``{New Half-lives of
  $r$-process Zn and Ga Isotopes Measured with Electromagnetic Separation},''
  {\sc \protect\JournalTitle{Phys. Rev. Lett.}} \textbf{109},  112501 (2012).

\bibitem{Warburton1991}
{E.K. Warburton}, ``First-forbidden $\beta$-decay in the lead region and
  mesonic enhancement of the weak axial current,'' {\sc
  \protect\JournalTitle{Physical Review C}} \textbf{44},  233 (1991).

\bibitem{BertschBroglia1994}
{G.F. Bertsch and R.A. Broglia}, \textit{ Oscillations in Finite Quantum
  Systems}.
\newblock New York: Cambridge Univ. Press (1994).

\bibitem{Borzov2018}
{I.N. Borzov}, ``{Delayed Multineutron Emission in the Region of Heavy Calcium
  Isotopes},'' {\sc \protect\JournalTitle{Physics of Atomic Nuclei}}
  \textbf{81},  680 (2018).

\bibitem{Engel1999}
J.~Engel, M.~Bender, J.~Dobaczewski, W.~Nazarewicz, and R.~Surman, ``Beta-decay
  rates of r-process waiting-point nuclei in a self-consistent approach,'' {\sc
  \protect\JournalTitle{Physical Review C}} \textbf{60},  014302 (1999).

\bibitem{Stone2014}
N.~Stone, ``{Table of Nuclear Magnetic Dipole and Electric Quadrupole
  Moments},'' {\sc \protect\JournalTitle{International Atomic Energy Agency
  (IAEA)}} \textbf{INDC(NDS)--0658} (2014).

\bibitem{Moeller1995}
P.~M{\"o}ller, J.~R. Nix, W.~D. Myers, and W.~J. Swiatecki, ``{Nuclear
  Ground-State Masses and Deformations},'' {\sc \protect\JournalTitle{Atomic
  Data and Nuclear Data Tables}} \textbf{59},  185 (1995).

\bibitem{Wilson02}
{W. Wilson, T. England}, ``{Delayed Neutron Study Using {ENDF/B-VI} Basic
  Nuclear Data},'' {\sc \protect\JournalTitle{Prog. Nucl. Energy}} \textbf{41},
   71 (2002).

\bibitem{Sarriguren2010}
P.~Sarriguren and J.~Pereira, ``{$\beta$-decay properties of neutron-rich Zr
  and Mo isotopes},'' {\sc \protect\JournalTitle{Phys. Rev. C}} \textbf{81},
  064314 (2010).

\bibitem{Wu2020}
J.~Wu, S.~Nishimura, P.~M\"oller,\textit{  {et~al.}},
  ``$\ensuremath{\beta}$-decay half-lives of 55 neutron-rich isotopes beyond
  the $n=82$ shell gap,'' {\sc \protect\JournalTitle{Phys. Rev. C}}
  \textbf{101},  042801 (2020).

\bibitem{Moon2017}
B.~Moon, C.-B. Moon, P.-A. S\"oderstr\"om,\textit{  {et~al.}}, ``Nuclear
  structure and $\ensuremath{\beta}$-decay schemes for {T}e nuclides beyond
  {N=82},'' {\sc \protect\JournalTitle{Phys. Rev. C}} \textbf{95},  044322
  (2017).

\bibitem{Xu2014}
Z.~Y. Xu, S.~Nishimura, G.~Lorusso,\textit{  {et~al.}},
  ``$\ensuremath{\beta}$-decay half-lives of $^{76,77}\mathrm{Co}$,
  $^{79,80}\mathrm{Ni}$, and $^{81}\mathrm{Cu}$: Experimental indication of a
  doubly magic $^{78}\mathrm{Ni}$,'' {\sc \protect\JournalTitle{Phys. Rev.
  Lett.}} \textbf{113},  032505 (2014).

\bibitem{Colo1998}
G.~Col\`o, H.~Sagawa, N.~Van~Giai, P.~F. Bortignon, and T.~Suzuki, ``Widths of
  isobaric analog resonances: A microscopic approach,'' {\sc
  \protect\JournalTitle{Phys. Rev. C}} \textbf{57},  3049 (1998).

\bibitem{Alshudifat2016}
{M.F. Alshudifat,}\textit{  {et~al.}}, ``{Reexamining Gamow-Teller decays near
  $^{78}${N}i},'' {\sc \protect\JournalTitle{Phys. Rev.}} \textbf{C93},  044325
  (2016).
\newblock [Addendum: Phys. Rev.C93,no.5,059903(2016)].

\bibitem{Lorusso2015}
G.~Lorusso, S.~Nishimura, Z.~Y. Xu,\textit{  {et~al.}}, ``$\beta$-decay
  half-lives of 110 neutron-rich nuclei across the {N=82} shell gap:
  Implications for the mechanism and universality of the astrophysical $r$
  process,'' {\sc \protect\JournalTitle{Physical Review Letters}} \textbf{114},
   192501 (2015).

\bibitem{Morales2015}
{A.I. Morales, \textit{et al.}}, ``{First measurement of the $\beta$-decay
  half-life of $^{206}$Au},'' {\sc \protect\JournalTitle{Europhysical Letters}}
  \textbf{111},  5 (2015).

\bibitem{CaballeroFolch2017}
R.~Caballero-Folch, I.~Dillmann, J.~Agramunt,\textit{  {et~al.}}, ``{First
  evidence of multiple $\beta$-delayed neutron emission for isotopes with A
  \textgreater 100},'' {\sc \protect\JournalTitle{Acta Physica Polonica B}}
  \textbf{48},  517 (2017).

\bibitem{Borzov2011}
{I.N. Borzov}, ``{Beta-decay of nuclei near the neutron shell N=126},'' {\sc
  \protect\JournalTitle{Physics of Atomic Nuclei}} \textbf{74},  1435 (2011).

\bibitem{Moeller1990}
P.~M\"{o}ller and J.~Randrup, ``New developments in the calculation of beta
  strength junctions,'' {\sc \protect\JournalTitle{Nuclear Physics A}}
  \textbf{514},  1 (1990).

\bibitem{Koning2003}
A.~Koning and J.-P. Delaroche, ``Local and global nucleon optical models from 1
  kev to 200 {M}ev,'' {\sc \protect\JournalTitle{Nuclear Physics A}}
  \textbf{713},  231 (2003).

\bibitem{Kopecky1993}
J.~Kopecky, M.~Uhl, and R.~Chrien, ``{Radiative strength in the compound
  nucleus $^{157}$Gd},'' {\sc \protect\JournalTitle{Phys. Rev. C}} \textbf{47},
   312 (1993).

\bibitem{Belgya2006}
{T. Belgya, O. Bersillon, R. Capote, T. Fukahori, G. Zhigang, S. Goriely, M.
  Herman, A.V. Ignatyuk, S. Kailas, A. Koning, P. Oblozinsky, V. Plujko and P.
  Young}, ``Handbook for calculations of nuclear reaction data,'' Tech. Rep.
  IAEA-TECDOC-1506, {IAEA, Vienna} (2006).

\bibitem{Gilbert1965}
A.~Gilbert and {A.G.W. Cameron}, ``A composite nuclear-level density formula
  with shell corrections,'' {\sc \protect\JournalTitle{Canadian journal of
  Physics}} \textbf{43},  1446 (1965).

\bibitem{SkO'}
P.-G. Reinhard, D.~J. Dean, W.~Nazarewicz, J.~Dobaczewski, J.~A. Maruhn, and
  M.~R. Strayer, ``{Shape coexistence and the effective nucleon-nucleon
  interaction},'' {\sc \protect\JournalTitle{Phys. Rev. C}} \textbf{60},
  014316 (1999).

\bibitem{Iwamoto2013}
O.~Iwamoto, ``{Extension of a nuclear reaction calculation code CCONE toward
  higher incident energies - multiple preequilibrium emission, and spectrum in
  laboratory system},'' {\sc \protect\JournalTitle{Journal of Nuclear Science
  and Technology}} \textbf{50},  409 (2013).

\bibitem{Capote2009}
R.~Capote, M.~Herman, P.~Oblozinsky,\textit{  {et~al.}}, ``Ripl – reference
  input parameter library for calculation of nuclear reactions and nuclear data
  evaluations,'' {\sc \protect\JournalTitle{Nuclear Data Sheets}} \textbf{118},
   396  (2014).

\bibitem{Mengoni1994}
A.~Mengoni and Y.~Nakajima, ``Fermi-gas model parametrization of nuclear level
  density,'' {\sc \protect\JournalTitle{Journal of Nuclear Science and
  Technology}} \textbf{31},  151 (1994).

\bibitem{Waldo81}
{R.W. Waldo, R.A. Karam, R.A. Meyer}, ``{Delayed Neutron Yields: Time Dependent
  Measurements and a Predictive Model},'' {\sc \protect\JournalTitle{Phys. Rev.
  C}} \textbf{23},  1113 (1981).

\bibitem{Saleh97}
{H.H. Saleh, T.A. Parish, S. Raman and N. Shinohara}, ``Measurements of
  delayed-neutron decay constants and fission yields from $^{235}${U},
  $^{237}${Np}, $^{241}${Am} and $^{243}${Am},'' {\sc
  \protect\JournalTitle{Nucl. Sci. Eng.}} \textbf{125},  51 (1997).

\bibitem{Charlton97}
{W.S. Charlton, T.A. Parish, S. Raman, N. Shinohara, and M. Andoh}, ``{Delayed
  Neutron Emission Measurements for Fast Fission of $^{235}$U, $^{237}$Np, and
  $^{243}$Am},''
\newblock Proc. Int. Conf. Nucl. Data for Sci. and Tech., Italian Physical
  Society, Trieste, Italy, p. 491 (May 19-24 1997).

\bibitem{Besant77}
{C.B. Besant, P.J. Challen, M.H. McTaggart, P. Tavoularidis, J.G. Williams},
  ``Absolute yields and group constants of delayed neutrons in fast fission of
  $^{235}${U}, $^{238}${U} and $^{239}${Pu},'' {\sc \protect\JournalTitle{J.
  Br. Nucl. Energy Soc.}} \textbf{16},  161 (1977).

\bibitem{Benedetti82}
{G. Benedetti, A. Cesana, V. Sangiust and M. Terrani}, ``Delayed neutron yields
  from fission of {U}ranium-233, {N}eptunium-237, {P}lutonium-238, -240, -241,
  and {A}mericium-241,'' {\sc \protect\JournalTitle{Nucl. Sci. Eng.}}
  \textbf{80},  379 (1982).

\bibitem{Synetos79}
{S. Synetos, J.G. Williams}, ``Delayed neutron yield and decay constants for
  thermal neutron induced fission of $^{235}${U},''
\newblock IAEA Report INDC(NDS)-107, Vienna, Austria, 1979.

\bibitem{Rose57}
{H. Rose, R.D. Smith}, ``Delayed neutron investigations with the zephyr fast
  reactor, part {II} - the delayed neutrons arising from fast fission in
  $^{235}${U}, $^{233}${U}, $^{238}${U}, $^{239}${Pu} and $^{232}${Th},'' {\sc
  \protect\JournalTitle{J. Nucl. Energy}} \textbf{4},  133 (1957).

\bibitem{Gudkov89}
{N.A. Gudkov}, {V.M. Zhivun}, {A.V. Zvonarev}, {V.V. Kovalenko}, {A.B.
  Koldobskii}, {Yu.F. Koleganov}, {S.V. Krivasheev}, {V.B. Pavlovich}, {N.S.
  Piven}, and {E.V. Semenova}, ``Measurements of the delayed neutron yields for
  the fission of $^{233}${U}, $^{236}${U}, $^{237}${Np}, $^{240}${Pu} and
  $^{241}${Pu} by fast neutron reactor spectrum,'' {\sc
  \protect\JournalTitle{Atomic Energy}} \textbf{66},  100 (1989).

\bibitem{Loaiza97}
{D.J. Loaiza, G. Brunson, R. Sanchez}, ``{Measurement of delayed neutron
  parameters for $^{235}$U},'' {\sc \protect\JournalTitle{Trans. Amer. Nucl.
  Soc.}} \textbf{76},  361 (1997).

\bibitem{Loaiza98}
{D.J. Loaiza, G. Brunson, R. Sanchez, K. Butterfield}, ``Measurements of
  absolute delayed neutron yield and group constants in the fast fission of
  $^{235}${U} and $^{237}${Np},'' {\sc \protect\JournalTitle{Nucl. Sci. Eng.}}
  \textbf{128},  270 (1998).

\bibitem{Masters69}
{C.F. Masters, \textit{et al.}}, ``The measurement of absolute delayed-neutron
  yields from 3.1- and 14.9-{MeV} fission,'' {\sc \protect\JournalTitle{Nucl.
  Sci. Eng.}} \textbf{36},  202 (1969).

\bibitem{Krick72}
{M.S. Krick and A.E. Evans}, ``{The Measurement of Total Delayed-Neutron Yields
  as a Function of the Energy of the Neutron Inducing Fission},'' {\sc
  \protect\JournalTitle{Nucl. Sci. Eng.}} \textbf{47},  311 (1972).

\bibitem{Cox68}
{S.A. Cox and E.D. Whiting}, ``Energy dependence of the delayed neutron yield
  from neutron induced fission of $^{232}${Th}, $^{235}${U} and $^{238}${U},''
\newblock Reactor Physics Division Annual Report ANL-7410, p. 27 (July 1, 1967
  to June 30, 1968).

\bibitem{Cox70}
{S.A. Cox and E.D. Whiting}, ``Energy dependence of the delayed neutron yield
  from neutron induced fission of $^{232}${Th}, $^{235}${U} and $^{238}${U},''
\newblock Report ANL-7610, p. 45 (1970).

\bibitem{Cox74}
{S.A. Cox}, ``{Delayed Neutron Data - Review and Evaluation},''
\newblock Rep. ANL/NDM-5, Argonne National Laboratory (1974).

\bibitem{Piksaikin06}
{V.M. Piksaikin, N.N. Semenova, V.I. Mil’shin, V.A. Roshchenko, G.G.
  Korolev}, ``Method and setup for studying the energy dependence of delayed
  neutron characteristics in nuclear fission induced by neutrons from the
  {T}$(p,n)$, {D}$(d,n)$, and {T}$(d,n)$ reactions,'' {\sc
  \protect\JournalTitle{Instrum Exp. Tech.}} \textbf{49},  765 (2006).

\bibitem{Fieg72}
G.~Fieg, ``Measurements of delayed fission neutron spectra of {$^{235}$U,
  $^{238}$U and $^{239}$Pu} with proton recoil proportional counters,'' {\sc
  \protect\JournalTitle{J. Nucl. Energy}} \textbf{26},  585 (1972).

\bibitem{Spriggs99}
{G.D. Spriggs, J.M. Campbell, V.M. Piksaikin}, ``An 8-group neutron model based
  on a consistent set of half-lives,''
\newblock {Report LA-UR-98-1619, Rev.3, Los Alamos National Laboratory (1999)}.

\bibitem{Spriggs02}
{G.D. Spriggs, J.M. Campbell, V.M. Piksaikin}, ``An 8-group delayed neutron
  model based on a consistent set of half-lives,'' {\sc
  \protect\JournalTitle{Prog. Nucl. Energy}} \textbf{41},  223 (2002).

\bibitem{Piksaikin02a}
{V.M. Piksaikin, L.E. Kazakov, S.G. Isaev, M.Z. Tarasko, V.A. Roshchenko, R.G.
  Tertytchnyi, G.D. Spriggs, J.M. Campbell}, ``Energy dependence of relative
  abundances and periods of delayed neutrons from neutron-induced fission of
  {$^{235}$U, $^{238}$U, $^{239}$Pu} in 6{-} and 8{-}group model
  representation,'' {\sc \protect\JournalTitle{Prog. Nucl. Energy}}
  \textbf{41},  203 (2002).

\bibitem{Piksaikin02b}
{V.M. Piksaikin, S.G. Isaev, and A.A. Goverdovski}, ``{Characteristics of
  Delayed Neutrons: Systematics and Correlation Properties},'' {\sc
  \protect\JournalTitle{Prog. Nucl. Energy}} \textbf{41},  361 (2002).

\bibitem{Piksaikin02c}
{V.M. Piksaikin, L.E. Kazakov, V.A. Roshchenko, S.G. Isaev, G.G. Korolev, A.A.
  Goverdovski, R.G. Tertytchnyi}, ``Experimental studies of the absolute total
  delayed neutron yields from neutron induced fission of $^{238}${U} in the
  energy range 1–5 {MeV},'' {\sc \protect\JournalTitle{Prog. Nucl. Energy}}
  \textbf{41},  135 (2002).

\bibitem{Roshchenko06}
{V.A. Roshchenko, V.M. Piksaikin, G.G. Korolev, Yu. F. Balakshev},
  ``Experimental studies of the absolute total delayed neutron yields from
  neutron induced fission of $^{236}${U} in the energy range 1-5 {MeV},''
\newblock Proc. XIVth Int. Seminar on Interaction of Neutrons with Nuclei, p.
  144 (24 May 2006), Dubna, Russia.

\bibitem{Piksaikin13}
{V.M. Piksaikin, A.S. Egorov, K.V. Mitrofanov}, ``The absolute total delayed
  neutron yields, relative abundances and half-life of delayed neutron groups
  from neutron induced fission of $^{232}${Th}, $^{233}${U}, $^{236}${U},
  $^{239}${Pu} and $^{241}${Am},''
\newblock {Report INDC(NDS)-0646, IAEA, Vienna, Austria (2013)}.

\bibitem{Maksyutenko58}
{B.P. Maksyutenko, \textit{et al.}}, ``Relative yields of delayed neutrons in
  fission of $^{238}${U}, $^{235}${U} and $^{232}${Th} by fast neutrons,'' {\sc
  \protect\JournalTitle{J. Exptl. Theoret. Phys. (USSR)}} \textbf{35},  815
  (1958).

\bibitem{Piksaikin02d}
{V.M. Piksaikin, \textit{et al.}}, ``Energy dependence of relative abundances
  and periods of delayed neutrons from fission of $^{239}${Pu} in the energy
  range of primary neutrons from 0.37 to 4.97 {MeV},'' {\sc
  \protect\JournalTitle{Atomic Energy}} \textbf{92},  233 (2002).

\bibitem{Piksaikin98}
{V.M. Piksaikin, Yu.F. Balakshev, S.G. Isaev, \textit{et al.}}, ``Measurement
  of the energy dependence of the parameters of delayed neutrons accompanying
  fission of $^{237}${Np} by fast neutrons,'' {\sc \protect\JournalTitle{Atomic
  Energy}} \textbf{85},  479 (1998).

\bibitem{Roshchenko06a}
{ V.A. Roshchenko, V.M. Piksaikin, L.E. Kazakov, and G. G. Korolev}, ``Relative
  yield of delayed neutrons and half-life of their precursor nuclei with
  fissioning of $^{239}$pu by 14.2 - 17.9 {MeV} neutrons,'' {\sc
  \protect\JournalTitle{Atomic Energy}} \textbf{101},  897 (2006).

\bibitem{Piksaikin02e}
{V.M. Piksaikin, L.E. Kazakov, \textit{et al.}}, ``Relative yield and period of
  individual groups of delayed neutrons in $^{233}${U}, $^{235}${U} and
  $^{239}${Pu} fission by epithermal neutrons,'' {\sc
  \protect\JournalTitle{Atomic Energy}} \textbf{92},  147 (2002).

\bibitem{Piksaikin97}
{V.M. Piksaikin, S.G. Isaev, \textit{et al.}}, ``Absolute calibration of
  neutron detector with $^{252}${Cf} neutron source, {M}onte {C}arlo
  calculations and activation technique,''
\newblock Proc. Int. Conf. on Nuclear Data for Science and Technology, Triest
  (19-24 May), p. 646 (1997).

\bibitem{Piksaikin99}
{V.M. Piksaikin, V.S. Shorin, R.G. Tertytchnyi}, ``Fission rate determination
  in delayed neutron emission measurements with {T}(p,n) and {D}(d,n)
  neutrons,''
\newblock {Report INDC(CCP)-0442, IAEA, Vienna, Austria (1999)}.

\bibitem{Piksaikin06a}
{V.M. Piksaikin, V.A. Roshchenko, G.G. Korolev}, ``A method for determining the
  intensity of concomitant neutron source {D}(d,n)$^{3}${He} when studying the
  characteristics of delayed neutrons from nuclear fission induced by neutrons
  from reactions {T}(d,n)$^{4}${He},'' {\sc \protect\JournalTitle{Instrum. Exp.
  Tech.}} \textbf{49},  43 (2006).

\bibitem{Piksaikin07}
{V.M. Piksaikin, V.A. Roshchenko, G.G. Korolev}, ``Relative yield of delayed
  neutrons and half-life of their precursor nuclei from $^{238}${U} fission by
  14.2-17.9 {M}e{V} neutrons,'' {\sc \protect\JournalTitle{Atomic Energy}}
  \textbf{102},  124 (2007).

\bibitem{Evans79}
{A.E. Evans and J.D. Brandenbergen}, ``High resolution fast neutron
  spectrometry without time-of-flight,'' {\sc \protect\JournalTitle{IEEE Trans.
  Nucl. Sci.}} \textbf{NS-26} (1979).

\bibitem{Tanczyn88}
{R.S. Tanczyn, Q. Sharfuddin, W.A. Schier, D.J. Pullen, M.H. Haghighi, L.
  Fisteag and G.P. Couchell}, ``Composite delayed neutron energy spectra for
  thermal fission of $^{235}${U},'' {\sc \protect\JournalTitle{Nucl. Sci.
  Eng.}} \textbf{94},  353 (1988).

\bibitem{Das94}
{S. Das}, ``Importance of delayed neutrons in nuclear research – a review,''
  {\sc \protect\JournalTitle{Prog. Nucl. Energy}} \textbf{28},  209 (1994).

\bibitem{Piksaikin17}
{V.M. Piksaikin, A.S. Egorov, A.A. Goverdovski, D.E. Gremyachkin, K.V.
  Mitrofanov}, ``High resolution measurements of time-dependent integral
  delayed neutron spectra from thermal neutron induced fission of
  $^{235}${U},'' {\sc \protect\JournalTitle{Ann. Nucl. Energy}} \textbf{102},
  408 (2017).

\bibitem{Vilani92}
{M.F. Vilani, G.P. Couchell, M.H. Haghighi, D.J. Pullen, W.A. Schier, Q.
  Sharfuddin}, ``{Six-Group Decomposition of Composite Delayed Neutron Spectra
  from $^{235}${U} Fission},'' {\sc \protect\JournalTitle{Nucl. Sci. Eng.}}
  \textbf{111},  422 (1992).

\bibitem{Santamarina12}
{A. Santamarina, P. Blaise, L. Erradi, P. Fougeras}, ``Calculation of {LWR}
  $\beta$eff kinetic parameters: Validation on the {MISTRAL} experimental
  program,'' {\sc \protect\JournalTitle{Ann. Nucl. Energy}} \textbf{48},  51
  (2012).

\bibitem{DAngelo02}
{A. D'Angelo}, ``Overview of the delayed neutron data activities and results
  monitored by the {NEA/WPEC} {S}ubgroup 6,'' {\sc \protect\JournalTitle{Prog.
  Nucl. Energy}} \textbf{41},  5 (2002).

\bibitem{Egorov17}
{A.S. Egorov, V.M. Piksaikin, A.A. Goverdovski, \textit{et al.}}, ``Measurement
  of the delayed neutron characteristics in the interaction of relativistic
  protons with $^{238}${U} nuclei,'' {\sc \protect\JournalTitle{Prog Nucl.
  Energy}} \textbf{97},  106 (2017).

\bibitem{Spriggs02a}
{G.D. Spriggs and J.M. Campbell}, ``Summary of measured delayed neutron
  parameters,'' {\sc \protect\JournalTitle{Prog Nucl. Energy}} \textbf{41},
  145 (2002).

\bibitem{Tuttle75}
{R.J. Tuttle}, ``Delayed neutron data for reactor physics analysis,'' {\sc
  \protect\JournalTitle{Nucl. Sci. Eng.}} \textbf{56},  37 (1975).

\bibitem{Piksaikin04}
{V.M. Piksaikin, V.A. Roshchenko, S.G. Isaev, L.E. Kazakov, G.G. Korolev, Yu.F.
  Balakshev, A.A. Goverdovsky}, ``Cumulative yields and average half-life of
  delayed neutron precursors from neutron induced fission of $^{233}${U},''
\newblock {Proc. XIIth Int. Seminar on Interaction of Neutrons with Nuclei
  Neutron Spectroscopy, Nuclear Structure, Related Topics, Dubna, Russia, p.
  342 (24-28 May 2004)}.

\bibitem{Isaev98}
{S.G. Isaev, V.M. Piksaikin, L.E. Kazakov, M.Z. Tarasko}, ``Energy dependence
  of average half-life of delayed neutron precursors in fast neutron induced
  fission of $^{235}${U} and $^{236}${U},''
\newblock {Proc. XIVth Int. Workshop on Nuclear Fission Physics, Obninsk,
  Russia, p. 257 (12-15 October 1998)}.

\bibitem{Piksaikin11}
{V.M. Piksaikin, A.S. Egorov, K.V. Mitrofanov, A.A. Goverdovski}, ``Relative
  abundances and half-lives of their precursors for fission of $^{241}${A}m
  nucleus by neutrons in the energy range from 1 to 5 {M}e{V},''
\newblock {Proc. of XIIth Int. Seminar on Interaction of Neutrons with Nuclei
  ``Neutron Spectroscopy, Nuclear Structure, Related Topics", Dubna, Russia, p.
  62 (24-28 May, 2011)}.

\bibitem{Gremyachkin17}
{D.E. Gremyachkin, V.M. Piksaikin, K.V. Mitrofanov, A.S. Egorov},
  ``Measurements of temporal characteristics of delayed neutrons from neutron
  induced fission of $^{237}${N}p in energy range from 14.2 to 18 {M}e{V},''
  {\sc \protect\JournalTitle{EPJ Web of Conferences}} \textbf{146},  04059
  (2017).

\bibitem{Gremyachkin17a}
{D.E. Gremyachkin, V.M. Piksaikin, A.S. Egorov, K.V. Mitrofanov}, ``Measurement
  of the temporal characteristics of delayed neutrons from neutron induced
  fission of $^{241}${Am} in the energy range from 14.2 to 18 {M}e{V},''
\newblock {to be published in Proceedings of the seminar ISINN-25, Dubna,
  Russian Federation 2017}.

\bibitem{Roshchenko10}
{V.A. Roshchenko, V.M. Piksaikin, \textit{et al.}}, ``Temporary characteristics
  of delayed neutrons and partial cross sections of emissive fission in fission
  of $^{232}${Th} by neutrons in the energy range from 3.2-17.9 {MeV},'' {\sc
  \protect\JournalTitle{Physics of At. Nuclei}} \textbf{73},  913 (2010).

\bibitem{Piksaikin97a}
{V.M. Piksaikin}, J.~Balakshev, S.~Isaev, L.~Kazakov, G.~Korolev, B.~Kuzminov,
  N.~Sergachev, and M.~Tarasko, ``Measurements of periods, relative abundances
  and absolute total yields of delayed neutrons from fast neutron induced
  fission of $^{235}${U} and $^{237}${Np},'' in \textit{ Proceedings of
  Conference on Nuclear Data for Science and Technology, Trieste, Italy, vol.
  59, p. 485} (1997).

\bibitem{Piksaikin99a}
{V.M. Piksaikin, S.G. Isaev, L.E. Kazakov, G.G. Korolev, B.D. Kuzminov, V.G.
  Pronyaev}, ``Features of the energy dependence of total delayed neutron
  yields for fast neutron induced fission of the $^{235}${U} and
  $^{237}${Np},'' {\sc \protect\JournalTitle{Phys. Atomic Nuclei}} \textbf{62},
   1279 (1999).
\newblock {\textit{Yad. Fiz.}, No. 6, p.1 (1999) (in Russian)}.

\bibitem{Shibata2011}
{K. Shibata, O. Iwamoto, T. Nakagawa, N. Iwamoto, A. Ichihara, S. Kunieda, S.
  Chiba, K. Furutaka, N. Otuka, T. Ohsawa, T. Murata, H. Matsunobu, A. Zukeran,
  S. Kamada, and J. Katakura}, ``{JENDL-4.0:} a new library for nuclear science
  and engineering,'' {\sc \protect\JournalTitle{Nucl. Sci. Technology}}
  \textbf{48},  1 (2011).

\bibitem{Piksaikin12}
{V.M. Piksaikin, A.S. Egorov, K.V. Mitrofanov, A.A. Goverdovski}, ``Total
  delayed neutron yields for neutron induced fission of $^{232}${Th} in energy
  range 3.2-4.9 {M}e{V},'' {\sc \protect\JournalTitle{Atomic Energy}}
  \textbf{112},  350 (2012).

\bibitem{Gremyachkin15}
{D.E. Gremyachkin, V.M. Piksaikin, K.V. Mitrofanov, A.S. Egorov},
  ``Verification of the evaluated fission product yields data from the neutron
  induced fission of $^{235}${U}, $^{238}${U} and $^{239}${Pu} based on the
  delayed neutron characteristics,'' {\sc \protect\JournalTitle{Prog. Nucl.
  Energy}} \textbf{83},  13 (2015).

\bibitem{dAngelo2002a}
{A. D'Angelo, J.L. Rowlands}, ``A review of delayed neutron data for
  calculating effective delayed neutron fractions,''
\newblock {PHYSOR 2002, Seoul, Korea, October 7-10 (2002)}.

\bibitem{Foligno19}
D.~Foligno, \textit{ New evaluation of delayed-neutron data and associated
  covariances}.
\newblock PhD thesis, Université Aix-Marseille (2019).

\bibitem{Tuttle79}
{R.J. Tuttle}, ``{Review of Delayed Neutron Yields in Nuclear Fission},''
\newblock {in: Proc. Consultants' Mtg. on Delayed Neutron Properties, Report
  INDC(NDS)-107, IAEA, Vienna, Austria, p. 29 (August 1979)}.

\bibitem{Manero72}
{F. Manero and V.A. Konshin}, ``{Status of the energy dependent
  $\bar{\nu}$-values for the heavy isotopes ($Z\ge90$) from thermal to 15 MeV,
  and of $\bar{\nu}$-values for spontaneous fission},''
\newblock INDC(NDS)-34/G, 10, 1 (1972).

\bibitem{Minato18}
{F. Minato}, ``Neutron energy dependence of delayed neutron yields and its
  assessments,'' {\sc \protect\JournalTitle{J. Nucl. Sci. Technol.}} (2018).

\bibitem{Nethaway74}
{D.R. Nethaway}, ``{Tables of Values of $Z_p$, The Most Probable Charge in
  Fission},'' (1974).
\newblock UCRL-51640, Lawrence Livermore Laboratory, 1974.

\bibitem{Roshchenko06b}
{V.A. Roshchenko, V.M. Piksaikin, S.G. Isaev, A.A. Goverdovski}, ``Energy
  dependence of nuclear charge distribution in neutron induced fission of
  $z$-even nuclei,'' {\sc \protect\JournalTitle{Phys. Rev. C}} \textbf{74},
  014607 (2006).

\bibitem{England94}
{T.R. England and B.F. Rider}, ``{Evaluation and Compilation of Fission Product
  Yields},''

\bibitem{Madland76}
{D.G. Madland and T.R. England}, ``{Distribution of Independent Fission-Product
  Yields to Isomeric States},''

\bibitem{Bobkov89}
{E.Yu. Bobkov}, {A.N. Gudkov}, {A.N. Dyumin}, {A.B. Koldobsky}, {M.Ya.
  Kondratko}, {S.V. Krivasheev}, {A.V. Mosesov}, {L.M. Nikitin}, {V.A. Smolin},
  and {A.A. Solonkin}, ``Measurement of delayed-neutron group yields following
  the fission of $^{235}${U},$^{236}${U},$^{238}${U},$^{237}${Np},$^{242}${Pu}
  by $14.7$ mev neutrons,'' {\sc \protect\JournalTitle{Atomnaya Energiya}}
  \textbf{67},  408 (1989).

\bibitem{Cesana80}
{A. Cesana, G. Sandrelli, V. Sangiust, M. Terrani}, ``Absolute total yields of
  delayed neutrons in the fission of {U}-233, {Np}-237, {Pu}-238, 240, 241,
  {Am}-241,'' {\sc \protect\JournalTitle{Energia Nucleare}} \textbf{26},  542
  (1979).

\bibitem{Cesana80b}
{G. Lammer and M. Lammer}, ``Progress in fission product nuclear data,''
\newblock IAEA Report INDC(NDS)-113/G+P, Vienna, Austria, 1980.

\bibitem{Darwin}
L.~{San-Felice}, R.~Eschbach, and P.~Bourdot, ``{Experimental Validation of the
  {DARWIN2.3} Package for Fuel Cycle Applications},'' {\sc
  \protect\JournalTitle{Nuclear Technology}} \textbf{184},  217 (2013).

\bibitem{MURE}
{O. Meplan, \textit{et al.}}, ``{MURE}, {MCNP} {U}tility for {R}eactor
  {E}volution,'' Tech. Rep. 0912, CNRS/in2p3/LPSC, Grenoble, France (2009).
\newblock Also at \url{http://www.nea.fr/tools/abstract/detail/nea-1845}.

\bibitem{Conrad}
P.~Archier, C.~{De Saint Jean}, O.~Litaize, G.~Noguere, L.~Berge, E.~Privas,
  and P.~Tamagno, ``{CONRAD Evaluation Code: Development Status and
  Perspectives},'' {\sc \protect\JournalTitle{Nuclear Data Sheets}}
  \textbf{118},  448 (2014).

\bibitem{ICSBEP}
``{International Handbook of Evaluated Criticality Safety Benchmark
  Experiments},'' Tech. Rep. NEA/NSC/DOC(95)03, NEA/OECD (2016).

\bibitem{MCNP611b}
{D.B. Pelowitz}, ``{MCNP} {U}ser's {M}anual, code version 6.1.1beta,'' Tech.
  Rep. LA-CP-14-00745, Los Alamos National Laboratory (2014).

\bibitem{Briggs2014}
{J.B. Briggs and J.D. Bess and J. Gulliford}, ``Integral benchmark data for
  nuclear data testing through the icsbep \& irphep,'' {\sc
  \protect\JournalTitle{Nuclear Data Sheets}} \textbf{118},  396  (2014).

\bibitem{Keepin1965}
{G.R. Keepin}, \textit{ Physics of Nuclear Kinetics}.
\newblock Addison-Wesley series in nuclear science and engineering,
  Addison-Wesley Publishing Company (1965).

\bibitem{DAngelo1993}
A.~D'Angelo and A.~Filip, ``The effective beta sensitivity to the incident
  neutron energy dependence of the absolute delayed neutron yields,'' {\sc
  \protect\JournalTitle{Nuclear Science and Engineering}} \textbf{114},  332
  (1993).

\bibitem{Romojaro2019}
P.~Romojaro, F.~Álvarez Velarde, and N.~García-Herranz, ``Sensitivity methods
  for effective delayed neutron fraction and neutron generation time with
  summon,'' {\sc \protect\JournalTitle{Annals of Nuclear Energy}} \textbf{126},
   410  (2019).

\bibitem{ICE}
{L. Koch, \textit{et al.}}, ``{The Isotope Correlation Experiment {ICE} - Final
  Report},'' Tech. Rep. 2/81 (KfK3737), ESARDA (1981).

\bibitem{ASTRID}
{J. Bess, \textit{et al.}}, ``{DOE-CEA Benchmark on SFR ASTRID Innovative Core:
  Neutronic and Safety Transients Simulation},'' in \textit{ Proc. Int. Conf.
  Fast Reactors and Related Fuel Cycles: Safe Technologies and Sustainable
  Scenarios (FR13, IAEACN-199/281), March 4-7, 2013, Paris, France} (2013).

\bibitem{MYRRHA}
J.~Engelen, {H.A. Abderrahim}, P.~Baeten, {D. De Bruyn}, and P.~Leysen,
  ``{MYRRHA}: preliminary front-end engineering design,'' {\sc
  \protect\JournalTitle{International Journal of Hydrogen Energy}} \textbf{40},
   15137 (2015).

\bibitem{GlaserRamana}
{A. Glaser and M.V. Ramana}, ``{Weapon-Grade Plutonium Production Potential in
  the Indian Prototype Fast Breeder Reactor},'' tech. rep., Princeton
  University (2007).

\bibitem{Cormon2012}
S.~Cormon, \textit{ Study of the potential of antineutrino detection for
  reactor monitoring and safeguards}.
\newblock PhD thesis, University of Nantes, Subatech Laboratory, CNRS/in2p3,
  Univ. of Nantes, IMTA, Nantes, FRANCE (2012).

\bibitem{Moscati62}
{G. Moscati, J. Goldemberg}, ``Delayed {N}eutron {Y}ields in the {P}hotofission
  of $^{238}${U} and $^{232}${Th},'' {\sc \protect\JournalTitle{Phys. Rev.}}
  \textbf{126},  1098 (1962).

\bibitem{Caldwell75}
{J.T. Caldwell and E.J. Dowdy}, ``Experimental determination of photofission
  neutron multiplicities for eight isotopes in the mass range 232 $\leq$ {A}
  $\leq$ 239,'' {\sc \protect\JournalTitle{Nucl. Sci. Eng.}} \textbf{56},  179
  (1975).

\bibitem{Pai76}
{H.L. Pai}, ``{On the (3Z-A) Dependence of Delayed Neutron Yield from Fission
  Fragments},'' {\sc \protect\JournalTitle{Ann. Nucl. Energy}} \textbf{3},  125
  (1976).

\bibitem{Ronen96}
{Y. Ronen}, ``{Correlations for Delayed Neutron Yields for Thermal, Fast and
  Spontaneous Fissions},'' {\sc \protect\JournalTitle{Ann. Nucl. Energy}}
  \textbf{23},  239 (1996).

\bibitem{Zemyatnin76}
{Yu.S. Zamyatnin, V.M. Gorbachev and A.A. Lbov}, ``Interaction of the radiation
  with heavy nuclei and nuclear fission,''
\newblock {in: Handbook, Moscow, Atomizdat (1976)}.

\bibitem{Rei17}
P.-G. Reinhard and W.~Nazarewicz, ``Toward a global description of nuclear
  charge radii: Exploring the {F}ayans energy density functional,'' {\sc
  \protect\JournalTitle{Phys. Rev. C}} \textbf{95},  064328 (2017).

\bibitem{Saperstein2016}
{E.E. Saperstein, I.N. Borzov, and S.V. Tolokonnikov}, ``Anomalous charge radii
  in heavy {C}a isotopes,'' {\sc \protect\JournalTitle{JETP Letters}}
  \textbf{104},  218 (2016).

\bibitem{Litvinova2014}
{E. Litvinova, B.A. Brown, D.-L. Fang, T. Marketin, R.G.T. Zegers},
  ``{Benchmarking nuclear models for {Gamow-Teller} response},'' {\sc
  \protect\JournalTitle{Physics Letters B}} \textbf{730},  307 (2014).

\end{thebibliography}

\end{document}